\documentclass[sigconf, nonacm]{acmart}

\usepackage{xr-hyper}

\usepackage{eccvabbrv}

\usepackage[capitalise]{cleveref}

\usepackage{myconfig}

\usepackage{longtable}
\usepackage{multirow}

\begin{document}
\title[\textsc{Plasma}: Layout-Aware Benchmark for Graph-based ANNS on GPU]{\textsc{Plasma}: A Layout-Aware Benchmark Reveals Memory Layout Matters for Graph-based ANNS on GPU}

\author{Yutaro Oguri}
\affiliation{%
  \institution{The University of Tokyo}
  \city{Tokyo}
  \country{Japan}
}
\email{}

\author{Mai Nishimura}
\affiliation{%
  \institution{OMRON SINIC X Corporation}
  \city{Tokyo}
  \country{Japan}
}
\email{}

\author{Yusuke Matsui}
\affiliation{%
  \institution{The University of Tokyo}
  \city{Tokyo}
  \country{Japan}
}
\email{}

\begin{abstract}
  We propose a \textbf{P}latform for
  \textbf{L}ayout-\textbf{A}ware
  \textbf{S}earch and \textbf{M}emory \textbf{A}rrangement (\textsc{Plasma}), a unified evaluation
  framework for graph-based Approximate Nearest Neighbor Search (ANNS) on GPU
  that isolates the effects of graph index topology and memory layout.
  Graph-based ANNS is essential in modern AI applications such as
  RAG, and GPU utilization is attracting attention for datasets of
  millions or more vectors.
  Our framework extracts the topology of arbitrary graph-based
  indices and enables execution under a unified, GPU-optimized search
  algorithm, specifying the correspondence between vertex
  IDs and positions on memory to allow arbitrary vertex orderings.
  Through comprehensive experiments, we demonstrate that vertex
  reordering yields up to \textbf{80\%} (typically \textbf{10--30\%}) QPS improvement while
  preserving search accuracy.
\end{abstract}


\maketitle

\begin{figure*}[t] %
  \centering
  \includegraphics[width=\textwidth]{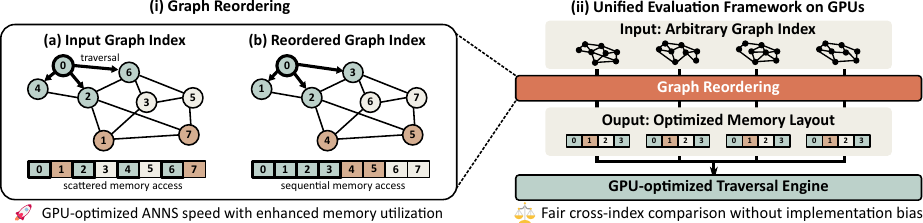}
  \caption{\textbf{Overview of Plasma}. (i) Graph Reordering transforms scattered memory access
    patterns into sequential ones by relocating co-accessed nodes to
    consecutive memory regions.
    (ii) Unified Execution Framework processes arbitrary graph
    topologies through optimized memory layouts and standardized GPU
    search algorithms, enabling fair performance evaluation across
  different graph-based indices.}
  \label{fig:teaser}
\end{figure*}

\section{Introduction}

The rise of foundation models has made large-scale vector search
a critical building block for fast retrieval, recommendation, and
generative AI systems, with Approximate Nearest Neighbor Search
(ANNS) serving as the indispensable core primitive at
scale~\cite{douze2024faiss_library,johnson2019billion_faiss}.
Modern AI applications such as Retrieval-Augmented Generation
(RAG)~\cite{cocom,kotoge-etal-2026-compact} demand billion-scale vector search, driving the
need for GPU-accelerated ANNS~\cite{cagra,bang}.
Among these, graph-based
ANNS~\cite{malkov2018efficient_HNSW,wang2021comprehensive_survey_graph_2021},
which traverses proximity graphs to locate neighbors,
stands out for its high accuracy and throughput.
In this work, we are particularly interested in improving GPU
search throughput by rearranging the memory layout of graph vertices without modifying graph structure.

While the field has advanced primarily through graph topology
and search algorithm
design~\cite{malkov2018efficient_HNSW,nsg,cagra}, memory layout
substantially affects performance on GPUs, where massive core-level
parallelism makes bandwidth and latency bottlenecks far more
prominent than on CPUs~\cite{bang,cagra}.
Graph reordering, which permutes node IDs, is the most direct
means of controlling memory layout because vertex IDs map
one-to-one to physical memory positions.
Reordering is known to improve CPU-based graph
ANNS~\cite{NEURIPS2022_reordering_anns_cpu}, but its effect on
GPU-based graph ANNS remains unexplored.

Isolating the effect of reordering is difficult because graph
structure, search algorithms, and implementation optimizations are
tightly coupled, preventing fair comparison across GPU graph
indices.
Conventional evaluations often use %
low-dimensional datasets that underestimate the impact of memory
layout, which grows with dimensionality.
Achieving high search throughput on GPUs for high-dimensional
embeddings from language and multimodal data %
is a pressing
need in modern AI applications.

To address these challenges, we propose \textsc{Plasma}, a platform
that decouples graph topology from memory layout, enabling fair
cross-index comparison on GPU (\cref{fig:teaser}).
Specifically, \textsc{Plasma} accepts the topology of any graph-based
index, applies vertex reordering, and evaluates it through a shared
GPU-optimized traversal engine, so that performance differences can
be attributed to graph structure or memory layout alone.
\textsc{Plasma} covers multiple graph indices, high-dimensional
datasets including modern %
multimodal embeddings,
and several reordering algorithms, allowing practitioners to
benchmark hardware-level performance.

Our hardware-level analysis attributes the speedup to improved DRAM
bandwidth utilization: reordering places co-accessed nodes in
consecutive memory, reducing random access and increasing effective
bandwidth; as dimensionality grows, cache hit ratio decreases and DRAM
bandwidth becomes the dominant bottleneck, matching our real-world results.

The main contributions of this study are summarized as follows:
\begin{itemize}
  \item We propose \textsc{Plasma}, an evaluation framework that
    decouples graph topology from memory layout by executing arbitrary
    graph-based indices under a unified GPU search algorithm.
    This enables controlled, fair comparison of these two factors,
    which are inseparable in conventional evaluations.
  \item Comprehensive experiments across multiple graph indices
    and datasets reveal that, surprisingly, topology
    differences are often minor on GPU; for example, NSG~\cite{nsg}
    can outperform CAGRA~\cite{cagra} by 1.2$\times$.
    Graph reordering yields up to \textbf{80\%} throughput improvement (typically 10--30\%),
    which we attribute to increased DRAM bandwidth utilization, an effect
    amplified in high-dimensional vectors.
  \item We will release the framework and reordering pipeline as
    open-source code with a \texttt{pixi}-based build
    system~\cite{vslamlab2025,pixi-whitepaper}, enabling
    single-command environment setup including CUDA dependencies
    to facilitate reproduction and extension.
\end{itemize}

\section{Related Work}
\subsection{ANNS on GPU}
ANNS has evolved through
distinct paradigms on CPU architectures, encompassing tree-based
structures~\cite{silpa2008optimised_KDTree,Echihabi2022-hercules-tree,Hu2024-DIDS-tree,muja2014scalableFLANN},
Locality-Sensitive Hashing
(LSH)~\cite{gionis1999similarityLSH,andoni2015practical_and_optimal_LSH_angular,Wei2024-det-LSH-vldb},
Inverted File indexing (IVF) combined with Product Quantization
(PQ)~\cite{Baranchuk2018RevisitingIVFHNSW,Jegou2011ProductQuantization,huijben2024residual_neural_quantization,ge2013optimized},
and graph-based
approaches~\cite{nsg,diskann,wang2021comprehensive_survey_graph_2021}.
While these methods are effective for many practical applications,
CPU-based approaches face fundamental performance constraints due to
memory bandwidth and parallelism limitations~\cite{cagra}.
\diff{Prior GPU-based methods each couple an index to a custom search
kernel: Faiss-GPU's IVF-PQ
$k$-selection~\cite{johnson2019billion_faiss}, CAGRA's warp-tuned
graph~\cite{cagra}, BANG's on-GPU DiskANN traversal~\cite{bang}, and
FusionANNS's CPU-GPU cooperation~\cite{fusionanns}.
Such entanglement prevents fair cross-index comparison and demands HPC
expertise~\cite{optimizing-graph-processing-on-gpus-2017}.}
\textsc{Plasma} instead separates index topology from memory layout
under one search engine, letting researchers exploit GPU performance
without such expertise.

\subsection{Graph-based ANNS}
Graph-based ANNS algorithms can be systematically organized from two
perspectives: \emph{Graph Construction} and \emph{Search Strategy}.

\emph{Graph construction} include three main approaches:
divide-and-conquer~\cite{chen2018sptag,HCNNG_2019},
refinement from base graphs
(\eg, KNN graph~\cite{paredes2005using_knn_graph},
RNG~\cite{jaromczyk1992relative_RNG}) to improve
navigability~\cite{nsg,diskann,fu2021high_NSSG,%
li2019approximate_DPG,fu2016efanna}, and incremental
insertion~\cite{malkov2014approximate_NSW,%
iwasaki2018optimization_NGT,malkov2018efficient_HNSW}.
In terms of topology, methods with multi-layer hierarchical
structures (\eg, HNSW~\cite{malkov2018efficient_HNSW},
HVS~\cite{hvs}) differ from single-layer flat graphs
(\eg, NN-Descent~\cite{dong2011efficient_Kgraph_NNDescent},
RNN-Descent~\cite{Ono2023-relative-nndescent},
NSG~\cite{nsg}, Vamana~\cite{diskann}).
Hierarchical traversal is inherently sequential and incompatible
with massive GPU parallelism.
We therefore focus on single-layer flat graph structures, as
adopted by CAGRA~\cite{cagra}, and exclude hybrid
approaches~\cite{azizi2023elpis,LSHAPG}.

\emph{Search strategies} include
hierarchical approaches (\eg,
HNSW~\cite{malkov2018efficient_HNSW}) and greedy
beam search on single-layer graphs (\eg, NSG,
DiskANN~\cite{nsg,diskann}) with various entry point
selection
methods~\cite{Oguri2024-theoretical-entrypoint,%
oguri2023generalSisap,ni2023diskann++}.
CAGRA~\cite{cagra} adopts fixed out-degree
single-layer graphs to uniformly parallelize search
and extract high GPU throughput.
\textsc{Plasma} unifies the search algorithm to CAGRA's
GPU-optimized beam search, enabling the first fair comparison
across graph indices on GPU.

\subsection{Graph Reordering}
Graph reordering optimizes memory access patterns by reassigning
vertex indices to improve spatial locality and cache
efficiency~\cite{hubsort,rcm,gorder,graph-reordering-real-world-2020},
but evaluations have been limited to traditional graph algorithm
benchmarks on CPUs (\eg, PageRank~\cite{PageRank},
SSSP~\cite{cormen2009introduction-to-algorithms-book},
SCC~\cite{scc-strong-connected-component}).
Recent work applies graph reordering to CPU-based
ANNS~\cite{NEURIPS2022_reordering_anns_cpu} and GNNs
training~\cite{Merkel2024-GNN-reordering},
reporting $10$--$40$\% speedup in CPU-based ANNS.
Reordering serves as a modular optimization that maintains
index structure while improving memory access.
However, the effectiveness of graph
reordering for GPU-based ANNS remains unexplored.
This study provides the first systematic evaluation of
graph reordering methods for multiple graph-based indices on
GPU.

\begin{table*}[t]
  \caption{%
    Comparison with existing ANNS benchmarks.
    Our benchmark is the first to introduce an
    architecture-aware evaluation perspective, analyzing the
    impact of memory layout and cache behavior on graph-based
    ANNS performance on GPU.
    \diff{\emph{HW profiling}: built-in hardware-counter collection
    (DRAM bandwidth, L1/L2 hit rates); \emph{Memory-layout aware}:
    varying physical vertex layout (reordering) independently of
    topology under a fixed search engine.}}
  \label{tab:benchmark-comparison}
  \centering
  \normalsize
  \setlength{\tabcolsep}{6pt}
  \begin{tabular}{@{}l|cccc|c@{}}
    \toprule
    & ANNBench.~\cite{aumuller2020ann-benchmarks}
    & BigANN~\cite{Simhadri2022NeuripsComp}
    & VDBBench~\cite{ziliz-vdbbench-dataset}
    & VIBE~\cite{VIBE-benchmark}
    & \textbf{\textsc{Plasma}} \\
    \midrule
    Platform                & \tblCPU   & \tblCPU       &  \tblCPU & \tblCPU+\tblGPU & \tblGPU \\
    Target                 & General   & General       & General         & General  & Graph index \\
    \midrule
    Modern embeddings & \tblNone  & \tblPartial & \tblPartial         & \tblFull & \tblFull \\
    HW profiling   & \tblNone & \tblNone      & \tblNone        & \tblNone & \tblFull \\
    Memory-layout aware     & \tblNone  & \tblNone      & \tblNone        & \tblNone & \tblFull \\
    \midrule
    Reproducible env.& \texttt{docker} & \texttt{docker}       & \texttt{docker} & \texttt{apptainer} & \texttt{pixi}~\cite{pixi-whitepaper} \\
    \bottomrule
  \end{tabular}
\end{table*}
\subsection{ANNS Benchmarks}
\Cref{tab:benchmark-comparison} compares \textsc{Plasma} against existing benchmarks.
Existing ANNS benchmarks serve complementary
purposes:
ANN-Benchmarks~\cite{aumuller2020ann-benchmarks} provides broad
algorithm coverage,
Big-ANN~\cite{Simhadri2022NeuripsComp,%
simhadri2024comp-resultsbigannneurips23} addresses large-scale
scenarios with filtered and out-of-distribution queries,
VIBE~\cite{VIBE-benchmark} targets modern LLM-era embeddings
including GPU methods,
VDBBench~\cite{ziliz-vdbbench-dataset} evaluates end-to-end
product performance.
Survey studies~\cite{wang2021comprehensive_survey_graph_2021,%
sigmod2025-graph-index-survey} offer systematic algorithmic
comparisons, but do not provide open benchmark.
\textsc{Plasma} adds an orthogonal axis, hardware memory
utilization: it can evaluate CPU-designed graph indices on
GPU and unlock their performance through memory layout
arrangement.

\section{Preliminaries}
\subsection{Approximate Nearest Neighbor Search (ANNS)}
Consider a database that contains a collection of $N$
$d$-dimensional vectors, which we denote by
$\mathcal{D} = \{\bm x_1, \bm x_2, \dots, \bm x_N\} \subset \mathbb R^d$.
Given a query vector $\bm q \in \mathbb{R}^d$, the vector
search problem seeks to identify the most similar vector in the database,
\begin{equation}
  n = \argmin_{i \in \{1,2, \dots, N\}} \| \bm q -  \bm x_i \|_2^2\,.
  \label{eq:vector-search}
\end{equation}
This brute-force approach must evaluate all $N$ vectors,
leading to $\mathcal{O}(Nd)$ computational complexity that becomes
impractical as datasets scale to billions of vectors.

To address this scalability challenge, ANNS employs index
structures~\cite{Jegou2011ProductQuantization,malkov2018efficient_HNSW,Pan2024-vector-db-survey}
that transform the feature space into efficiently searchable representations.
ANNS pursues the trade-off between search efficiency and accuracy by
approximating \cref{eq:vector-search}.

\subsection{Graph-based Index}
Among ANNS methods, graph-based ANNS constructs a graph structure
(graph-based index) on a vector set and \diff{performs efficient
approximate nearest neighbor search by traversing the
graph}~\cite{wang2021comprehensive_survey_graph_2021,sigmod2025-graph-index-survey}.
A graph-based index models the vector space as a proximity graph $G =
(\mathcal V, \mathcal E)$.
Here, each vertex $v_i \in \mathcal V$ corresponds to a
$d$-dimensional vector $\bm x_i \in \mathcal D$ in the database.
We construct each edge $(v_i, v_j) \in \mathcal E$ based on
heuristics designed to enhance subsequent search efficiency~\cite{nsg,diskann}.
We refer to the connectivity structure defined by $G$
as the \emph{graph topology}.
We perform search by traversing the constructed graph using beam-search-based algorithms~\cite{nsg,cagra} to efficiently identify
neighbors of the query.
The typical search algorithm maintains a candidate queue of size at most $L$
and explores the graph by visiting neighbors of the nearest unvisited
node until no further improvements are found (see Supp. \Cref{app:search-algorithm} for the pseudocode).

\subsection{Graph Reordering}
Given a graph-based index $G = (\mathcal V, \mathcal E)$ with
vertices $\mathcal{V} = \{v_1, v_2, \ldots, v_N\}$, graph reordering
seeks a permutation $\pi$ that reassigns each vertex $v_i$ to a new
index $\pi(i)$ so that vertices frequently co-accessed during
traversal receive nearby IDs.
Because vertex IDs directly index the underlying arrays, this ID
reassignment determines physical memory layout
(\cref{sec:memory-layout}).
\Cref{fig:teaser}~(i) illustrates the effect: the transition from
(a) to (b) collocates vertices at similar hop distances from the
entry point, converting scattered memory accesses into sequential
ones.
Since most graph-based ANNS methods~\cite{diskann,nsg,cagra} employ
similar beam-search-based traversal, the reordering benefits examined
here generalize to other indices with comparable access patterns.

\section{\textsc{Plasma}: Platform for Layout-Aware Search and Memory Arrangement of Graph-based ANNS on GPU}

\textsc{Plasma} is an evaluation platform for graph-based ANNS on
GPU that decouples graph topology from memory layout.
\textsc{Plasma} comprises three components:
(i)~a \textbf{graph adapter},
(ii)~a \textbf{graph reordering} module, and
(iii)~a \textbf{unified graph traversal engine}.
This pipeline enables controlled comparison across all combinations
$(G, \pi, \mathcal{D})$ of graph topologies $G$, reordering functions $\pi$,
and datasets $\mathcal{D}$.
The following subsections detail each component.

\subsection{Graph Adapter}
The graph adapter bridges diverse index formats and
\textsc{Plasma}.
It accepts graph-based indices in various formats
(\eg Faiss~\cite{douze2024faiss_library},
DiskANN~\cite{diskann}), extracts their graph topology, and converts
them into Faiss's GPU index
format~\cite{douze2024faiss_library}.
This topology-preserving conversion enables any single-layer
(flat) graph index, including those originally designed for CPU
execution, to run on GPU under a shared search algorithm.
The adapter is extensible to arbitrary CPU indices by extracting
adjacency information from a custom format.

This conversion is an implementation-level mechanism, yet it is
essential for fair evaluation.
Conventional graph indices are tied to their own libraries, each with
distinct data structures and search algorithms. Consequently, when
one library's index outperforms another's, the advantage may stem
from the search algorithm rather than the graph topology itself.
By converting all topologies into a common format and executing them
with a single search kernel, the adapter isolates graph topology as
the sole variable, enabling comparisons that were previously
confounded by implementation differences.

\vspace{-12pt}
\subsection{Search Kernel}
Given a GPU index produced by the adapter, \textsc{Plasma}
executes a highly optimized beam-search
kernel~\cite{cagra,cuvs-library} on GPU.
A fast search implementation is a prerequisite for meaningful
evaluation: benchmarking under a slow kernel would underestimate
throughput and obscure the relative impact of graph topology
and memory layout.
Equally important, applying the same kernel to every condition
eliminates search-algorithm variation as a confounding factor,
so that observed performance differences reflect only graph topology
and memory layout.

The kernel partitions data structures across GPU shared memory
and global memory (DRAM).
Shared memory holds per-block working state: candidate index and
distance buffers, a visited hashmap, a parent node list, and a
top-$k$ sorting workspace.
Global memory stores the vector dataset $\mathcal{D}$ and neighbor
graph $G(\mathcal{V}, \mathcal{E})$ in separate regions, which are
read during distance computation.
The kernel's runtime parameter $L$ controls candidate set size,
enabling consistent evaluation of heterogeneous indices, including
CPU-native NSG and Vamana, under diverse memory layouts~$\pi$.
The physical layout of these memory regions and how vertex reordering
transforms it are detailed in~\cref{sec:memory-layout}.

\begin{figure}[t]
  \centering
  \includegraphics[width=0.9\linewidth]{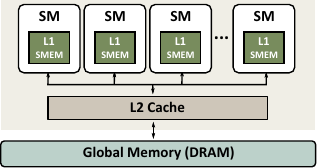}
  \caption{GPU Memory Hierarchy: global memory provides high capacity but low bandwidth. L1 cache and shared memory offer high bandwidth with limited capacity.}
  \label{fig:gpu-arch}
\end{figure}

\begin{figure}[t]
  \centering
  \includegraphics[width=\linewidth]{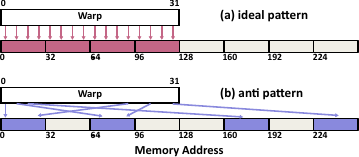}
  \caption{Coalesced Memory Access: top, coalesced access where threads access consecutive memory addresses. bottom, worst access patterns with scattered memory requests.}
  \label{fig:gpu-access}
\end{figure}

\subsection{Memory Layout Arrangement}
\label{sec:memory-layout}
GPUs have a hierarchical memory system
(\cref{fig:gpu-arch}:
global memory (DRAM) provides large capacity but high latency,
while L1/L2 caches and shared memory offer low latency with limited
capacity.
GPUs execute instructions in groups of 32 threads called warps.
When threads in a warp access consecutive global-memory addresses,
the hardware combines them into fewer transactions, a pattern known
as \emph{coalesced access} (\cref{fig:gpu-access})
Because graph traversal involves irregular access patterns,
coalescing efficiency and cache utilization in global memory can
dominate overall search throughput
(see Supp. \Cref{app:gpu-architecture} for details).

\Cref{fig:dataset-graph-memory-layout} shows how \textsc{Plasma}
stores the dataset and graph in GPU global memory.
The dataset region consists of $N$ vectors in row-major
order. Each vector occupies $d + p$ elements, where padding $p$
enforces 16-byte alignment.
The graph region consists of $N$ neighbor lists in row-major
order. Each list stores the
IDs of up to $K$ neighbor nodes without padding, with a stride of
$K$ elements per node.
Graph reordering applies a permutation $\pi$ to vertex IDs,
changing the physical positions of both vectors and neighbor lists in
these regions.
\textsc{Plasma} supports various reordering algorithms as $\pi$,
including
Gorder~\cite{gorder}, RCM~\cite{rcm}, and methods tailored for ANNS
graphs~\cite{NEURIPS2022_reordering_anns_cpu}.
Vertices frequently co-accessed during beam-search traversal receive
consecutive IDs under $\pi$, promoting coalesced access and boosting cache reuse, thereby reducing DRAM traffic.

\begin{figure}[t]
  \centering
  \begin{subfigure}[b]{0.5\textwidth}
    \centering
    \resizebox{\linewidth}{!}{\begin{tikzpicture}[
    font=\small\ttfamily,
    node distance=0.5cm and 0.5cm,
    cell/.style={
      draw, minimum width=1.1cm, minimum height=0.65cm,
      anchor=center, inner sep=0pt, align=center,
      text=black, line width=\memLayoutLineWidth
    },
    dataset_cell/.style={ cell, fill=datasetCellLight },
    dataset_null_cell/.style={ cell, fill=datasetCellDark, text=white },
    labelnode/.style={ font=\footnotesize\sffamily, anchor=east, align=right, text=black },
    collabel/.style={ font=\scriptsize\sffamily, text=black, above=2pt },
  ]

  \node[font=\sffamily\bfseries, text=black] (dtitle) {Dataset Memory Region};

  \matrix (dataset) [
    matrix of nodes,
    nodes in empty cells,
    row sep=-\pgflinewidth,
    column sep=-\pgflinewidth,
    nodes={dataset_cell},
    below=0.5cm of dtitle
  ] {
    0.12 & -1.5 & $\cdots$ & 3.14 & |[dataset_null_cell]| null &
    |[dataset_null_cell]| $\cdots$ & |[dataset_null_cell]| null \\
    -0.8 & 2.71 & $\cdots$ & -0.1 & |[dataset_null_cell]| null &
    |[dataset_null_cell]| $\cdots$ & |[dataset_null_cell]| null \\
    2.3 & -0.3 & $\cdots$ & 1.2 & |[dataset_null_cell]| null &
    |[dataset_null_cell]| $\cdots$ & |[dataset_null_cell]| null \\
    |[draw=none, fill=none]| $\vdots$ & |[draw=none, fill=none]|
    $\vdots$ & |[draw=none, fill=none]| & |[draw=none, fill=none]|
    $\vdots$ & |[draw=none, fill=none]| $\vdots$ & |[draw=none,
    fill=none]| & |[draw=none, fill=none]| $\vdots$ \\
    1.61 & -0.5 & $\cdots$ & 0.0 & |[dataset_null_cell]| null &
    |[dataset_null_cell]| $\cdots$ & |[dataset_null_cell]| null \\
  };

  \node[collabel] at (dataset-1-1.north) {$0$};
  \node[collabel] at (dataset-1-2.north) {$1$};
  \node[collabel] at (dataset-1-3.north) {$\cdots$};
  \node[collabel] at (dataset-1-4.north) {$d-1$};
  \node[collabel] at (dataset-1-5.north) {$0$};
  \node[collabel] at (dataset-1-6.north) {$\cdots$};
  \node[collabel] at (dataset-1-7.north) {$p-1$};

  \node[labelnode, left=0.3cm of dataset-1-1.west] {Vector $\bm{x}_0$:};
  \node[labelnode, left=0.3cm of dataset-2-1.west] {Vector $\bm{x}_1$:};
  \node[labelnode, left=0.3cm of dataset-3-1.west] {Vector $\bm{x}_2$:};
  \node[labelnode, left=0.3cm of dataset-4-1.west] {Vector $\bm{x}_{N-1}$:};

  \draw[<-, >=latex, line width=\memLayoutLineWidth, black]
    (dataset-1-1.north west) -- ++(-0.6, 0.6) coordinate (dptr_start);
  \node[anchor=south east, font=\footnotesize\ttfamily, text=black] at (dptr_start) {dataset\_ptr};

  \draw[decorate, decoration={brace, amplitude=5pt, mirror, raise=5pt},
    line width=\memLayoutLineWidth, black]
  (dataset-5-1.south west) -- (dataset-5-7.south east)
  node[midway, below=15pt, font=\footnotesize\sffamily, text=black]
  {Dimension of Dataset ($d$) with Padding ($p$)};
\end{tikzpicture}}
    \caption{Dataset memory region}
    \label{fig:dataset-memory-region}
  \end{subfigure}
  \hfill
  \begin{subfigure}[b]{0.5\textwidth}
    \centering
    \resizebox{\linewidth}{!}{\begin{tikzpicture}[
    font=\small\ttfamily,
    node distance=0.5cm and 0.5cm,
    cell/.style={
      draw, minimum width=1.1cm, minimum height=0.65cm,
      anchor=center, inner sep=0pt, align=center,
      text=black, line width=\memLayoutLineWidth
    },
    graph_cell/.style={ cell, fill=graphCellLight },
    graph_null_cell/.style={ cell, fill=graphCellDark, text=white },
    labelnode/.style={ font=\footnotesize\sffamily, anchor=east, align=right, text=black },
    collabel/.style={ font=\scriptsize\sffamily, text=black, above=2pt },
  ]

  \node[font=\sffamily\bfseries, text=black] (gtitle) {Graph Memory Region};

  \matrix (graph) [
    matrix of nodes,
    nodes in empty cells,
    row sep=-\pgflinewidth,
    column sep=-\pgflinewidth,
    nodes={graph_cell},
    below=0.5cm of gtitle
  ] {
    12 & 5 & 25 & $\cdots$ & 84 & 145 & 3124 \\
    749 & 0 & 100 & $\cdots$ & 9512 & |[graph_null_cell]| null & |[graph_null_cell]| null \\
    422 & 22 & 1000 & $\cdots$ & 1512 & 6 & |[graph_null_cell]| null \\
    |[draw=none, fill=none]| $\vdots$ & |[draw=none, fill=none]|
    $\vdots$ & |[draw=none, fill=none]| & |[draw=none, fill=none]|
    $\vdots$ & |[draw=none, fill=none]| $\vdots$ & |[draw=none, fill=none]| $\vdots$ \\
    912 & 234 & 139 & $\cdots$ & 4 & 12 & 4482 \\
  };

  \node[collabel] at (graph-1-1.north) {$0$};
  \node[collabel] at (graph-1-2.north) {$1$};
  \node[collabel] at (graph-1-3.north) {$2$};
  \node[collabel] at (graph-1-4.north) {$\cdots$};
  \node[collabel] at (graph-1-5.north) {$K-3$};
  \node[collabel] at (graph-1-6.north) {$K-2$};
  \node[collabel] at (graph-1-7.north) {$K-1$};

  \node[labelnode, left=0.3cm of graph-1-1.west] {Node $v_0$:};
  \node[labelnode, left=0.3cm of graph-2-1.west] {Node $v_1$:};
  \node[labelnode, left=0.3cm of graph-3-1.west] {Node $v_2$:};
  \node[labelnode, left=0.3cm of graph-5-1.west] {Node $v_{N-1}$:};

  \draw[<-, >=latex, line width=\memLayoutLineWidth, black]
    (graph-1-1.north west) -- ++(-0.6, 0.6) coordinate (gptr_start);
  \node[anchor=south east, font=\footnotesize\ttfamily, text=black] at (gptr_start) {graph\_ptr};

  \draw[decorate, decoration={brace, amplitude=5pt, mirror, raise=5pt},
    line width=\memLayoutLineWidth, black]
  (graph-5-1.south west) -- (graph-5-7.south east)
  node[midway, below=15pt, font=\footnotesize\sffamily, text=black]
  {Max Graph Degree ($K$)};
\end{tikzpicture}}
    \caption{Graph memory region}
    \label{fig:graph-memory-region}
  \end{subfigure}
  \caption{Memory layout of the dataset and graph structure in GPU
    global memory.
    (\subref{fig:dataset-memory-region}) The dataset memory region
    consists of vectors, each containing $d$ data elements plus padding
    $p$ to satisfy 16-byte alignment.
    (\subref{fig:graph-memory-region}) The graph memory region
    consists of nodes, where each node stores the IDs of up to $K$
    neighbor nodes without padding.}
  \label{fig:dataset-graph-memory-layout}
\end{figure}

\subsection{Dependency Management}
\label{sec:dependency-management}
Reproducing GPU-based ANNS benchmarks requires coordinating
C++/CUDA compilers, GPU-specific libraries such as
cuVS~\cite{cuvs-library}, and matrix computation libraries such as
MKL and OpenBLAS across exact versions.
\textsc{Plasma} manages these dependencies with
\texttt{pixi}~\cite{pixi-whitepaper}, a modern multi-language package
manager whose declarative lockfile pins the entire dependency graph.
With a single \texttt{pixi install} command followed by
\texttt{pixi run build}, users can set up the environment and
build the project without manually resolving version conflicts.
This lightweight virtual environment, which encompasses the full
CUDA stack, lowers the barrier to extending \textsc{Plasma} to
broader applications such as integration with foundation models
and facilitates future research on GPU-based graph indices.

\section{Experimental Results}
\subsection{Experimental Setup}
\label{sec:exp-setup}
We provide an overview of the experimental setup; details including
software versions are given in the Supp. \Cref{app:exp-setup-details}.

\subsubsection{Execution Environment}
We conduct experiments primarily using AMD EPYC 7742 (64 cores) CPU and NVIDIA A100 (VRAM 80GB) GPU.
For some experiments described later, we also use Intel(R) Xeon(R) w5-3435X (16 cores) CPU and NVIDIA RTX 6000 Ada (VRAM 48GB) GPU.
Note that we use a single GPU for all experiments, which is a standard configuration, and CAGRA~\cite{cagra} also targets single-GPU execution.
For speed measurements, we perform one warm-up run followed by five
measurements and take the average.

Our environment consists of CUDA 12.0 and cuVS 25.08.00~\cite{cuvs-library}.
We use our customized implementation, extended from Faiss 1.10.0~\cite{douze2024faiss_library}.

\subsubsection{Datasets}
\label{sec:exp-dataset}
Using our evaluation framework, we evaluate various graph-based
indices on multiple datasets.
These datasets include not only those used in conventional benchmarks
but also datasets designed for modern applications.

As traditional datasets widely used in image feature-based
benchmarks, we use SIFT~\cite{Jegou2011ProductQuantization},
GIST~\cite{Jegou2011ProductQuantization}, and
Deep~\cite{babenko2016efficientDeep1M}, which consists of image
embeddings generated by GoogLeNet~\cite{szegedy2015going_googlenet}.
These are relatively low-dimensional datasets and have been standardly used in
conventional ANNS evaluations.

Additionally, as modern datasets reflecting recent AI applications,
we use Yandex Text-to-Image (Yandex
T2I)~\cite{yandex2021yandext2i-dataset} with a cross-modal search
setting, OpenAI
Embedding~\cite{Matsui_LotusFilter,Oguri2024-theoretical-entrypoint,simhadri2024comp-resultsbigannneurips23}
consisting of text embeddings of WikiText generated by OpenAI's
model, Wikipedia text embeddings by Cohere
V2~\cite{ziliz-vdbbench-dataset}, text embeddings of the BioASQ
question answering corpus by
OpenAI~\cite{Tsatsaronis2015-bioasq1,Krithara2023-bioasq-2,ziliz-vdbbench-dataset},
and text embeddings of the Colossal Clean Crawled Corpus (C4)
generated by Cohere V3~\cite{C4Dataset2020-JMLR,ziliz-vdbbench-dataset}.
In particular, Yandex T2I is a setting where the database and queries
come from different distributions, which causes
performance degradation in ANNS~\cite{jaiswal2022ood_diskann}.

\subsubsection{Graph-based Indices}
\label{sec:exp-graph-index}
We evaluate the following graph-based indices.
\begin{itemize}
  \item \textbf{CAGRA}~\cite{cagra}: GPU-optimized index that constructs a
    $k$NN graph via NN-Descent and optimizes it with reachability
    considerations (cuVS~\cite{cuvs-library}).
  \item \textbf{NSG}~\cite{nsg}: CPU-oriented index approximating MRNG with
    low out-degree and short search paths (Faiss~\cite{douze2024faiss_library}).
  \item \textbf{Vamana}~\cite{diskann}: DiskANN's graph-based
    index\footnote{Vamana was originally designed for SSD-resident
    data; we use it as an on-memory graph.}
    (DiskANN~\cite{diskann-github}).
  \item \textbf{NN-Descent}~\cite{dong2011efficient_Kgraph_NNDescent}:
    Iterative $k$NN graph construction, widely used as a component in
    NSG and CAGRA (Faiss~\cite{douze2024faiss_library}).
\end{itemize}
CAGRA is a state-of-the-art index designed for GPU execution. NSG, Vamana, NN-Descent are CPU-oriented methods.
We focus on the single layer graph indices.
We exclude hierarchical indices like HNSW~\cite{malkov2018efficient_HNSW} and
focus on single-layer indices as adopted by CAGRA~\cite{cagra}.
In general, HNSW performs well and is a typical choice in practice. However, previous work has shown that NSG shows a similar or equivalent performance to HNSW~\cite{oguri2023generalSisap,Oguri2024-theoretical-entrypoint,cvpr23_tutorial_neural_search}.

\begin{table*}[tb]
  \caption{Dataset descriptions. L2 and IP denote Euclidean and
  inner product distance, respectively.}
  \label{table:datasets}
  \centering
  \normalsize
  \begin{minipage}[t]{\linewidth}
    \centering
    \subcaption{Classical datasets}
    \label{table:datasets-classical}
    \begin{tabular}{@{}lccclp{8.44cm}@{}}
      \toprule[1pt]
      Dataset & \#Vec. & Dim. & \#Query & Metric & Description \\
      \midrule
      SIFT~\cite{Jegou2011ProductQuantization} & 1M & 128 & 10,000 & L2 & Image feature descriptors \\
      GIST~\cite{Jegou2011ProductQuantization} & 1M & 960 & 1,000 & L2 & Image feature descriptors \\
      Deep~\cite{babenko2016efficientDeep1M} & 1M, 10M, 40M & 96 & 10,000 & L2 & Image embeddings \\
      \bottomrule[1pt]
    \end{tabular}
  \end{minipage}

  \vspace{10pt}

  \begin{minipage}[t]{\linewidth}
    \centering
    \subcaption{Modern datasets}
    \label{table:datasets-modern}
    \begin{tabular}{@{}lcccll@{}}
      \toprule[1pt]
      Dataset & \#Vec. & Dim. & \#Query & Metric & Description \\
      \midrule
      Yandex T2I~\cite{yandex2021yandext2i-dataset} & 1M & 200 & 100,000 & IP & Text-to-image embeddings \\
      OpenAI Embed.~\cite{simhadri2024comp-resultsbigannneurips23} & 1M & 1536 & 10,000 & L2 & Text embeddings of WikiText by OpenAI's model \\
      Wikipedia~\cite{ziliz-vdbbench-dataset} & 1M, 10M & 768 & 1,000 & IP & Text embeddings from Wikipedia (Cohere V2) \\
      BioASQ~\cite{ziliz-vdbbench-dataset} & 1M, 10M & 1024 & 3,106 & IP & Text embeddings from a question-answering corpus \\
      C4~\cite{ziliz-vdbbench-dataset} & 5M & 1536 & 1,000 & IP & Text embeddings from a large-scale web corpus (Cohere V3) \\
      \bottomrule[1pt]
    \end{tabular}
  \end{minipage}
\end{table*}

\subsubsection{Reordering Algorithms}
\label{sec:exp-reordering}
We evaluate diverse graph reordering algorithms.
All methods operate only on the order of graph node placement in memory
without changing the graph topology.%
\begin{itemize}
  \item \textbf{Degree Sort}~\cite{degreesort}: Sorts nodes by degree.
    We consider \textbf{Indegree Sort} and \textbf{Outdegree Sort} variants.
  \item \textbf{Hub Sort}~\cite{hubsort}: Clusters high-degree hub nodes
    together.
  \item \textbf{GOrder}~\cite{gorder}: Window-based method maximizing
    graph locality via shared neighbors. We adopt the window size $w=5$ following the previous work~\cite{NEURIPS2022_reordering_anns_cpu}.
  \item \textbf{RCM}~\cite{rcm}: BFS-based approach with degree-sorted insertion.
\end{itemize}

\subsubsection{Evaluation Metrics}
\label{sec:exp-metrics}
We evaluate the search accuracy using Recall@$k$.
Recall@$k$ is an index that averages the match rate between the true
top-$k$ nearest neighbor set $R^*(\bm{q})$ and the set
$\widehat{R}(\bm{q})$ returned by the algorithm over the query set
$\mathcal{Q}$, defined as
  $\mathrm{Recall@}k
  = \frac{1}{\lvert\mathcal{Q}\rvert}\sum_{\bm{q}\in\mathcal{Q}}
  \frac{\lvert R^*(\bm{q})\cap \widehat{R}(\bm{q})\rvert}{k}$.
We also evaluate the search speed using Queries Per Second (QPS).
QPS is a throughput metric representing the number of queries that
can be processed per unit time.
A higher value indicates faster search.

\subsection{Performance Evaluation of Various Graph-based Indices on GPU}
\label{sec:exp-baseline}

\begin{figure*}[t]
  \centering
  \setlength{\tabcolsep}{2pt}
  \renewcommand{\arraystretch}{0}
  \begin{tabular}{cccc}
    \subcaptionbox{SIFT 1M \label{fig:base_eval_sift}}[0.23\textwidth]
    {\includegraphics[width=0.23\textwidth]{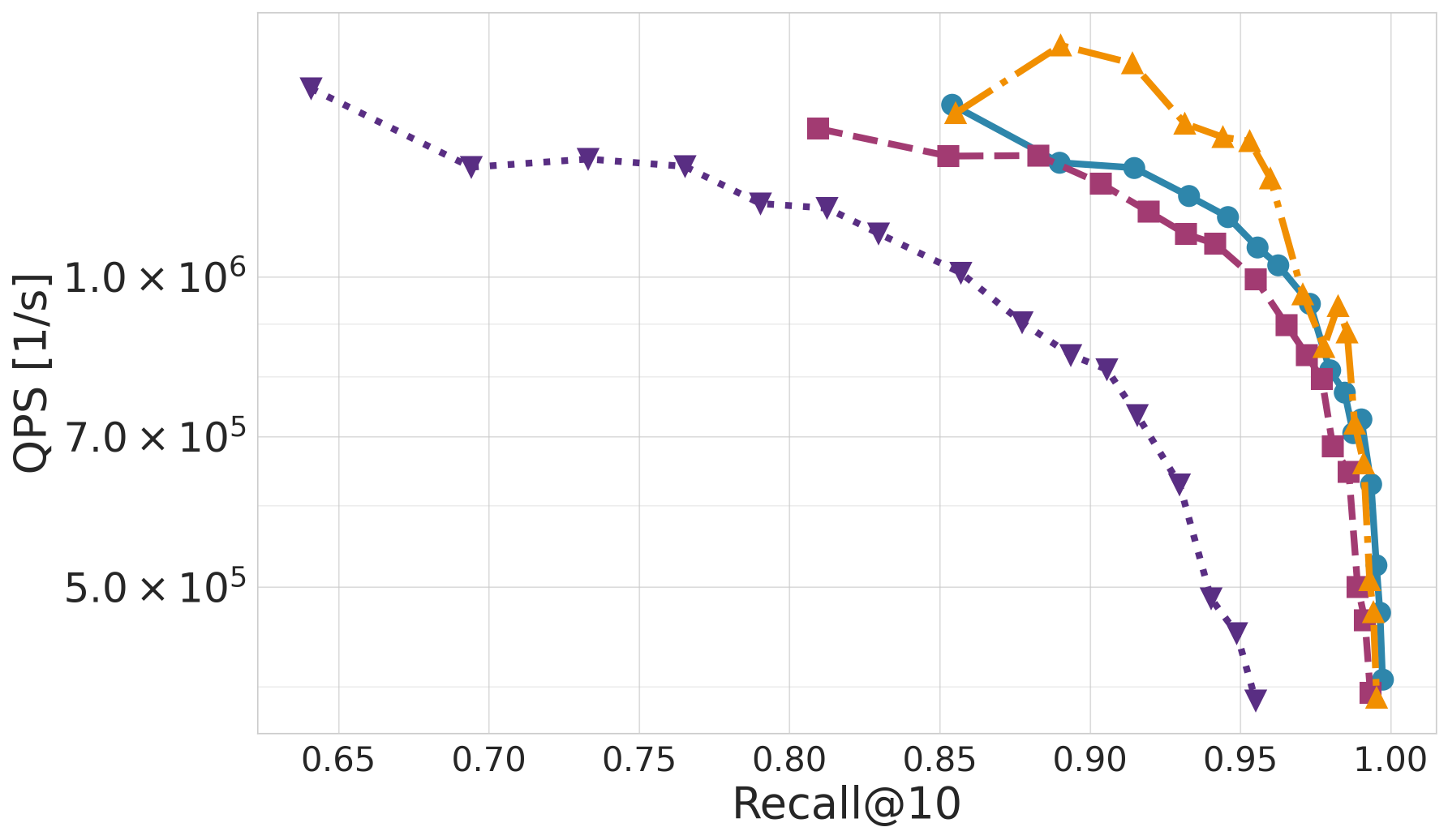}}
    &
    \subcaptionbox{GIST 1M \label{fig:base_eval_gist}}[0.23\textwidth]
    {\includegraphics[width=0.23\textwidth]{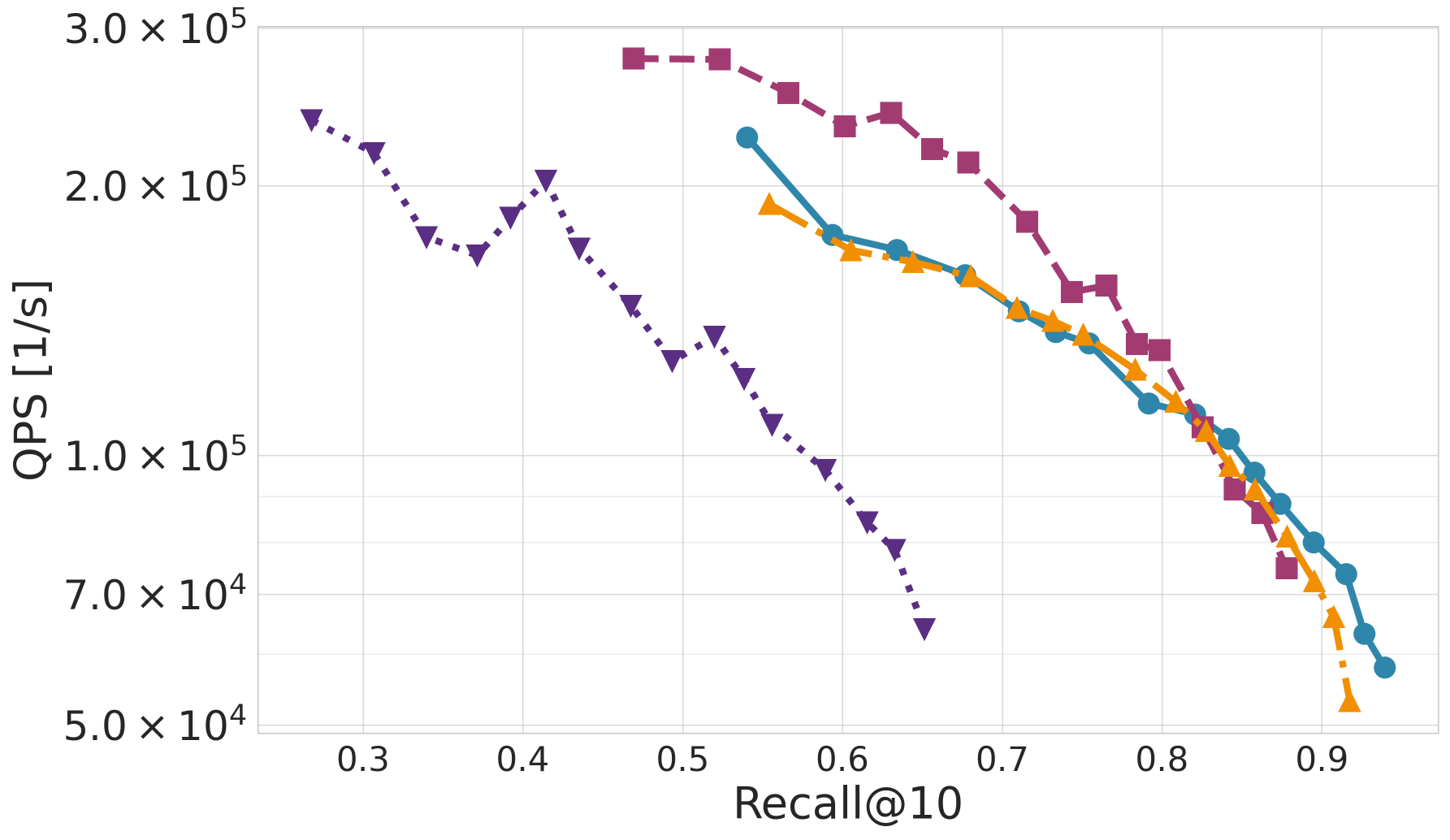}}
    &
    \subcaptionbox{Deep 1M \label{fig:base_eval_deep1m}}[0.23\textwidth]
    {\includegraphics[width=0.23\textwidth]{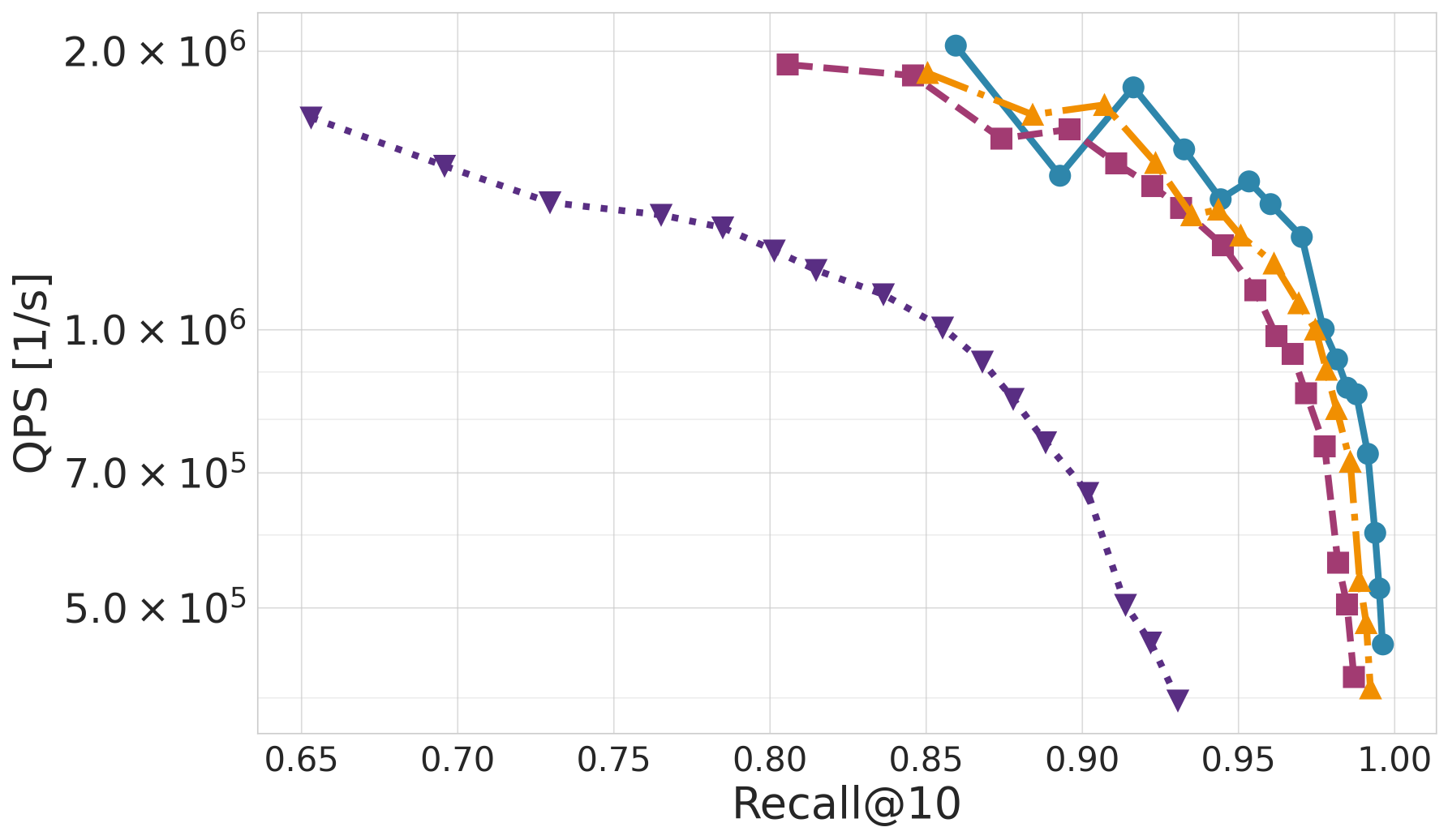}}
    &
    \subcaptionbox{Deep 10M \label{fig:base_eval_deep10m}}[0.23\textwidth]
    {\includegraphics[width=0.23\textwidth]{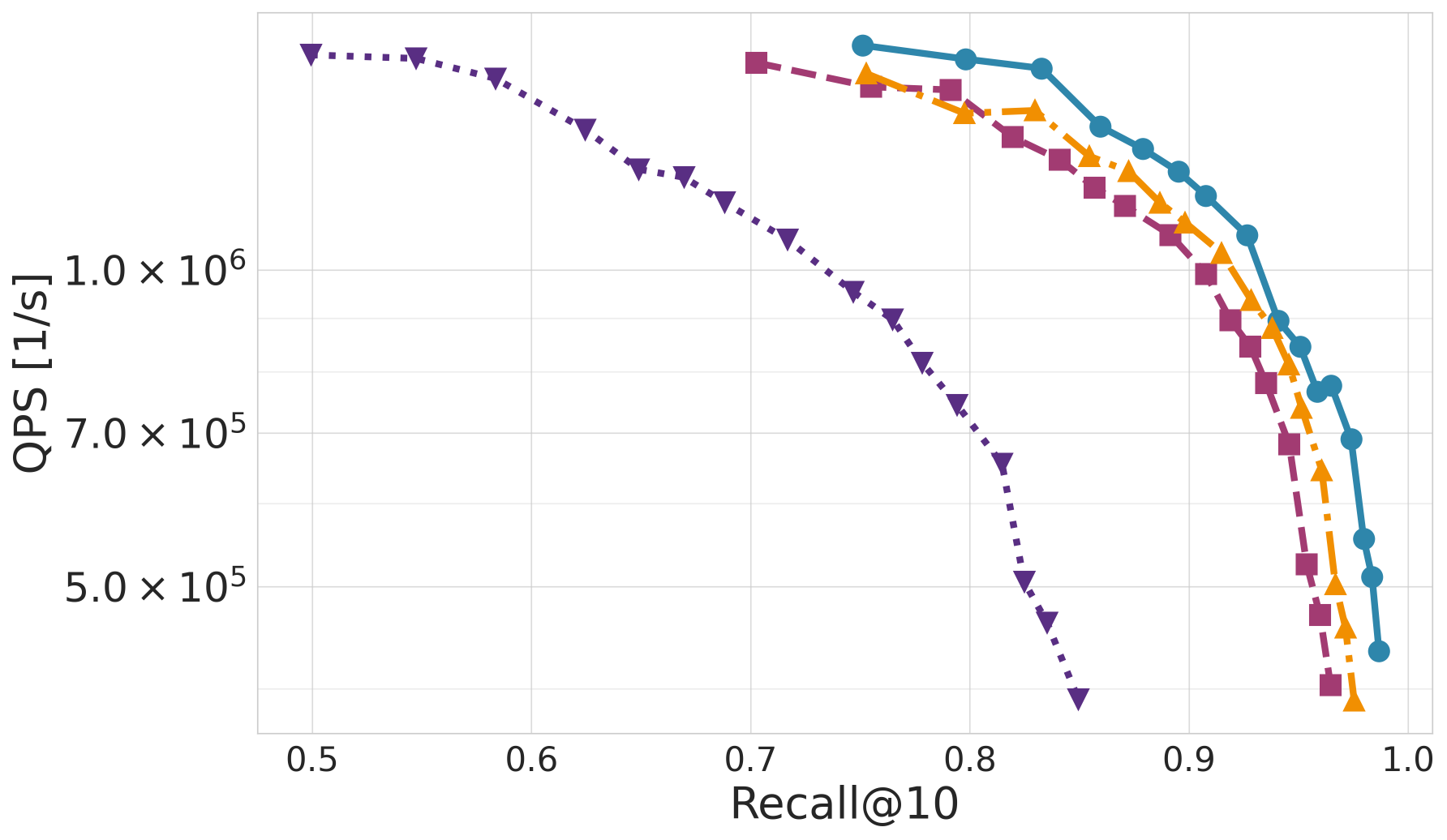}}
    \\
    \subcaptionbox{Deep 40M \label{fig:base_eval_deep40m}}[0.23\textwidth]
    {\includegraphics[width=0.23\textwidth]{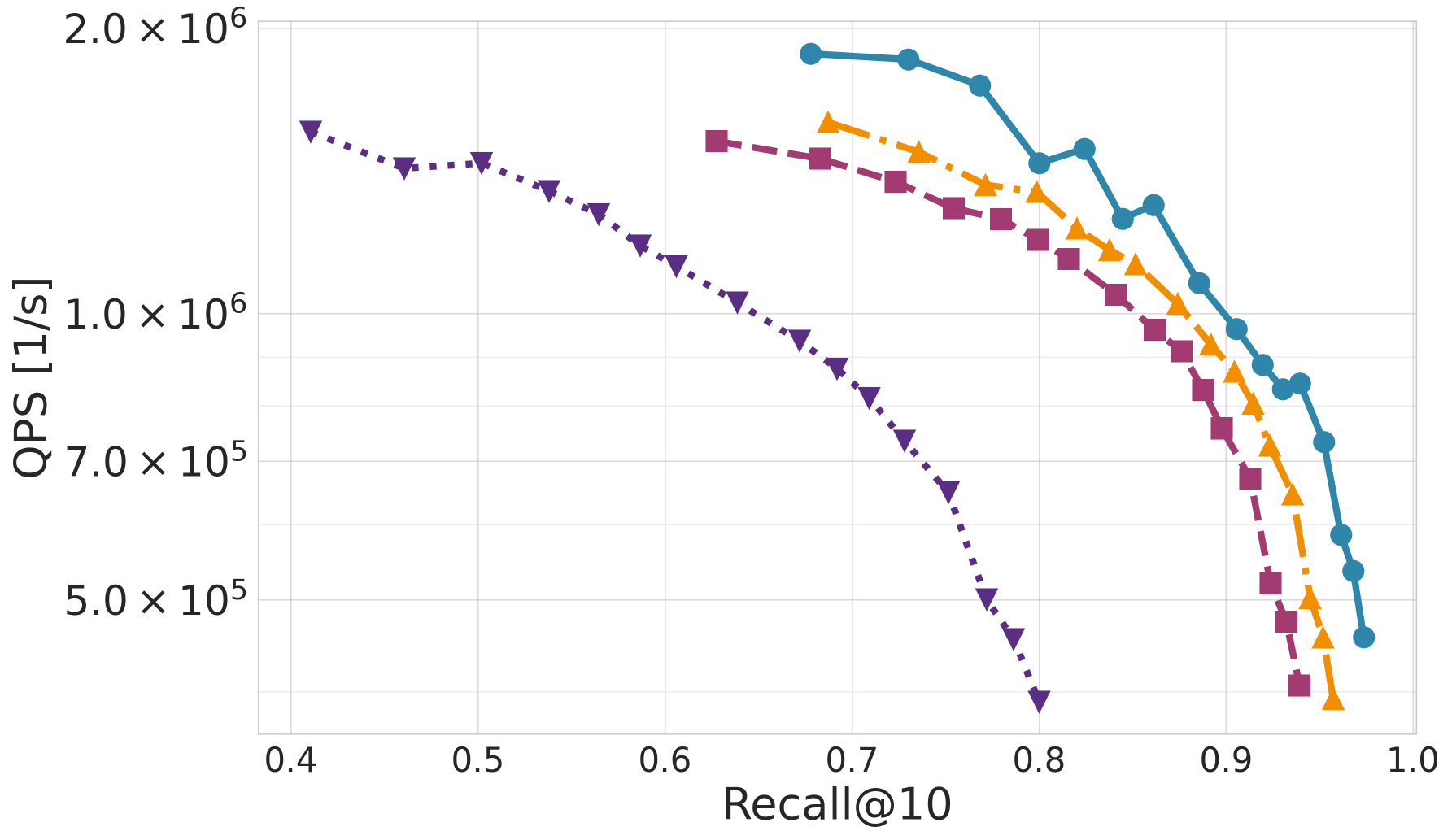}}
    &
    \subcaptionbox{Yandex T2I 1M \label{fig:base_eval_yandex}}[0.23\textwidth]
    {\includegraphics[width=0.23\textwidth]{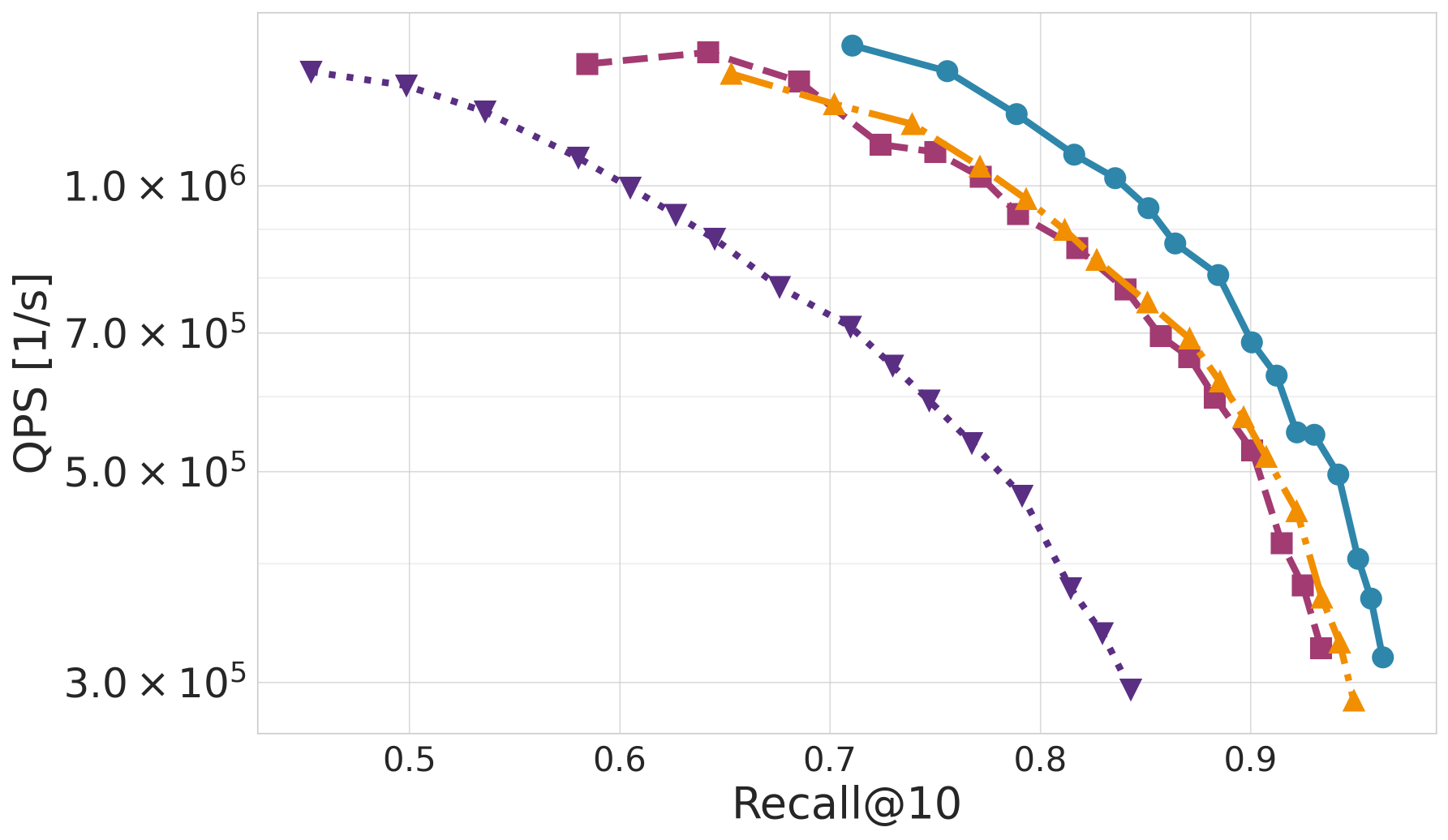}}
    &
    \subcaptionbox{OpenAI Embed.\ 1M
    \label{fig:base_eval_openai}}[0.23\textwidth]
    {\includegraphics[width=0.23\textwidth]{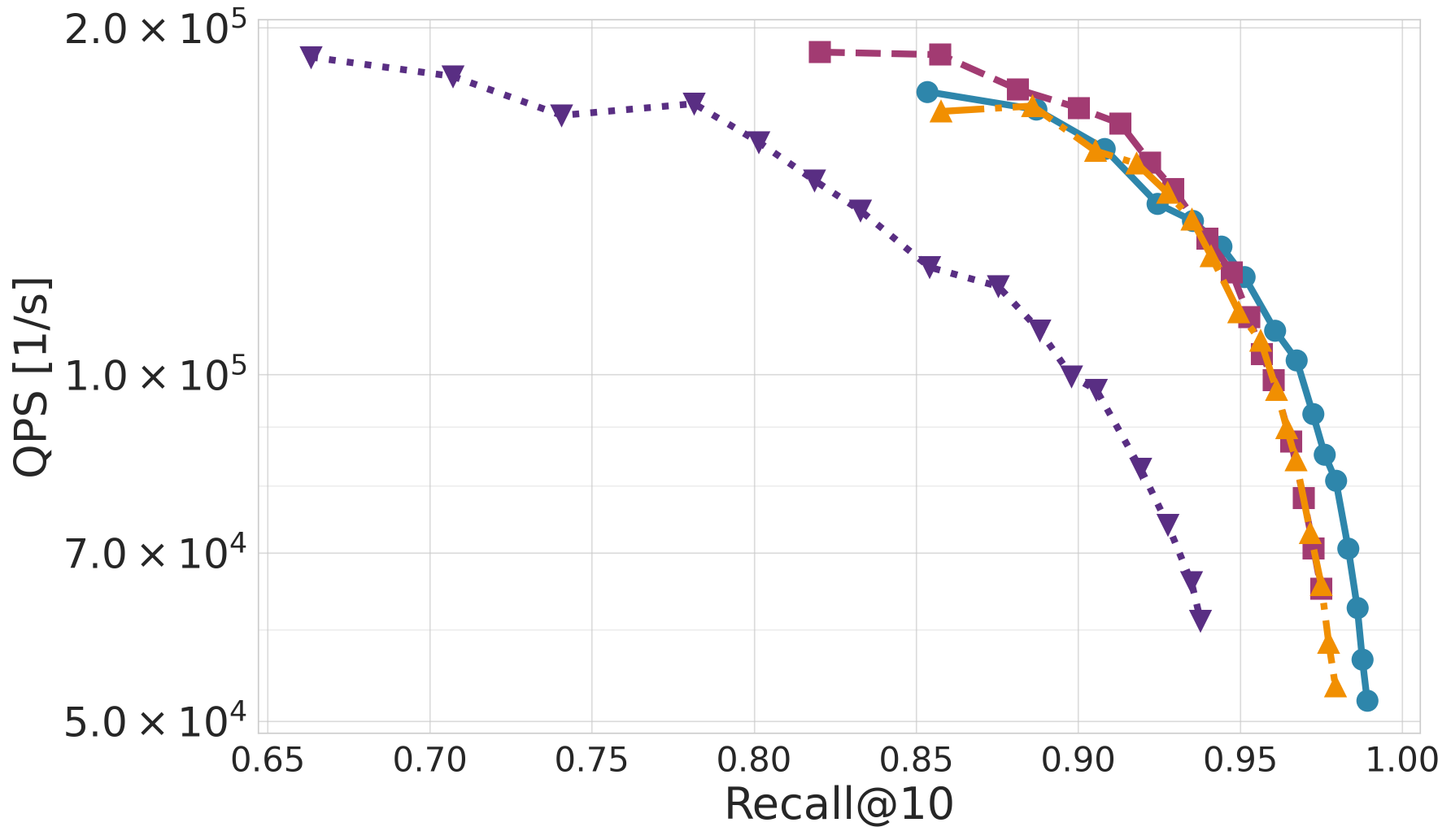}}
    &
    \subcaptionbox{Wikipedia 1M
    \label{fig:base_eval_wikipedia1m}}[0.23\textwidth]
    {\includegraphics[width=0.23\textwidth]{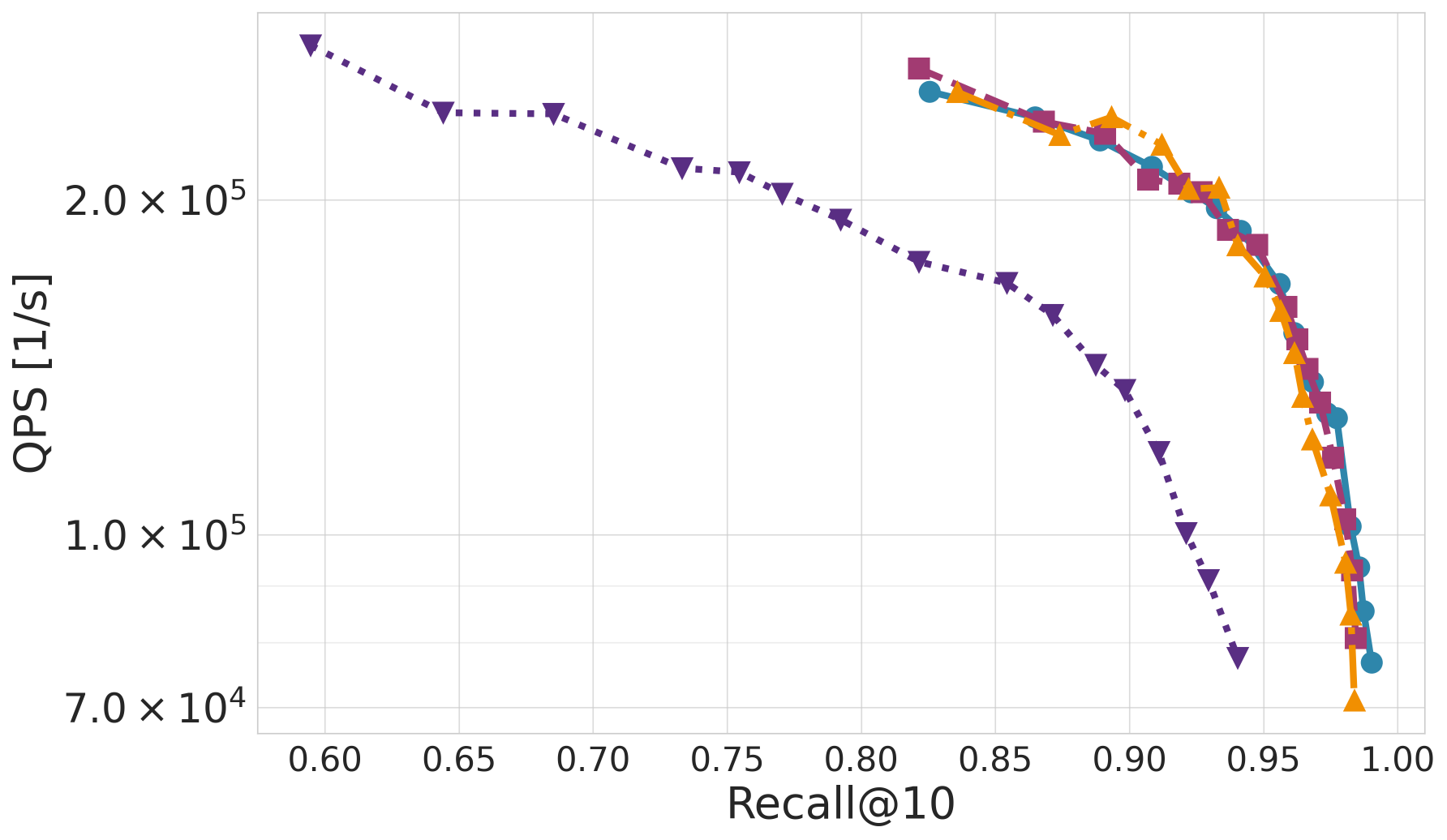}}
    \\
    \subcaptionbox{Wikipedia 10M
    \label{fig:base_eval_wikipedia10m}}[0.23\textwidth]
    {\includegraphics[width=0.23\textwidth]{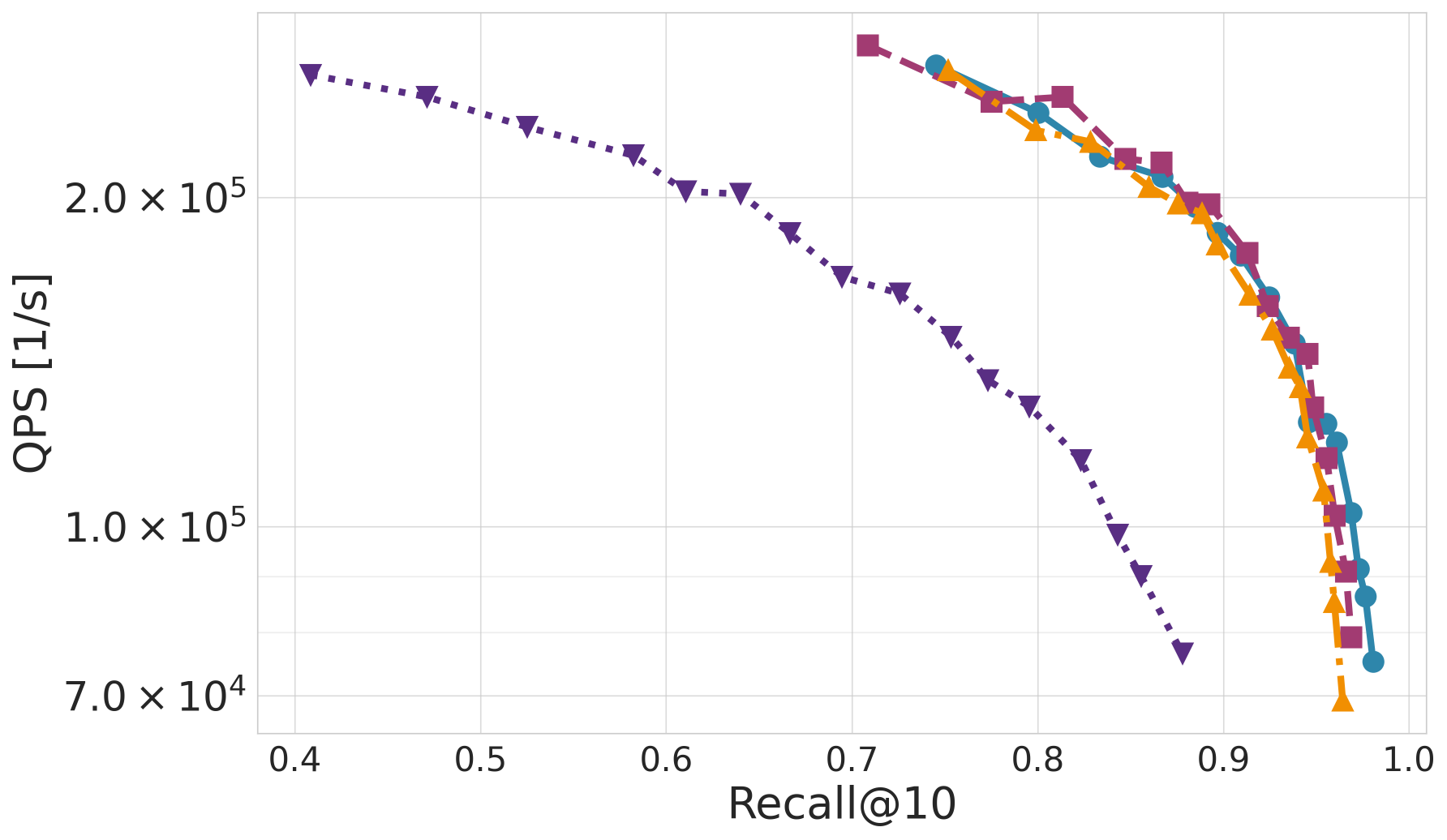}}
    &
    \subcaptionbox{BioASQ 1M \label{fig:base_eval_bioasq1m}}[0.23\textwidth]
    {\includegraphics[width=0.23\textwidth]{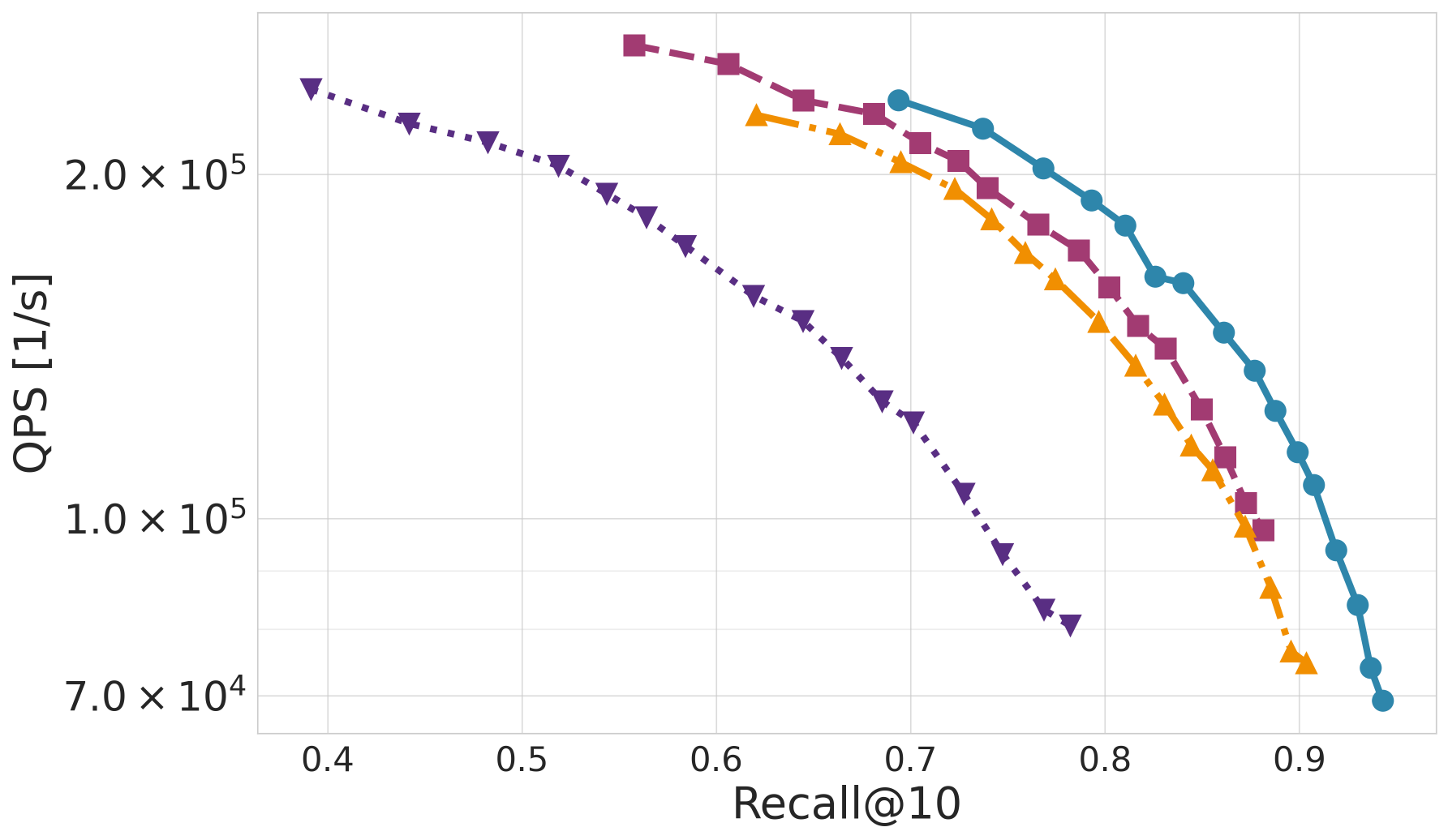}}
    &
    \subcaptionbox{BioASQ 10M \label{fig:base_eval_bioasq10m}}[0.23\textwidth]
    {\includegraphics[width=0.23\textwidth]{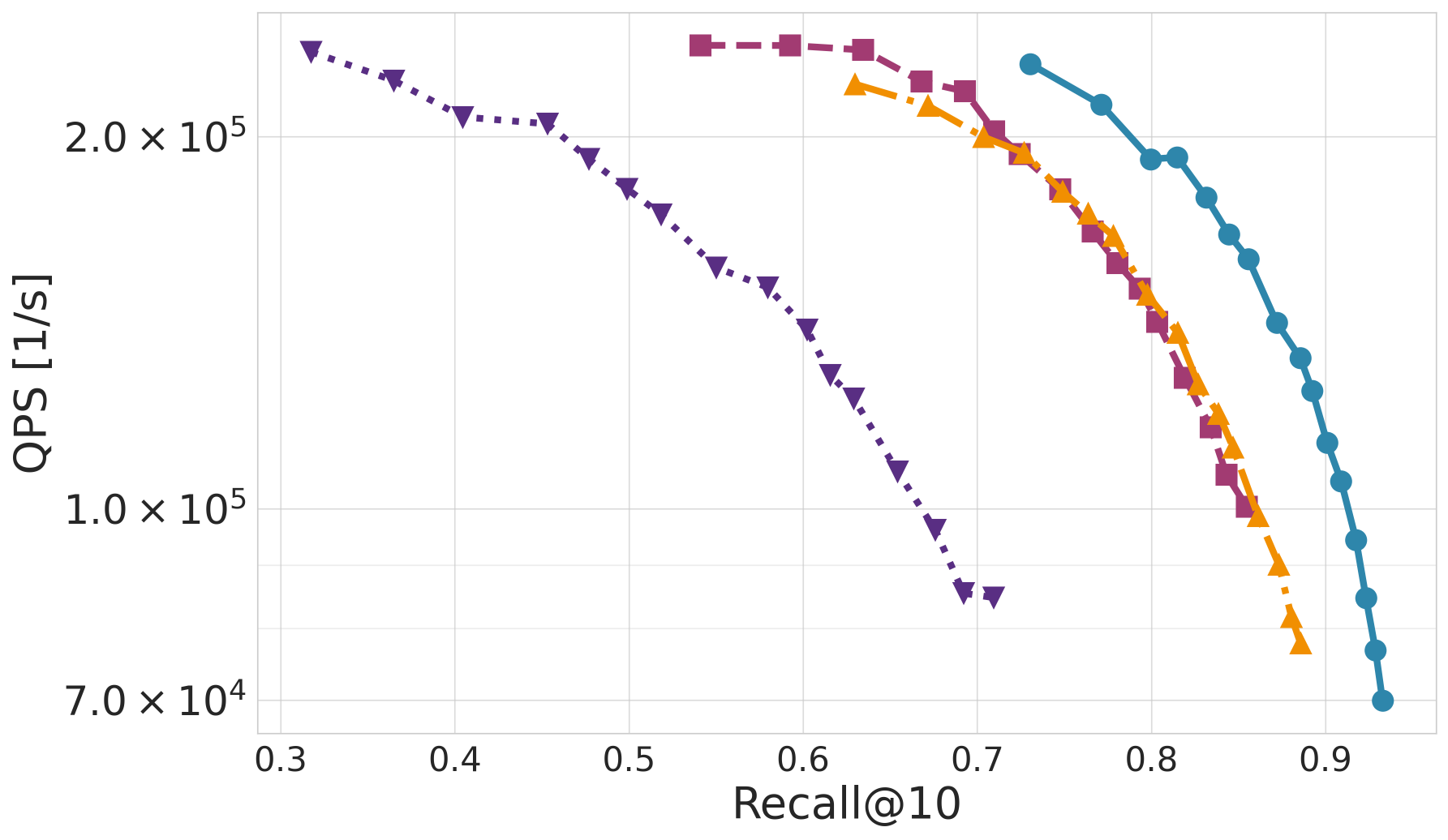}}
    &
    \subcaptionbox{C4 5M \label{fig:base_eval_c4}}[0.23\textwidth]
    {\includegraphics[width=0.23\textwidth]{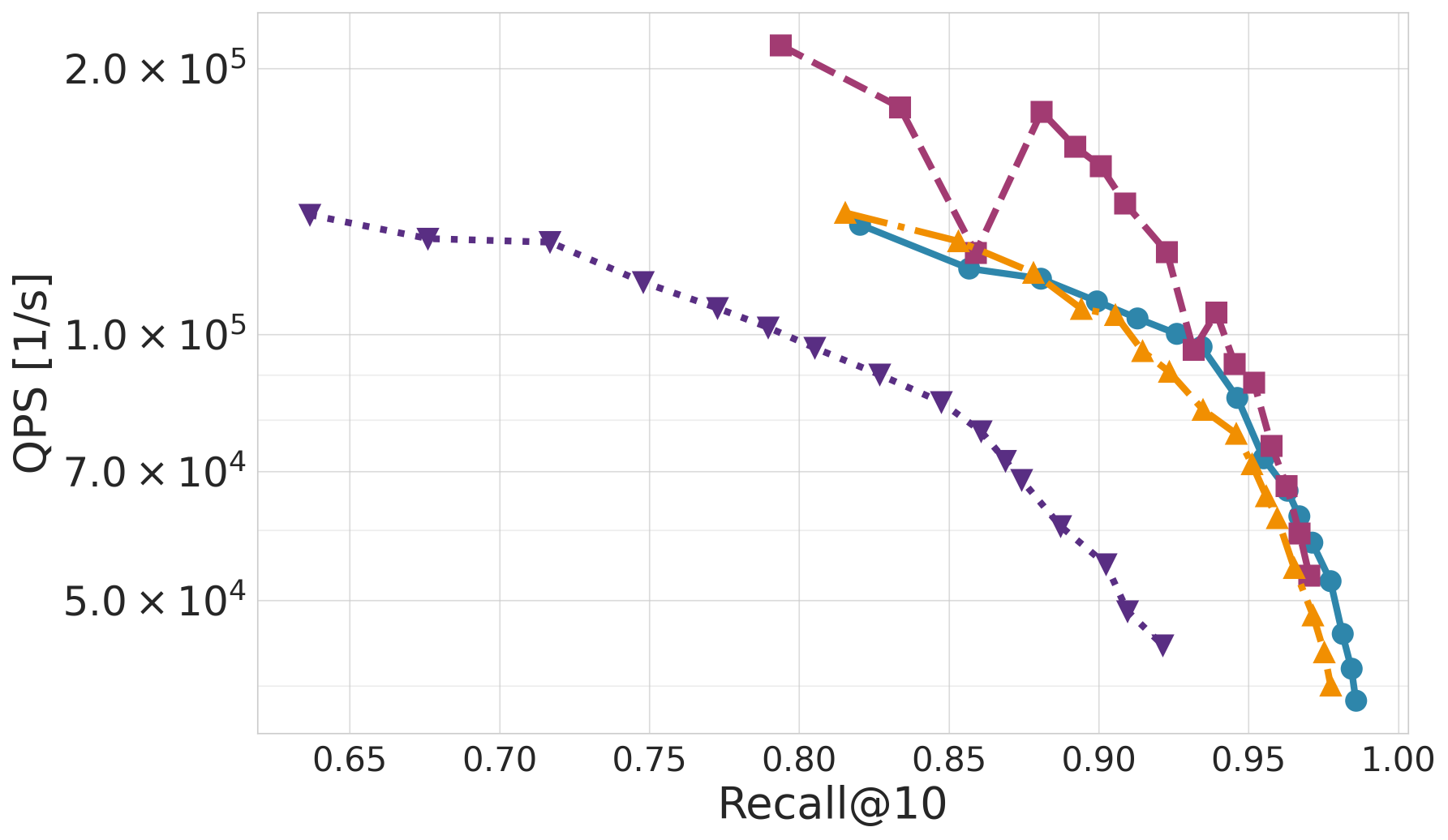}}
  \end{tabular}
  \vspace{0.5em}
  \begin{tabular}{c}
    \includegraphics[width=0.62\textwidth]{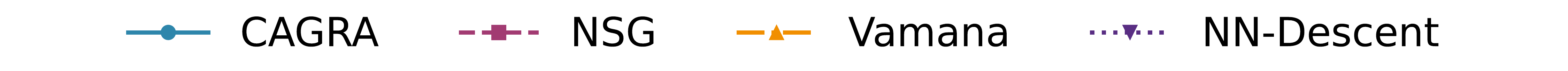}
  \end{tabular}
  \caption{Recall-QPS trade-off curves for four graph-based indices
    (CAGRA, NSG, Vamana, NN-Descent) across all 12 datasets.
  Upper-right is better.
  The full figure is also provided in the Supp. \Cref{app:full-baseline}.}
  \label{fig:base_eval}
\end{figure*}

Using our evaluation framework, we systematically benchmark the
graph-based indices introduced in \cref{sec:exp-graph-index} across
12 datasets on a single GPU.
We evaluate performance using Recall@$10$ and QPS.
We test 16 values of the runtime parameter $L$:
20, 25, 30, 35, 40, 45, 50, 60, 70, 80, 90, 100, 120, 140, 160, and 180.
\diffCR{All indices here are evaluated under their original
construction-time vertex ordering, without any reordering applied, so
that this comparison reflects topology alone; the effect of reordering
is isolated in \cref{sec:exp-reordering-results}.}
To the best of our knowledge, this is the first work to run CPU-oriented indices such as NSG,
Vamana, and NN-Descent on a GPU.

\Cref{fig:base_eval} shows the performance trade-off curves across
all 12 datasets.
Overall, NN-Descent consistently underperforms compared to the other
indices across all datasets, as expected since it constructs a
$k$NN graph without any optimization for search efficiency.
Surprisingly, NSG and Vamana, although originally designed for CPU
execution, perform as well as or better than the GPU-optimized CAGRA
on several datasets.
In particular, on C4 5M (\cref{fig:base_eval}~(l)), NSG
outperforms CAGRA by up to $1.2$ times.
On classical datasets such as SIFT and Deep, performance differences
among CAGRA, NSG, and Vamana are relatively small, while on modern
high-dimensional datasets such as Wikipedia 10M
(\cref{fig:base_eval}~(i)) and BioASQ 10M
(\cref{fig:base_eval}~(k)), the differences tend to be more pronounced.
We obtain these results on NVIDIA A100 GPU, and we also confirm similar trends on a mid-range GPU (NVIDIA RTX 6000 Ada).
See Supp. \Cref{app:full-baseline} for the full results.

\subsection{Effectiveness Measurement of Graph Reordering}
\label{sec:exp-reordering-results}

\begin{figure*}[t]
  \centering
  \setlength{\tabcolsep}{0.5pt}
  \renewcommand{\arraystretch}{0}
  \begin{tabular}{cccc}
    \subfigbox{subfiggood}{CAGRA|Deep10M}{fig:reordering_cagra_good1}{0.24\textwidth}{\includegraphics[width=0.24\textwidth]{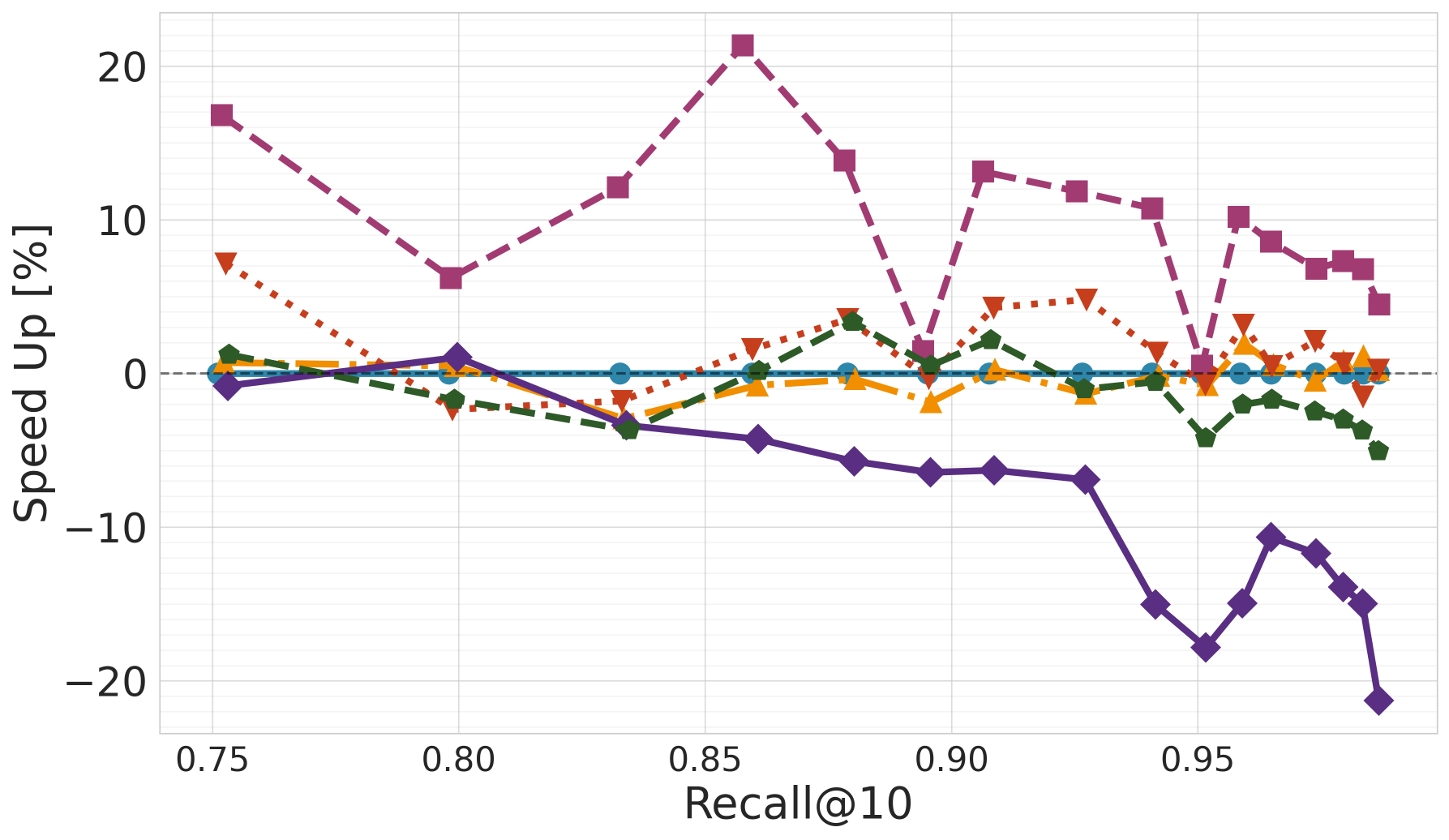}}
    &
    \subfigbox{subfiggood}{CAGRA|C4 5M}{fig:reordering_cagra_good2}{0.24\textwidth}{\includegraphics[width=0.24\textwidth]{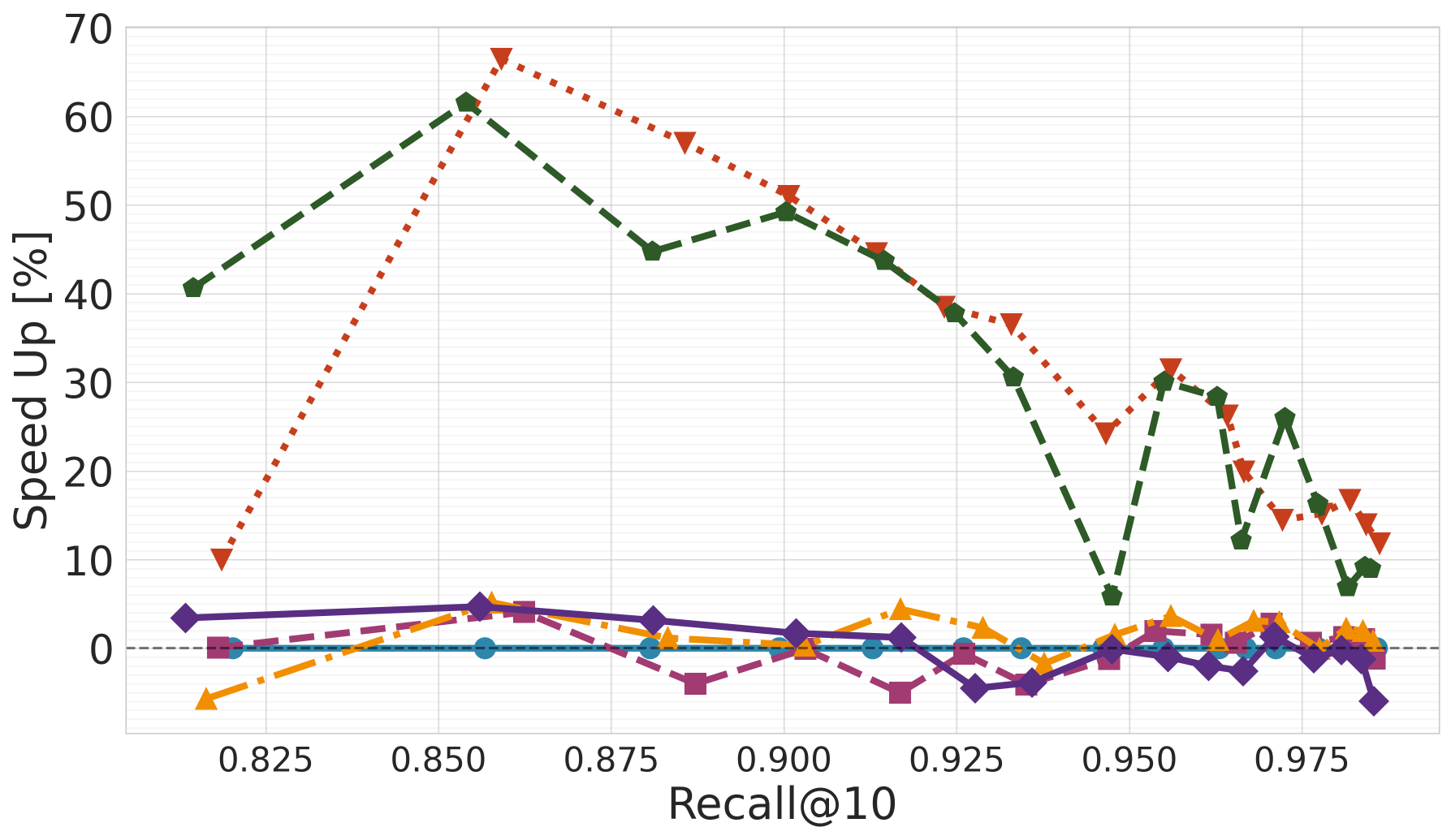}}
    &
    \subfigbox{subfiggood}{NSG|SIFT 1M}{fig:reordering_nsg_good1}{0.24\textwidth}{\includegraphics[width=0.24\textwidth]{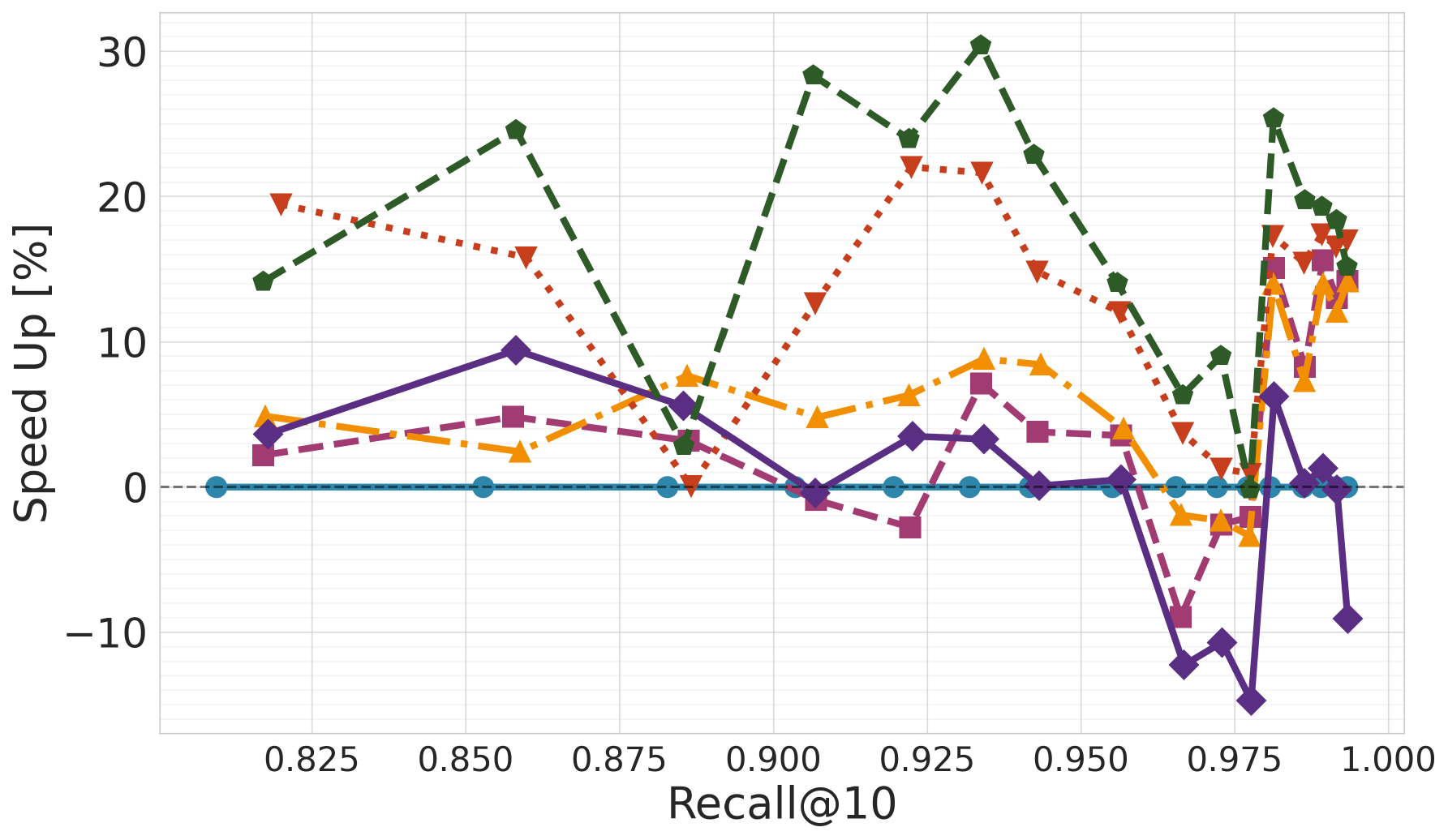}}
    &
    \subfigbox{subfiggood}{NSG|Deep10M}{fig:reordering_nsg_good2}{0.24\textwidth}{\includegraphics[width=0.24\textwidth]{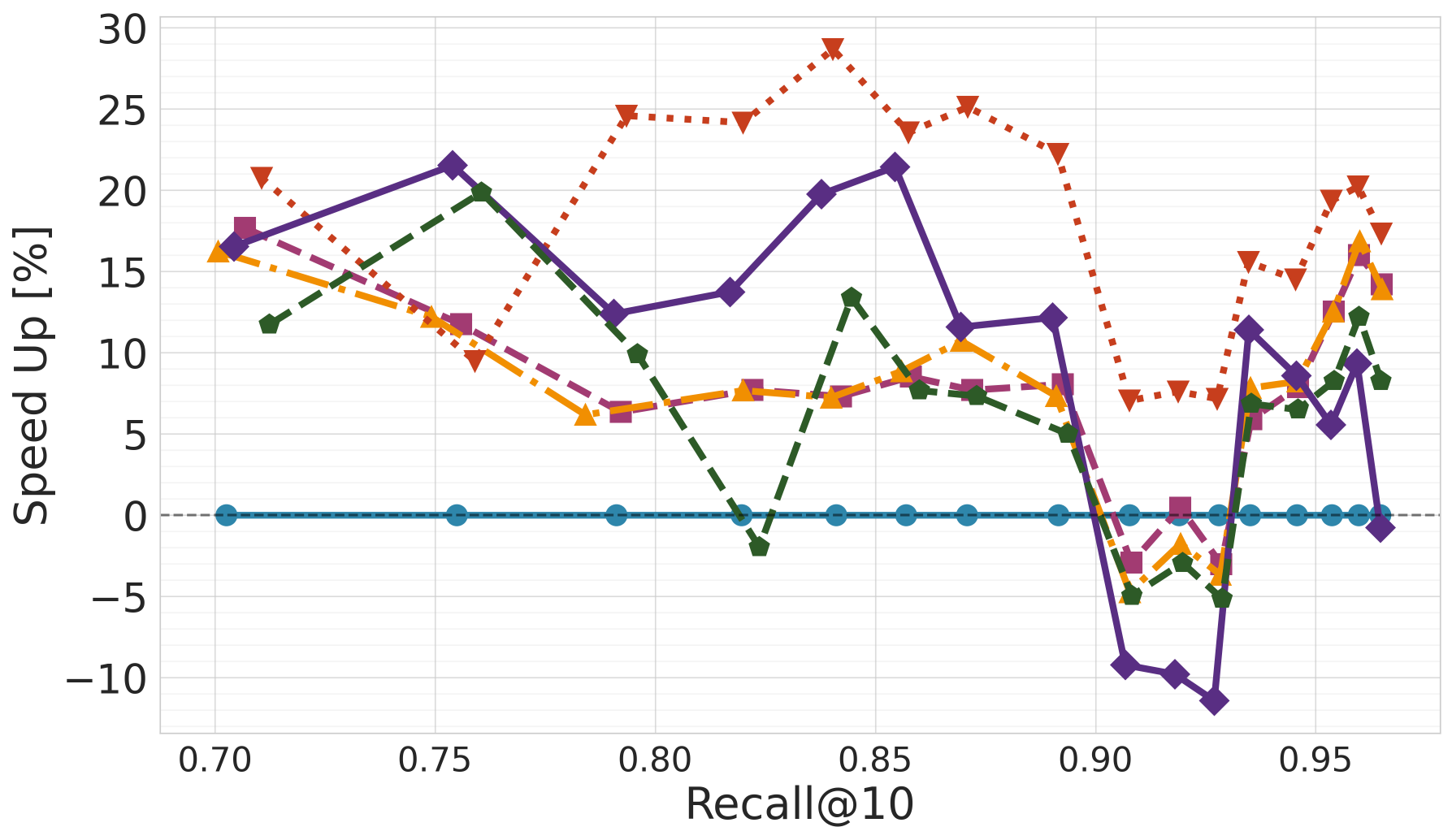}}
    \\
    \subfigbox{subfiggood}{Vamana|GIST 1M}{fig:reordering_diskann_good1}{0.24\textwidth}{\includegraphics[width=0.24\textwidth]{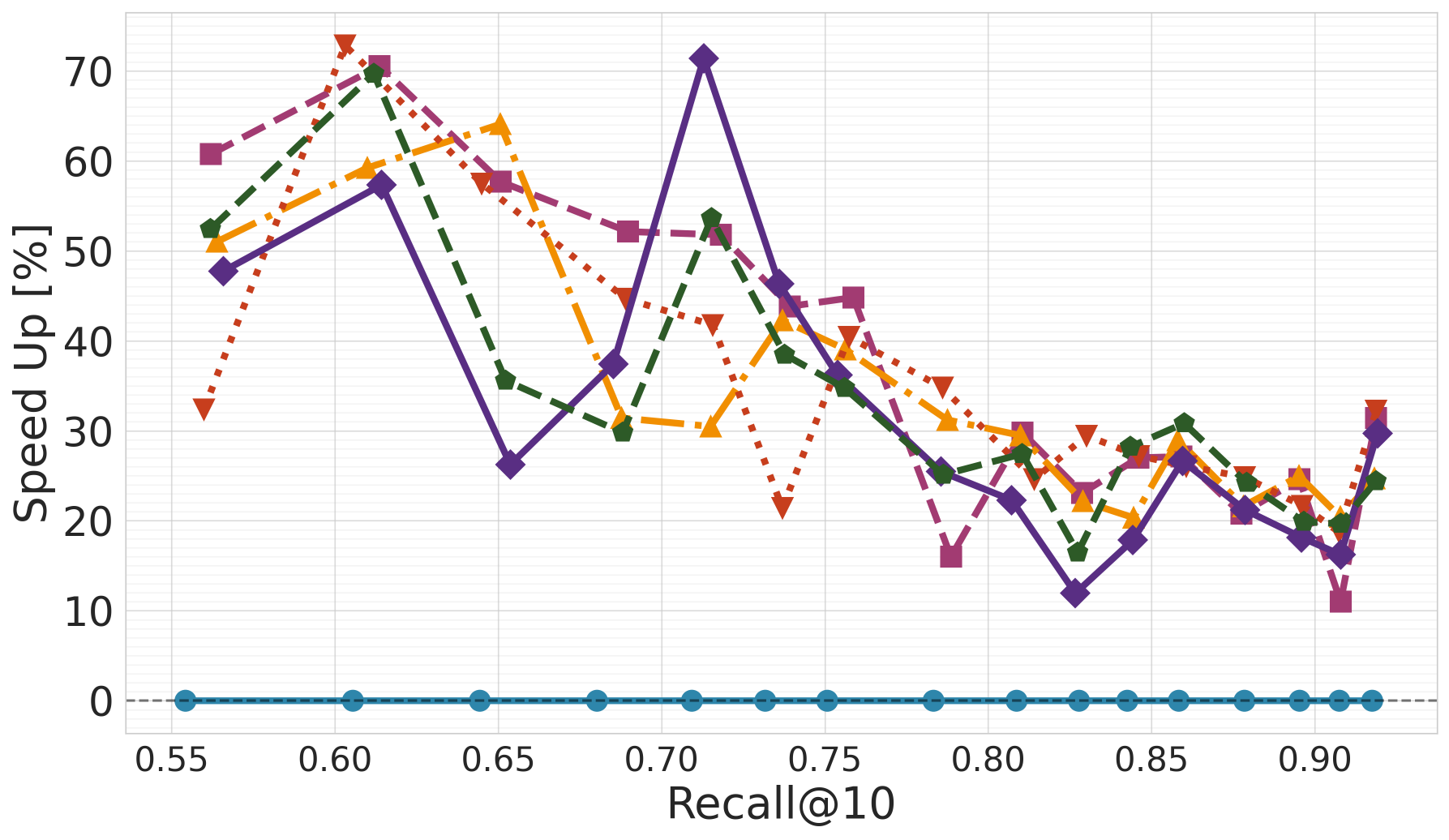}}
    &
    \subfigbox{subfiggood}{Vamana|Deep10M}{fig:reordering_diskann_good2}{0.24\textwidth}{\includegraphics[width=0.24\textwidth]{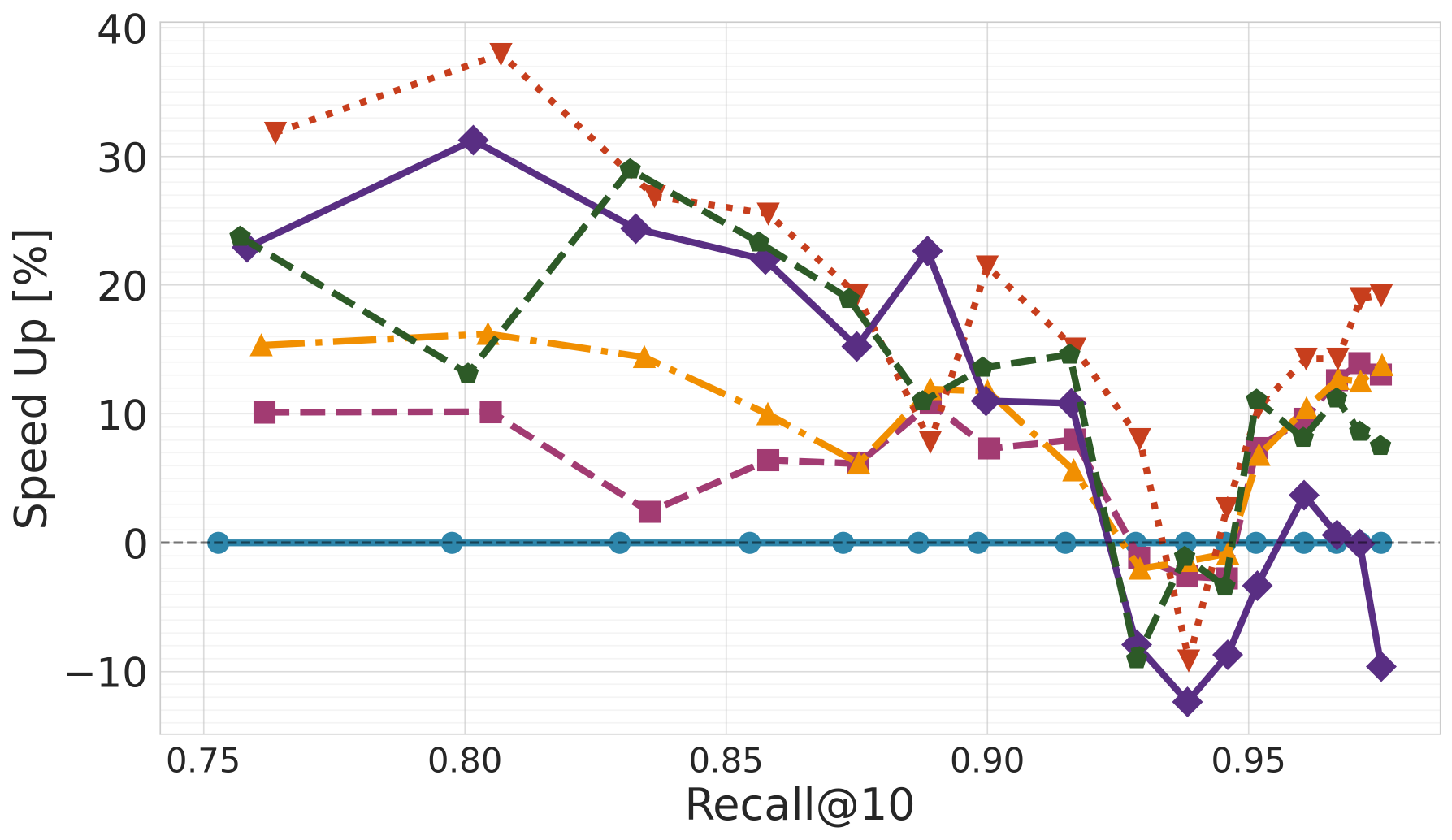}}
    &
    \subfigbox{subfiggood}{NNDes.|Wikipedia 1M}{fig:reordering_nndescent_good1}{0.24\textwidth}{\includegraphics[width=0.24\textwidth]{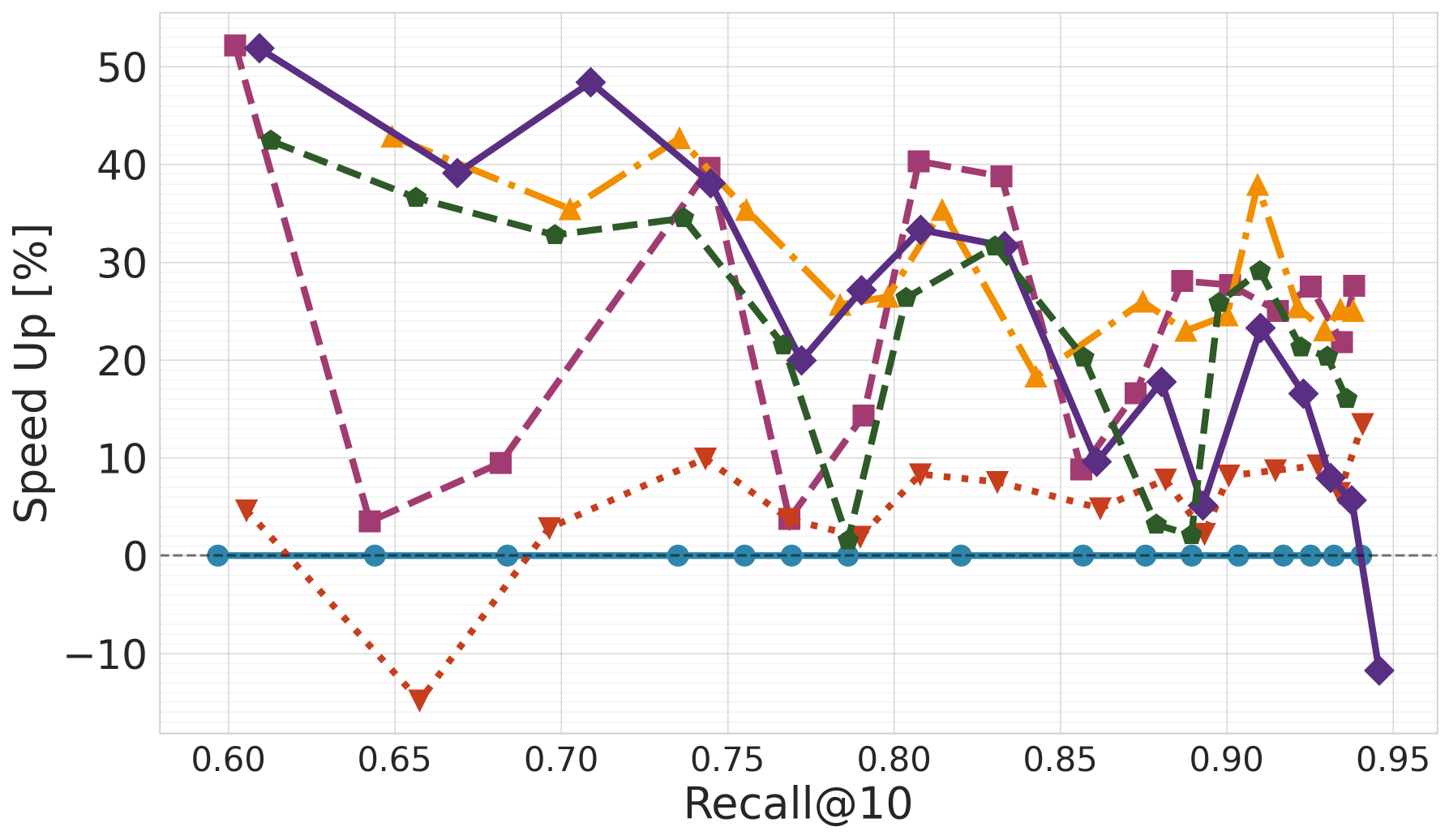}}
    &
    \subfigboxsmall{subfiggood}{NNDes.|BioASQ 1M}{fig:reordering_nndescent_good2}{0.24\textwidth}{\includegraphics[width=0.24\textwidth]{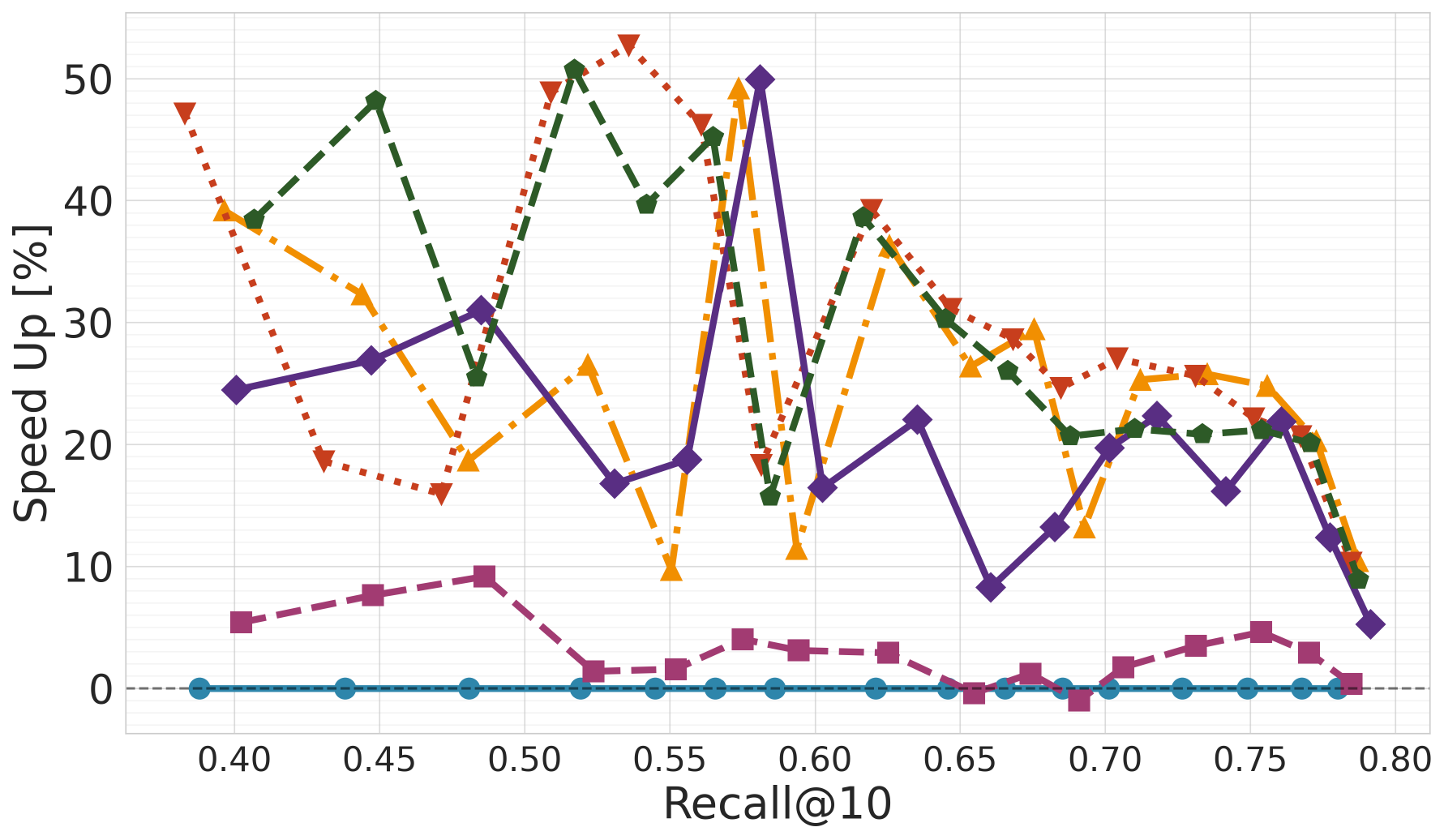}}
  \end{tabular}
  \vspace{0.1in}
  \begin{tabular}{c}
    \includegraphics[width=0.9\textwidth]{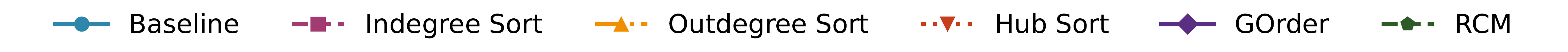}
  \end{tabular}
  \vspace{-0.1in}
  \caption{
    \colorbox{subfiggood!60}{Positive cases}: performance impact of \diffCR{graph reordering} for eight
    index--dataset combinations where reordering yields substantial QPS
    improvement (label format: graph index\,|\,dataset).
    The full results are in the Supp. \Cref{app:full-reordering}.}
  \label{fig:reordering_eval_positive}
\end{figure*}

\begin{figure*}[t]
  \centering
  \setlength{\tabcolsep}{0.5pt}
  \renewcommand{\arraystretch}{0}
  \begin{tabular}{cccc}
    \subfigbox{subfigbad}{CAGRA|SIFT 1M}{fig:reordering_cagra_bad1}{0.24\textwidth}{\includegraphics[width=0.24\textwidth]{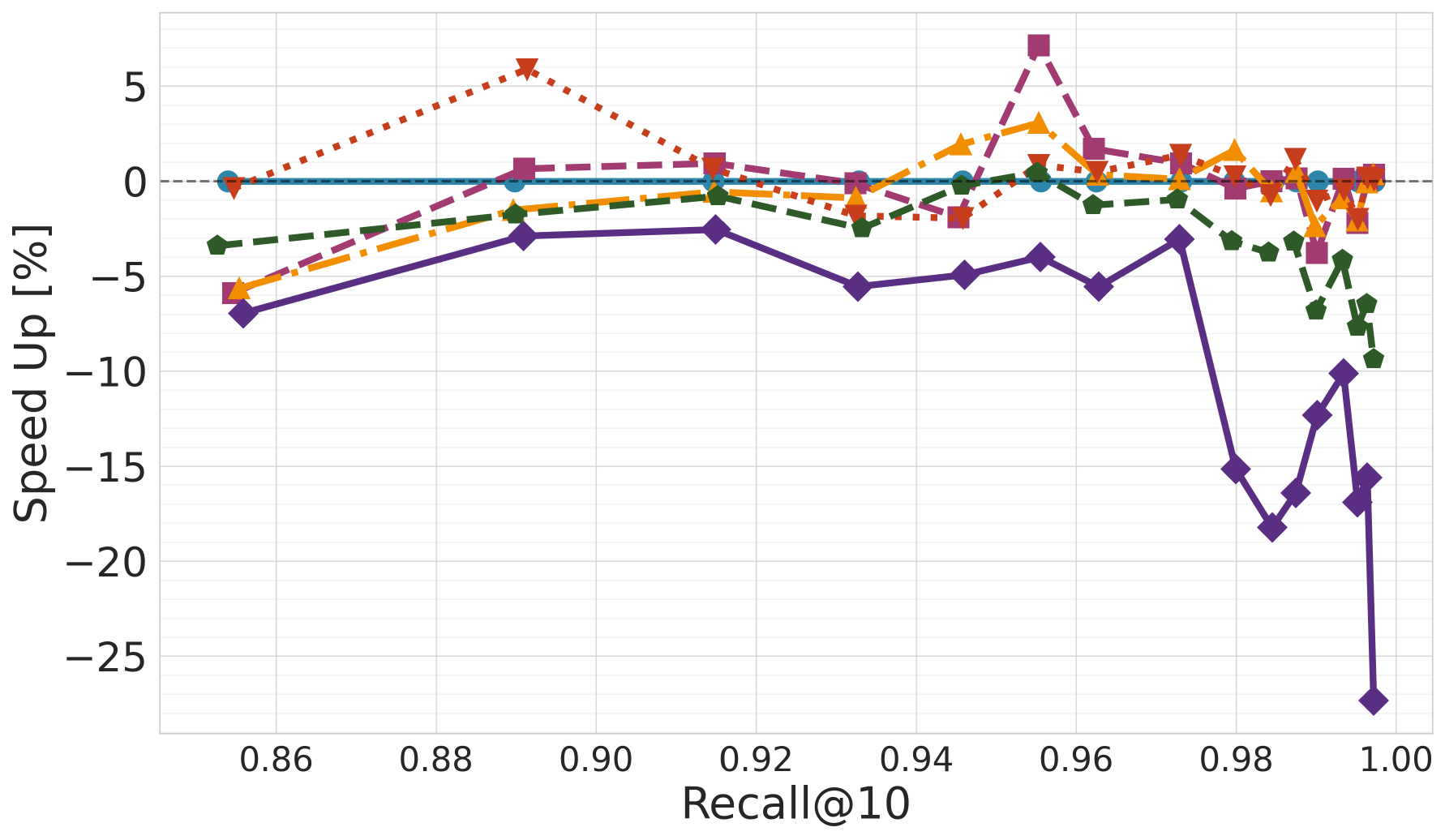}}
    &
    \subfigbox{subfigbad}{NSG|Deep 1M}{fig:reordering_nsg_bad1}{0.24\textwidth}{\includegraphics[width=0.24\textwidth]{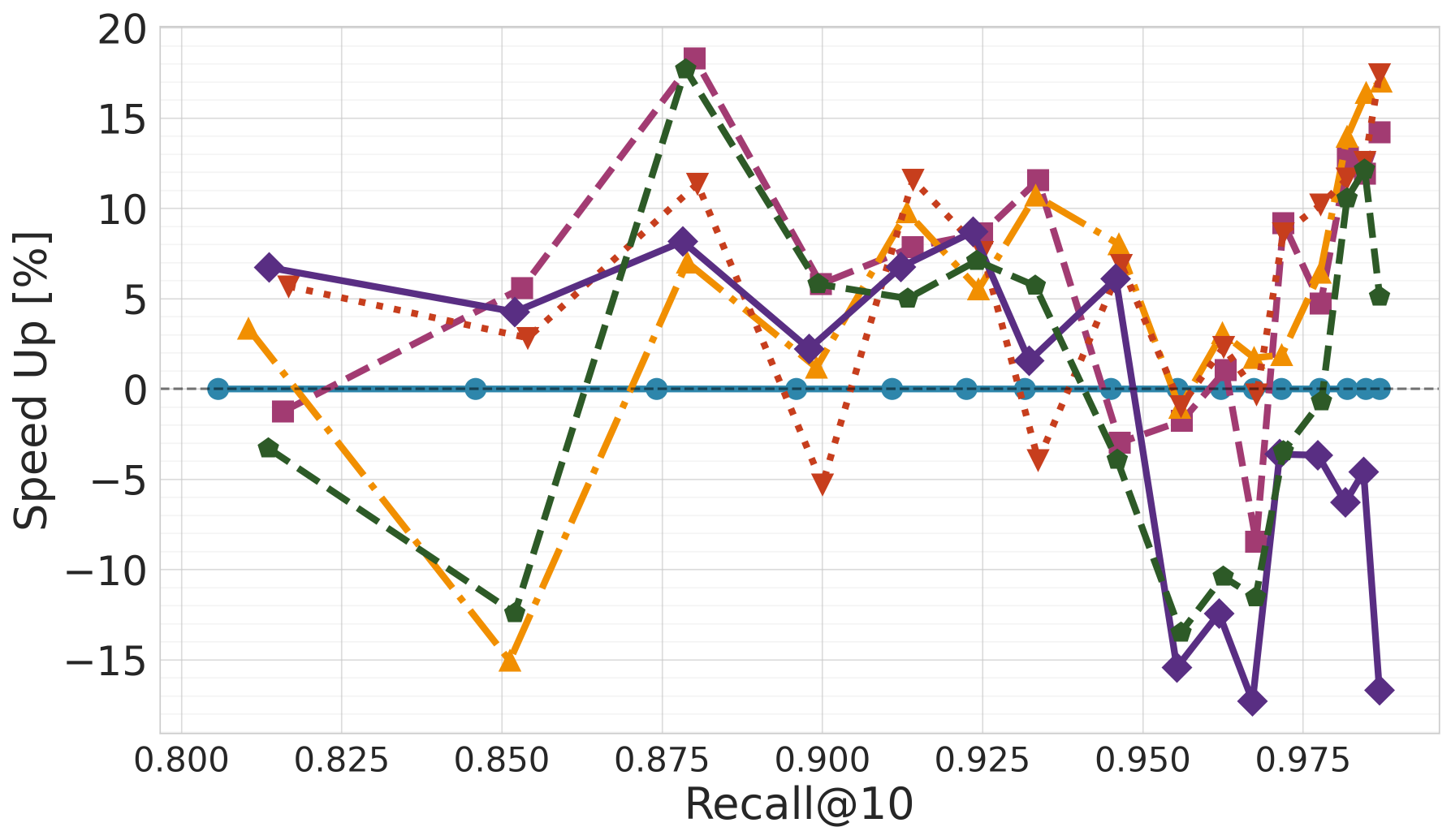}}
    &
    \subfigbox{subfigbad}{Vamana|SIFT 1M}{fig:reordering_diskann_bad1}{0.24\textwidth}{\includegraphics[width=0.24\textwidth]{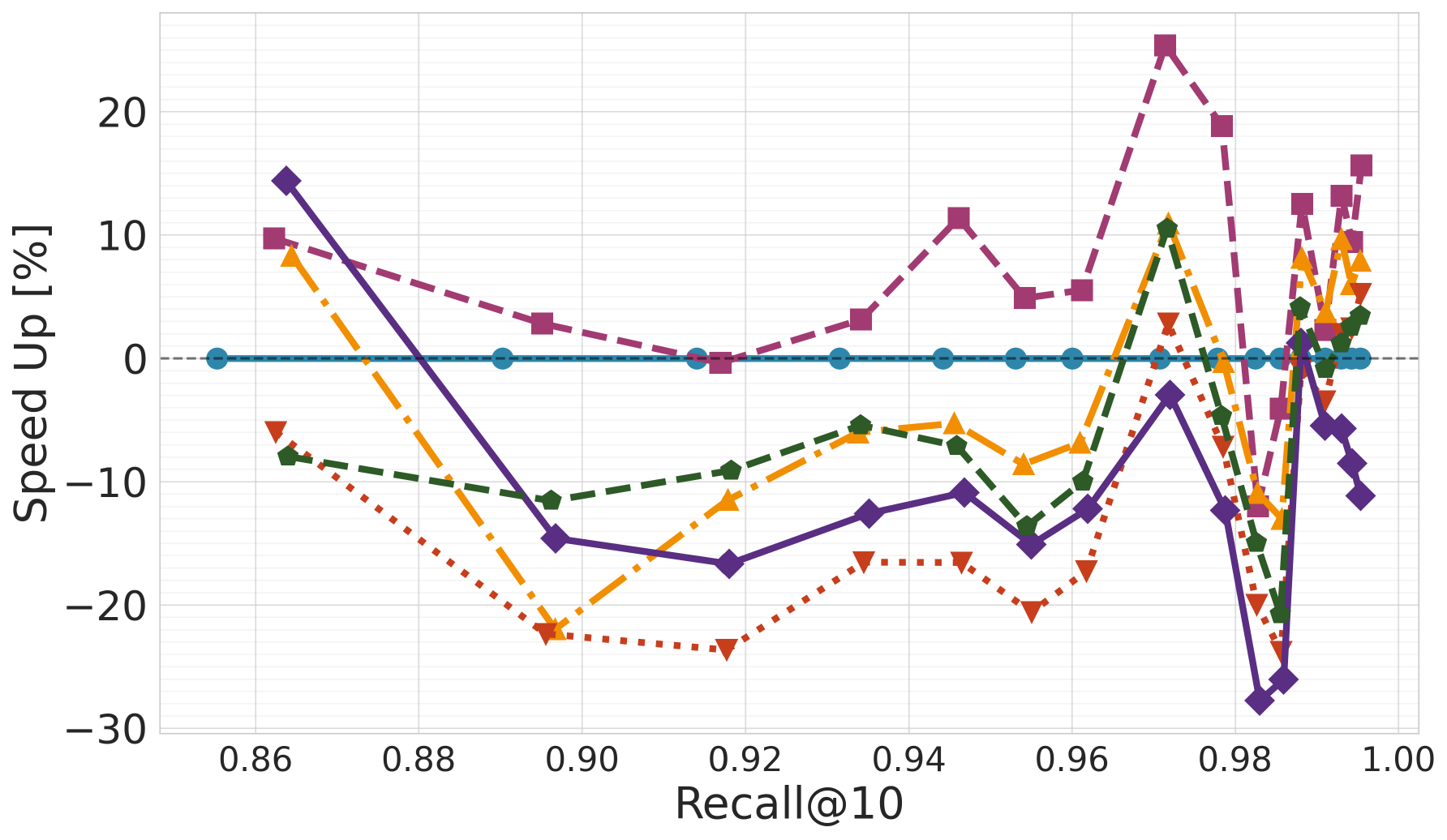}}
    &
    \subfigboxsmall{subfigbad}{NNDes.|BioASQ 10M}{fig:reordering_nndescent_bad1}{0.24\textwidth}{\includegraphics[width=0.24\textwidth]{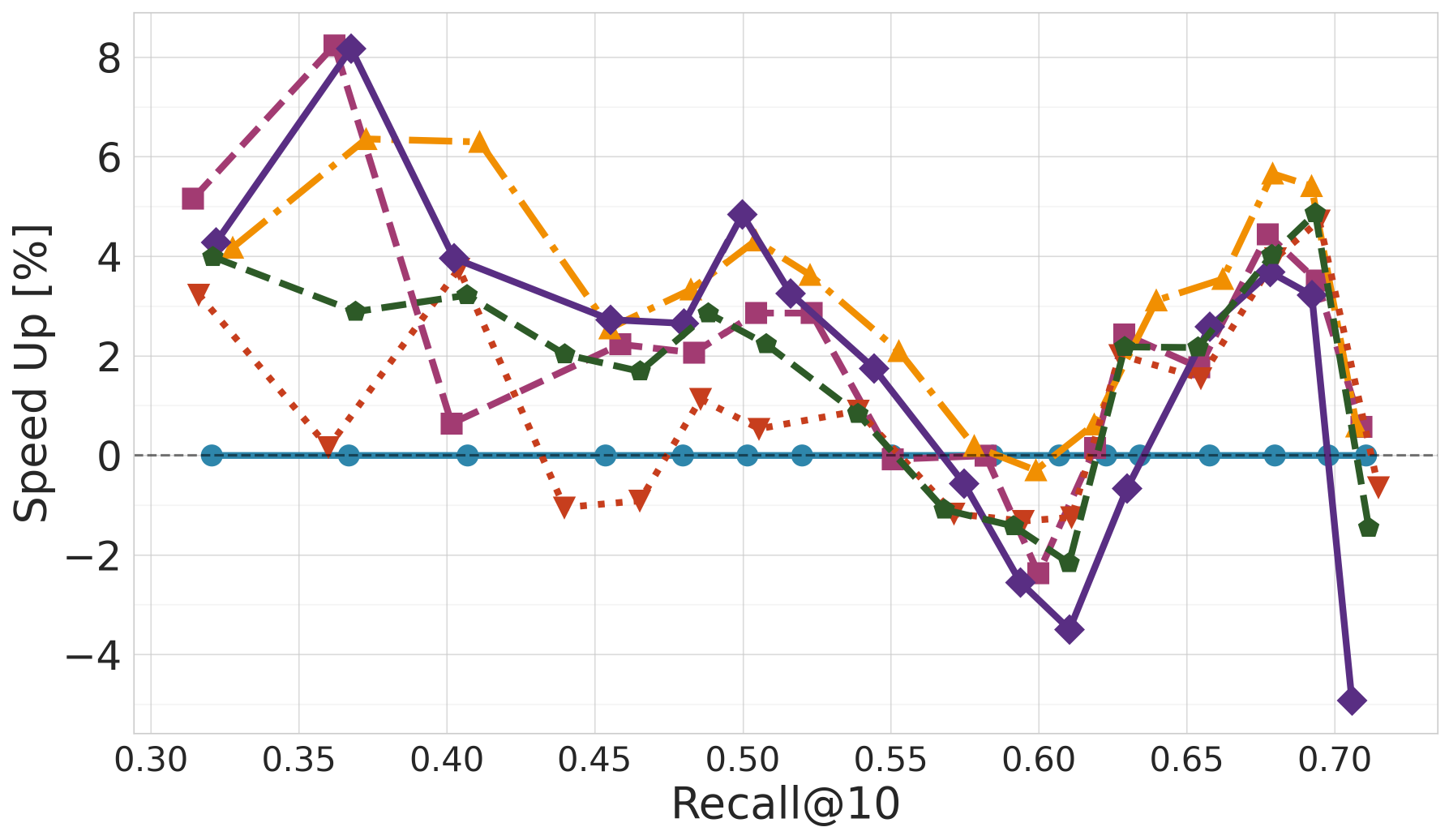}}
  \end{tabular}
  \caption{
    \colorbox{subfigbad!60}{Failure cases}: four index--dataset combinations with minimal or no
    QPS improvement from reordering.}
  \label{fig:reordering_eval_failure}
\end{figure*}

We apply the reordering algorithms introduced in
\cref{sec:exp-reordering} to four graph-based indices and evaluate
the QPS improvement across 12 datasets compared to the baseline
without reordering.
The runtime parameter settings are the same as in \cref{sec:exp-baseline}.
\Cref{fig:reordering_eval_positive} shows eight positive cases;
\Cref{fig:reordering_eval_failure} shows four representative failure cases.
The full combinations are in the Supp. \Cref{app:full-reordering}.
In some combinations, we observe up to 80\% QPS improvement while preserving search accuracy, with typical gains of 10--30\% in positive cases.
For CAGRA, Deep 10M and C4 5M achieve 20--70\% speedup
(\cref{fig:reordering_eval_positive}~(a--b)), while SIFT 1M shows minimal
improvement (\cref{fig:reordering_eval_failure}~(a)); Yandex T2I 1M also
shows minimal improvement.
For NSG, SIFT 1M and Deep 10M show 10--40\% speedup
(\cref{fig:reordering_eval_positive}~(c--d)), while Deep 1M shows
limited gains (\cref{fig:reordering_eval_failure}~(b)); BioASQ 1M also
shows limited gains.
For Vamana, GIST 1M and Deep 10M benefit most
(\cref{fig:reordering_eval_positive}~(e--f)), while SIFT 1M shows
limited improvement (\cref{fig:reordering_eval_failure}~(c)); BioASQ 1M
also shows limited improvement.
For NN-Descent, speedup rates are large on 1M-scale datasets
(\cref{fig:reordering_eval_positive}~(g--h))
but sometimes remain below 10\% on larger datasets
(\cref{fig:reordering_eval_failure}~(d)).
Notably, for BioASQ 1M, the speedup characteristics of
NN-Descent (\cref{fig:reordering_eval_positive}~(h)) and Vamana
(see Supp. \Cref{app:full-reordering}) exhibit highly similar shapes,
suggesting that for certain dataset--index combinations,
the effectiveness of reordering at each runtime parameter
follows similar patterns.
See Supp. \Cref{app:full-reordering} for full results.

At the dataset level, C4 5M shows remarkable performance improvement
for all indices, with maximum improvements of 60--80\%.
On the other hand, BioASQ 1M and 10M show improvements of less than
10\% except for NN-Descent, making them relatively ineffective
datasets for reordering.
Among reordering algorithms, GOrder consistently shows
unstable performance, with a tendency for speedup rates to drop
sharply in high-recall regions, making it unsuitable for applications
that require high recall.
For the other algorithms, no consistent performance trends are
observed across datasets and indices, suggesting that the optimal
choice of reordering algorithm may depend on specific use cases.
\diff{Preprocessing cost also varies: the lightweight methods finish
within minutes even on the largest datasets, whereas GOrder is far
more expensive; Supp.~\Cref{app:reordering-time} reports full
wall-clock reordering times.}

We conduct our primary experiments on NVIDIA A100 GPU, where many
cases show substantial speed improvements.
Furthermore, results are more sensitive to the runtime
parameter configuration, and there is no improvement in many cases.
Supp. \Cref{app:full-reordering} provides results on RTX 6000 Ada GPU.

\subsection{Hardware Metrics Measurement}
\label{sec:exp-hardware}
To quantitatively analyze the effectiveness of reordering on GPU, we
use NVIDIA's profiler Nsight Compute to measure L1/L2 cache hit rates
and DRAM bandwidth utilization across all combinations of four
graph-based indices, multiple datasets, and five reordering methods
with $L=120$.

\begin{figure}[t]
  \centering
  \setlength{\tabcolsep}{6pt}
  \renewcommand{\arraystretch}{0}
  \begin{tabular}{cc}
    \subcaptionbox{
      Overall execution time
      \label{fig:dram_total_time}
    }{
      \includegraphics[width=0.47\linewidth]{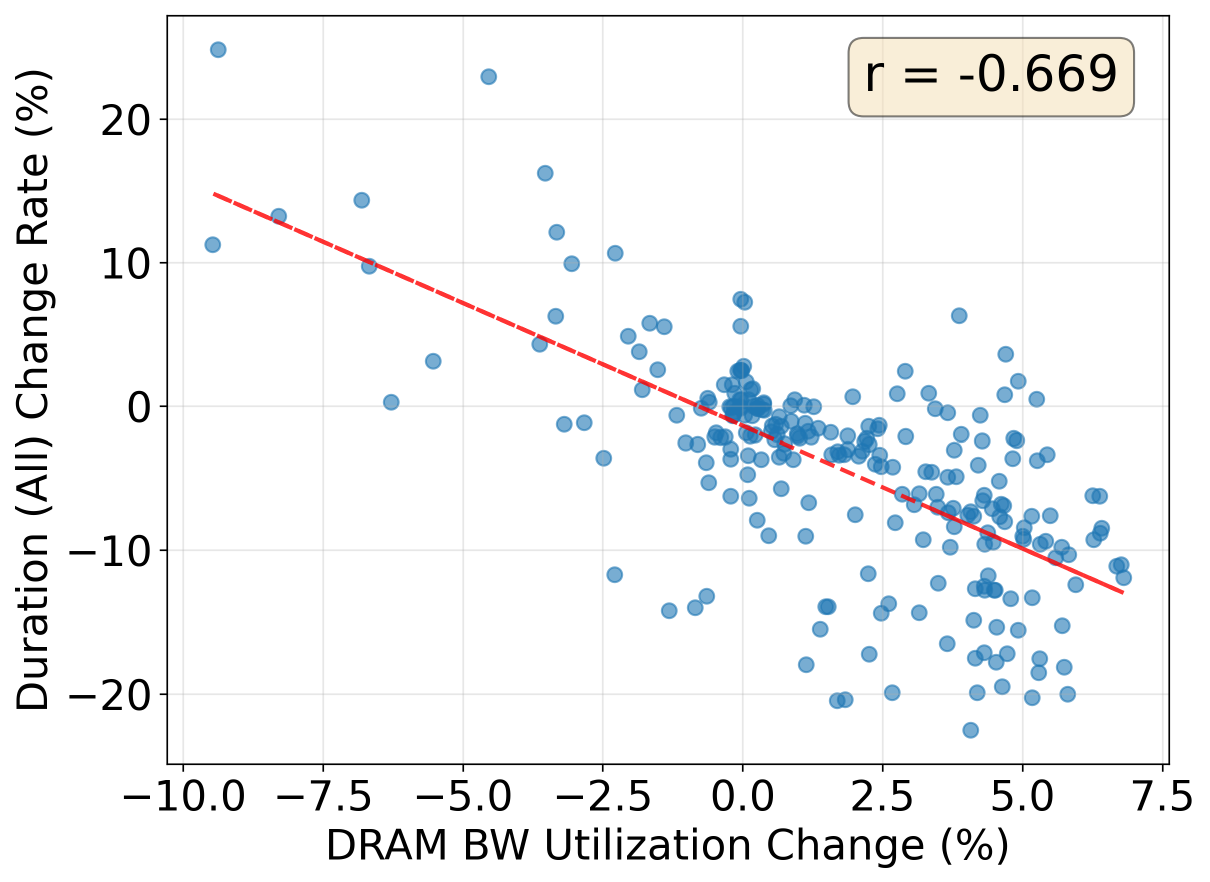}
    } &
    \subcaptionbox{
      Search execution time
      \label{fig:dram_kernel_time}
    }{
      \includegraphics[width=0.47\linewidth]{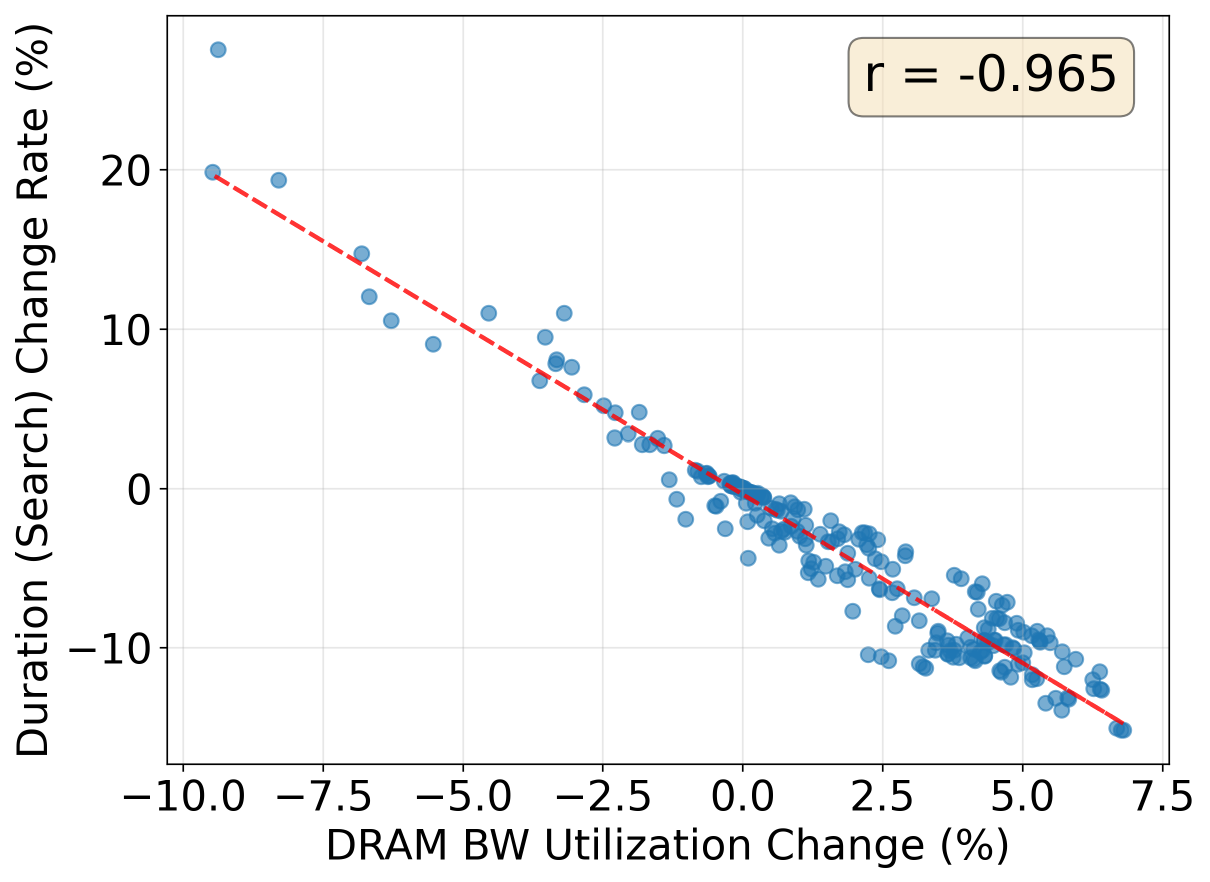}
    } \\
  \end{tabular}
  \caption{
    Correlation between DRAM bandwidth utilization change
    and execution time change.
  }
  \label{fig:dram-correlation}
\end{figure}

Only DRAM bandwidth utilization showed correlation with performance,
with correlation coefficients of $-0.669$ for overall execution time
and $-0.965$ for search kernel execution time
(\cref{fig:dram-correlation}).
This suggests that the larger the improvement in DRAM bandwidth
utilization before and after reordering, the more reordering reduces
execution time.
See Supp. \Cref{app:hardware-details} for detailed per-configuration measurements
and analysis of L1/L2 cache and DRAM behavior.
Examining the measured metrics reveals the characteristics of memory
access in graph-based ANNS.
L1 cache hit rates are consistently near zero across all
configurations, typically ranging from $0.007$\% to $0.025$\%.
For example, in CAGRA on Deep 1M, L1 hit rates remain around
$0.018$\% regardless of reordering method (baseline: $0.0178$\%, Hub
Sort: $0.0177$\%, RCM: $0.0183$\%).
These consistently near-zero L1 hit rates suggest that L1 cache is
effectively bypassed in graph-based ANNS workloads.

L2 cache hit rates are also low, typically ranging from $8$\% to
$28$\% depending on the index and dataset.
For instance, NSG on Deep 1M shows L2 hit rates around $18$ to $20$\%, while CAGRA on the same
dataset shows approximately $16$\%.
On higher-dimensional datasets, L2 hit rates tend to be even lower;
for example, CAGRA on OpenAI Embedding 1M shows L2 hit rates around $8.4$\%.
More importantly, reordering induces only minimal changes
in the L2 cache hit rates, typically less than $1$ percentage point.
We can conclude that the speedup by graph reordering is not
primarily achieved by improving L1 and L2 cache utilization.

In contrast, DRAM bandwidth utilization shows substantial changes
with reordering.
For example, in NSG on SIFT 1M, DRAM bandwidth increases from
$39.4$\% (baseline) to $44.2$\% (Hub Sort), a $4.8$\%
improvement that corresponds to roughly $15$\% QPS improvement.
Similarly, in Vamana on SIFT 1M, DRAM bandwidth increases from $47.9$\%
(baseline) to $52.6$\% (Hub Sort), a $4.7$\% improvement.
This indicates that memory bandwidth utilization, rather than cache
utilization, dominates the effectiveness of graph reordering on GPU.

\FloatBarrier

Randomly generated datasets with varying dimensions confirm that DRAM bandwidth becomes the dominant bottleneck as dimensionality increases (Supp. \Cref{app:gaussian-datasets}).

\FloatBarrier

\section{Discussion and Limitation}
\label{sec:discussion}

\subsection{Memory Layout Matters more than Graph Topology}
A surprising aspect of our results is that CPU-oriented indices (NSG,
Vamana) achieve performance comparable to or better than
GPU-optimized CAGRA under a unified algorithm.
Under conditions where search algorithms and implementations are
unified, we observe multiple cases where performance differences
between CAGRA, NSG, and Vamana are small or reversed
(\cref{fig:base_eval}).
This suggests that in GPU environments, the impact of memory access
characteristics is strong, and topological differences become
relatively small.
For high-dimensional modern datasets, the interaction between memory
layout and GPU-specific behavior (memory coalescing, warp-level
parallelism) appears to dominate search efficiency.

\subsection{Performance Improvement Factors via Hardware Metrics}
Our analysis reveals an extremely strong correlation ($r=-0.965$)
between search kernel execution time reduction and DRAM bandwidth
utilization improvement (\cref{fig:dram-correlation}).
In contrast, L1 and L2 cache hit rates show no clear correlation with
performance.
The strong correlation indicates that graph search on GPU is
memory bandwidth-limited, consistent with prior analysis attributing
approximately 40\% of kernel execution time to memory
latency from uncoalesced accesses~\cite{bang}.
Graph traversal accesses scattered neighborhoods irregularly, keeping
cache hit rates low regardless of layout.

Vertex reordering transforms scattered accesses into more coalesced
memory transactions.
When frequently accessed vertices are placed consecutively, warp-level
accesses can be combined into fewer, wider transactions, increasing
effective bandwidth utilization.
DRAM bandwidth utilization improvement reflects three underlying
factors: DRAM row buffer hit rate improvement, partial memory coalescing
improvement, and TLB hit rate improvement~\cite{walker2024_a100_tuning}.

\subsection{Impact of GPU Architecture}
Reordering effectiveness varies across GPU architectures.
On NVIDIA A100 (HBM2e, 40\,MB unified L2), improvements up to 80\% are
observed, while on NVIDIA RTX 6000 Ada (GDDR6, \diffCR{96}\,MB L2),
improvements remain below 15\%.
\diffCR{This indicates that reordering is most beneficial on
high-bandwidth-memory architectures: the A100's HBM2e provides
2{,}039\,GB/s of DRAM bandwidth, offering substantial headroom for
coalescing-driven improvements, whereas the RTX 6000 Ada's GDDR6 at
960\,GB/s limits the achievable gains.}
See Supp. \Cref{app:full-reordering} for the detailed evaluation.

\subsection{Limitation}
This study focuses on graph-based indices on GPU, and does
not cover other GPU-accelerable index families (such as IVF or
hash-based methods) or CPU execution.
\diff{We target GPU because vertex layout matters most
there: GPU throughput is dominated by warp-level
coalesced memory transactions and is far more bandwidth-bound than
CPU, where prefetching and large caches partly mask layout effects.}
Extending the framework to CPU execution and non-graph indices, and
amortizing reordering cost under frequent index updates, remain
natural future directions.

\section{Conclusion}
\textsc{Plasma} is a unified evaluation framework that decouples graph
topology from memory layout, enabling the first systematic analysis of
graph reordering for GPU-based ANNS.
Our experiments show that memory layout can dominate search throughput more
than topology: reordering yields up to 80\% QPS improvement (typically
10--30\%) while preserving accuracy, and CPU-oriented indices match or
surpass GPU-oriented ones once layout is optimized.
Profiling attributes this to GPU search being memory-bandwidth-limited,
where reordering improves DRAM bandwidth utilization.

\begin{acks}
We thank Prof.\ Hiroaki Shiokawa for his valuable comments and insightful discussions.
This work was supported by JST, Japan (AIP Acceleration Research JPMJCR23U2, PRESTO JPMJPR2518).
\end{acks}

\clearpage

\bibliographystyle{ACM-Reference-Format}
\bibliography{main}

\appendix

\section{Experimental Setup (Details)}
\label{app:exp-setup-details}
All experiments run on a single GPU.
Primary experiments use an AMD EPYC 7742 (64-core) CPU and an
NVIDIA A100 (80\,GB HBM2e) GPU.
Additional experiments use an NVIDIA RTX 6000 Ada (48\,GB GDDR6) GPU.
The software stack is \texttt{CUDA}~12.0, \texttt{cuVS}~25.08.00, and \texttt{Faiss}~1.10.0.
Hardware metrics are measured with NVIDIA Nsight Compute~2024.1.1.

\section{Search Algorithm Pseudocode}
\label{app:search-algorithm}
Our unified search engine executes any graph-based index under a
single algorithm, enabling fair comparison across indices with
different construction strategies.
Algorithm~\ref{alg:graph_search} presents the search procedure.
The runtime parameter $L$ (Queue Size Limit) controls the maximum
number of candidate nodes maintained in the search queue during
traversal.
Larger $L$ increases search accuracy at the cost of visiting more
vertices, directly affecting which vertices are accessed and in what
order. We vary $L$ across $16$ values to plot performance curves.
\begin{algorithm}[h]
  \caption{\textsc{Search on Graph-based Index}}
  \label{alg:graph_search}
  \begin{algorithmic}[1]
    \STATE {\bfseries Input:} Graph-based Index $G(\mathcal{V},
    \mathcal{E})$, Entry Point $v_s \in \mathcal{V}$, Query $\bm{q}
    \in \mathbb{R}^d$, Queue Size Limit $L \in \mathbb{Z}$
    \STATE {\bfseries Output:} Approximate Nearest Node to $\bm{q}$
    \STATE Initialize search queue $Q \gets \{v_s\}$
    \STATE Initialize visited set $T \gets \{v_s\}$
    \STATE $updated \gets \text{true}$
    \WHILE{$updated$}
    \STATE $updated \gets \text{false}$
    \STATE $u \gets Q$.popNearestNode()
    \FOR{$v \in G$.getNeighbors($u$)}
    \IF{$v \in T$}
    \STATE \textbf{continue}
    \ENDIF
    \STATE $Q$.enqueue($v$)
    \STATE $T \gets T \cup \{v\}$
    \STATE $updated \gets \text{true}$
    \IF{$|Q| > L$}
    \STATE $Q$.popFarthestNode()
    \ENDIF
    \ENDFOR
    \ENDWHILE
  \end{algorithmic}
\end{algorithm}

\section{Index Construction Hyperparameters}
\label{app:memory-layout}

All indices use maximum degree $K=32$.
CAGRA is constructed natively on GPU with intermediate graph degree~128.
NSG, Vamana, and NN-Descent are built on CPU and converted to the unified GPU format.
NSG uses NN-Descent for initial $k$-NN graph construction ($GK=32$, $S=10$, $R=100$, $L=GK{+}50$, iter$=10$), followed by final graph pruning to $K=32$.
Vamana uses $\alpha=1.2$.
NN-Descent uses $S=10$, $R=100$, $L=GK{+}50$, iter$=10$.
For each query batch, only the search queue length $L$ is varied; all other parameters remain fixed.

\section{GPU Architecture and Reordering Effect}
\label{app:gpu-architecture}

Table~\ref{tab:supp-gpu-specs} summarizes the key hardware
specifications of the two GPUs used in our experiments, extracted from
the official NVIDIA datasheets with a focus on elements impactful to
the proposed framework.
The highlighted rows in Table~\ref{tab:supp-gpu-specs} capture the memory characteristics most relevant to graph-based ANNS.
GPUs employ a three-level memory hierarchy.
Global Memory (DRAM) provides large capacity but high latency.
L2 cache serves as a unified intermediate layer shared across all SMs.
L1 cache and Shared Memory are architecturally unified, allowing flexible allocation between caching and per-SM scratchpad usage.

\begin{table}[t]
  \caption{Hardware specifications of the GPUs used in our experiments,
  extracted from the official NVIDIA datasheets.
  \myhl{tblFullBg!70}{Highlighted values} indicate metrics most relevant to
  the proposed framework.}
  \label{tab:supp-gpu-specs}
  \centering
  \small
  \begin{tabular}{@{}l|c|c@{}}
    \toprule[1pt]
    NVIDIA GPU & A100 & RTX 6000 Ada \\
    \midrule
    Architecture        & GA100 (\texttt{Ampere})       & AD102 (\texttt{Ada Lovelace}) \\
    Process             & TSMC 7\,nm           & TSMC 4N \\
    SMs                 & 108                  & 142 \\
    CUDA Cores          & 6{,}912              & 18{,}176 \\
    \midrule
    L1 / Shared Memory  & 20.25\,MB            & 17.75\,MB \\
    L2 Cache (total)    & \myhl{tblFullBg!70}{40\,MB}  & \myhl{tblFullBg!70}{96\,MB} \\
    \midrule
    DRAM                & \myhl{tblFullBg!70}{80\,GB}   & \myhl{tblFullBg!70}{48\,GB} \\
    Memory Type         & \myhl{tblFullBg!70}{HBM2e}    & \myhl{tblFullBg!70}{GDDR6} \\
    Memory Bandwidth    & \myhl{tblFullBg!70}{2{,}039\,GB/s} & \myhl{tblFullBg!70}{960\,GB/s} \\
    Memory Bus          & \myhl{tblFullBg!70}{5{,}120-bit}         & \myhl{tblFullBg!70}{384-bit} \\
    \midrule
    TDP                 & 400\,W               & 300\,W \\
    \bottomrule[1pt]
  \end{tabular}
\end{table}

In graph-based ANNS, vector datasets far exceed cache capacity (e.g., Deep 1M: 1M vectors $\times$ 96 dimensions $\times$ 4\,bytes $\approx$ 384\,MB, far exceeding the A100's 40\,MB L2 cache), making the workload inherently DRAM-bandwidth-bound.
Vertex reordering places frequently co-accessed vertices in adjacent memory locations, increasing the 
degree of coalescing in memory accesses within a warp and thereby improving effective DRAM bandwidth utilization.
Consequently, GPUs with higher memory bandwidth benefit more from reordering: the A100's HBM2e (2{,}039\,GB/s) provides greater headroom than the RTX~6000~Ada's GDDR6 (960\,GB/s), as HBM's architecture more effectively exploits improvements in DRAM row buffer hit rates and TLB hit rates~\cite{HBM-FPGA-benchmark}.

\section{Baseline Evaluation Results on Different GPU Architecture}
\label{app:full-baseline}

\begin{figure*}[t]
  \centering
  \setlength{\tabcolsep}{2pt}
  \renewcommand{\arraystretch}{0}
  \begin{tabular}{ccc}
    \subcaptionbox{SIFT
    1M}[0.32\textwidth]{\includegraphics[width=0.32\textwidth]{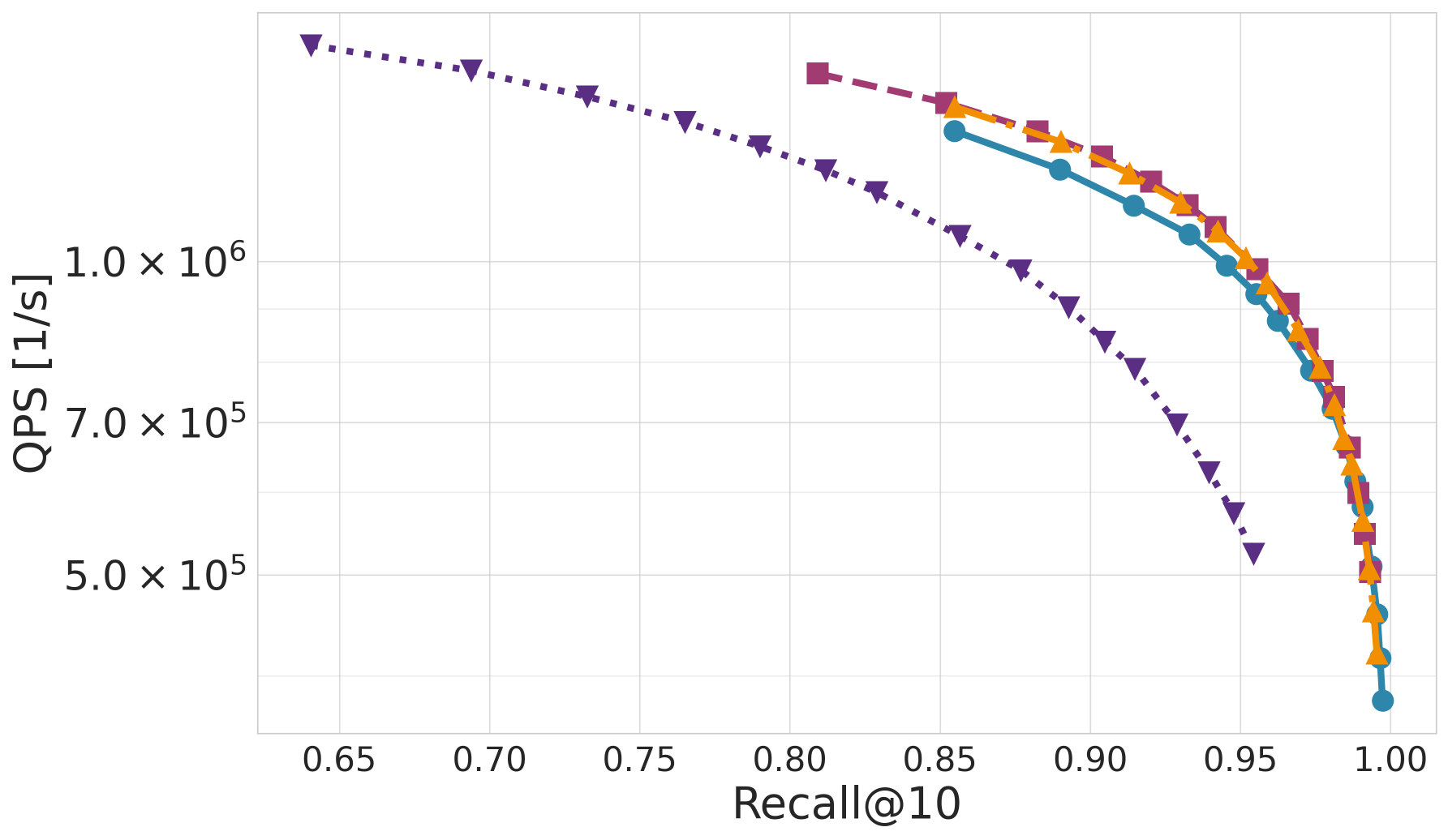}}
    &
    \subcaptionbox{GIST
    1M}[0.32\textwidth]{\includegraphics[width=0.32\textwidth]{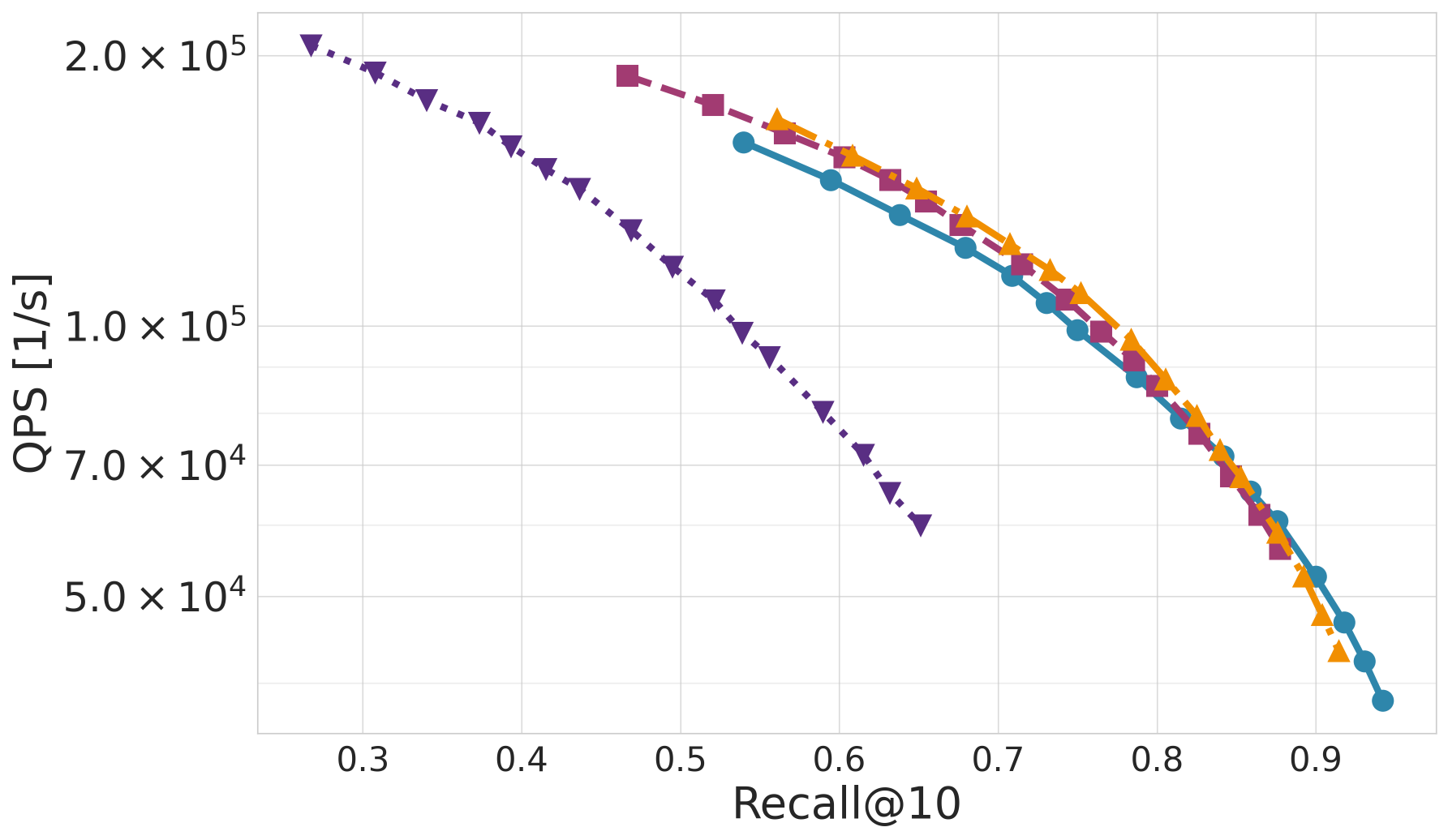}}
    &
    \subcaptionbox{Deep
    1M}[0.32\textwidth]{\includegraphics[width=0.32\textwidth]{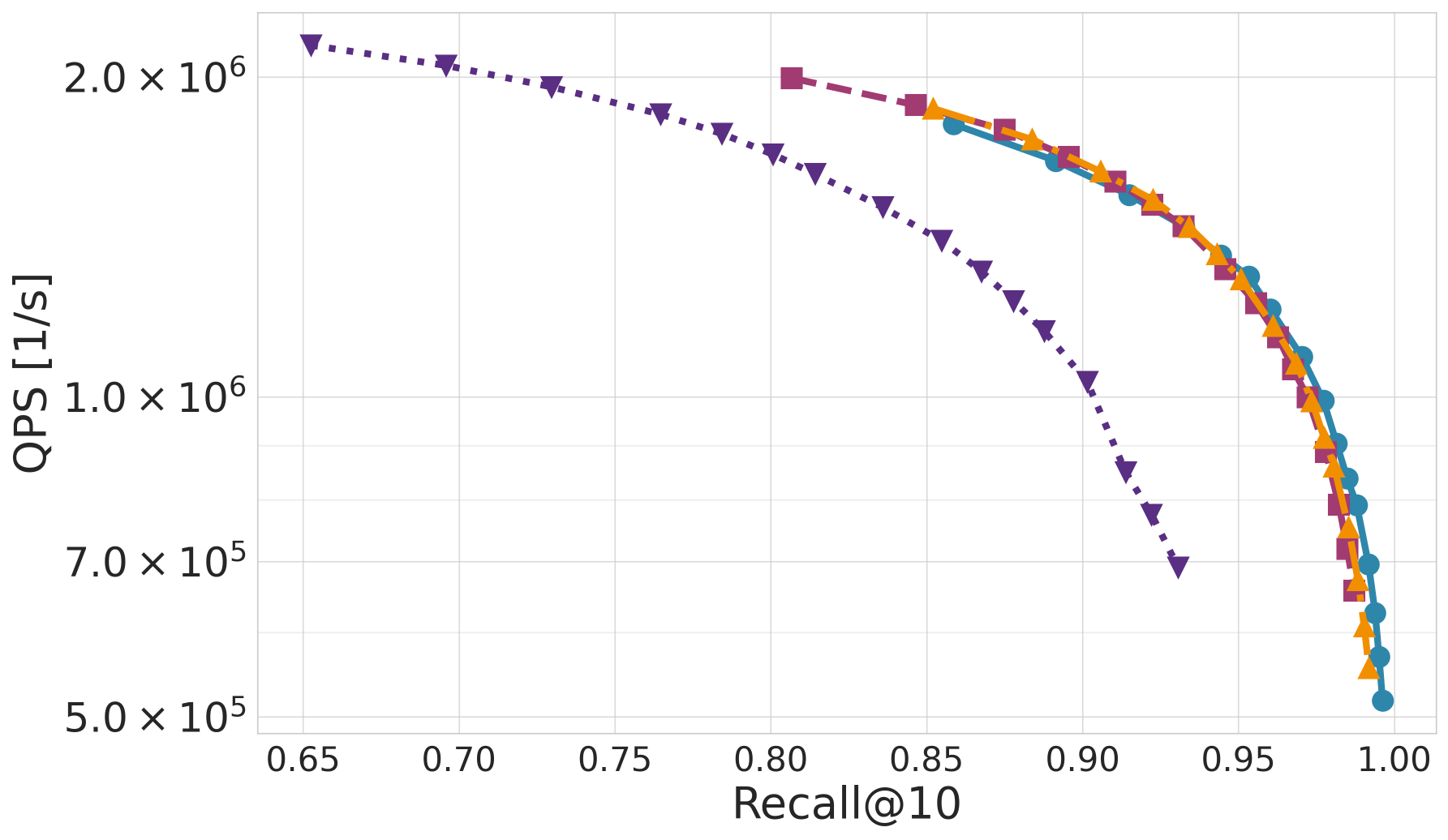}}
    \\
    \subcaptionbox{Deep
    10M}[0.32\textwidth]{\includegraphics[width=0.32\textwidth]{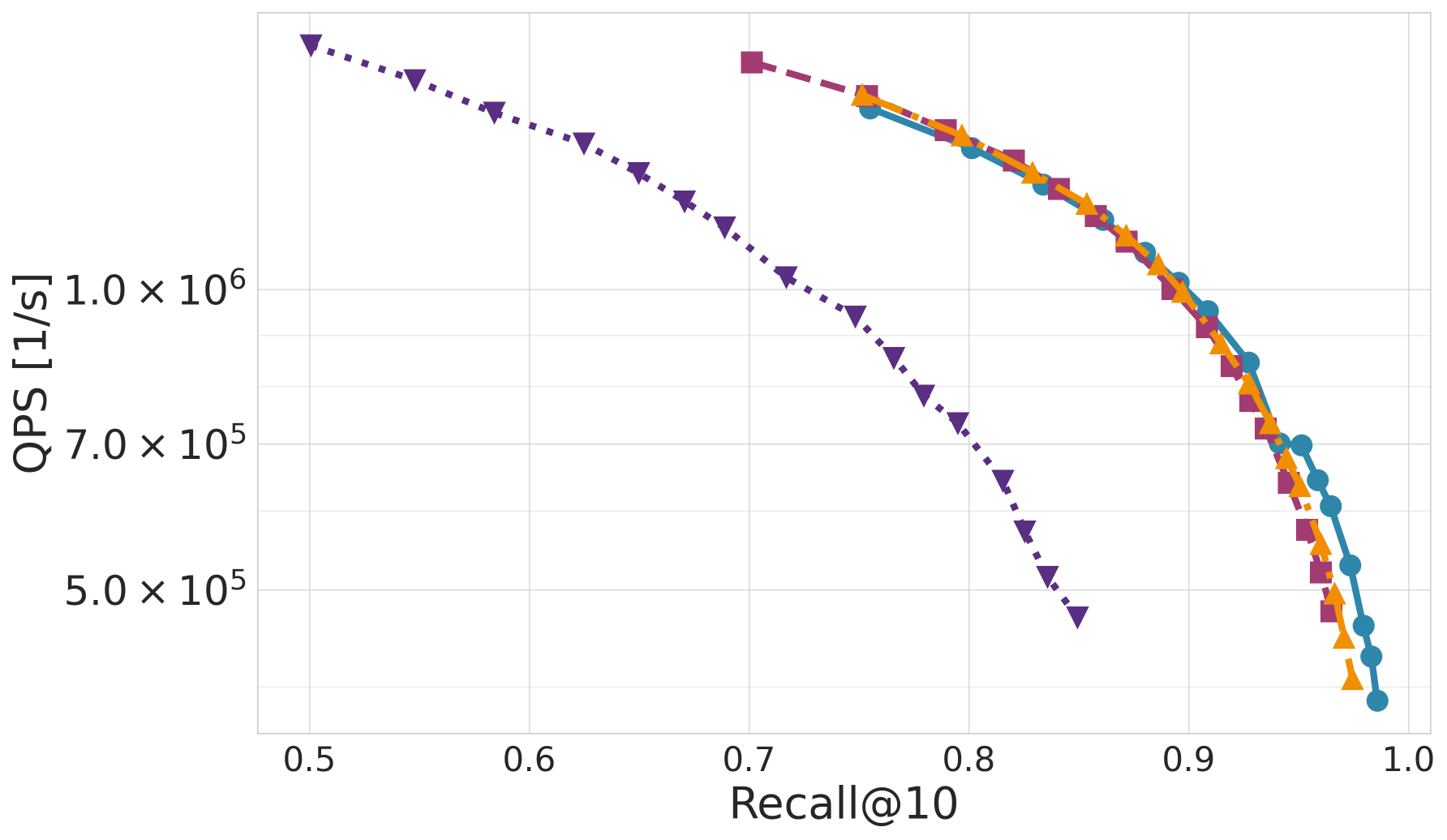}}
    &
    \subcaptionbox{Deep
    40M}[0.32\textwidth]{\includegraphics[width=0.32\textwidth]{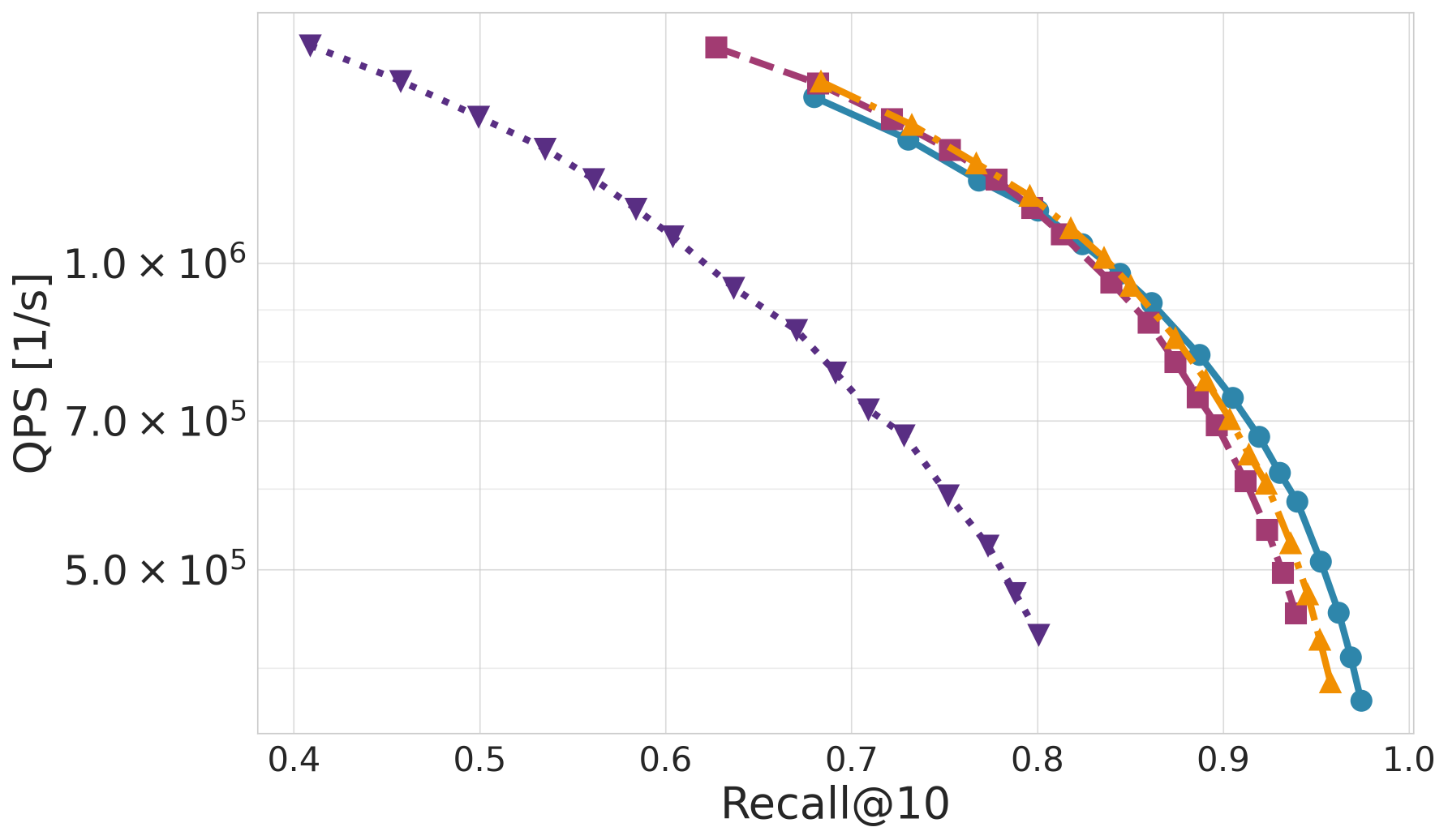}}
    &
    \subcaptionbox{Yandex T2I
    1M}[0.32\textwidth]{\includegraphics[width=0.32\textwidth]{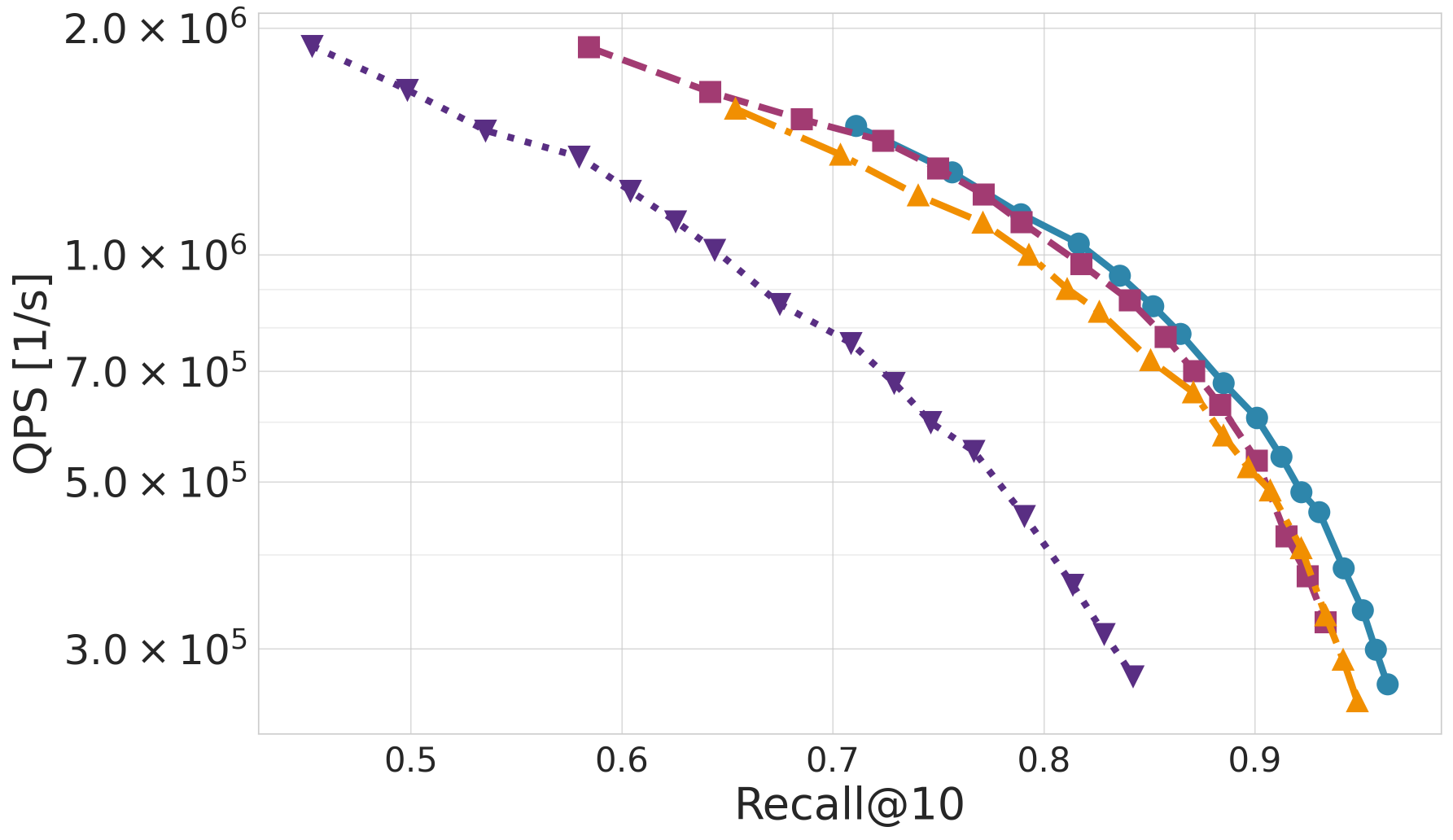}}
    \\
    \subcaptionbox{OpenAI
    Embed.\ 1M}[0.32\textwidth]{\includegraphics[width=0.32\textwidth]{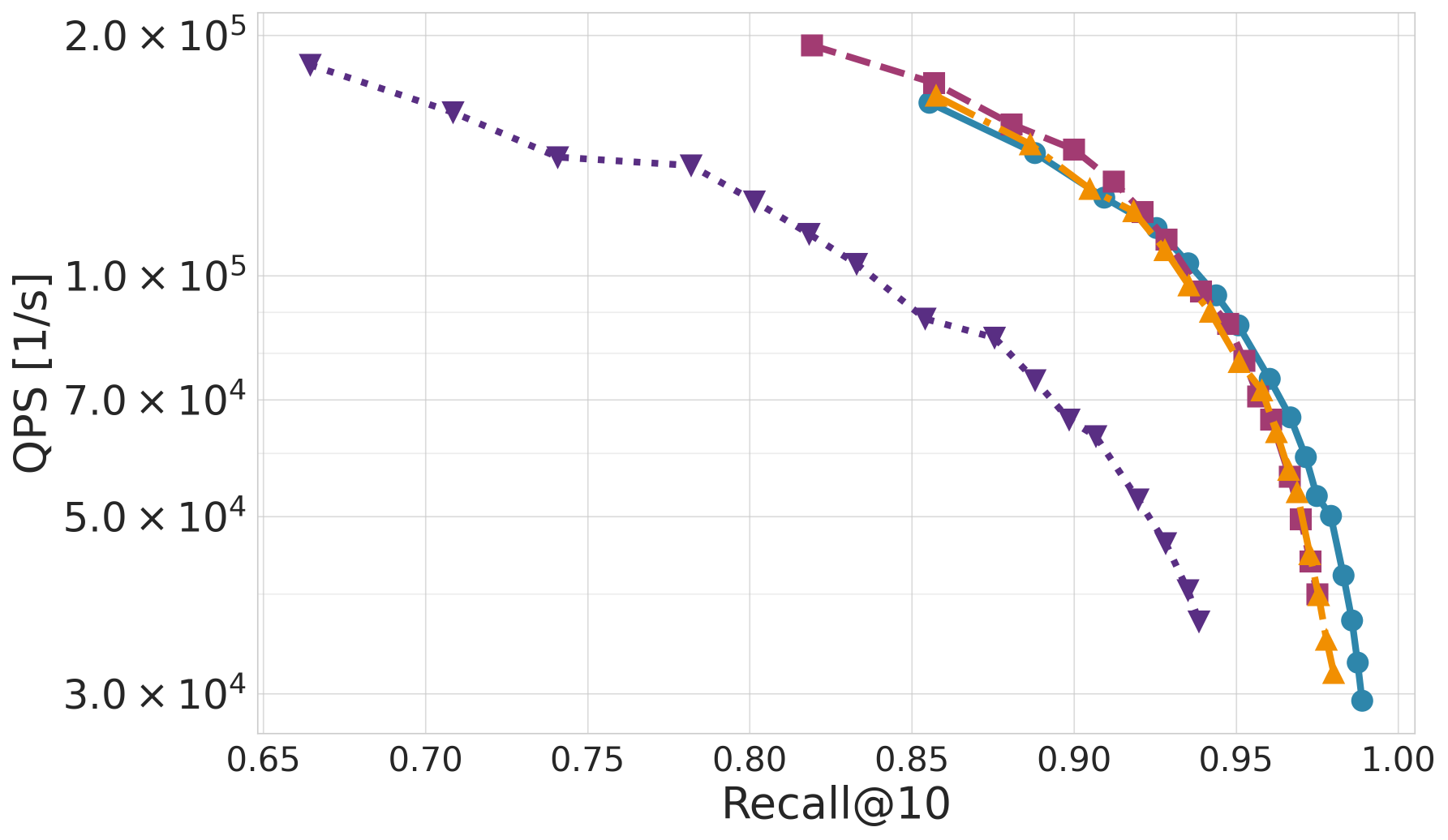}}
    &
    \subcaptionbox{Wikipedia
    1M}[0.32\textwidth]{\includegraphics[width=0.32\textwidth]{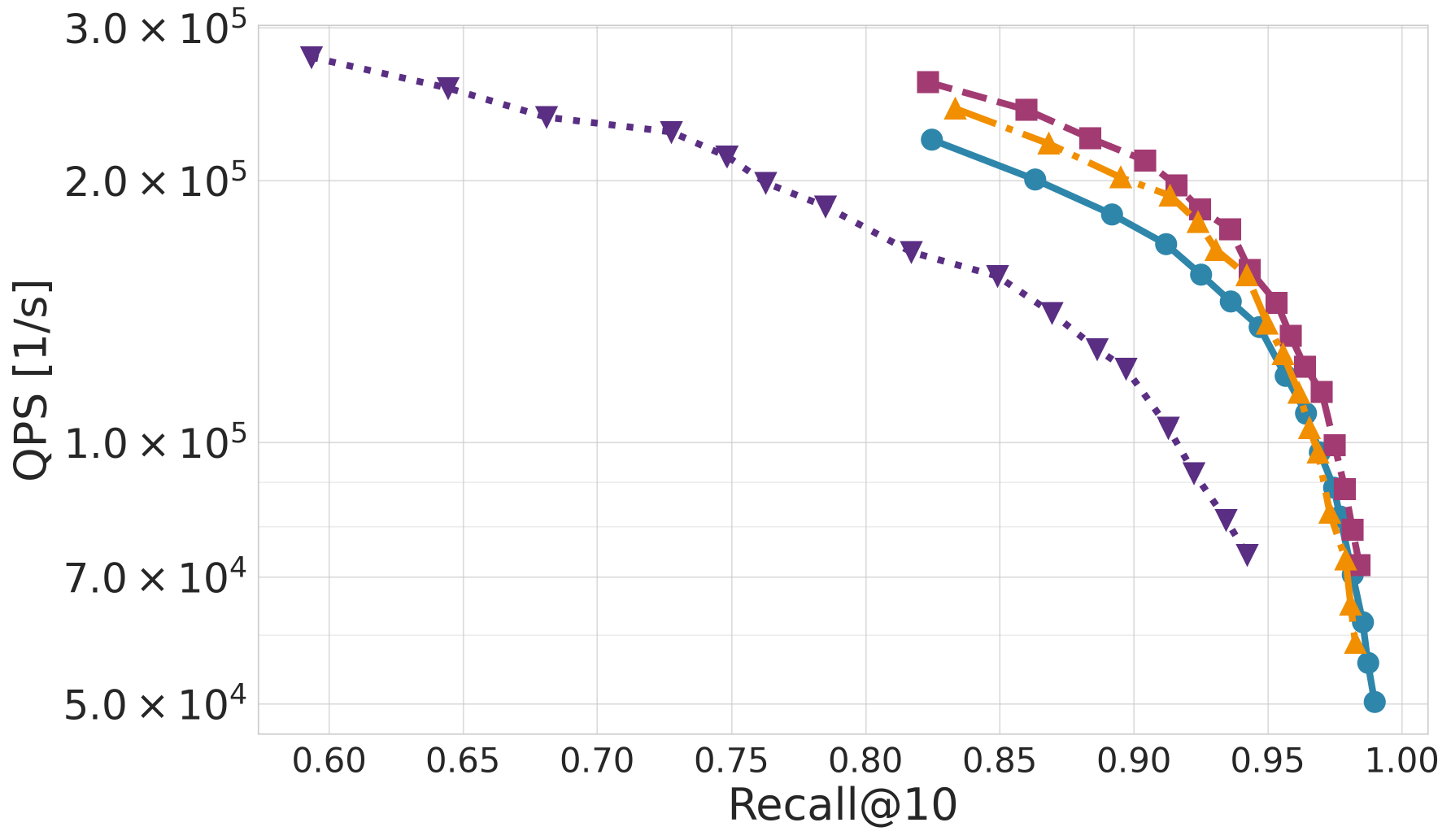}}
    &
    \subcaptionbox{Wikipedia
    10M}[0.32\textwidth]{\includegraphics[width=0.32\textwidth]{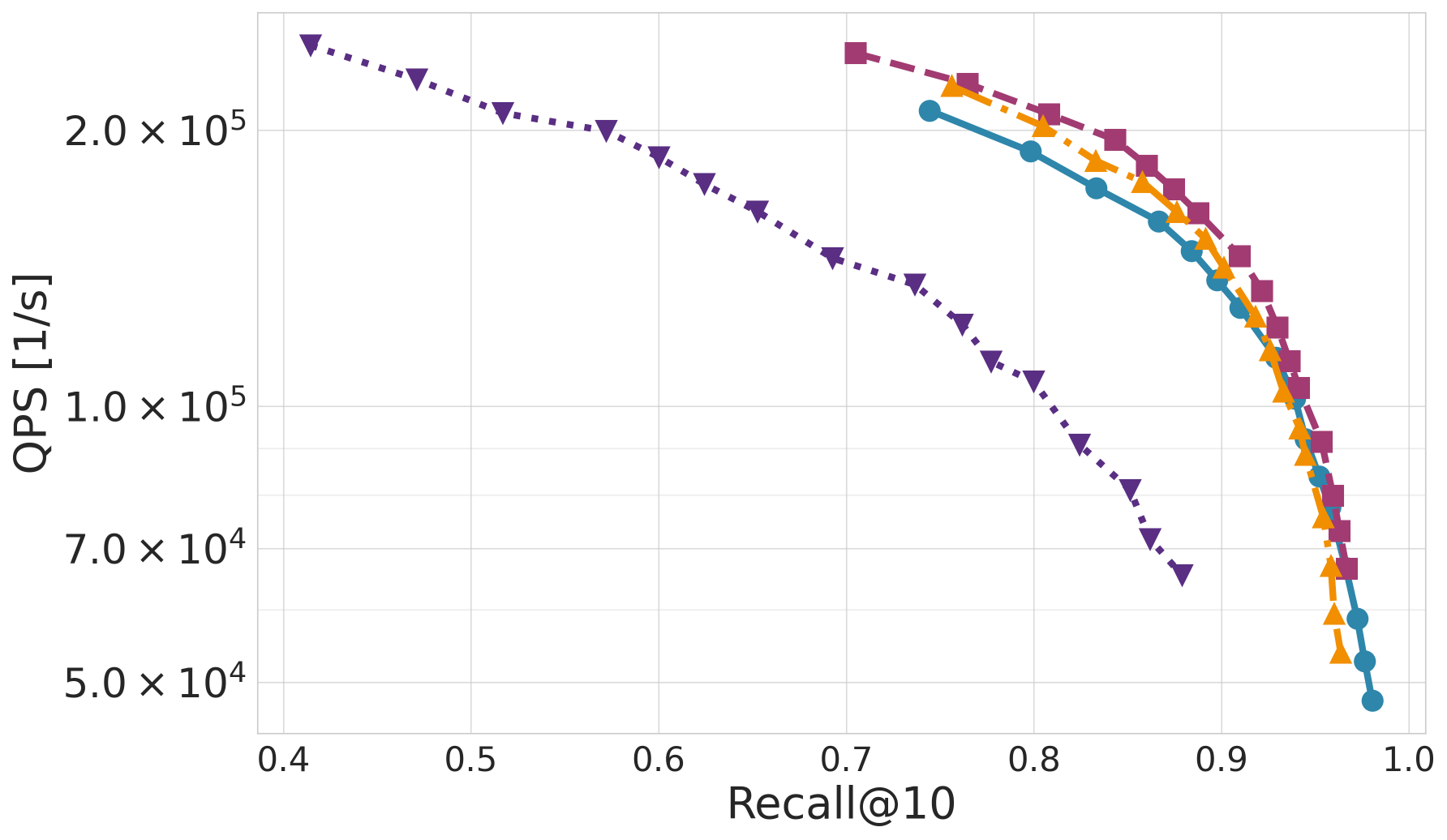}}
    \\
    \subcaptionbox{BioASQ
    1M}[0.32\textwidth]{\includegraphics[width=0.32\textwidth]{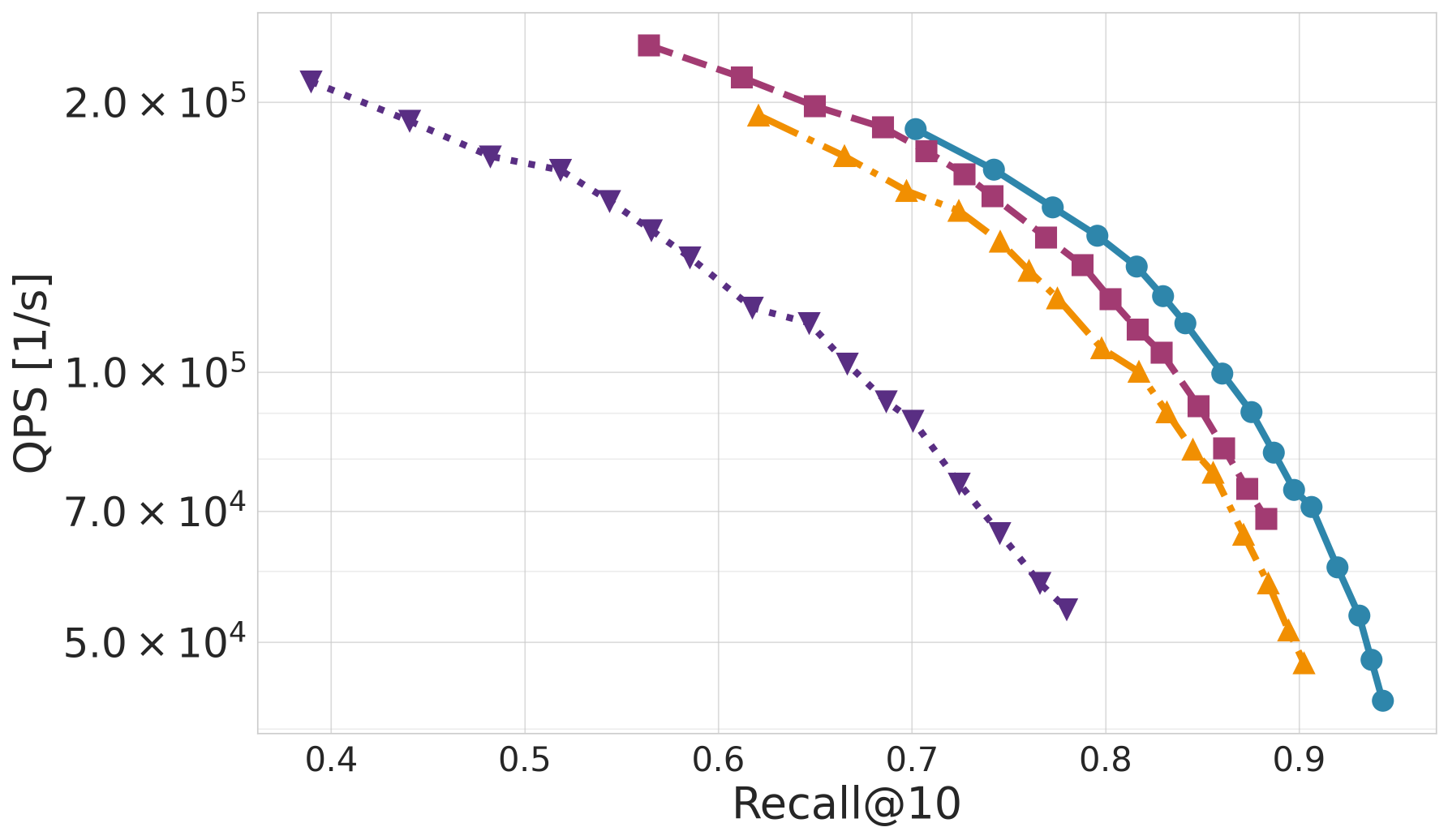}}
    &
    \subcaptionbox{BioASQ
    10M}[0.32\textwidth]{\includegraphics[width=0.32\textwidth]{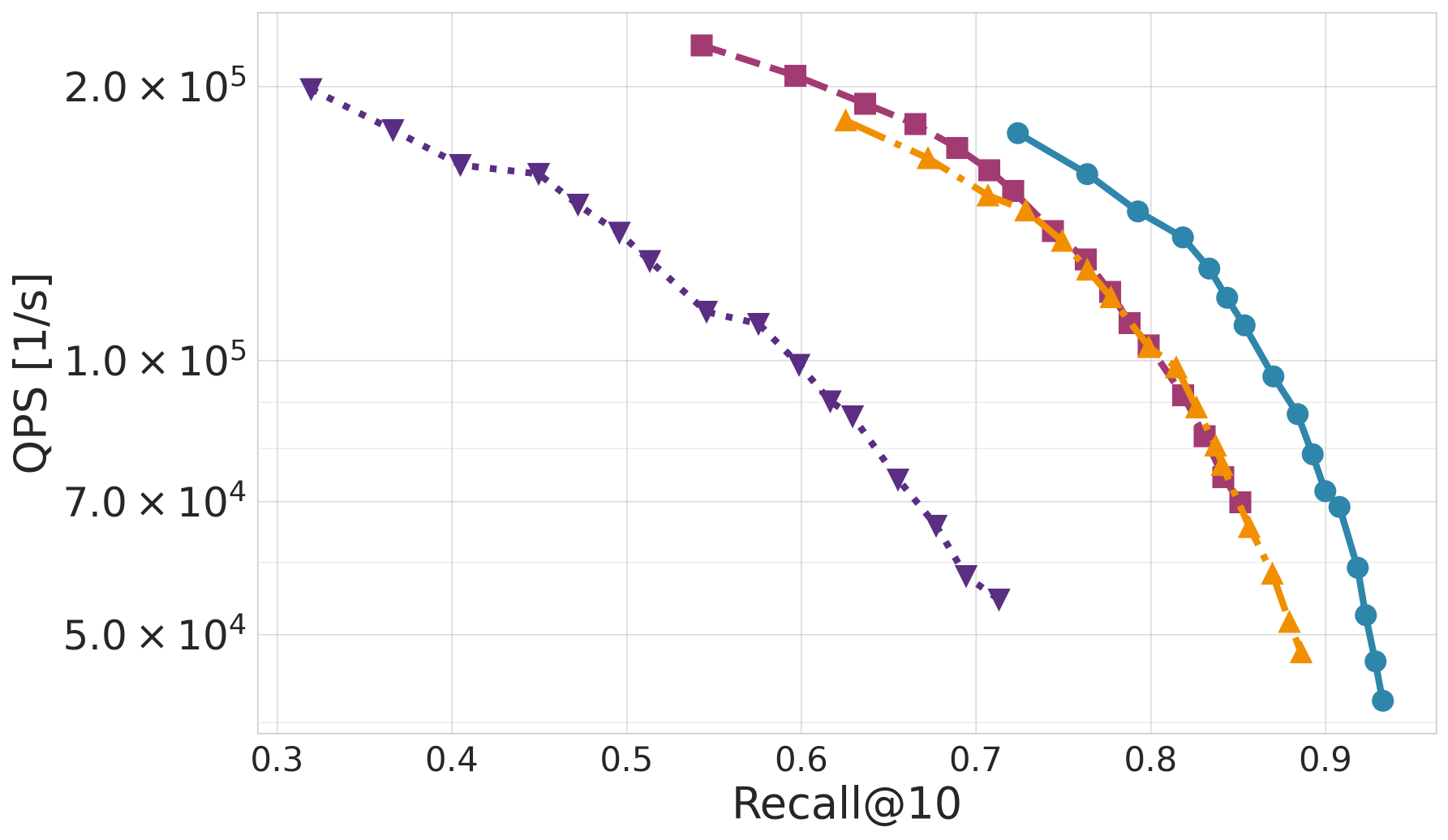}}
    &
    \subcaptionbox{C4
    5M}[0.32\textwidth]{\includegraphics[width=0.32\textwidth]{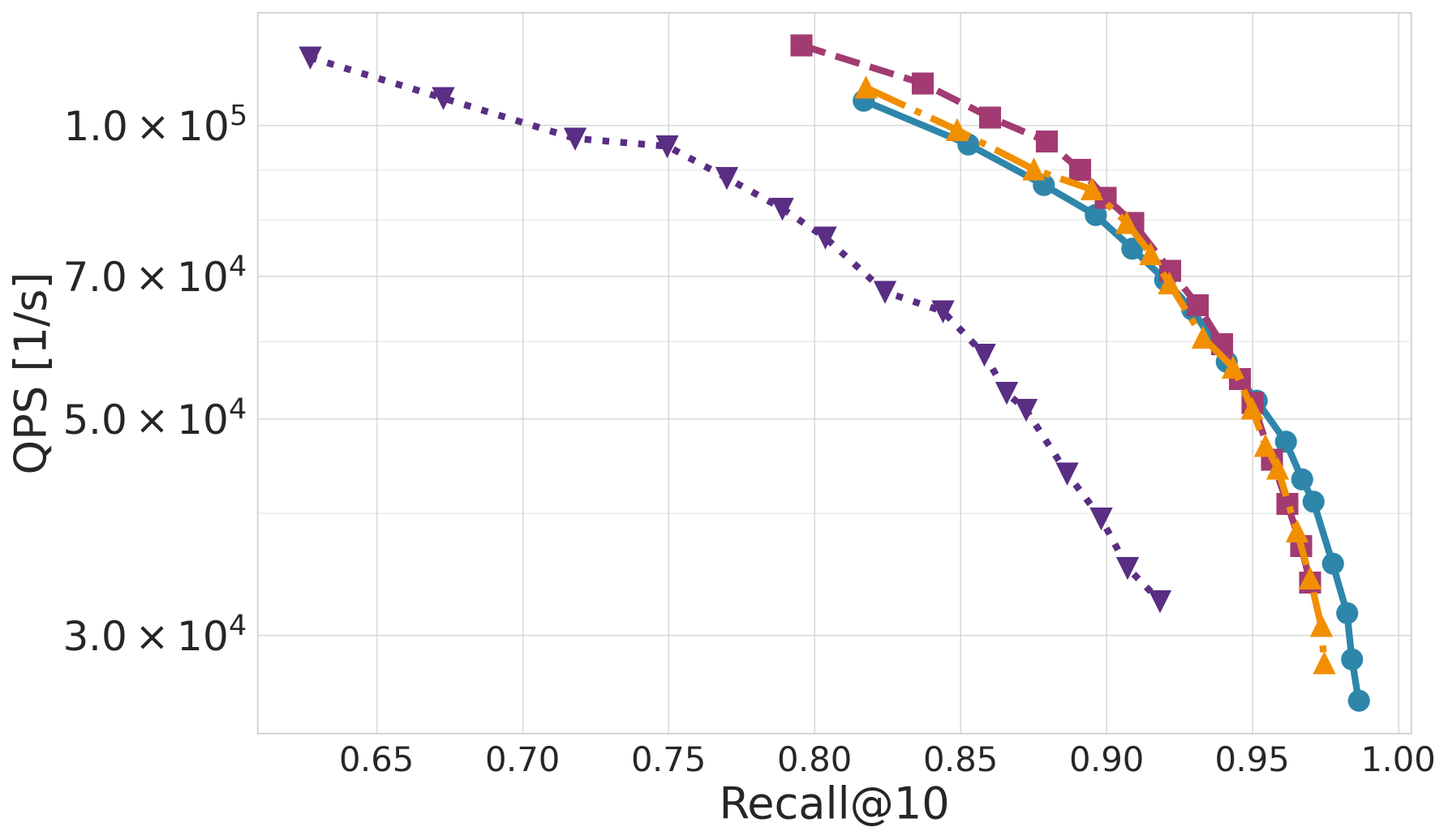}}
  \end{tabular}
  \vspace{0.5em}
  \begin{tabular}{c}
    \includegraphics[width=0.7\textwidth]{fig/baseline_algorithms_legend.pdf}
  \end{tabular}
  \caption{Full Recall-QPS trade-off curves for four graph-based indices
    (CAGRA, NSG, Vamana, NN-Descent) across all 12 datasets on NVIDIA
  RTX 6000 Ada. Upper-right is better.}
  \label{fig:supp-base-eval-A6000}
\end{figure*}

Figure~\ref{fig:supp-base-eval-A6000} shows the full baseline evaluation
results across all 12 datasets and four graph-based indices on NVIDIA
RTX 6000 Ada.

\section{Full Reordering Evaluation Results}
\label{app:full-reordering}
\begin{figure*}[h]
  \centering
  \setlength{\tabcolsep}{2pt}
  \renewcommand{\arraystretch}{0}
  \begin{tabular}{cccc}
    \subcaptionbox{CAGRA, Deep
    1M}[0.24\linewidth]{\includegraphics[width=\linewidth]{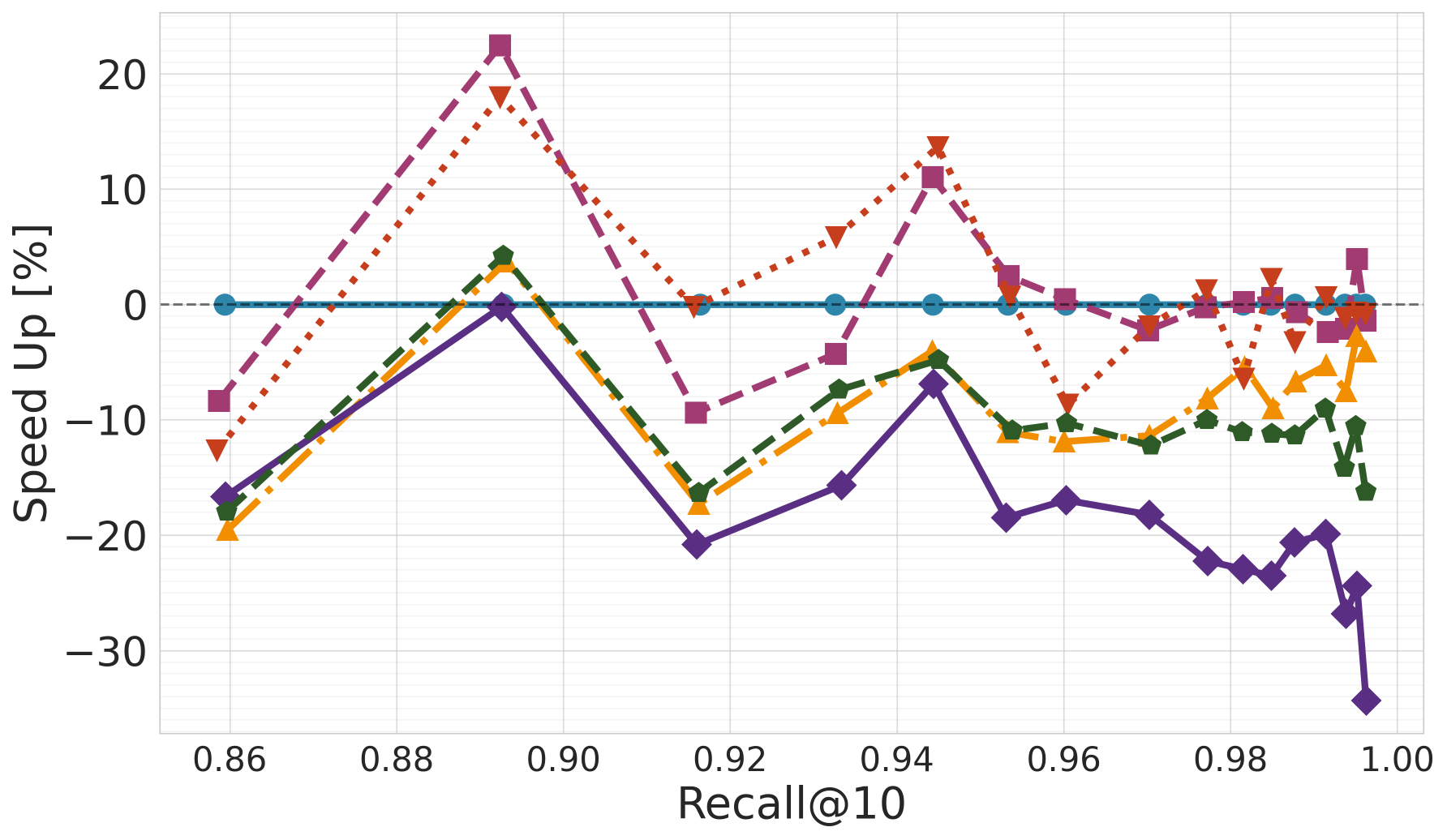}}
    &
    \subcaptionbox{NSG, Deep
    1M}[0.24\linewidth]{\includegraphics[width=\linewidth]{fig/A100-result/recall_qps_comparison_K32_gpu_deep1m-96-euclidean_nsg_reordering.pdf}}
    &
    \subcaptionbox{Vamana, Deep
    1M}[0.24\linewidth]{\includegraphics[width=\linewidth]{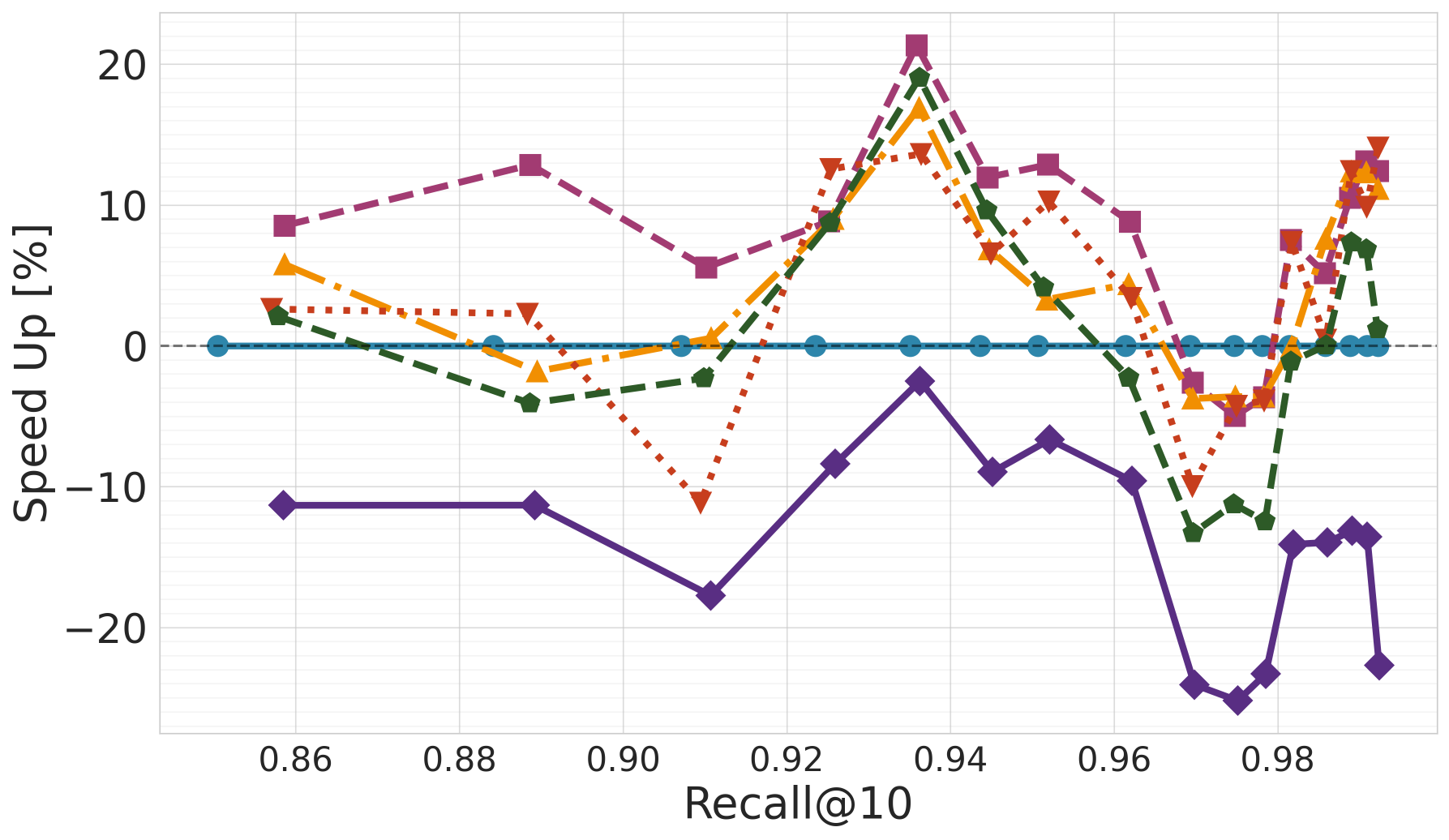}}
    &
    \subcaptionbox{NN-Descent, Deep
    1M}[0.24\linewidth]{\includegraphics[width=\linewidth]{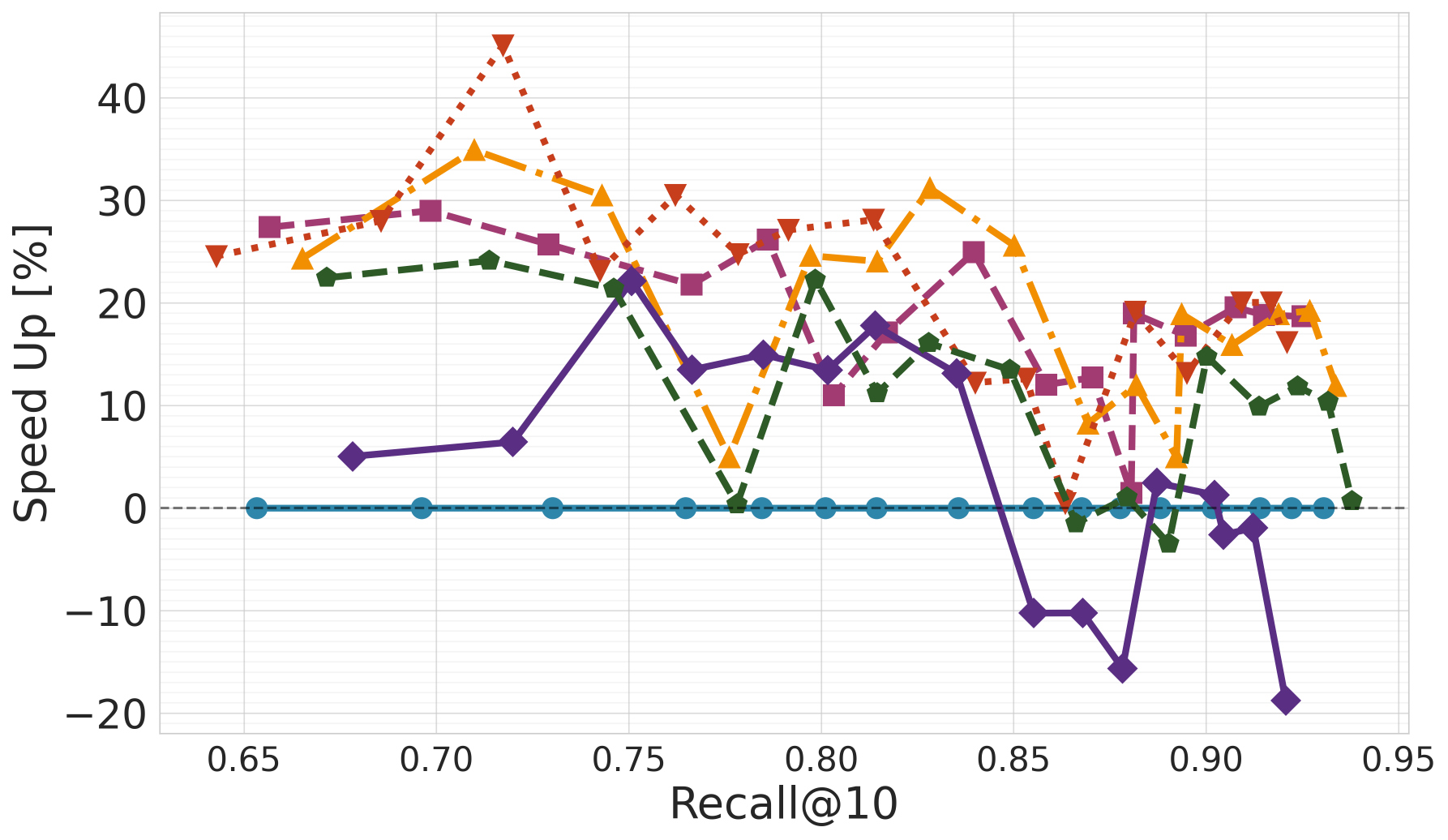}}
    \\
    \subcaptionbox{CAGRA, SIFT
    1M}[0.24\linewidth]{\includegraphics[width=\linewidth]{fig/A100-result/recall_qps_comparison_K32_gpu_sift-128-euclidean_cagra_reordering.pdf}}
    &
    \subcaptionbox{NSG, SIFT
    1M}[0.24\linewidth]{\includegraphics[width=\linewidth]{fig/A100-result/recall_qps_comparison_K32_gpu_sift-128-euclidean_nsg_reordering.pdf}}
    &
    \subcaptionbox{Vamana, SIFT
    1M}[0.24\linewidth]{\includegraphics[width=\linewidth]{fig/A100-result/recall_qps_comparison_K32_gpu_sift-128-euclidean_diskann_reordering.pdf}}
    &
    \subcaptionbox{NN-Descent, SIFT
    1M}[0.24\linewidth]{\includegraphics[width=\linewidth]{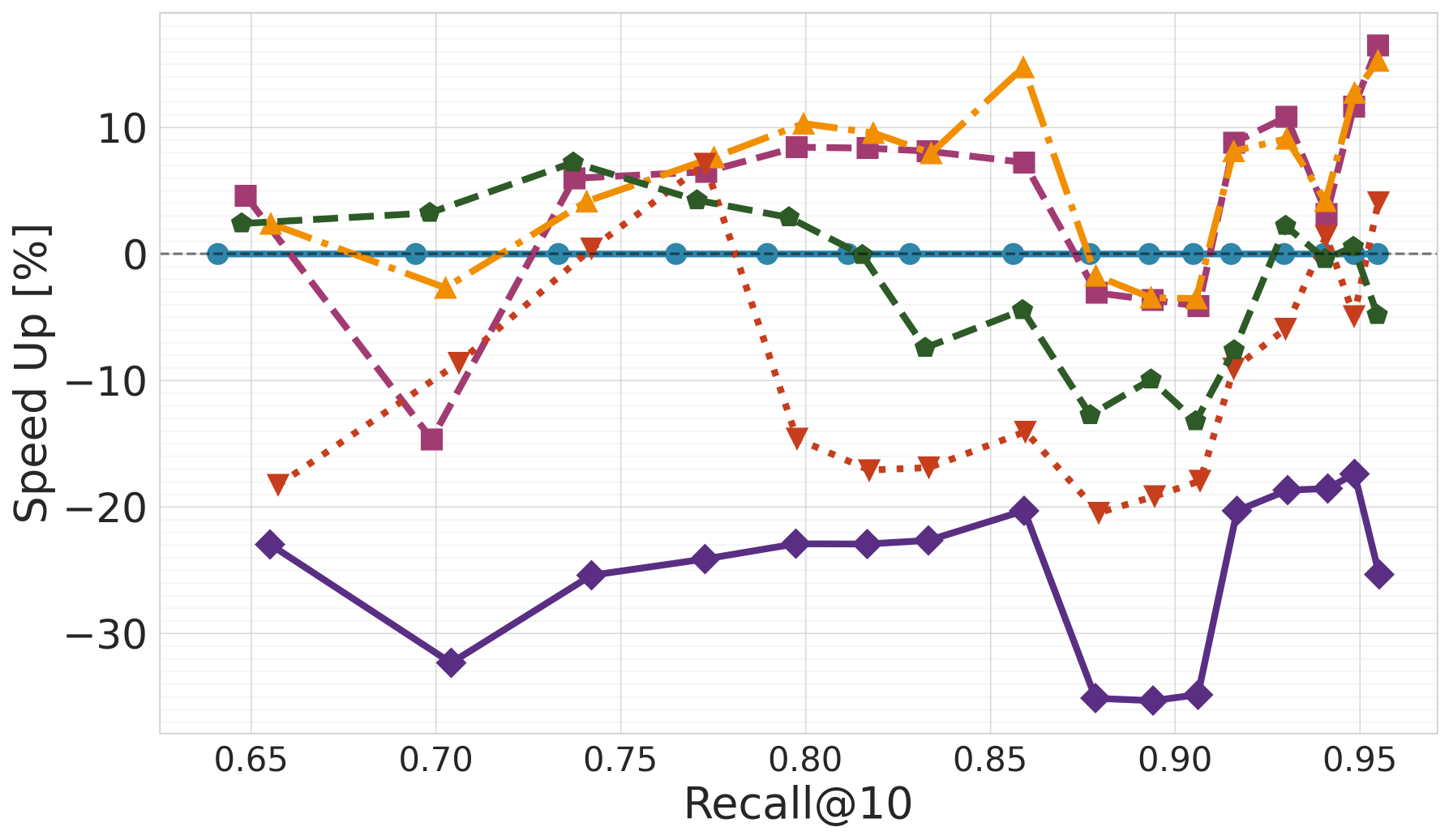}}
    \\
    \subcaptionbox{CAGRA, GIST
    1M}[0.24\linewidth]{\includegraphics[width=\linewidth]{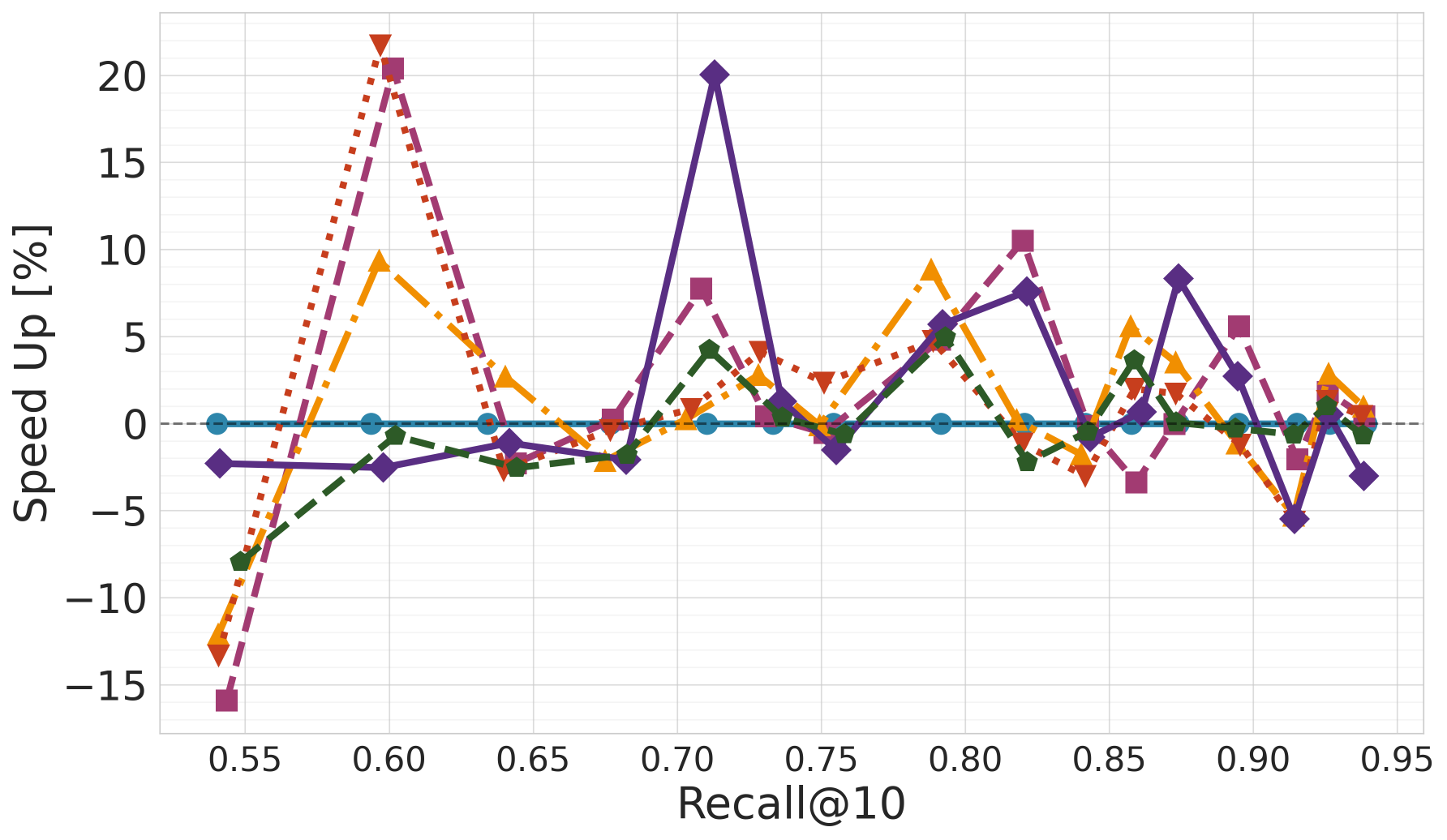}}
    &
    \subcaptionbox{NSG, GIST
    1M}[0.24\linewidth]{\includegraphics[width=\linewidth]{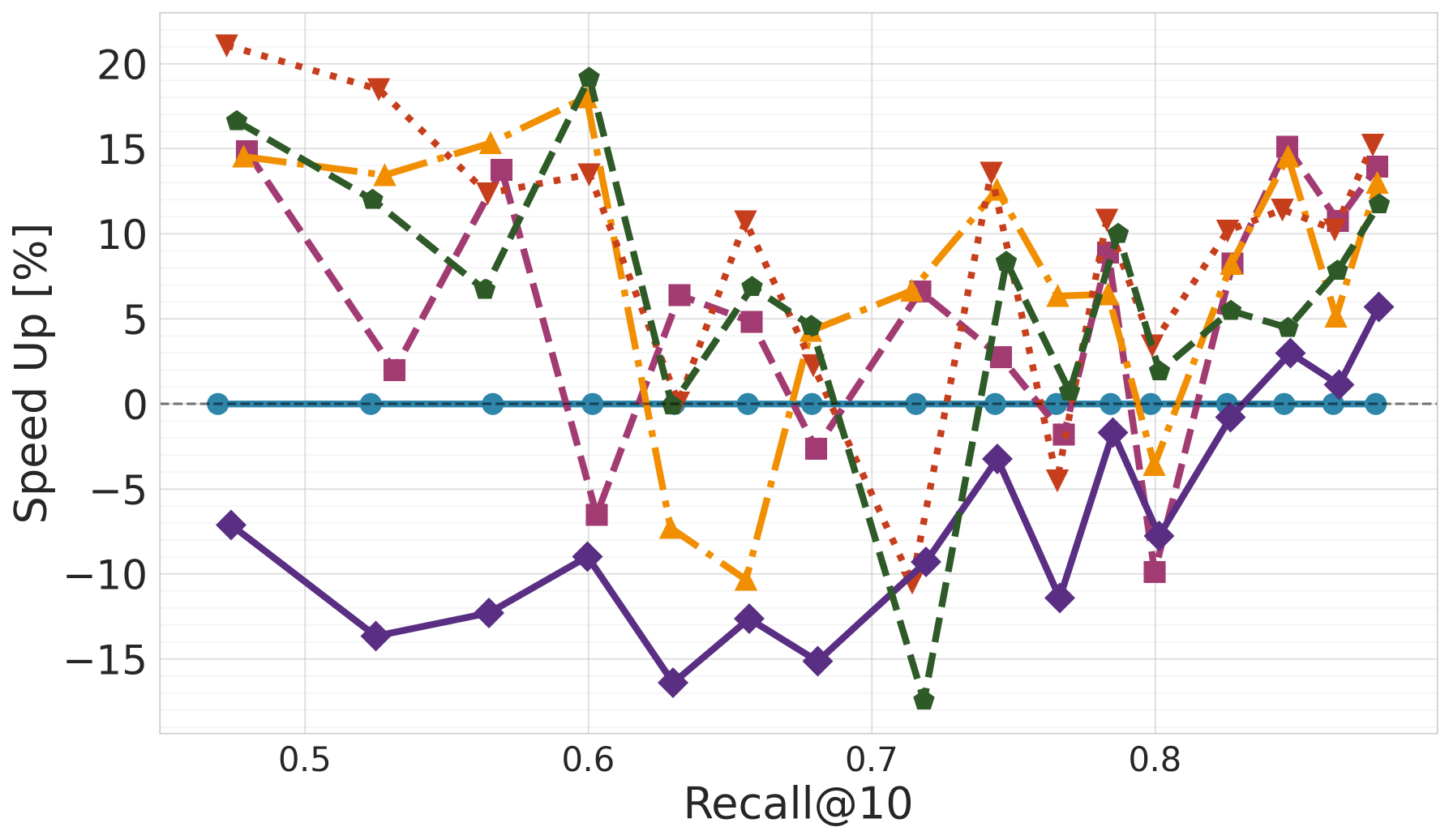}}
    &
    \subcaptionbox{Vamana, GIST
    1M}[0.24\linewidth]{\includegraphics[width=\linewidth]{fig/A100-result/recall_qps_comparison_K32_gpu_gist-960-euclidean_diskann_reordering.pdf}}
    &
    \subcaptionbox{NN-Descent, GIST
    1M}[0.24\linewidth]{\includegraphics[width=\linewidth]{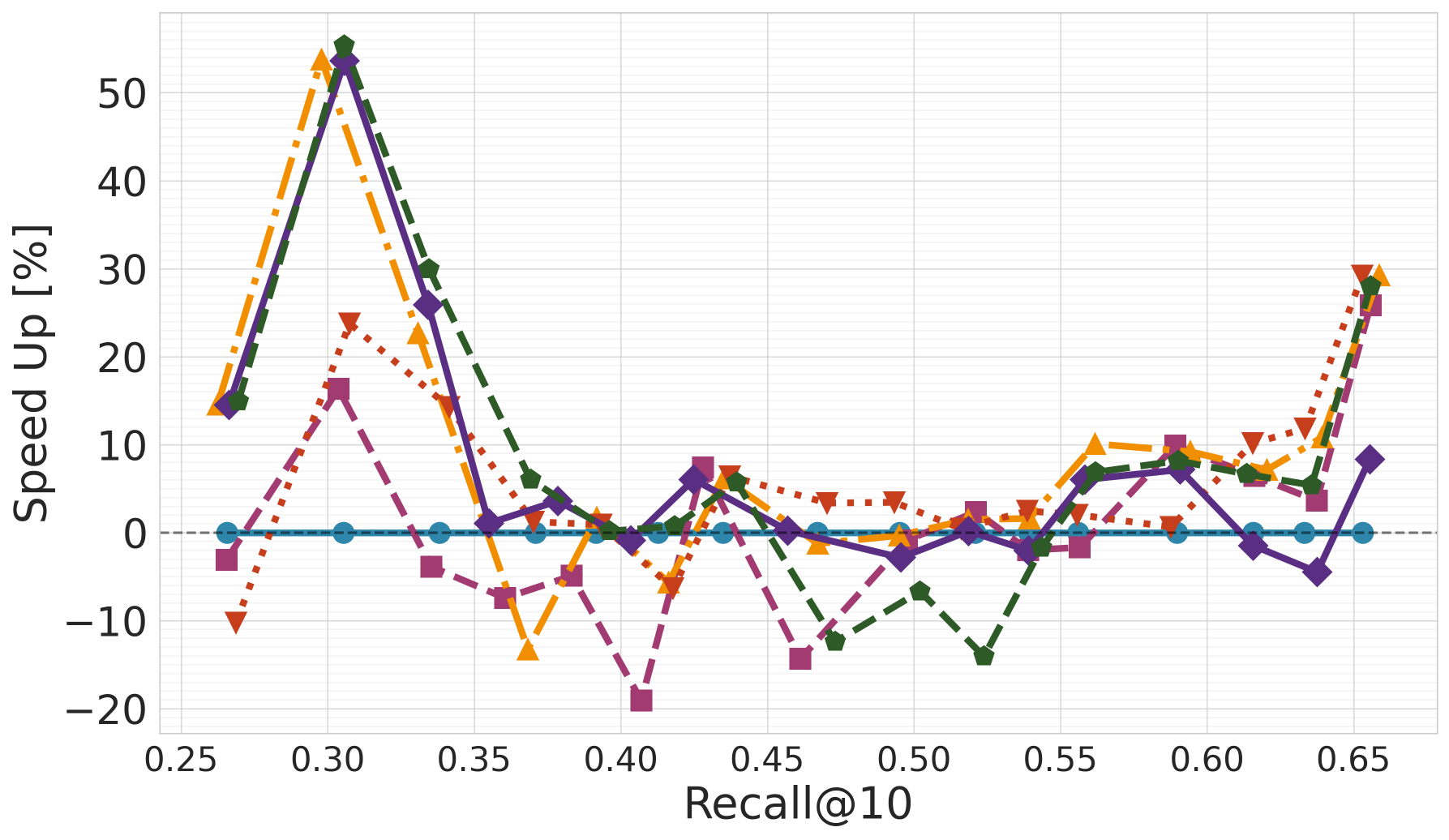}}
    \\
    \subcaptionbox{CAGRA, OpenAI
    1M}[0.24\linewidth]{\includegraphics[width=\linewidth]{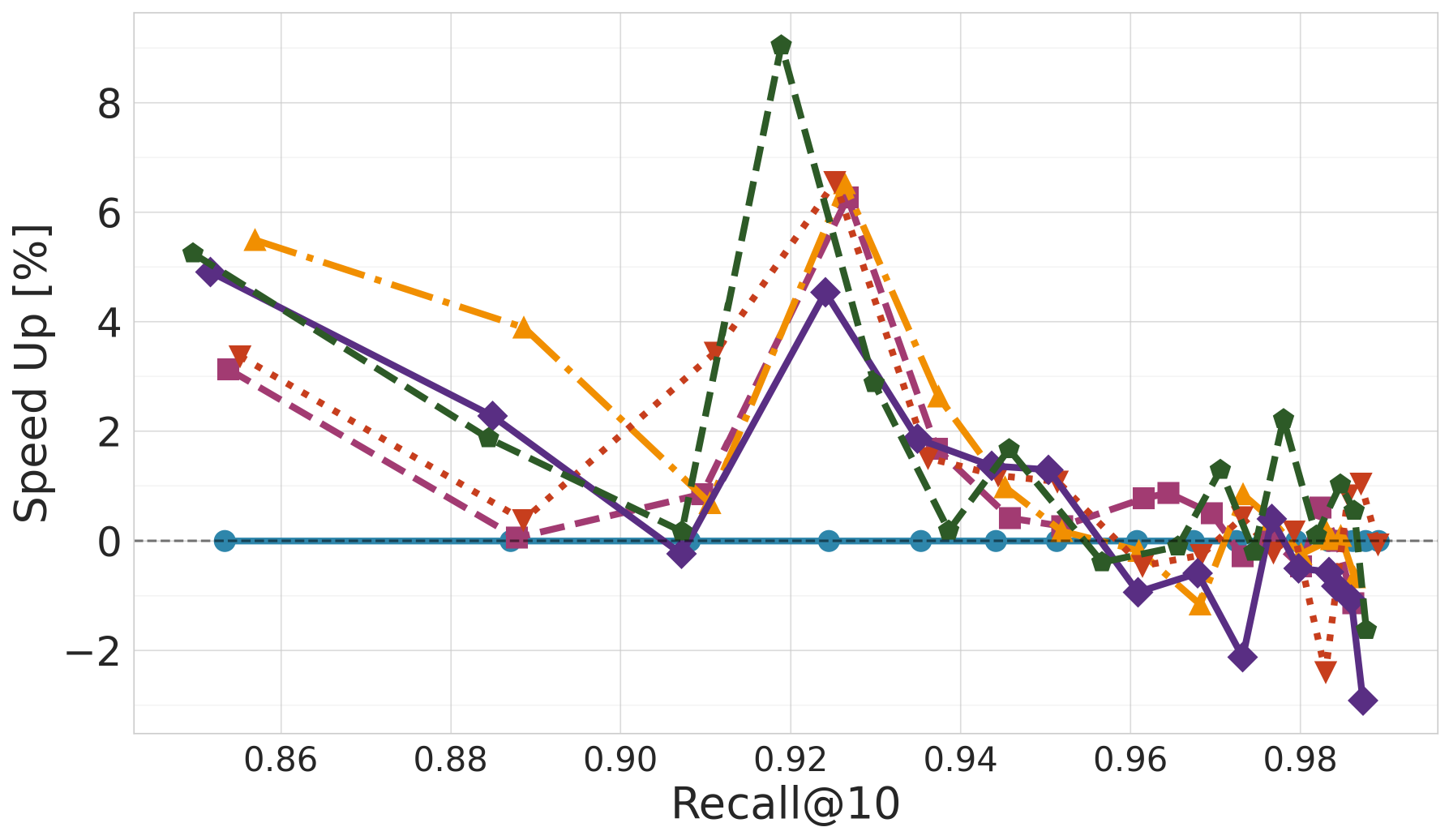}}
    &
    \subcaptionbox{NSG, OpenAI
    1M}[0.24\linewidth]{\includegraphics[width=\linewidth]{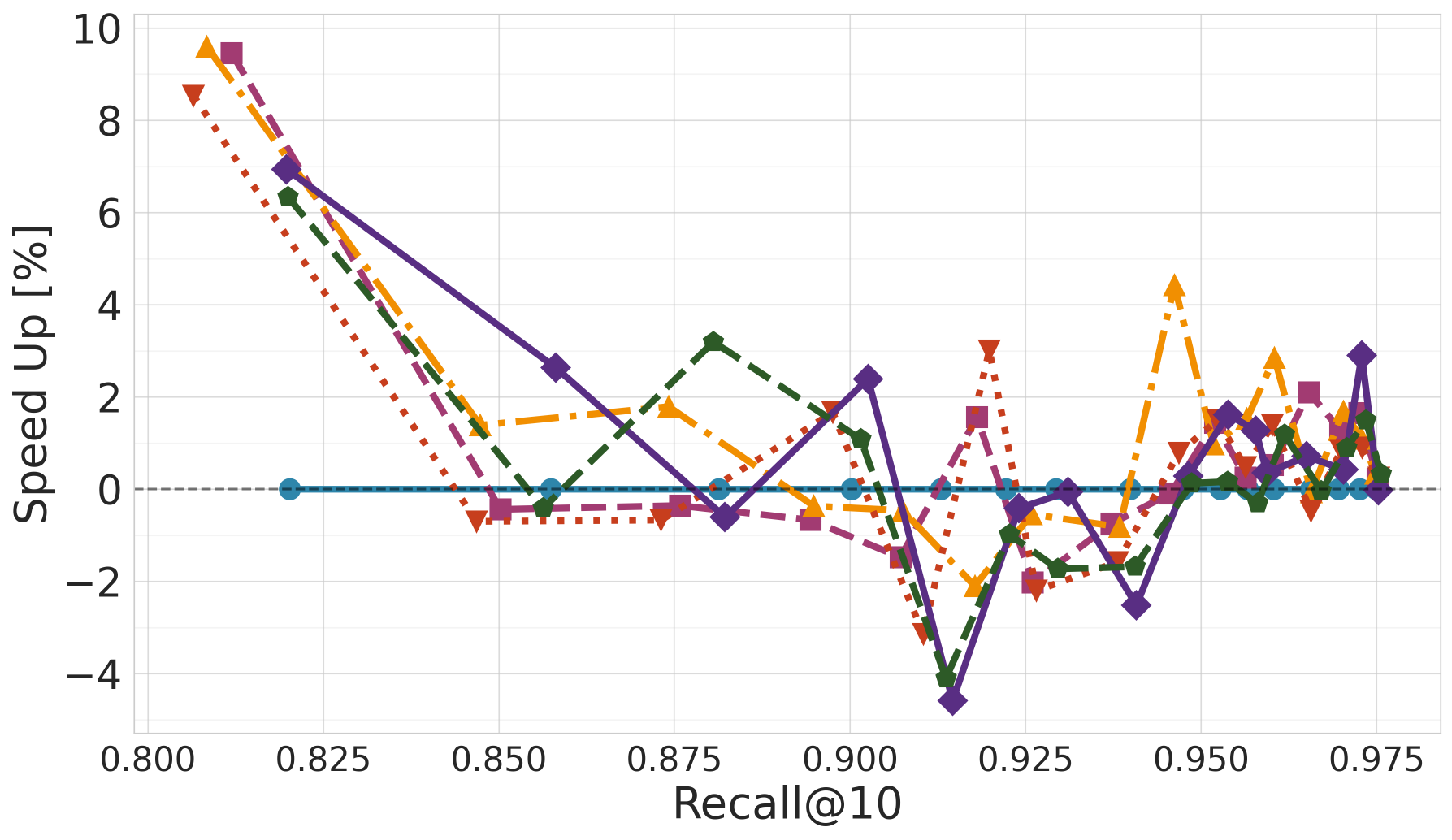}}
    &
    \subcaptionbox{Vamana, OpenAI
    1M}[0.24\linewidth]{\includegraphics[width=\linewidth]{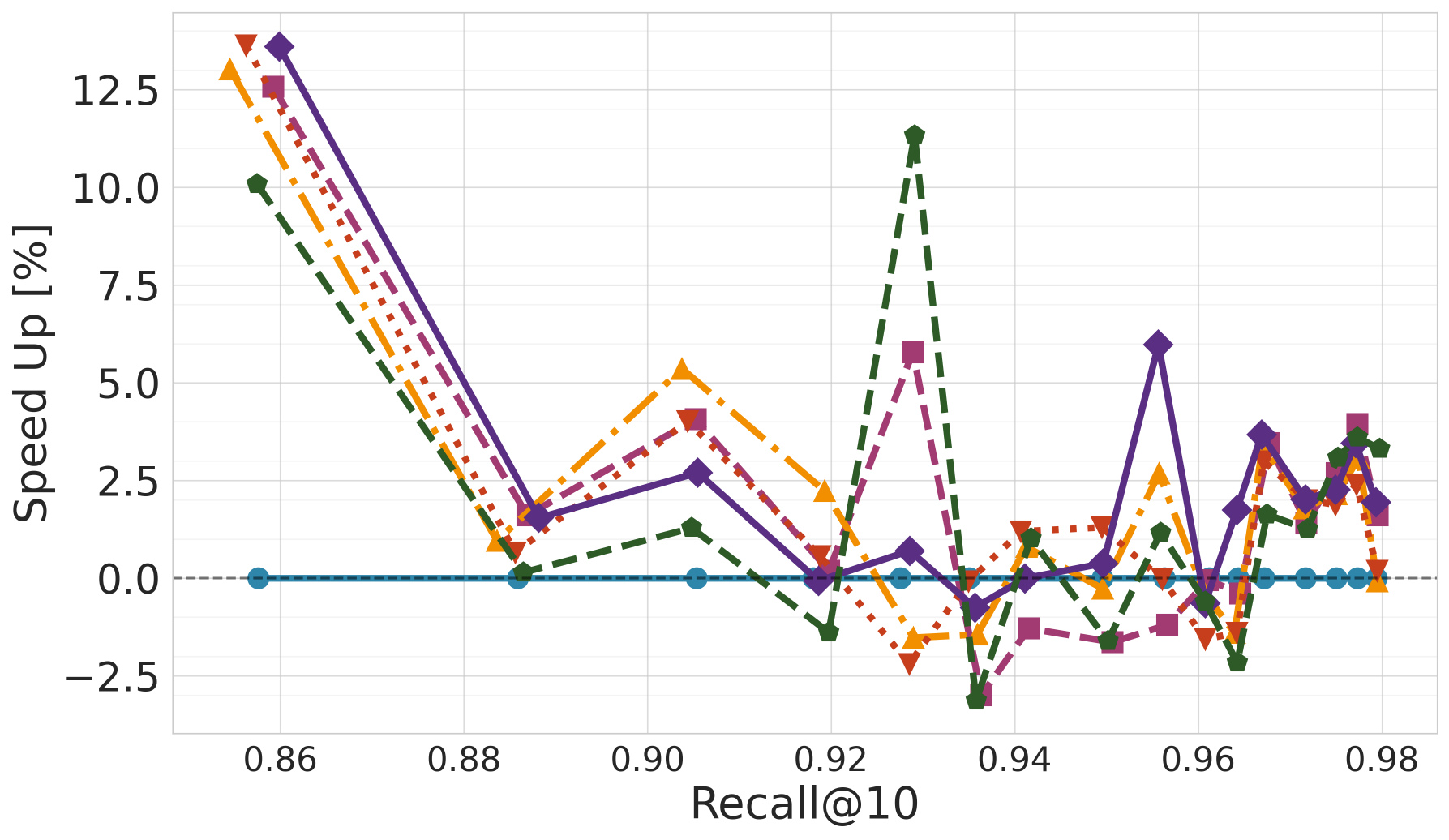}}
    &
    \subcaptionbox{NN-Descent, OpenAI
    1M}[0.24\linewidth]{\includegraphics[width=\linewidth]{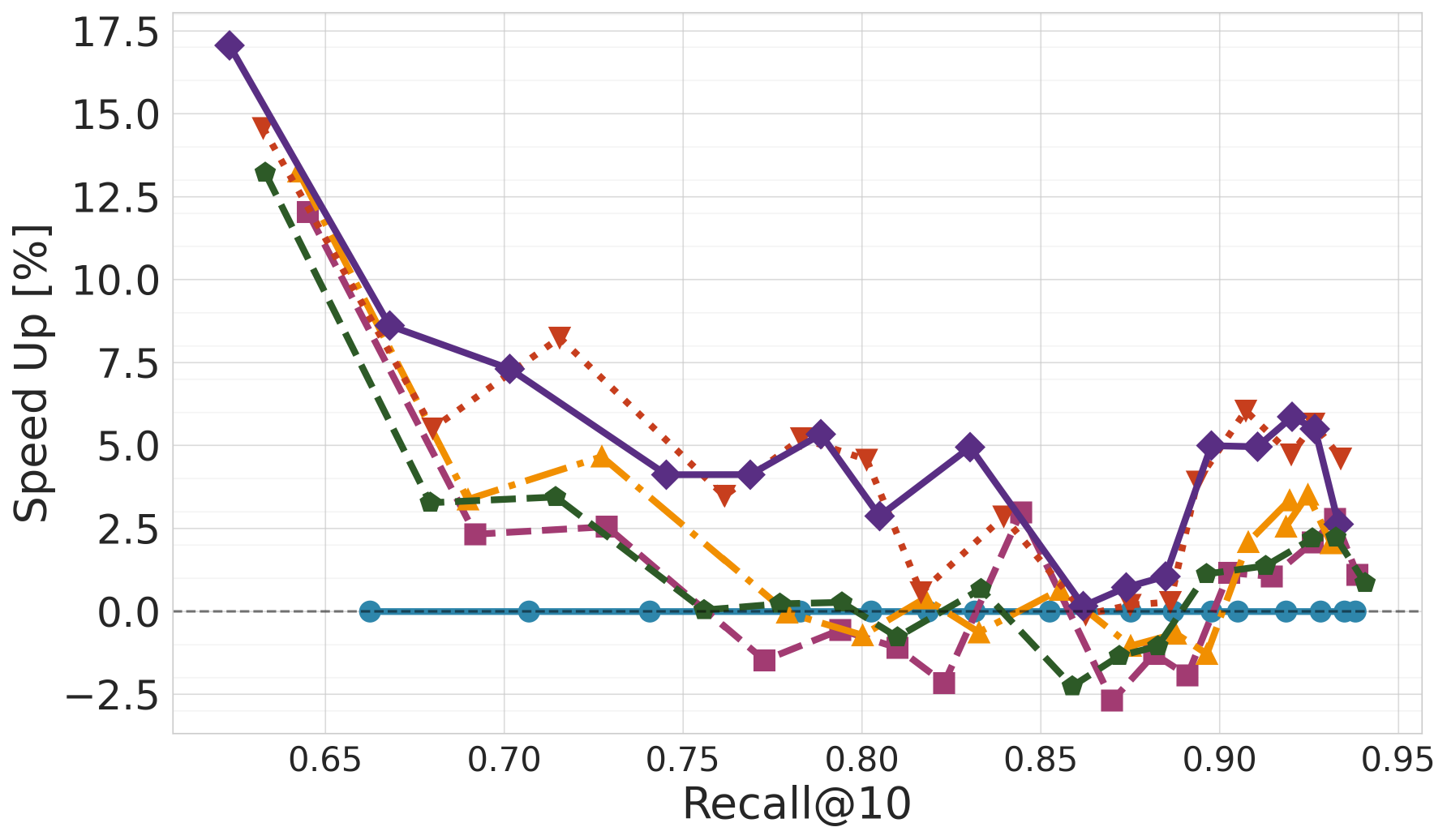}}
    \\
    \subcaptionbox{CAGRA, Yandex T2I
    1M}[0.24\linewidth]{\includegraphics[width=\linewidth]{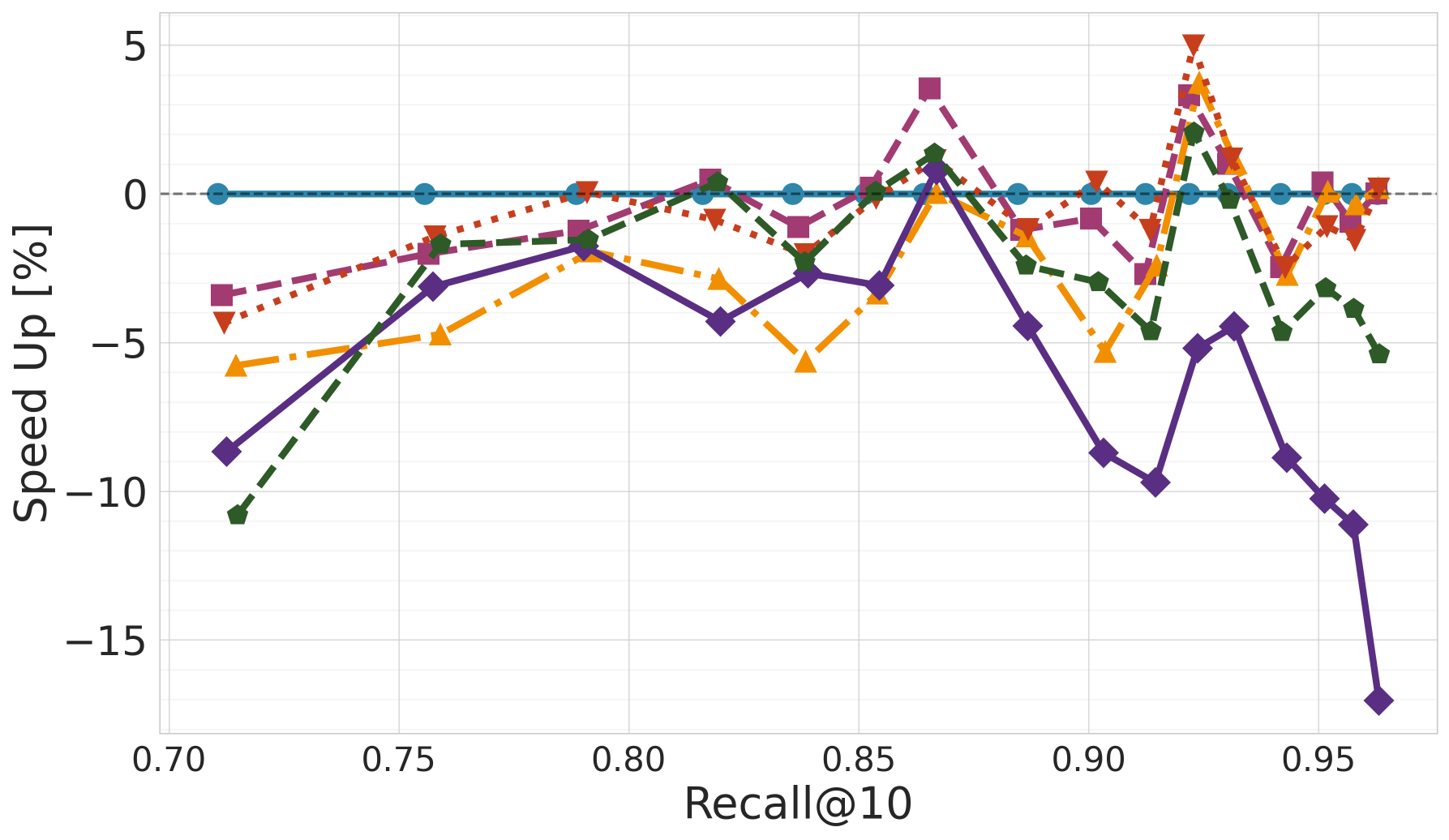}}
    &
    \subcaptionbox{NSG, Yandex T2I
    1M}[0.24\linewidth]{\includegraphics[width=\linewidth]{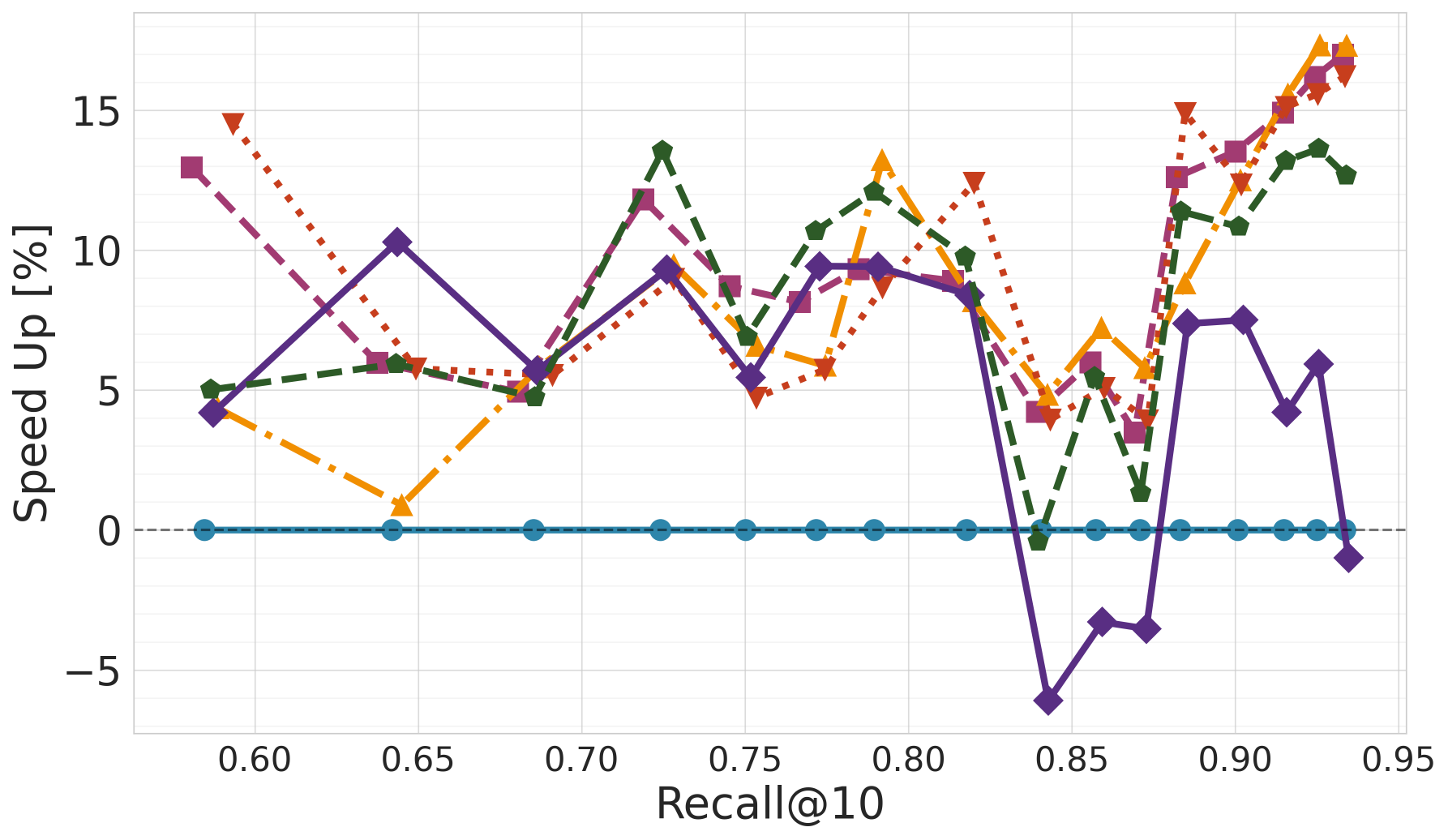}}
    &
    \subcaptionbox{Vamana, Yandex T2I
    1M}[0.24\linewidth]{\includegraphics[width=\linewidth]{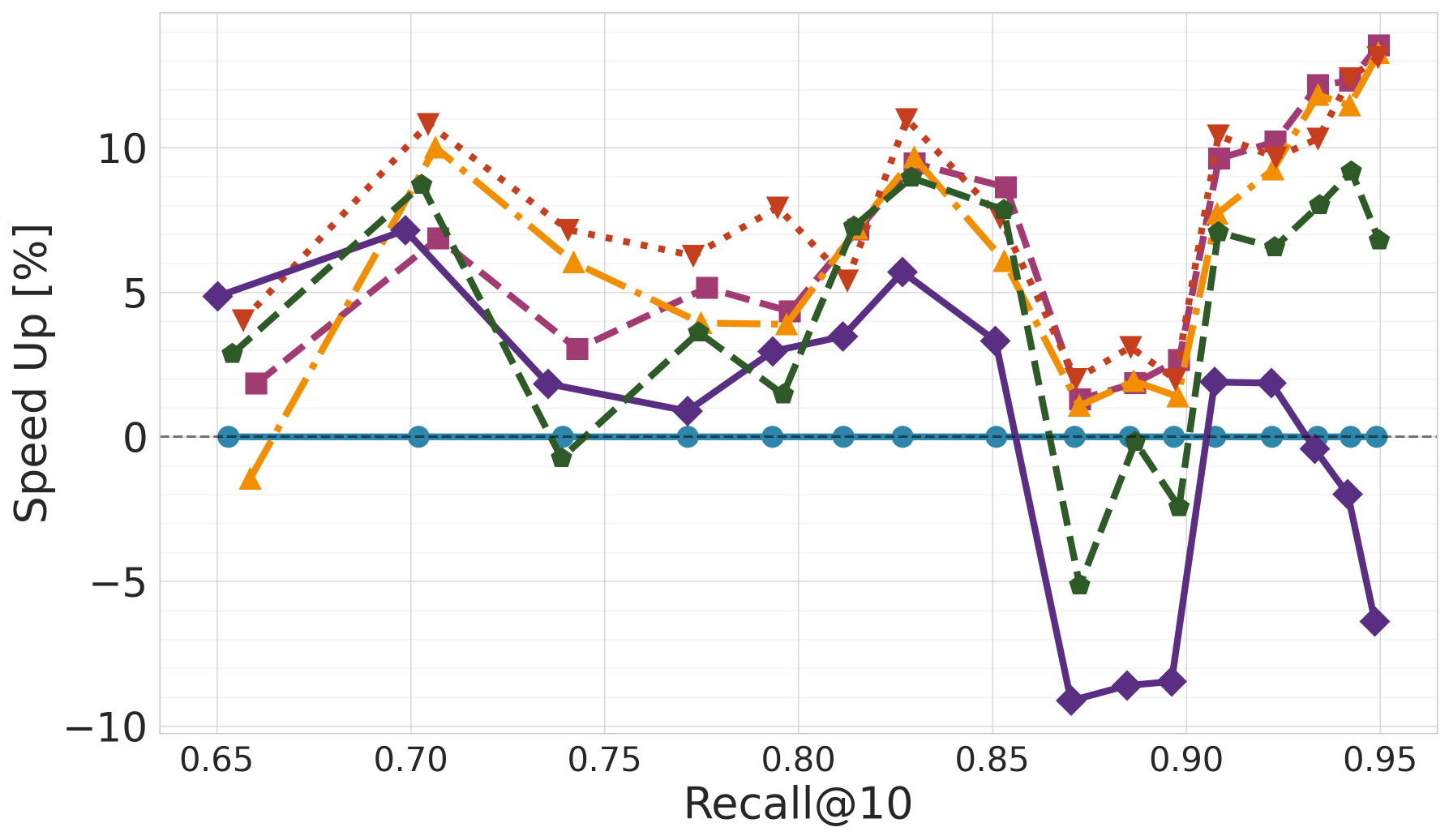}}
    &
    \subcaptionbox{NN-Descent, Yandex T2I
    1M}[0.24\linewidth]{\includegraphics[width=\linewidth]{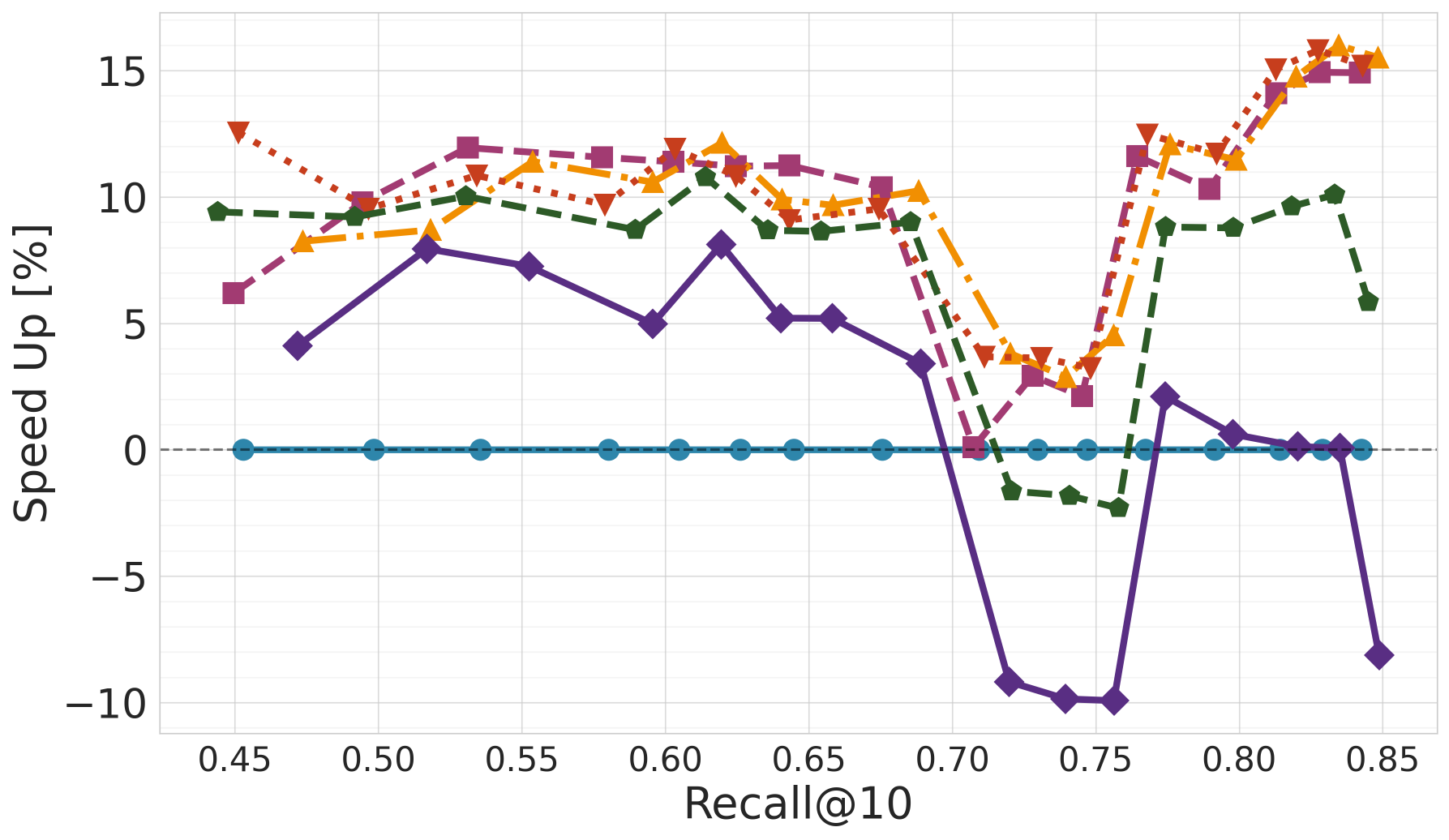}}
    \\
    \subcaptionbox{CAGRA, Deep
    10M}[0.24\linewidth]{\includegraphics[width=\linewidth]{fig/A100-result/recall_qps_comparison_K32_gpu_deep10m_cagra_reordering.pdf}}
    &
    \subcaptionbox{NSG, Deep
    10M}[0.24\linewidth]{\includegraphics[width=\linewidth]{fig/A100-result/recall_qps_comparison_K32_gpu_deep10m_nsg_reordering.pdf}}
    &
    \subcaptionbox{Vamana, Deep
    10M}[0.24\linewidth]{\includegraphics[width=\linewidth]{fig/A100-result/recall_qps_comparison_K32_gpu_deep10m_diskann_reordering.pdf}}
    &
    \subcaptionbox{NN-Descent, Deep
    10M}[0.24\linewidth]{\includegraphics[width=\linewidth]{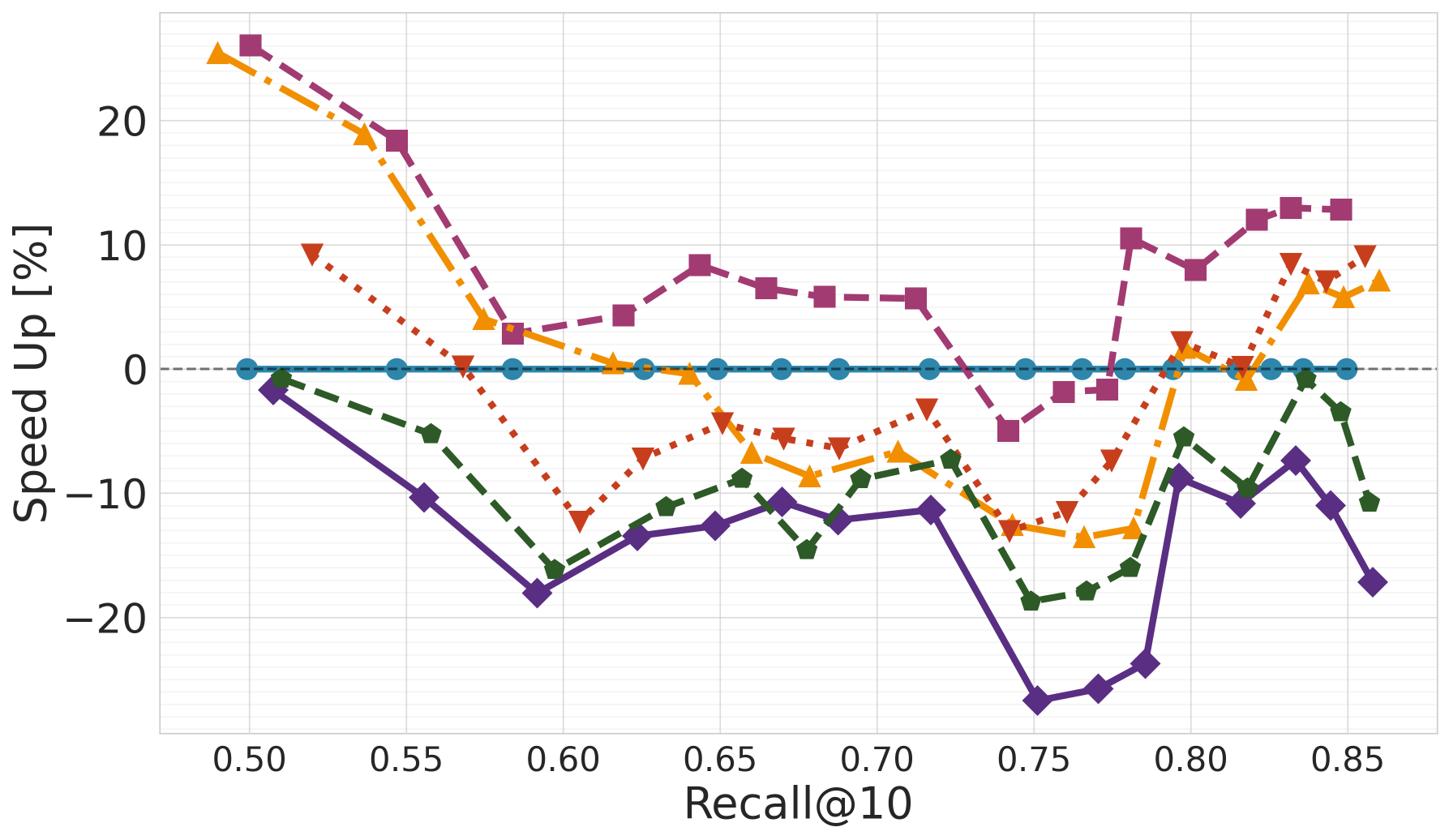}}
  \end{tabular}
  \vspace{0.1in}
  \begin{tabular}{c}
    \includegraphics[width=\textwidth]{fig/reordering_algorithms_legend.pdf}
  \end{tabular}
  \caption{Recall--QPS trade-off with reordering on NVIDIA A100 (Part
  I: traditional and medium-scale datasets).}
  \label{fig:supp-reordering-A100-part1}
\end{figure*}

\begin{figure*}[h]
  \centering
  \setlength{\tabcolsep}{2pt}
  \renewcommand{\arraystretch}{0}
  \begin{tabular}{cccc}
    \subcaptionbox{CAGRA, Wikipedia
    1M}[0.24\linewidth]{\includegraphics[width=\linewidth]{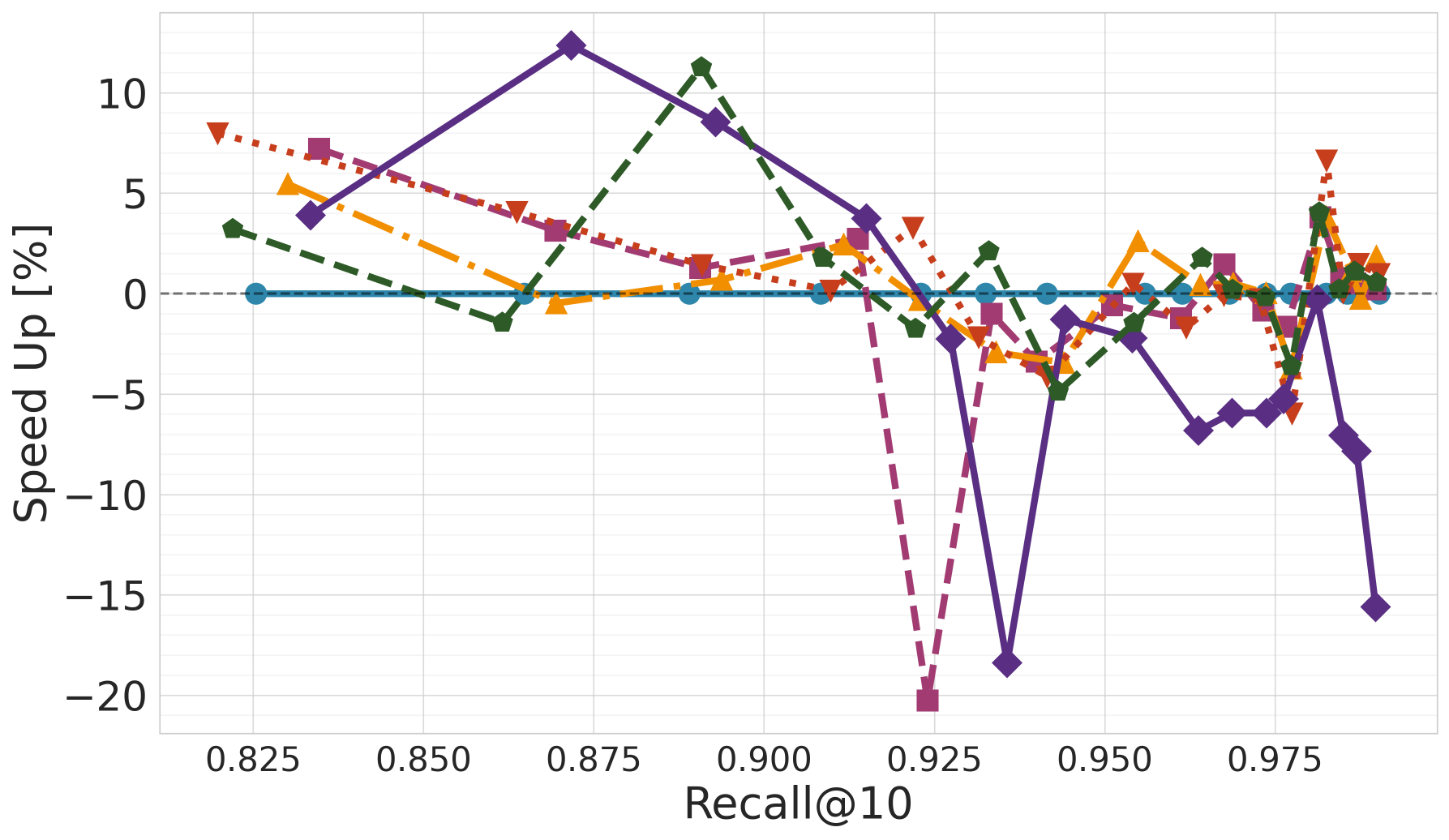}}
    &
    \subcaptionbox{NSG, Wikipedia
    1M}[0.24\linewidth]{\includegraphics[width=\linewidth]{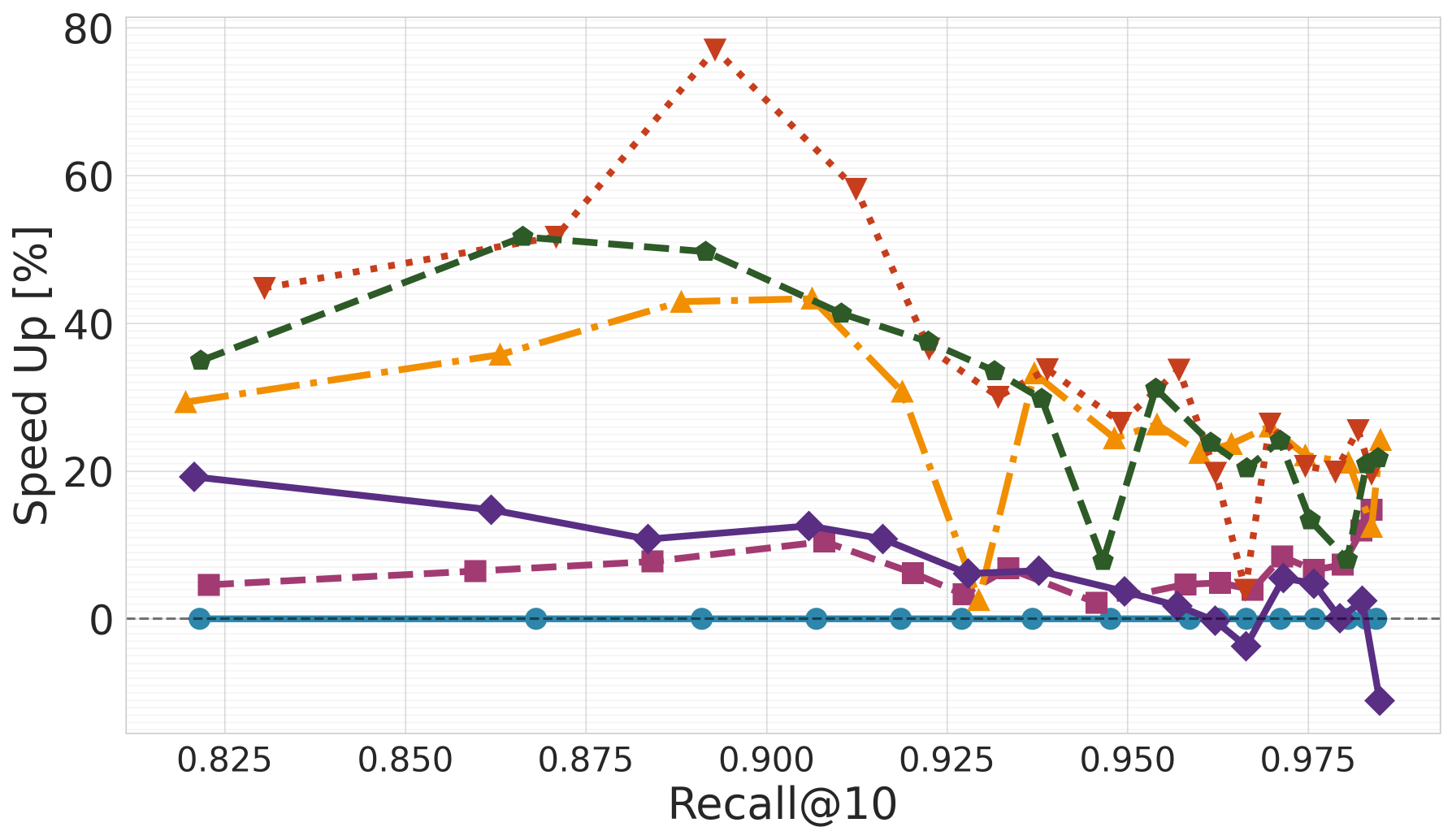}}
    &
    \subcaptionbox{Vamana, Wikipedia
    1M}[0.24\linewidth]{\includegraphics[width=\linewidth]{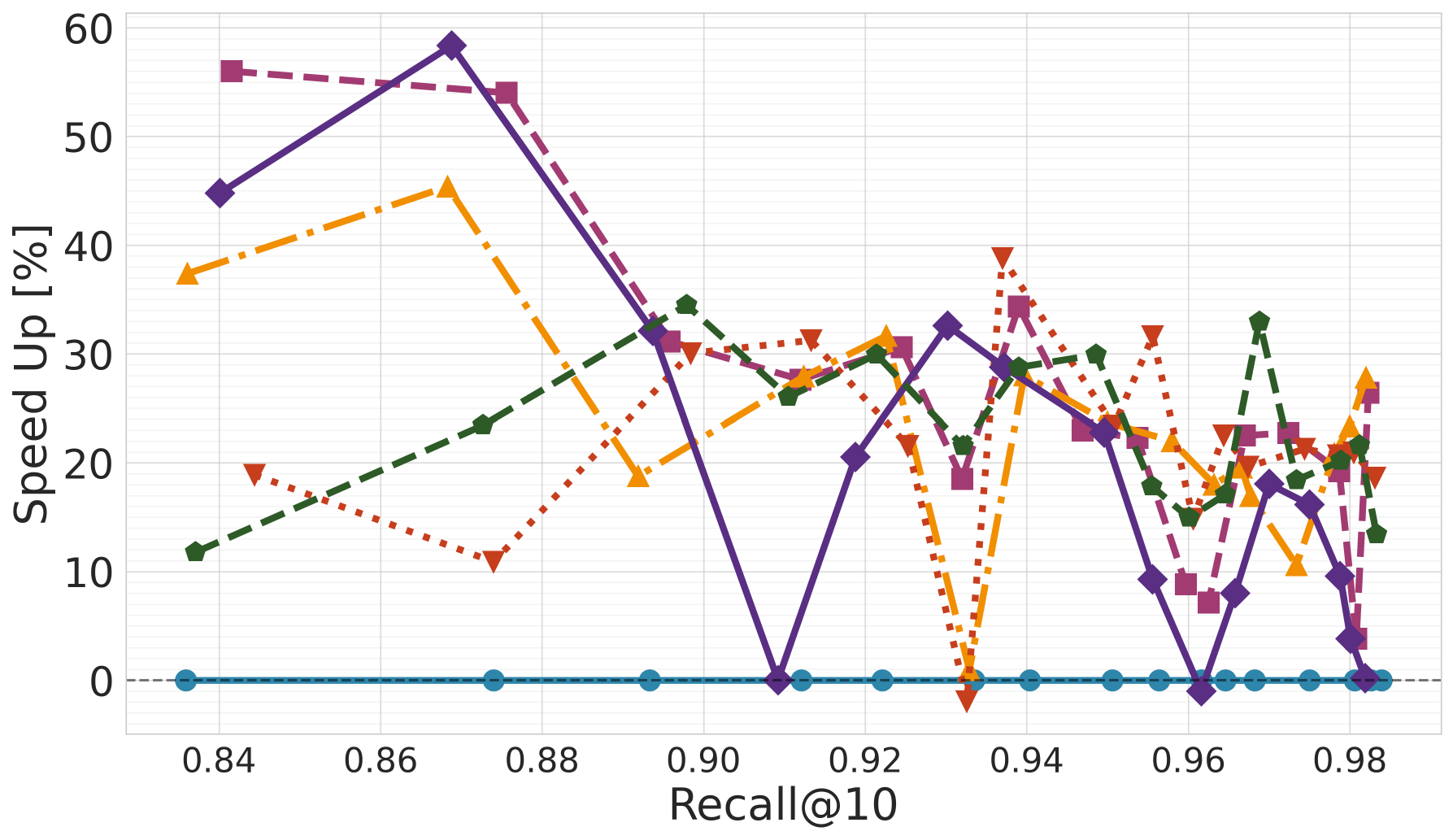}}
    &
    \subcaptionbox{NN-Descent, Wikipedia
    1M}[0.24\linewidth]{\includegraphics[width=\linewidth]{fig/A100-result/recall_qps_comparison_K32_gpu_wikipedia1m-768-ip_nndescent_reordering.pdf}}
    \\
    \subcaptionbox{CAGRA, BioASQ
    1M}[0.24\linewidth]{\includegraphics[width=\linewidth]{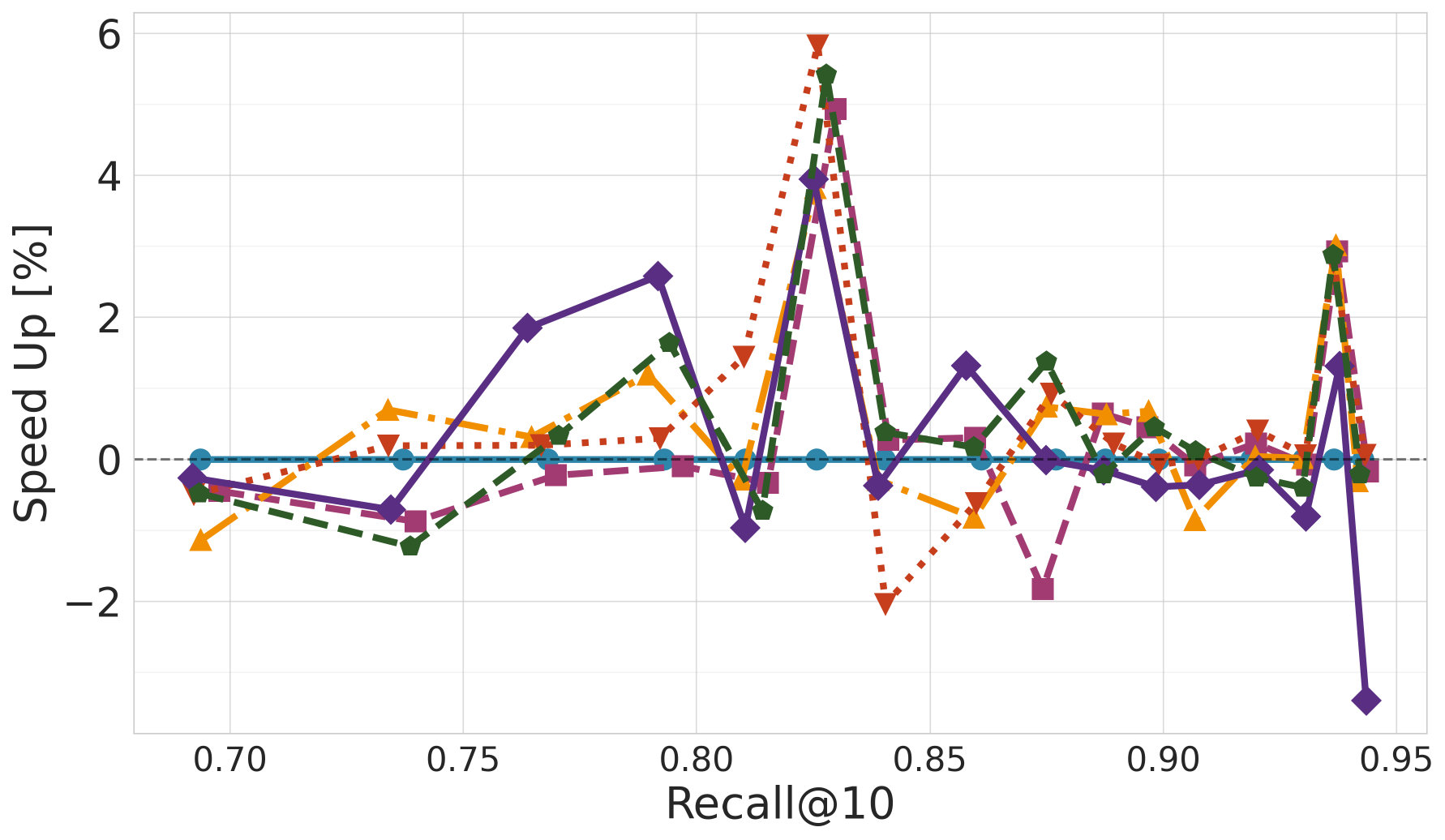}}
    &
    \subcaptionbox{NSG, BioASQ
    1M}[0.24\linewidth]{\includegraphics[width=\linewidth]{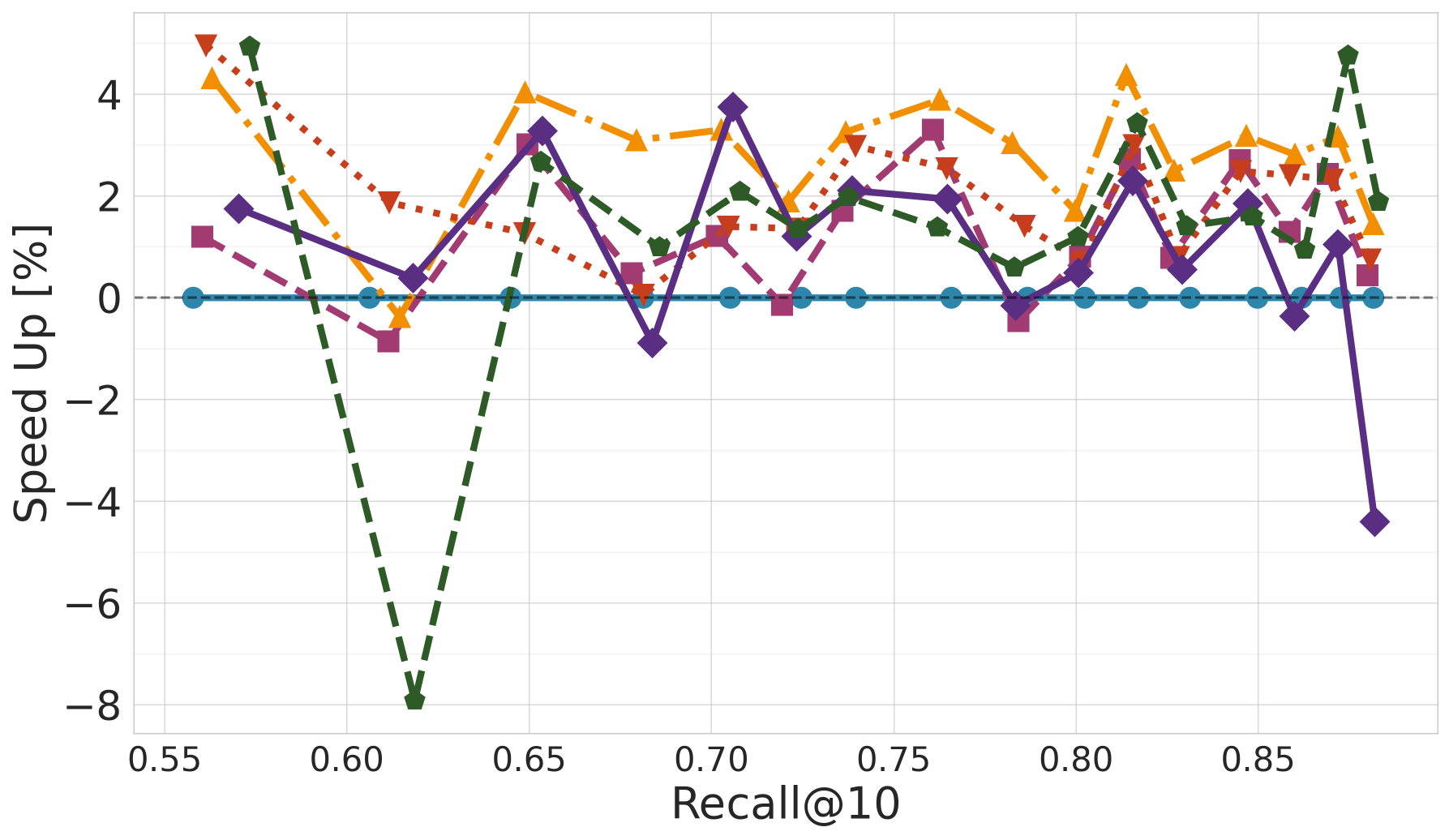}}
    &
    \subcaptionbox{Vamana, BioASQ
    1M}[0.24\linewidth]{\includegraphics[width=\linewidth]{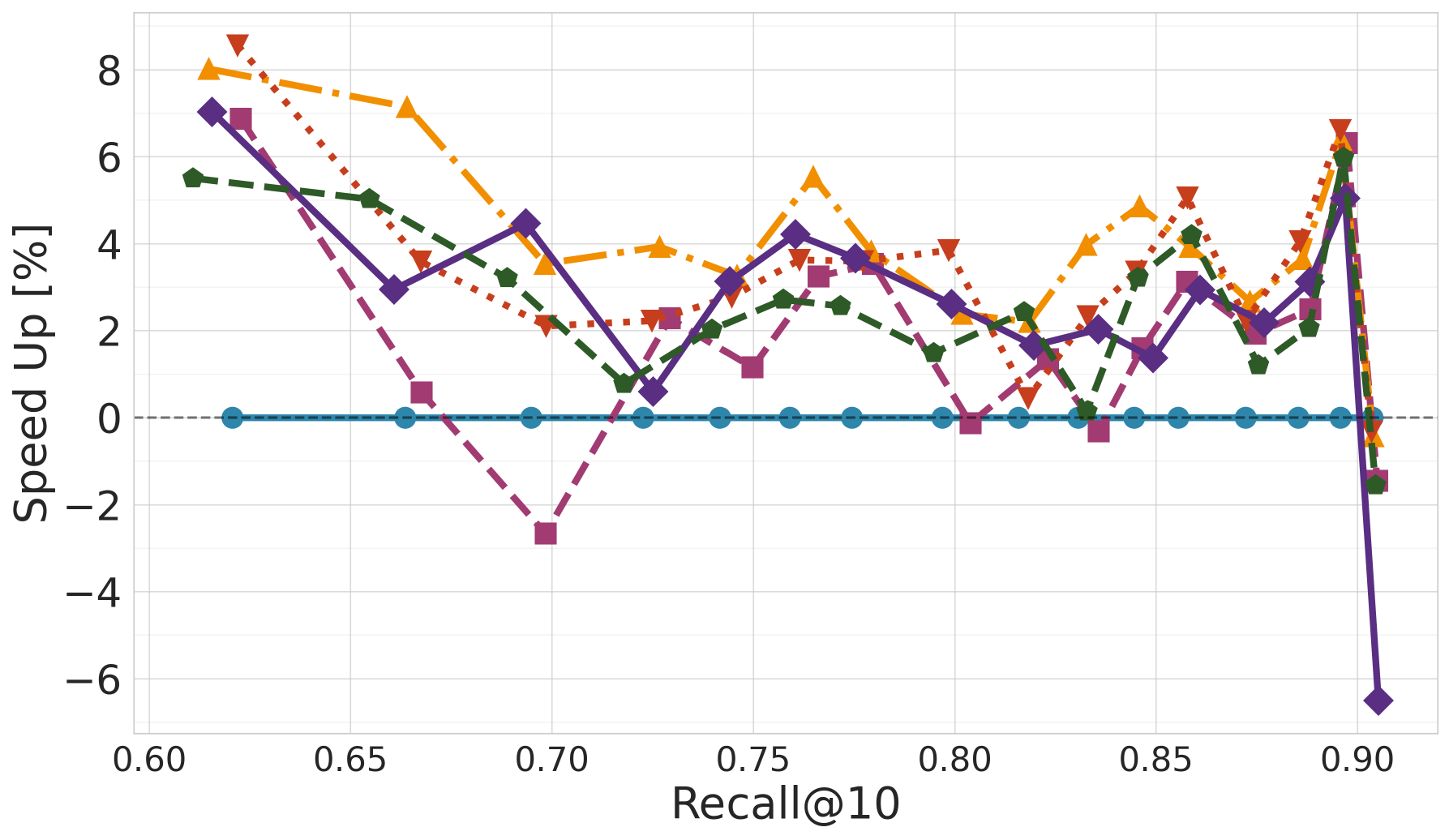}}
    &
    \subcaptionbox{NN-Descent, BioASQ
    1M}[0.24\linewidth]{\includegraphics[width=\linewidth]{fig/A100-result/recall_qps_comparison_K32_gpu_bioasq1m-1024-ip_nndescent_reordering.pdf}}
    \\
    \subcaptionbox{CAGRA, Wikipedia
    10M}[0.24\linewidth]{\includegraphics[width=\linewidth]{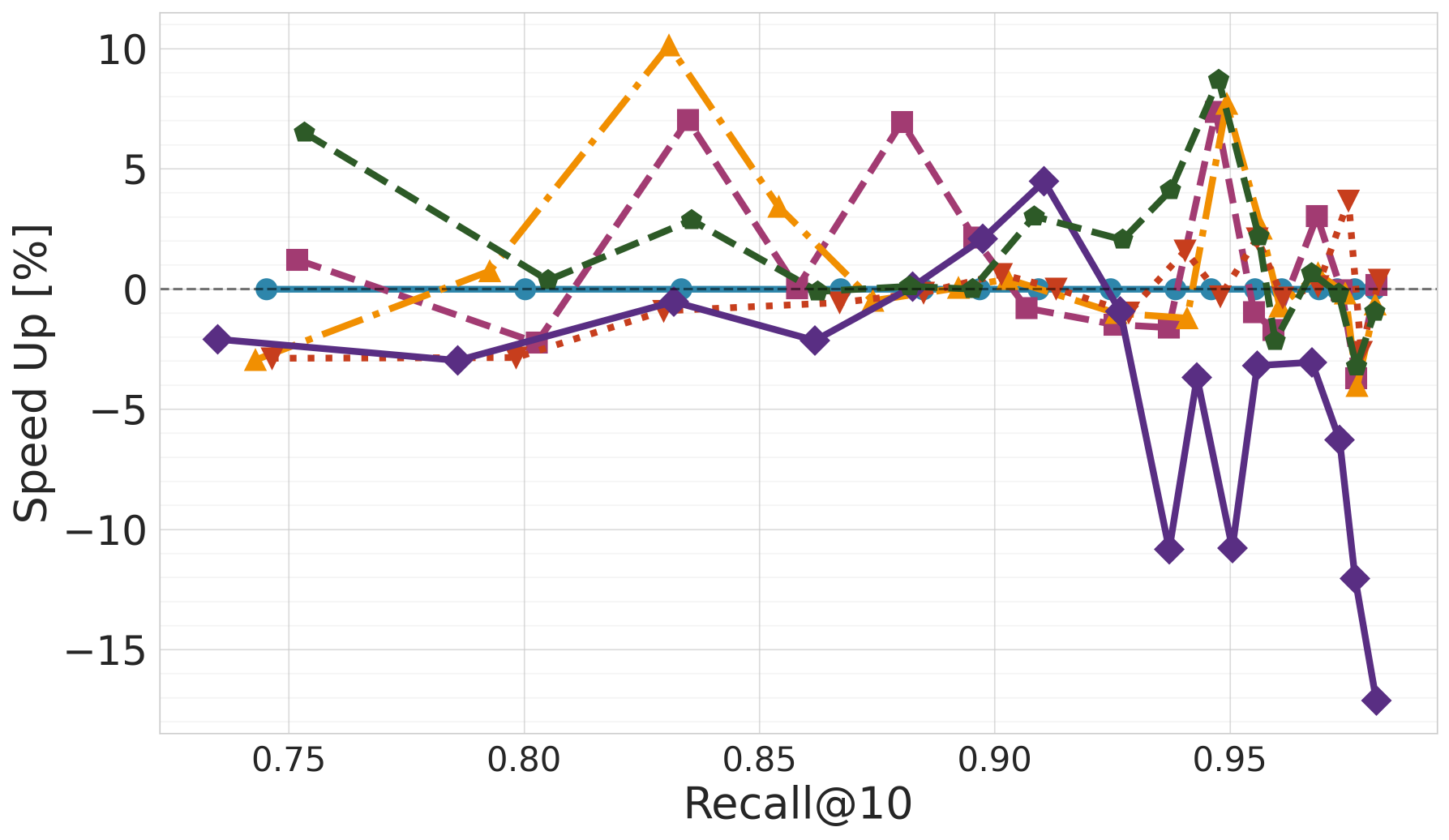}}
    &
    \subcaptionbox{NSG, Wikipedia
    10M}[0.24\linewidth]{\includegraphics[width=\linewidth]{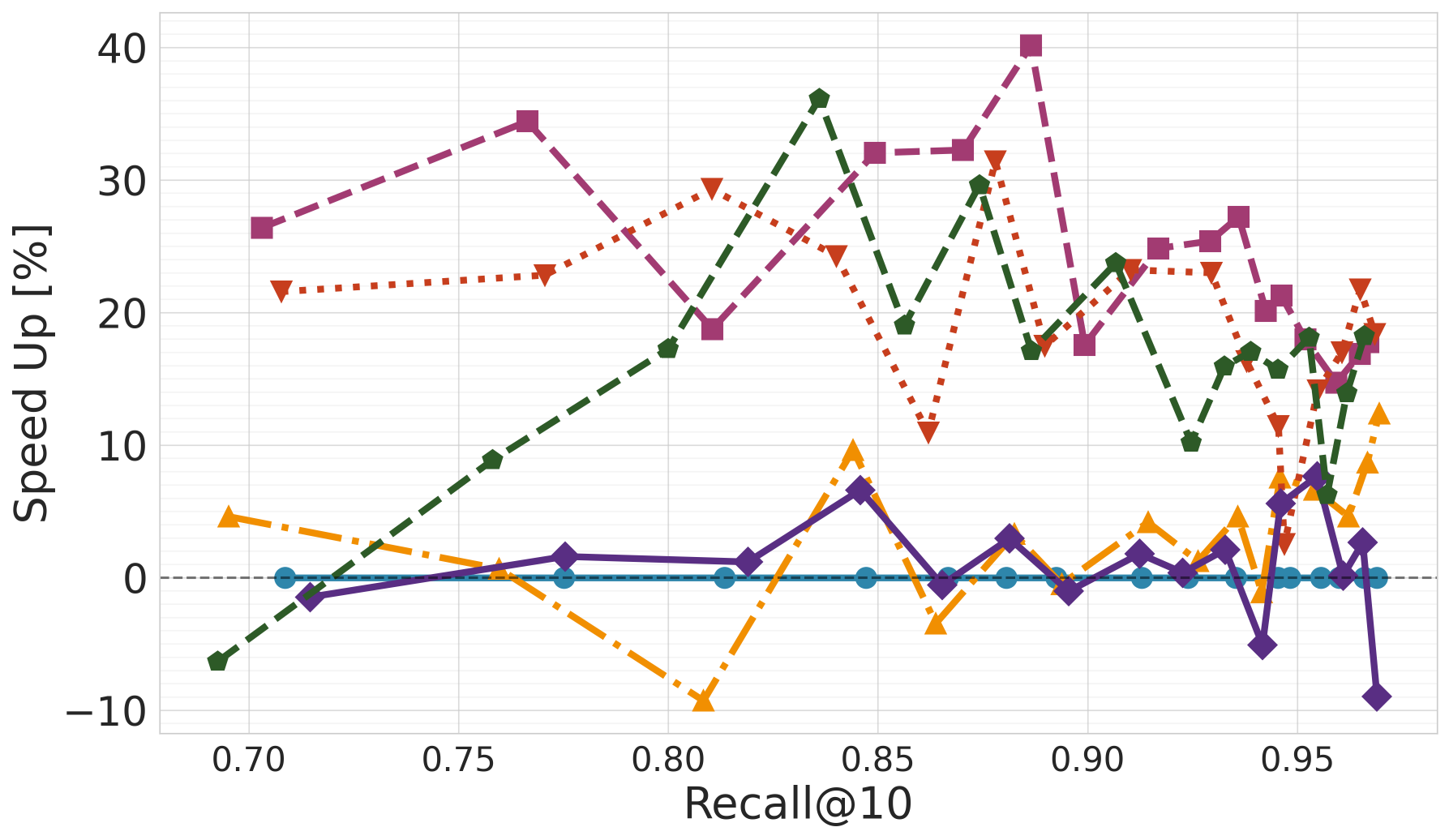}}
    &
    \subcaptionbox{Vamana, Wikipedia
    10M}[0.24\linewidth]{\includegraphics[width=\linewidth]{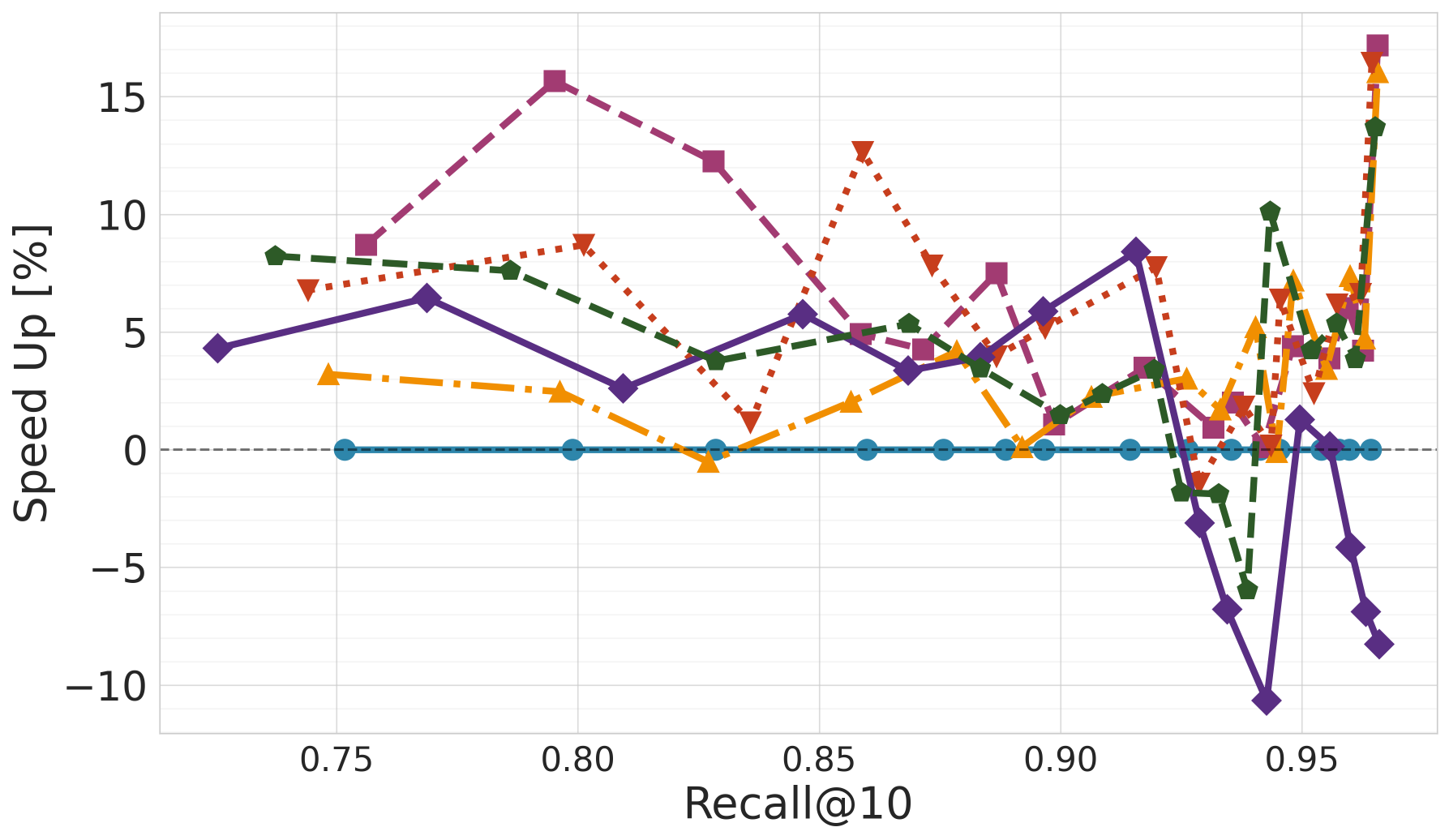}}
    &
    \subcaptionbox{NN-Descent, Wikipedia
    10M}[0.24\linewidth]{\includegraphics[width=\linewidth]{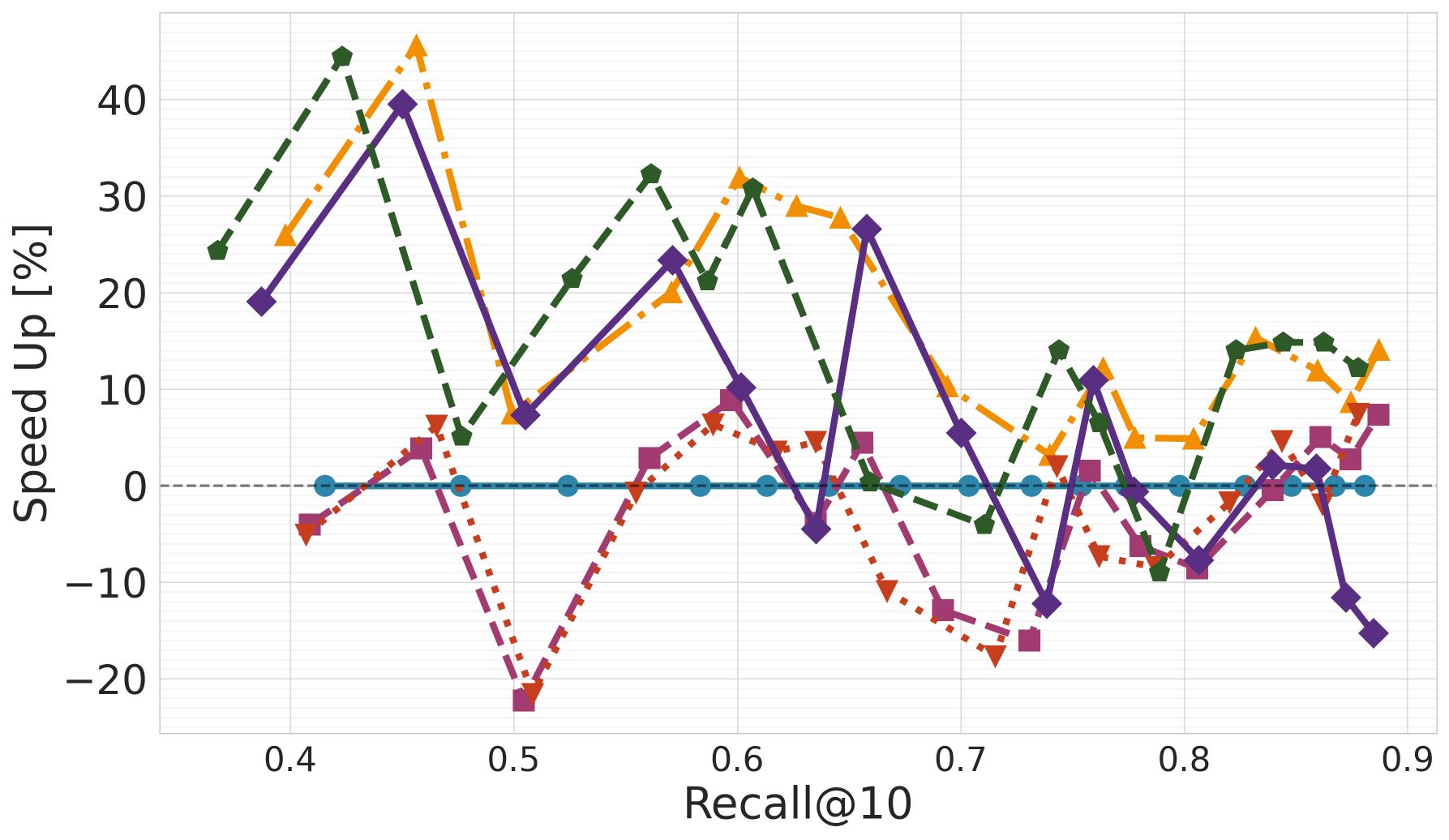}}
    \\
    \subcaptionbox{CAGRA, BioASQ
    10M}[0.24\linewidth]{\includegraphics[width=\linewidth]{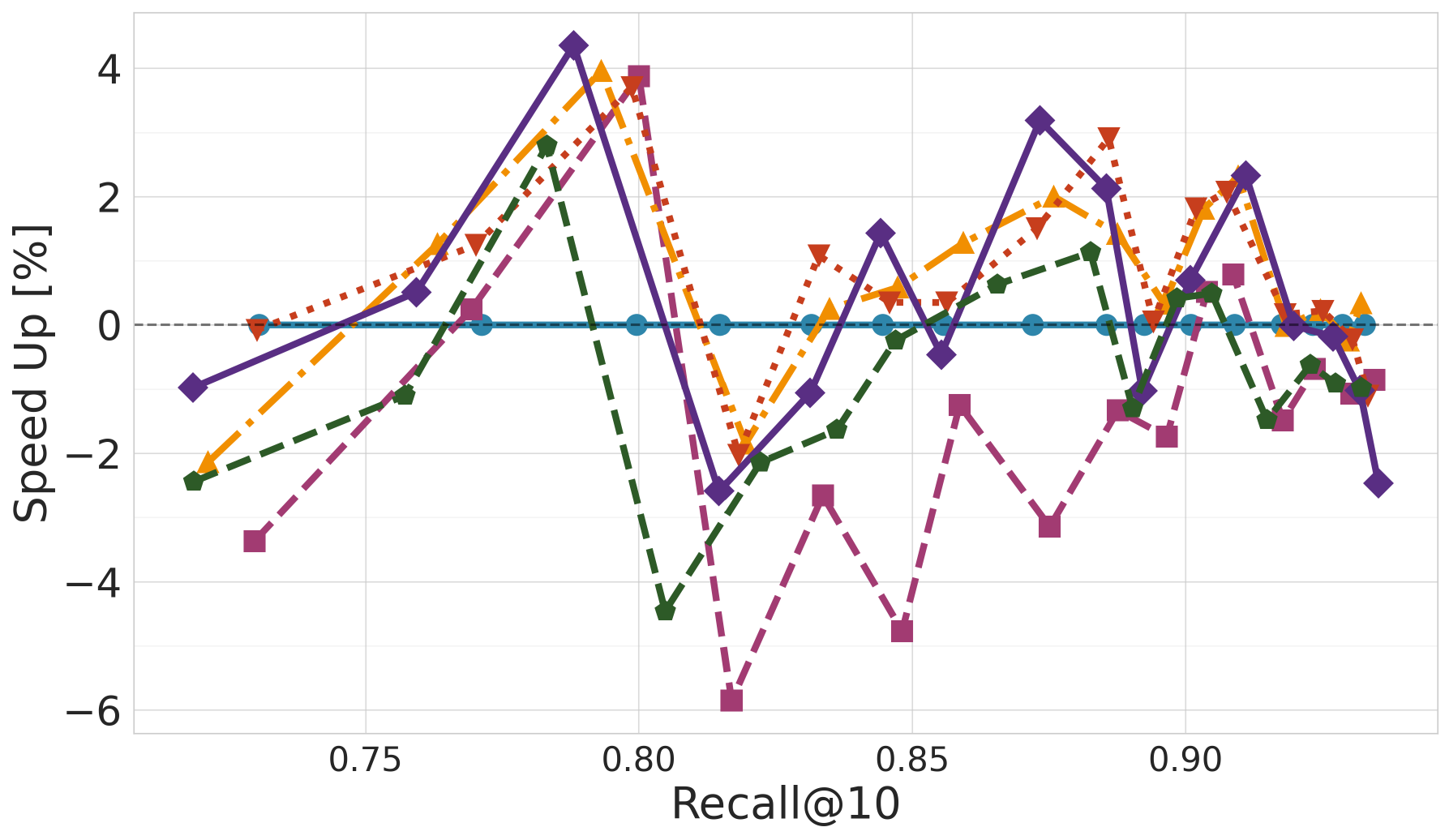}}
    &
    \subcaptionbox{NSG, BioASQ
    10M}[0.24\linewidth]{\includegraphics[width=\linewidth]{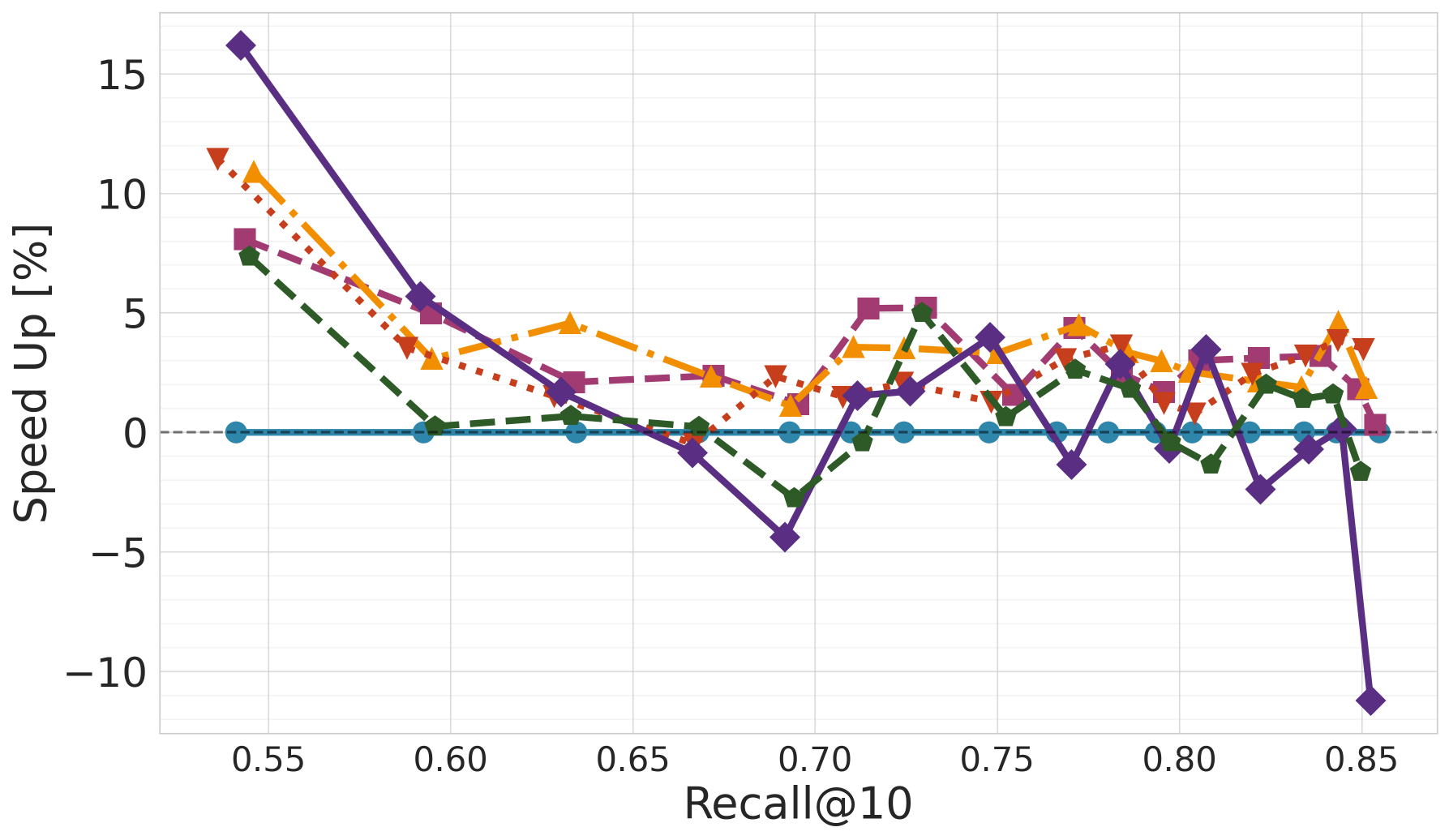}}
    &
    \subcaptionbox{Vamana, BioASQ
    10M}[0.24\linewidth]{\includegraphics[width=\linewidth]{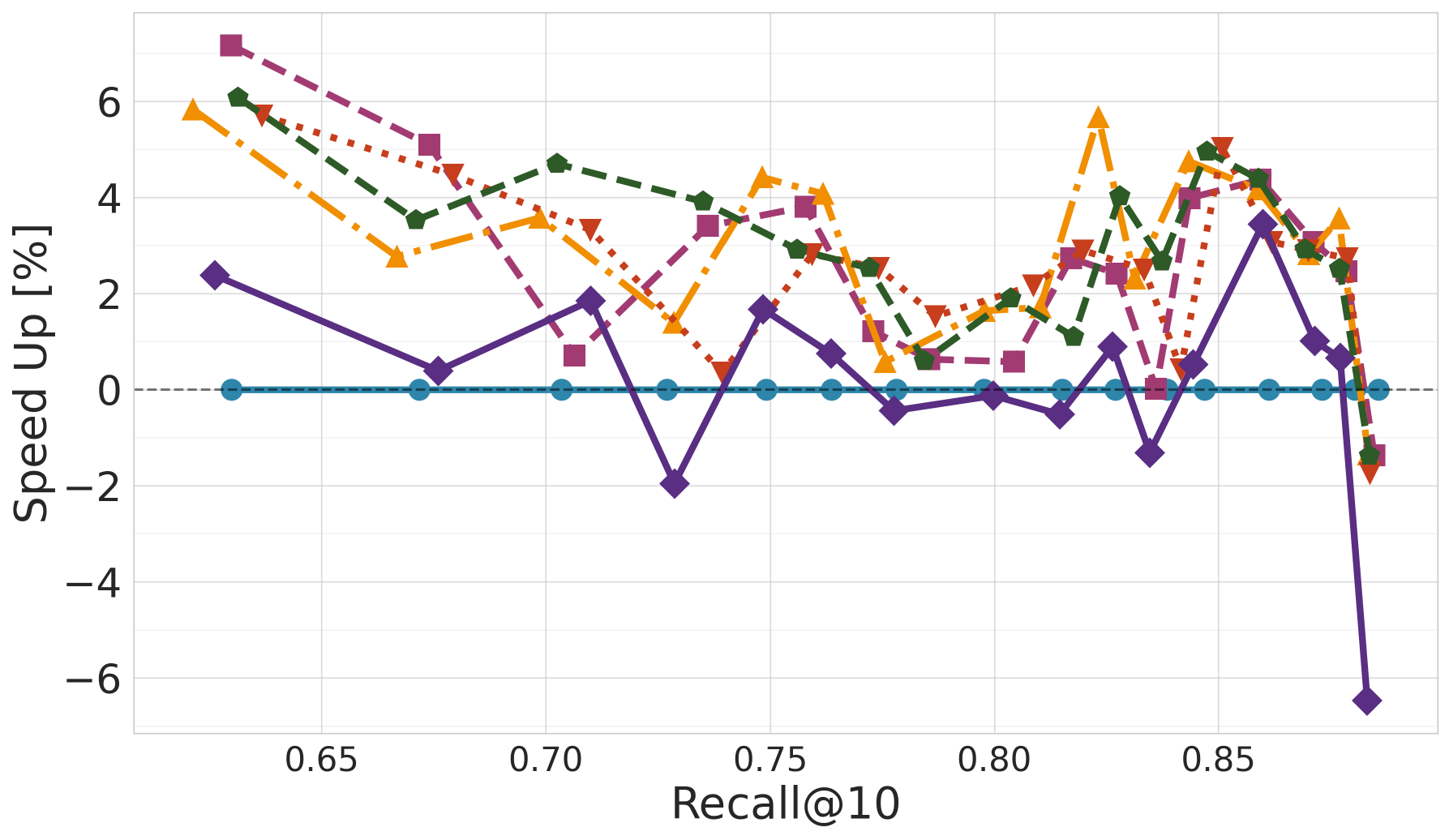}}
    &
    \subcaptionbox{NN-Descent, BioASQ
    10M}[0.24\linewidth]{\includegraphics[width=\linewidth]{fig/A100-result/recall_qps_comparison_K32_gpu_bioasq10m-1024-ip_nndescent_reordering.pdf}}
    \\
    \subcaptionbox{CAGRA, C4
    5M}[0.24\linewidth]{\includegraphics[width=\linewidth]{fig/A100-result/recall_qps_comparison_K32_gpu_c45m-1536-ip_cagra_reordering.pdf}}
    &
    \subcaptionbox{NSG, C4
    5M}[0.24\linewidth]{\includegraphics[width=\linewidth]{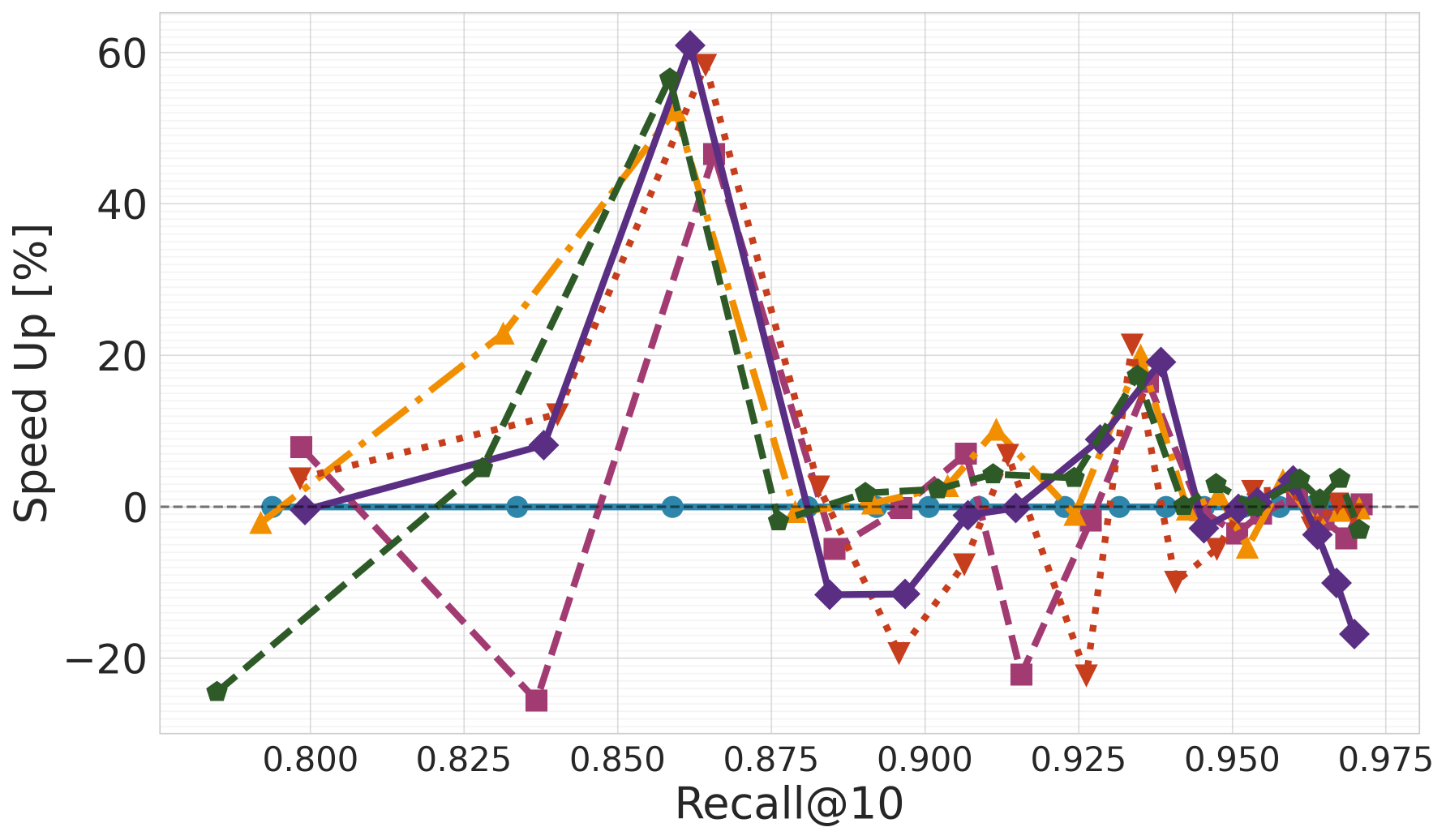}}
    &
    \subcaptionbox{Vamana, C4
    5M}[0.24\linewidth]{\includegraphics[width=\linewidth]{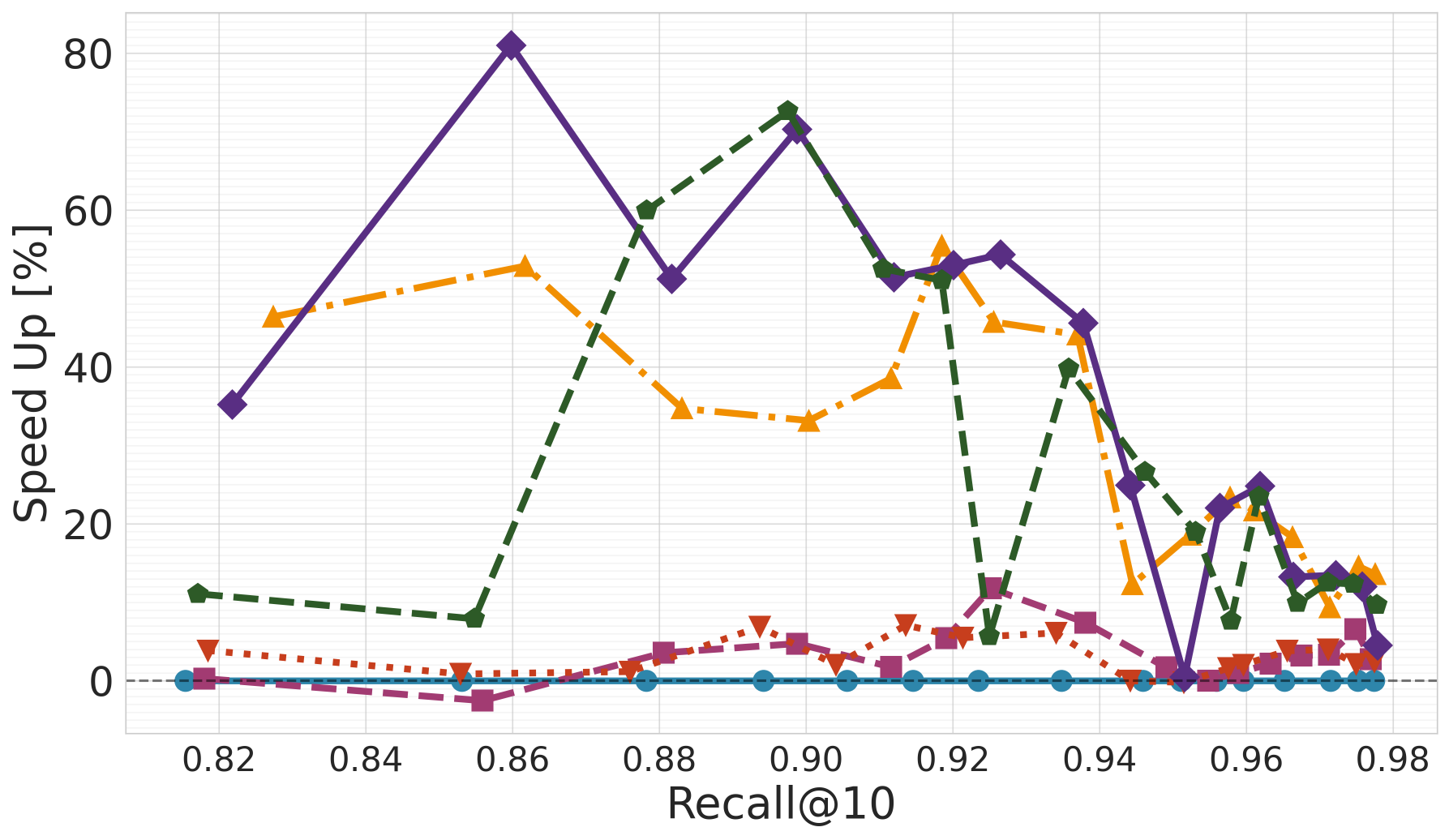}}
    &
    \subcaptionbox{NN-Descent, C4
    5M}[0.24\linewidth]{\includegraphics[width=\linewidth]{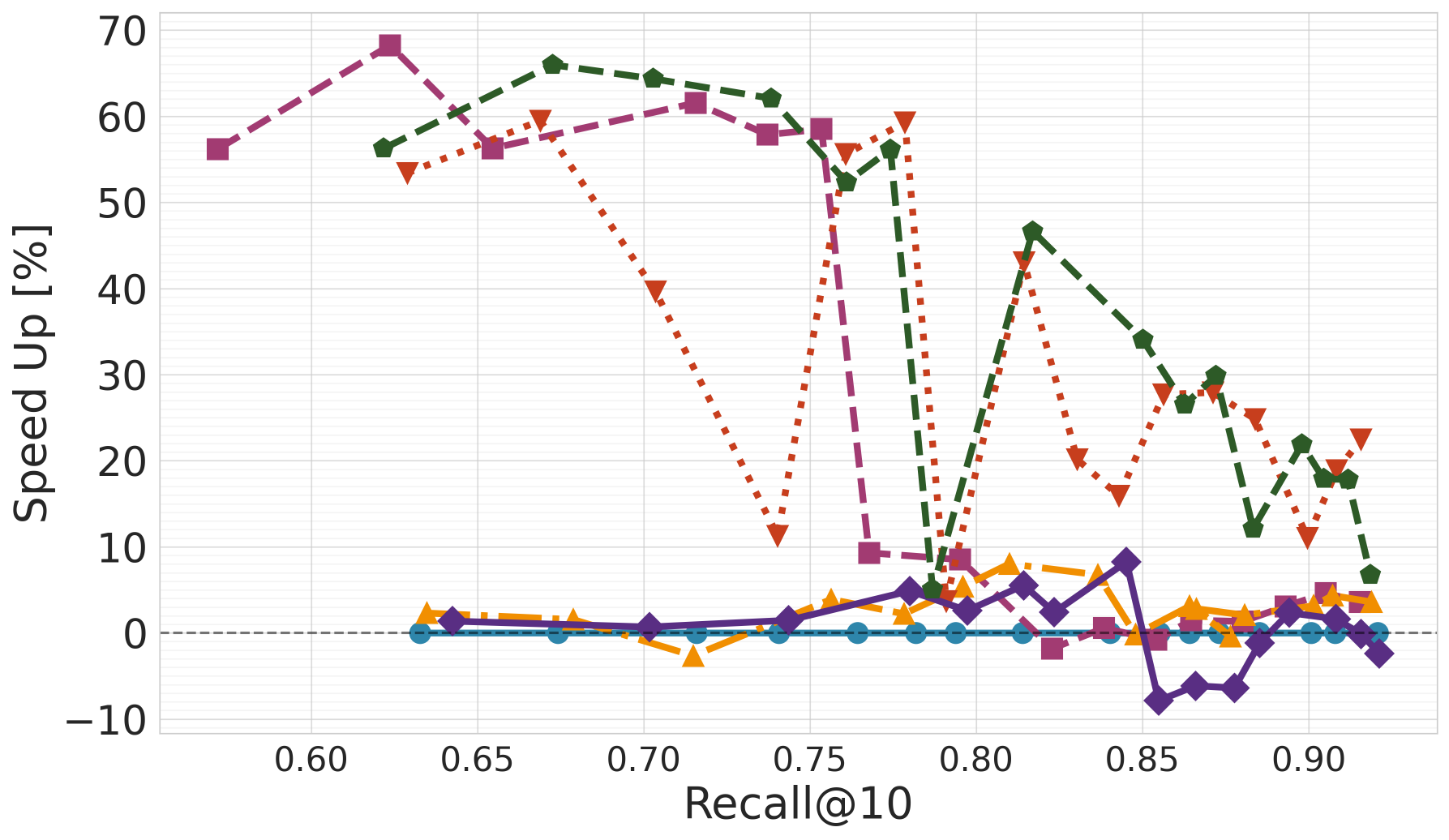}}
    \\
    \subcaptionbox{CAGRA, Deep
    40M}[0.24\linewidth]{\includegraphics[width=\linewidth]{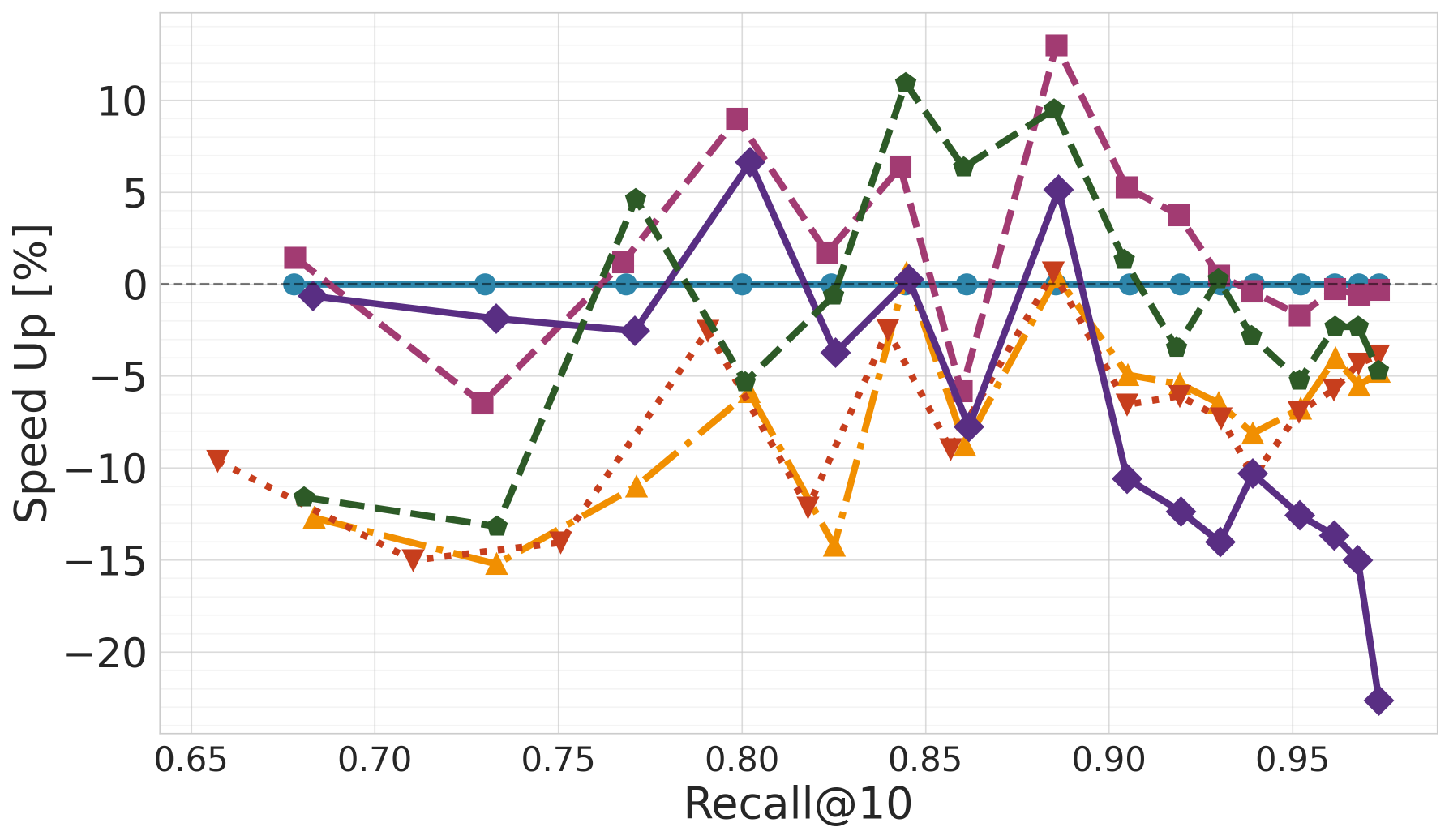}}
    &
    \subcaptionbox{NSG, Deep
    40M}[0.24\linewidth]{\includegraphics[width=\linewidth]{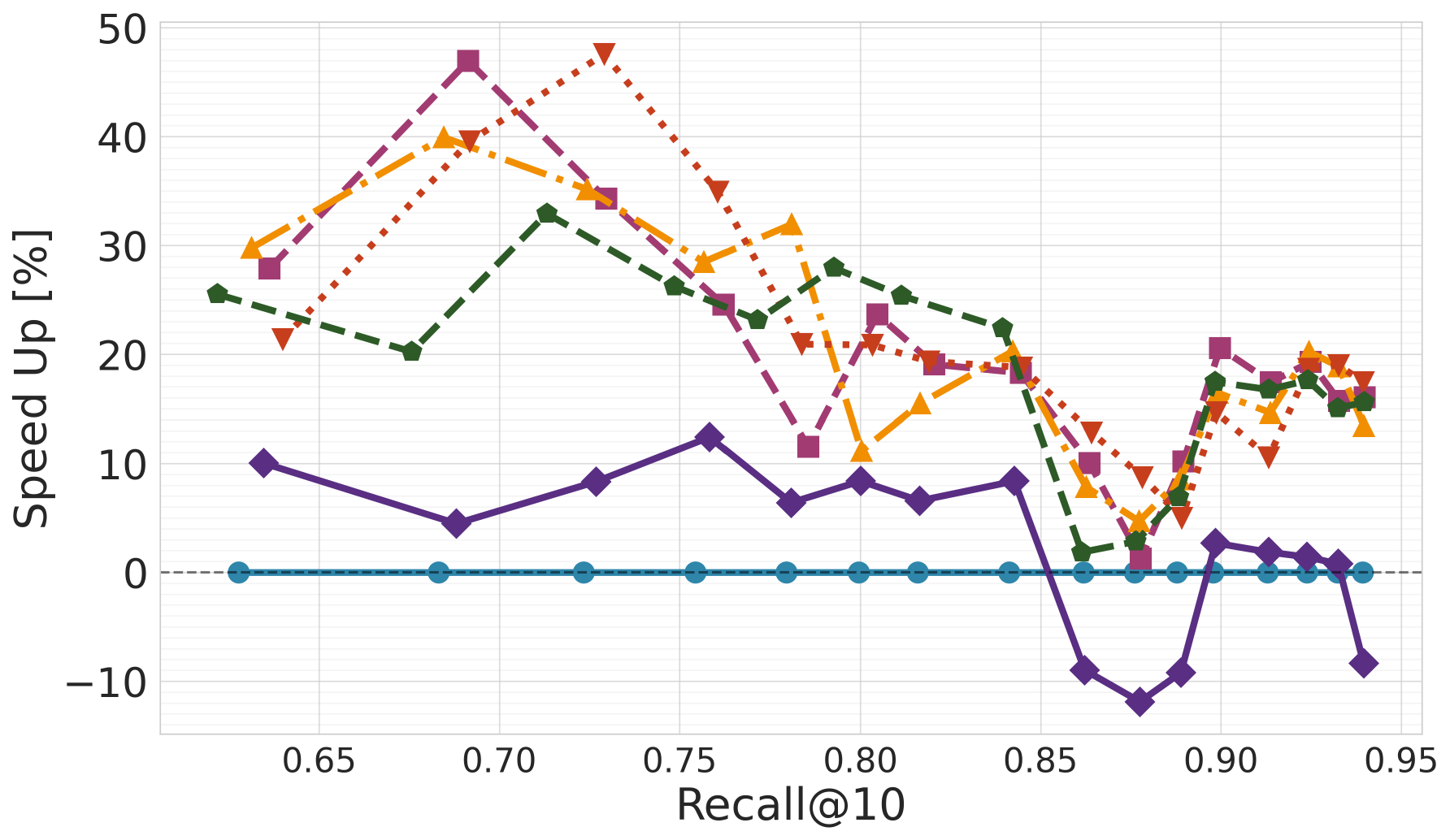}}
    &
    \subcaptionbox{Vamana, Deep
    40M}[0.24\linewidth]{\includegraphics[width=\linewidth]{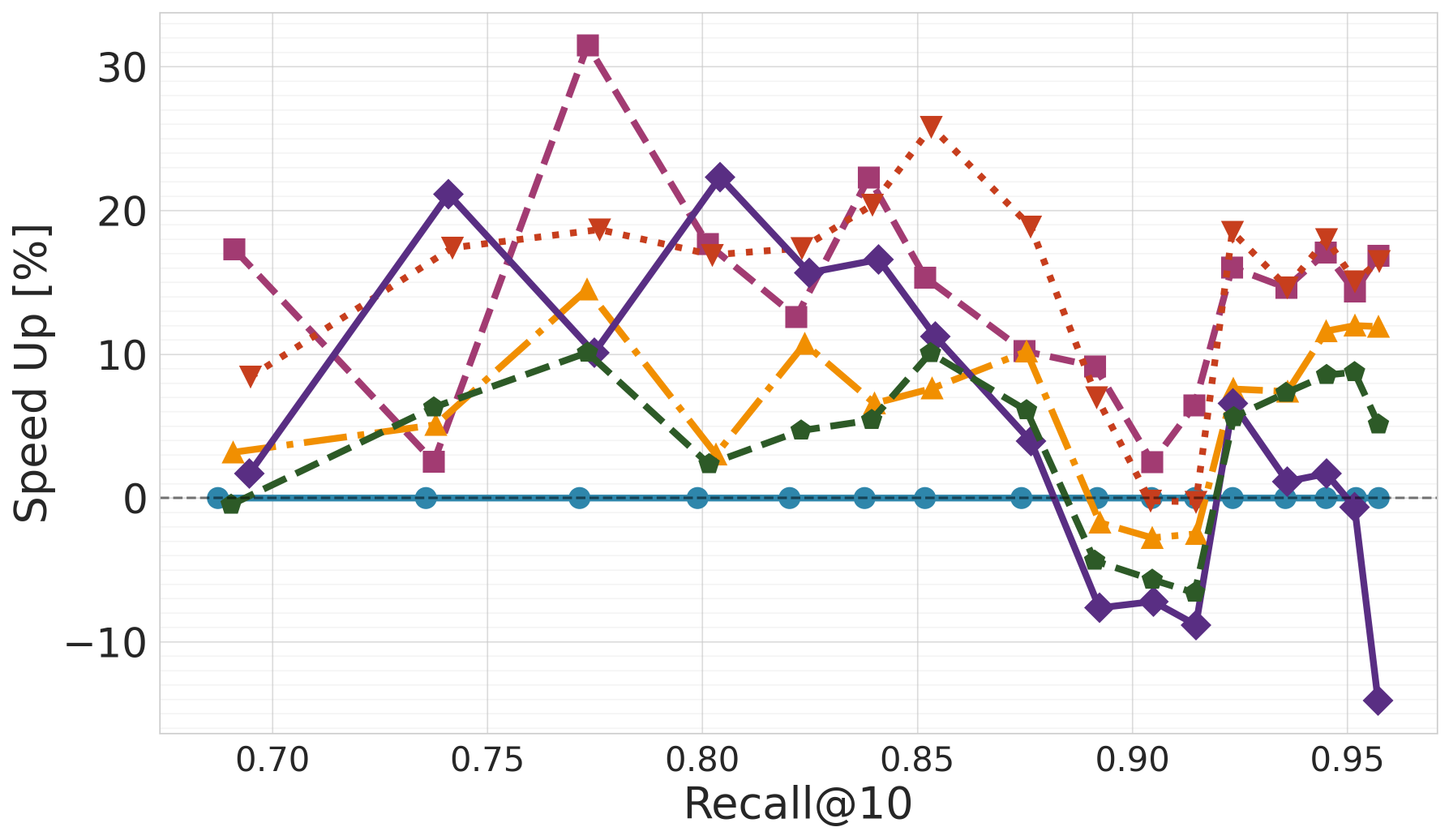}}
    &
    \subcaptionbox{NN-Descent, Deep
    40M}[0.24\linewidth]{\includegraphics[width=\linewidth]{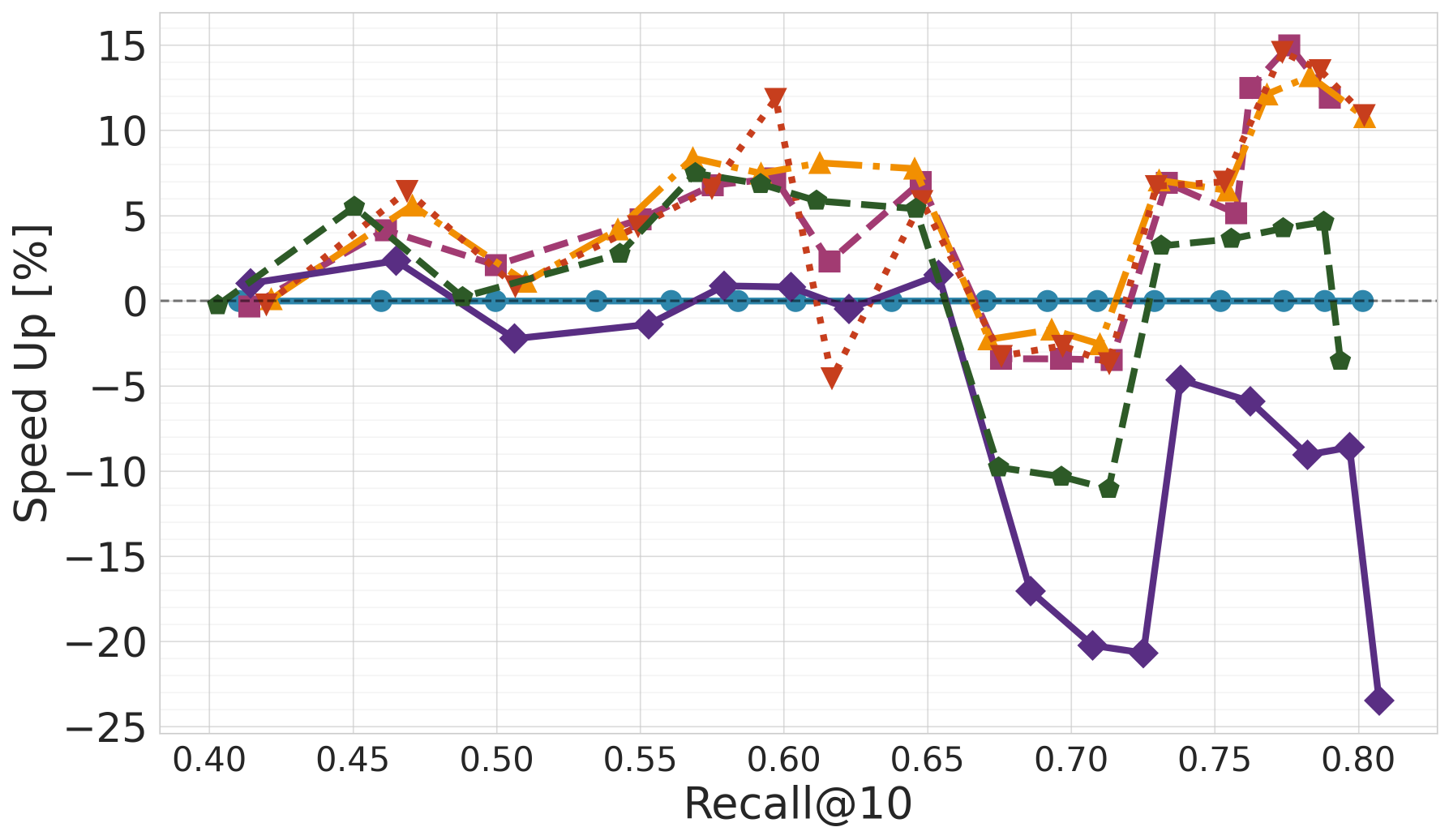}}
  \end{tabular}
  \vspace{0.1in}
  \begin{tabular}{c}
    \includegraphics[width=\textwidth]{fig/reordering_algorithms_legend.pdf}
  \end{tabular}
  \caption{Recall--QPS trade-off with reordering on NVIDIA A100 (Part
  II: large-scale and modern datasets).}
  \label{fig:supp-reordering-A100-part2}
\end{figure*}

\begin{figure*}[h]
  \centering
  \setlength{\tabcolsep}{2pt}
  \renewcommand{\arraystretch}{0}
  \begin{tabular}{cccc}
    \subcaptionbox{CAGRA, Deep
    1M}[0.24\linewidth]{\includegraphics[width=\linewidth]{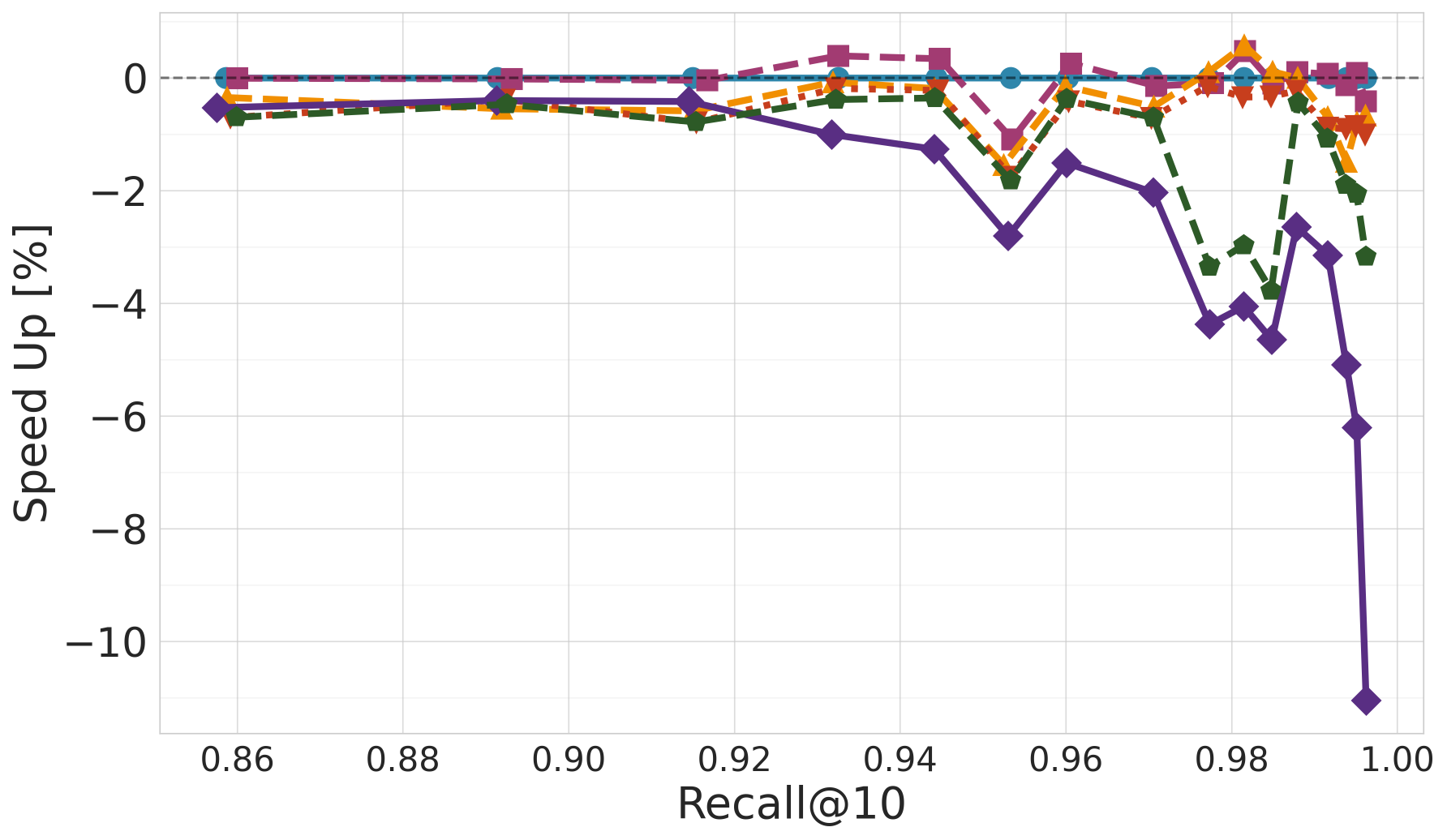}}
    &
    \subcaptionbox{NSG, Deep
    1M}[0.24\linewidth]{\includegraphics[width=\linewidth]{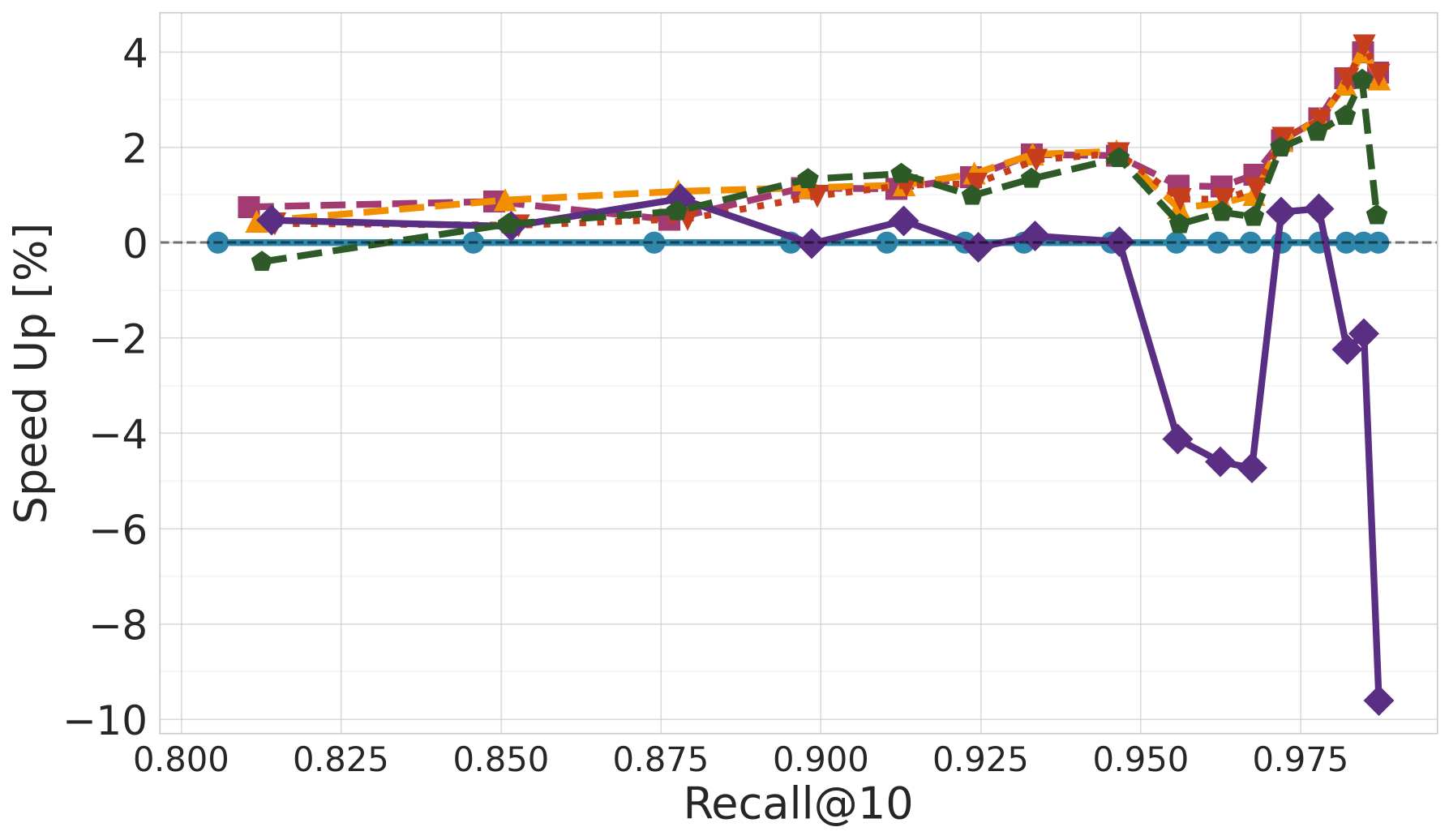}}
    &
    \subcaptionbox{Vamana, Deep
    1M}[0.24\linewidth]{\includegraphics[width=\linewidth]{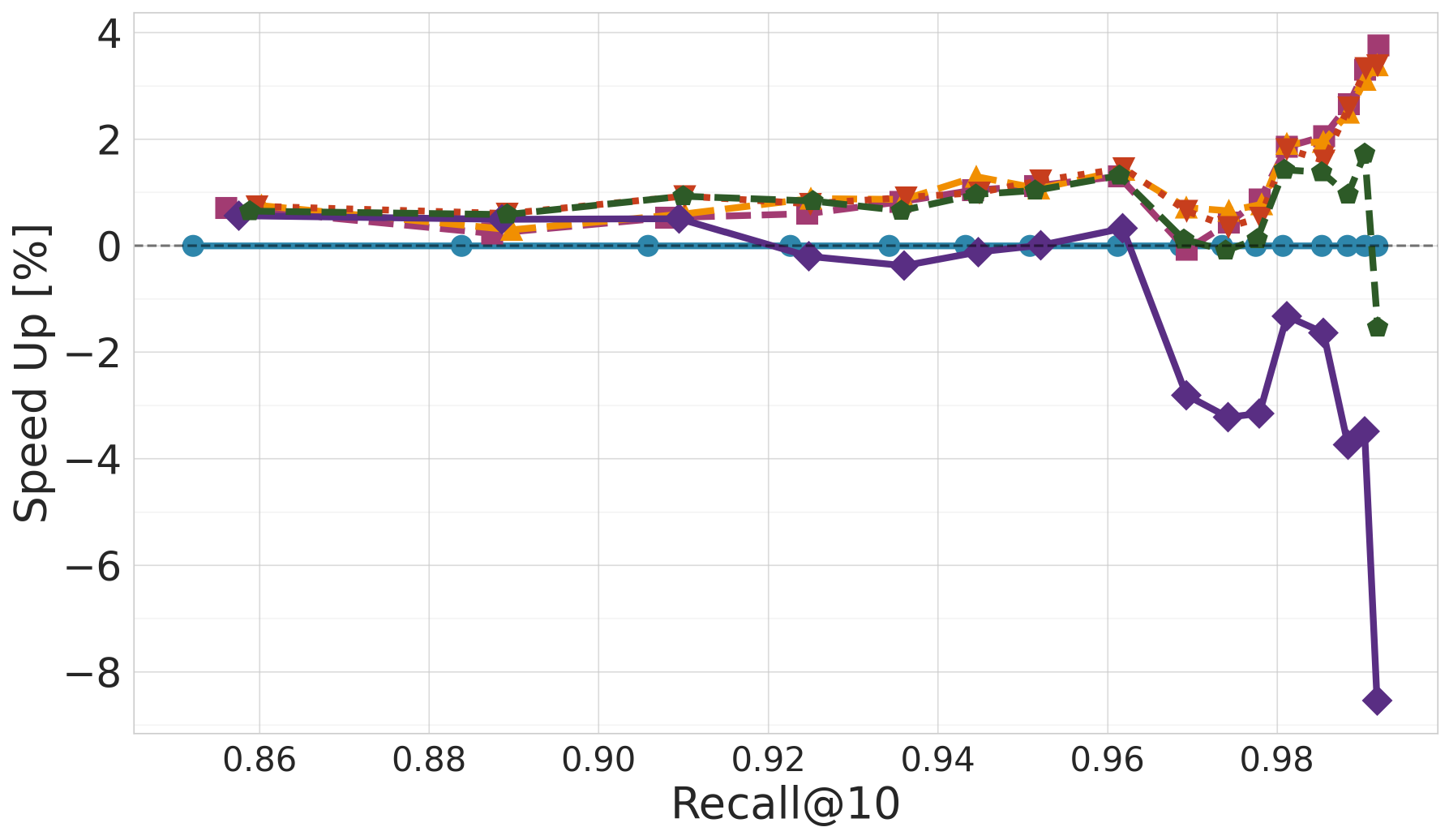}}
    &
    \subcaptionbox{NN-Descent, Deep
    1M}[0.24\linewidth]{\includegraphics[width=\linewidth]{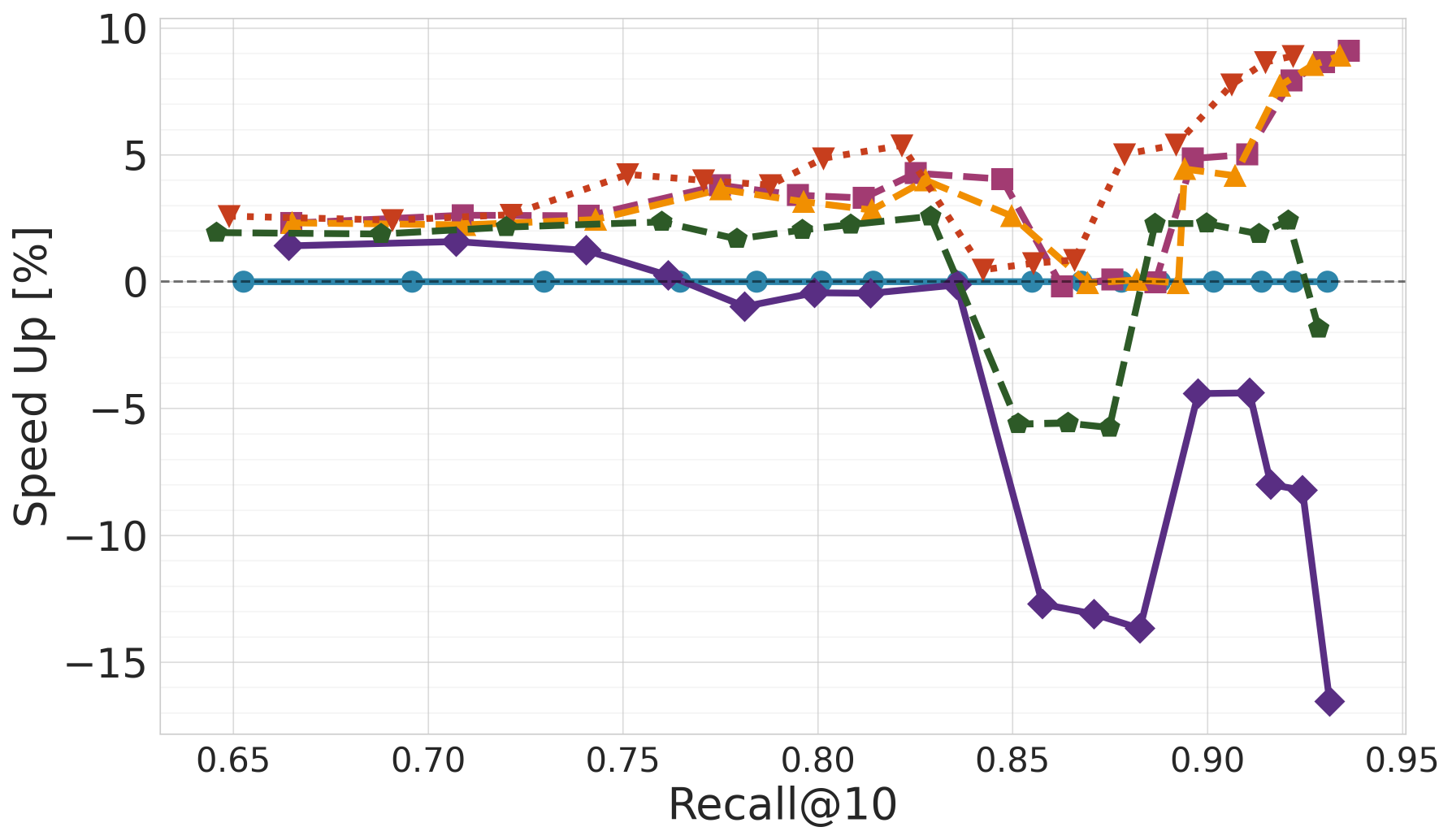}}
    \\
    \subcaptionbox{CAGRA, SIFT
    1M}[0.24\linewidth]{\includegraphics[width=\linewidth]{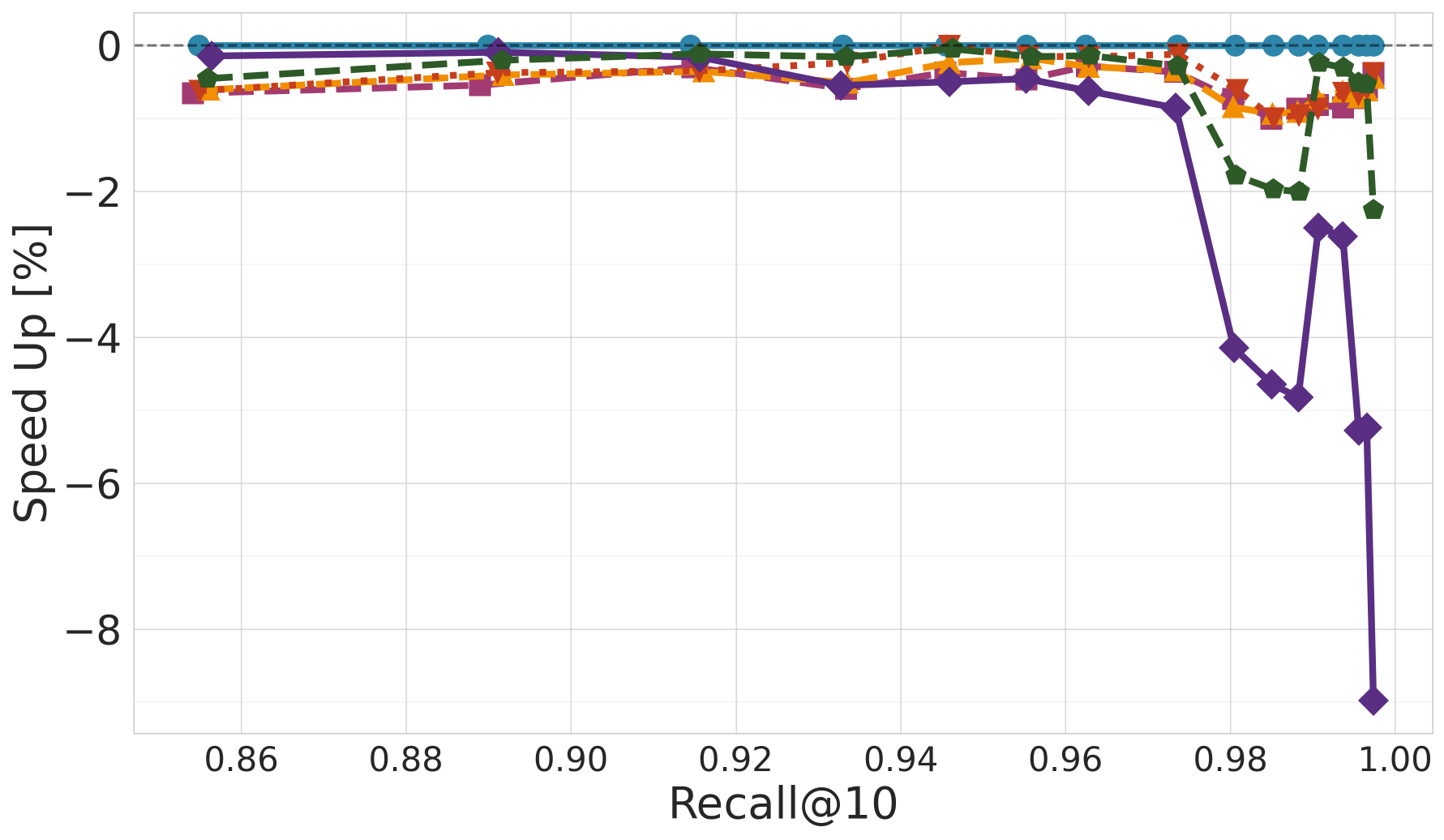}}
    &
    \subcaptionbox{NSG, SIFT
    1M}[0.24\linewidth]{\includegraphics[width=\linewidth]{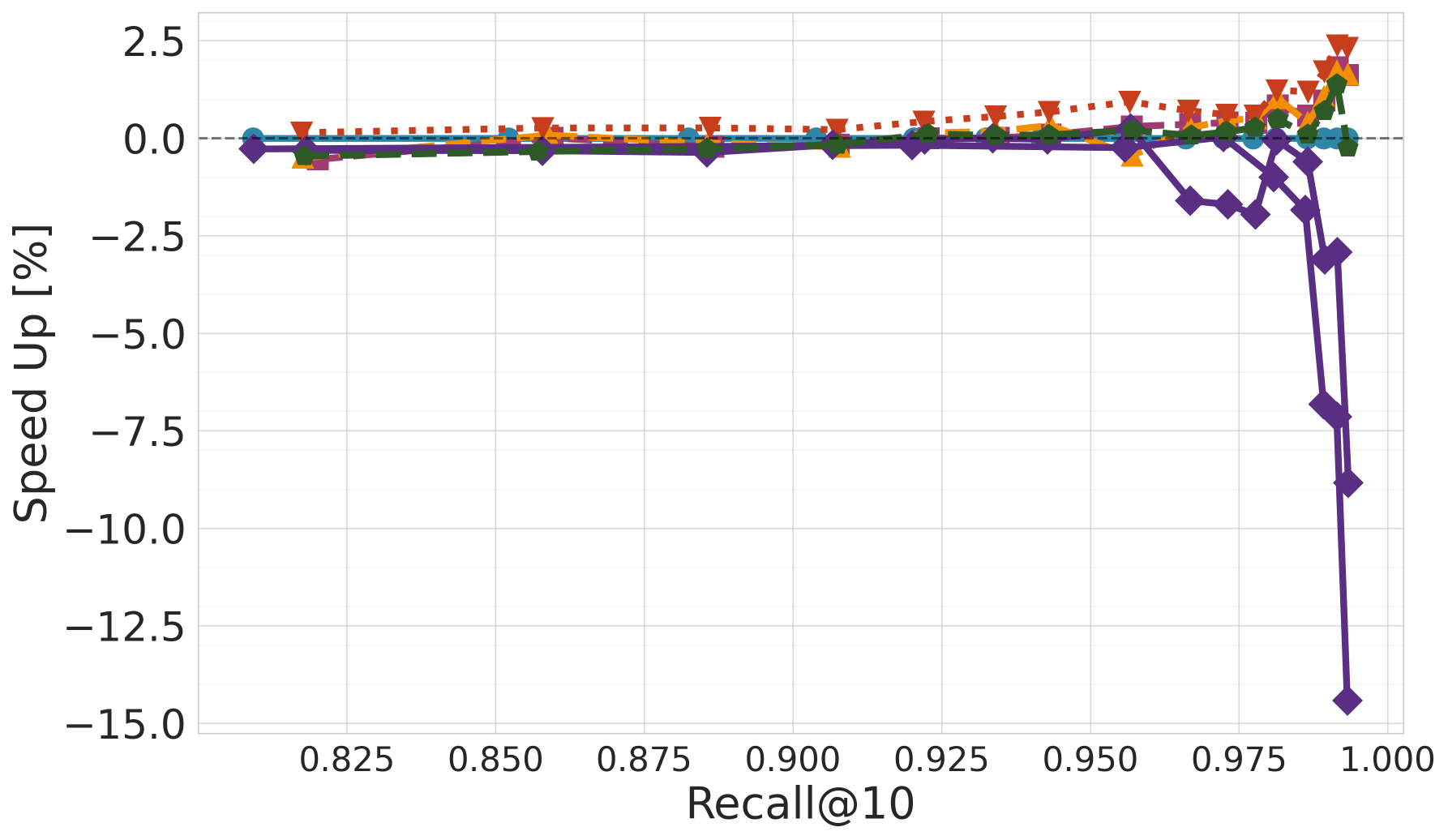}}
    &
    \subcaptionbox{Vamana, SIFT
    1M}[0.24\linewidth]{\includegraphics[width=\linewidth]{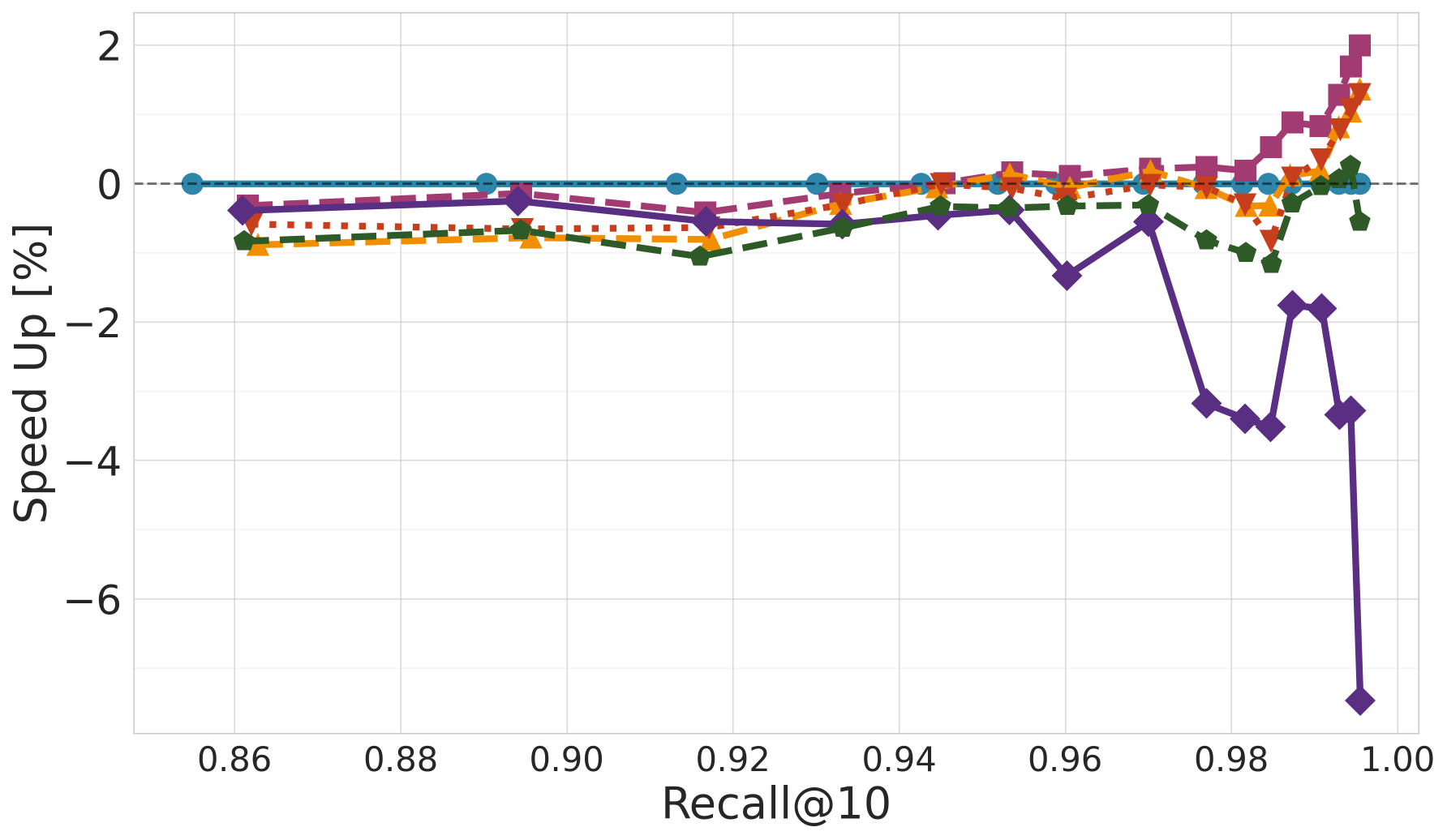}}
    &
    \subcaptionbox{NN-Descent, SIFT
    1M}[0.24\linewidth]{\includegraphics[width=\linewidth]{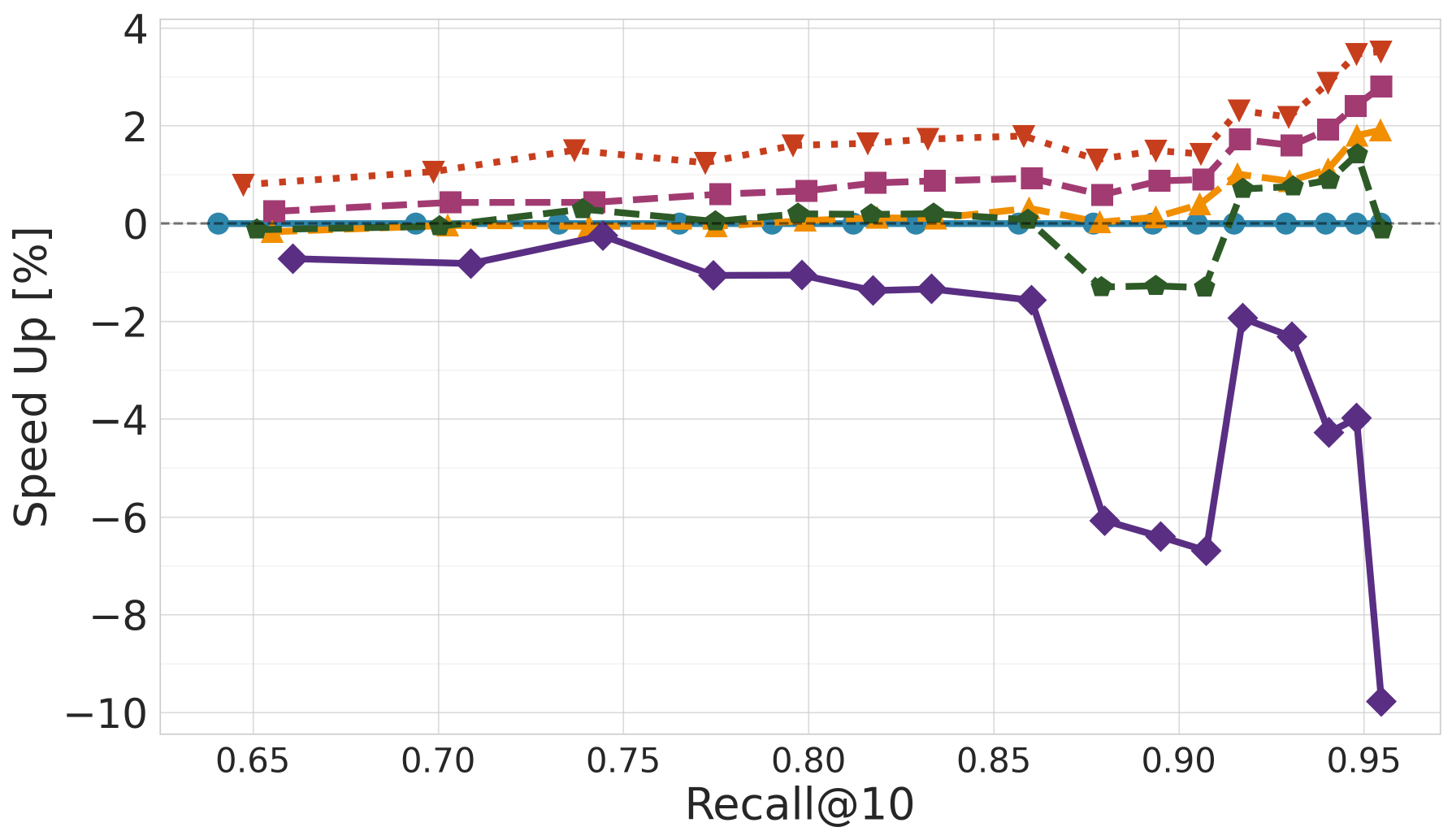}}
    \\
    \subcaptionbox{CAGRA, GIST
    1M}[0.24\linewidth]{\includegraphics[width=\linewidth]{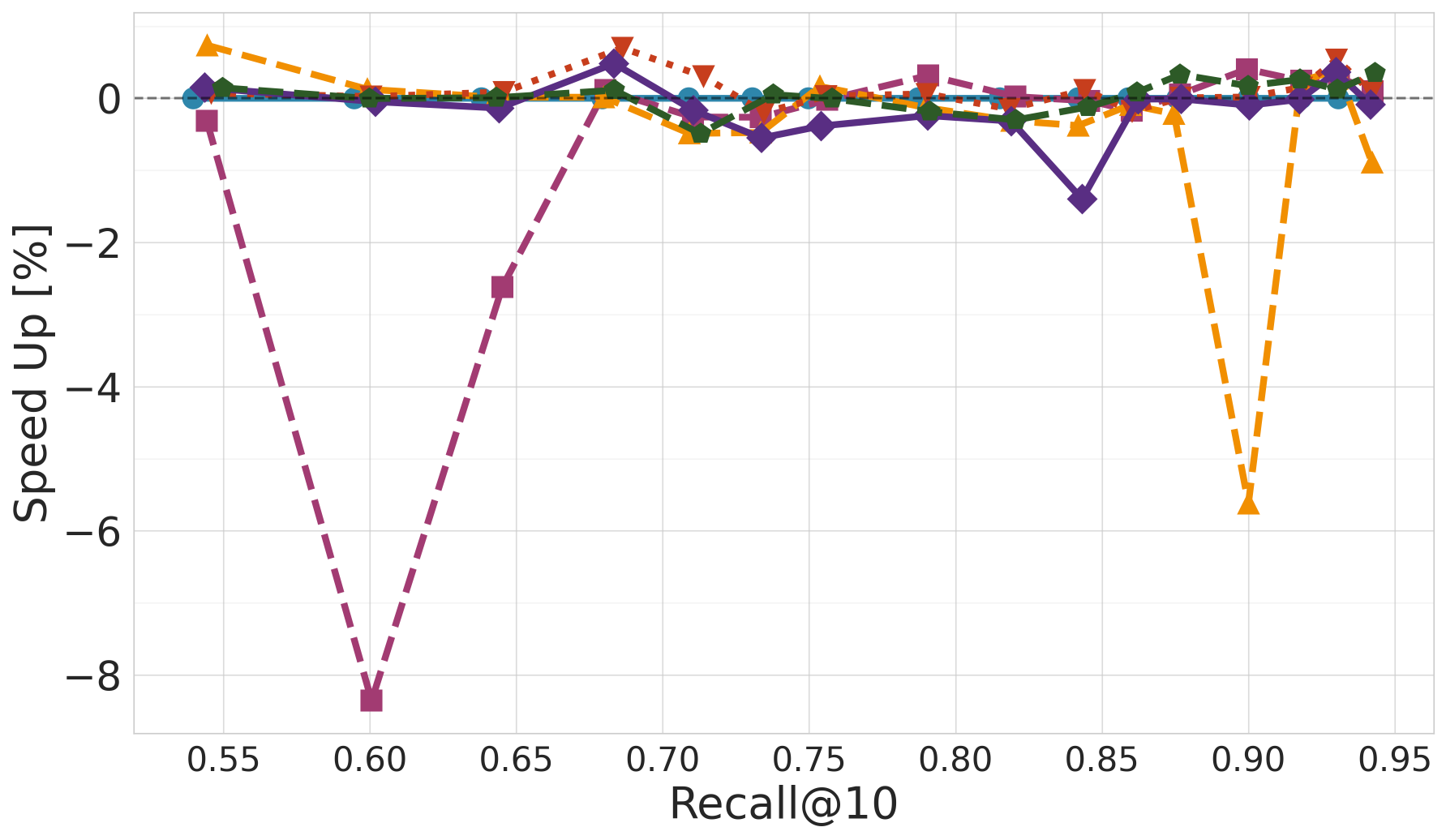}}
    &
    \subcaptionbox{NSG, GIST
    1M}[0.24\linewidth]{\includegraphics[width=\linewidth]{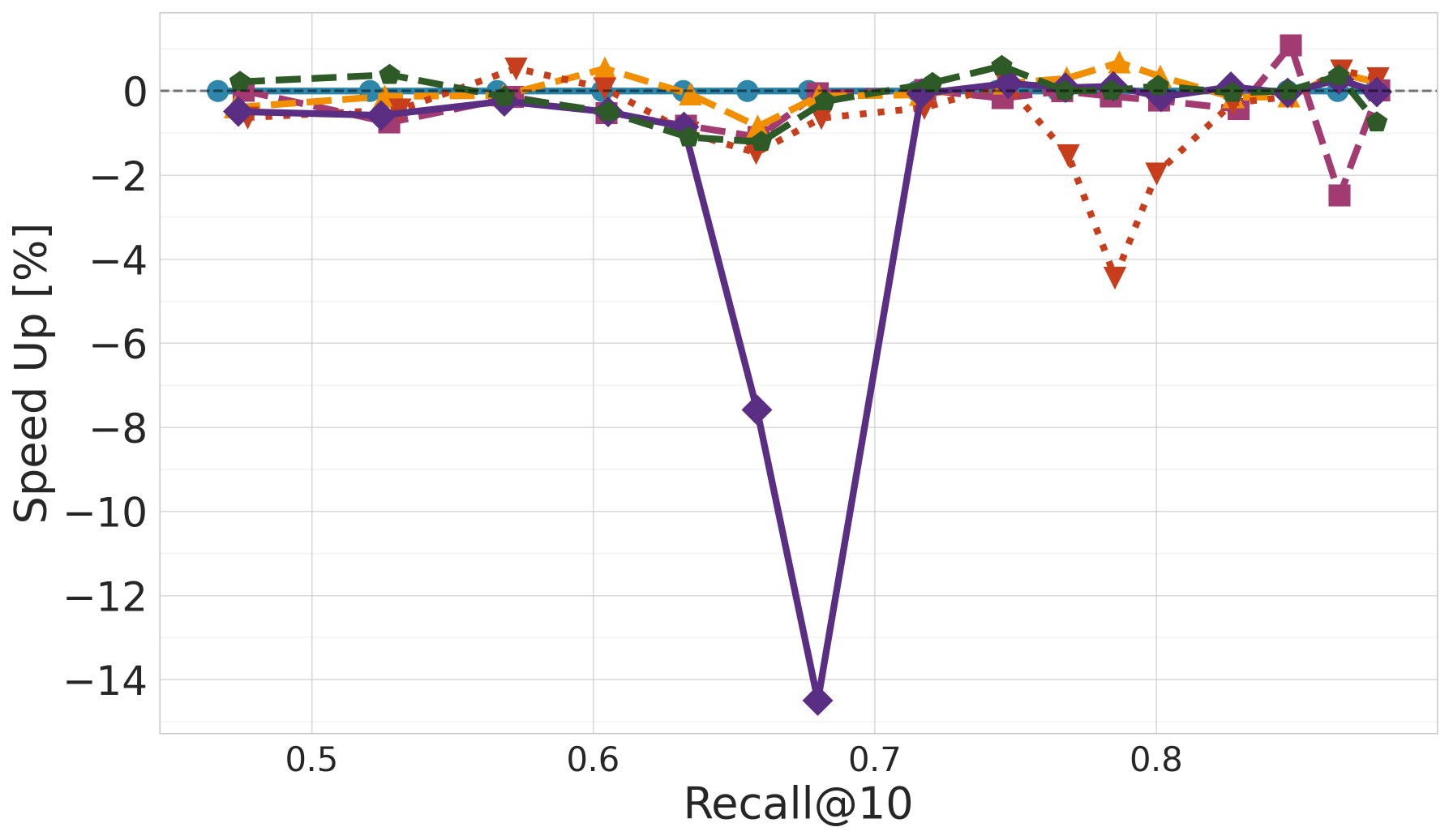}}
    &
    \subcaptionbox{Vamana, GIST
    1M}[0.24\linewidth]{\includegraphics[width=\linewidth]{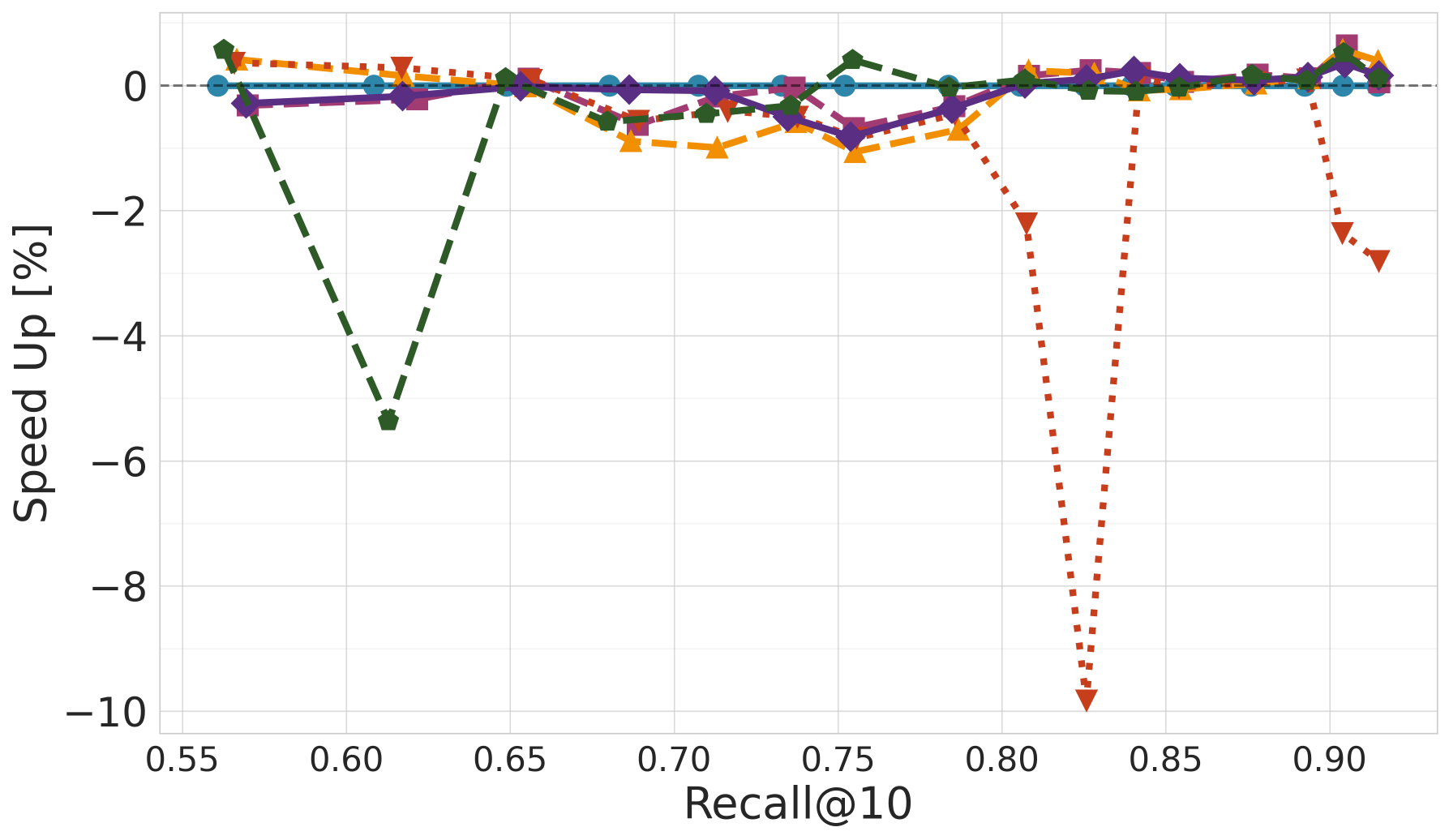}}
    &
    \subcaptionbox{NN-Descent, GIST
    1M}[0.24\linewidth]{\includegraphics[width=\linewidth]{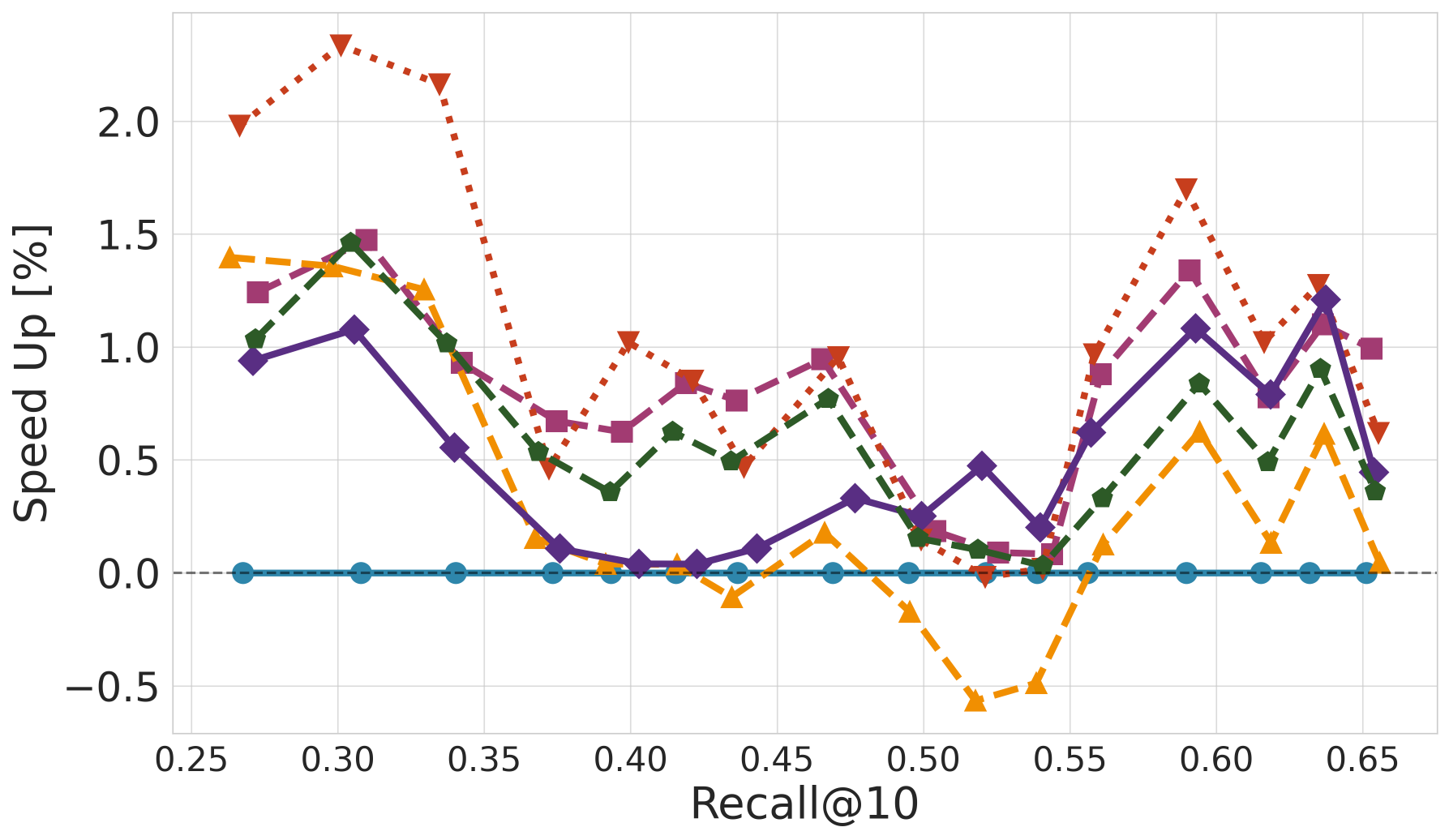}}
    \\
    \subcaptionbox{CAGRA, OpenAI
    1M}[0.24\linewidth]{\includegraphics[width=\linewidth]{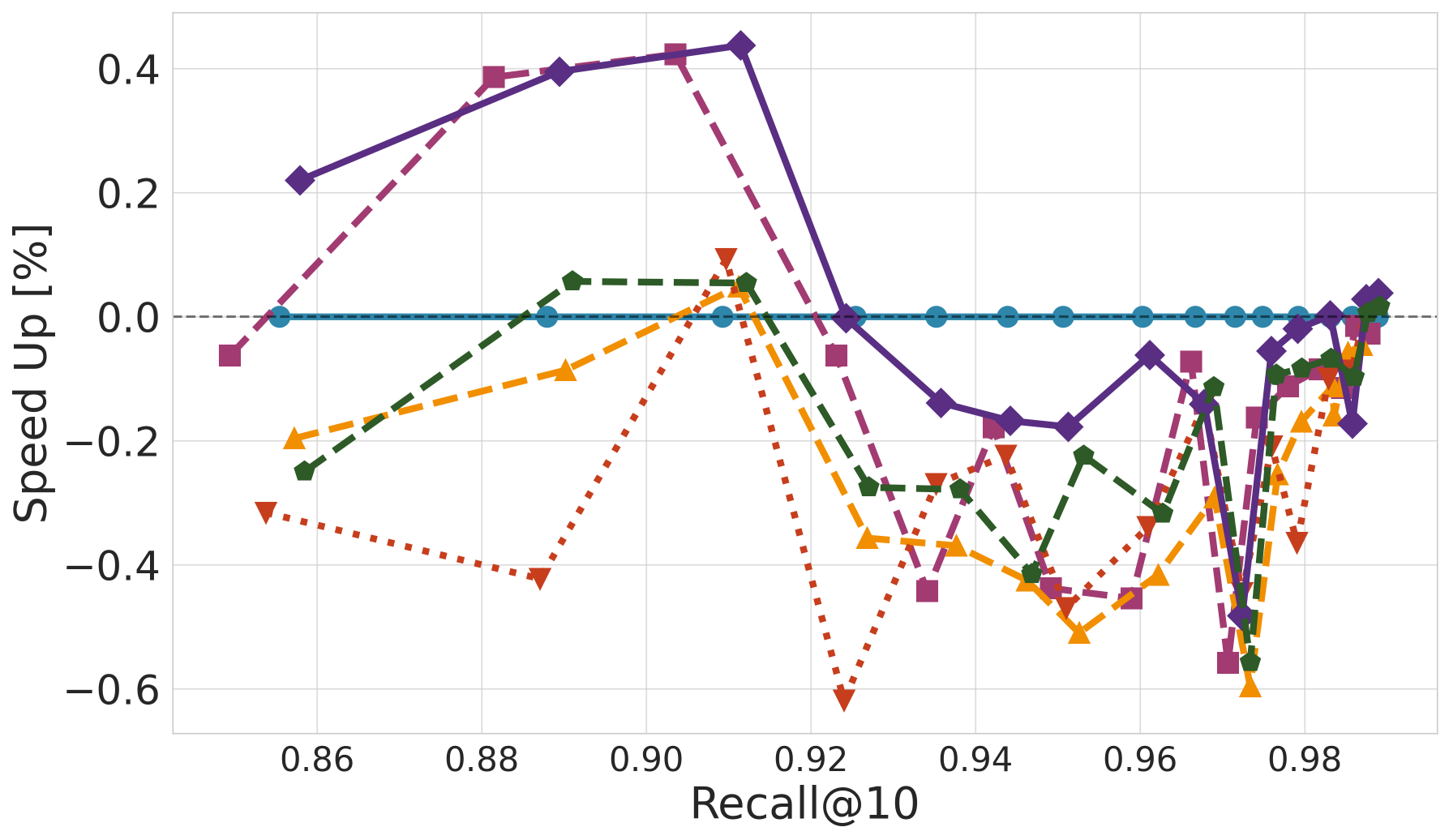}}
    &
    \subcaptionbox{NSG, OpenAI
    1M}[0.24\linewidth]{\includegraphics[width=\linewidth]{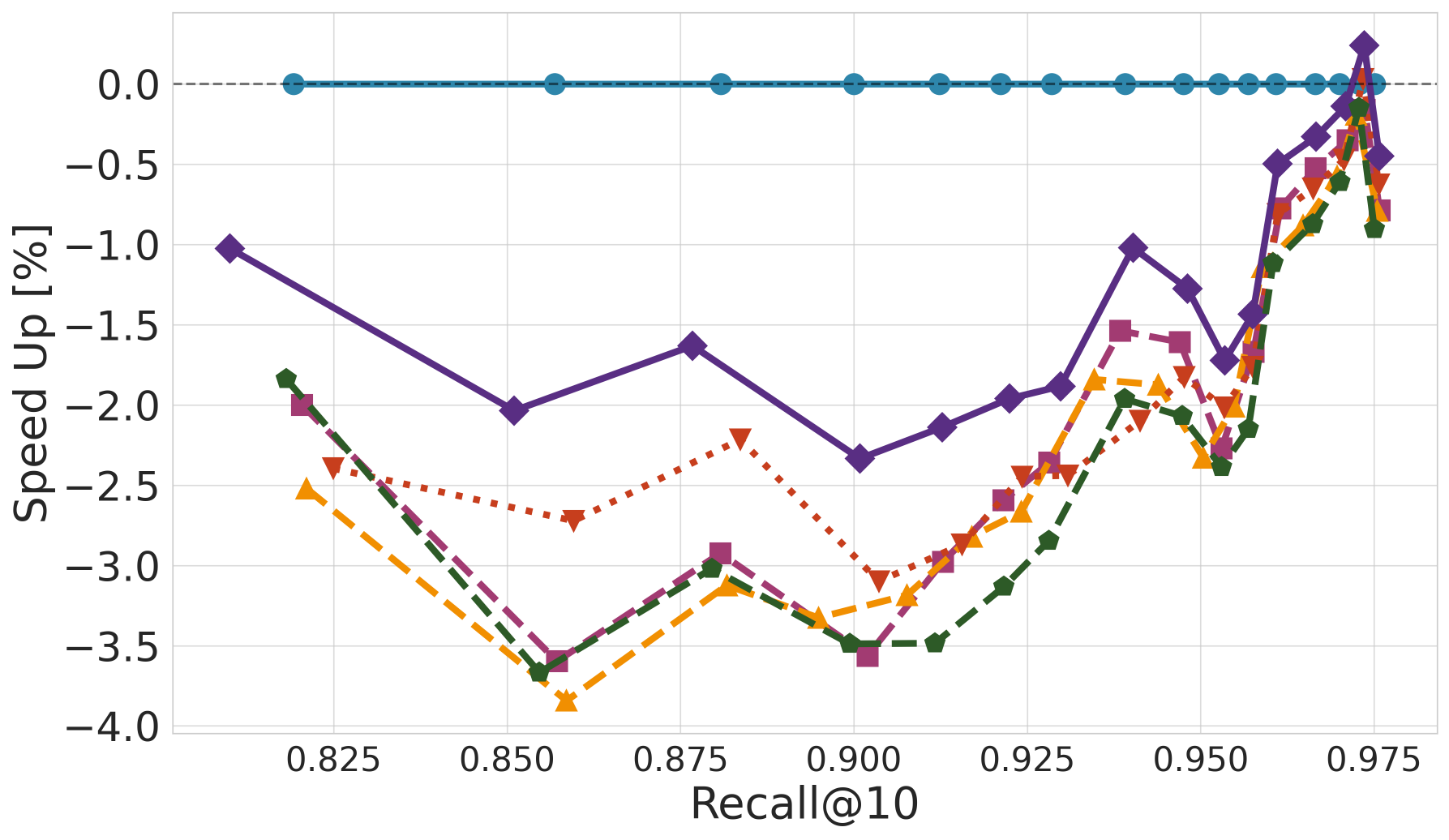}}
    &
    \subcaptionbox{Vamana, OpenAI
    1M}[0.24\linewidth]{\includegraphics[width=\linewidth]{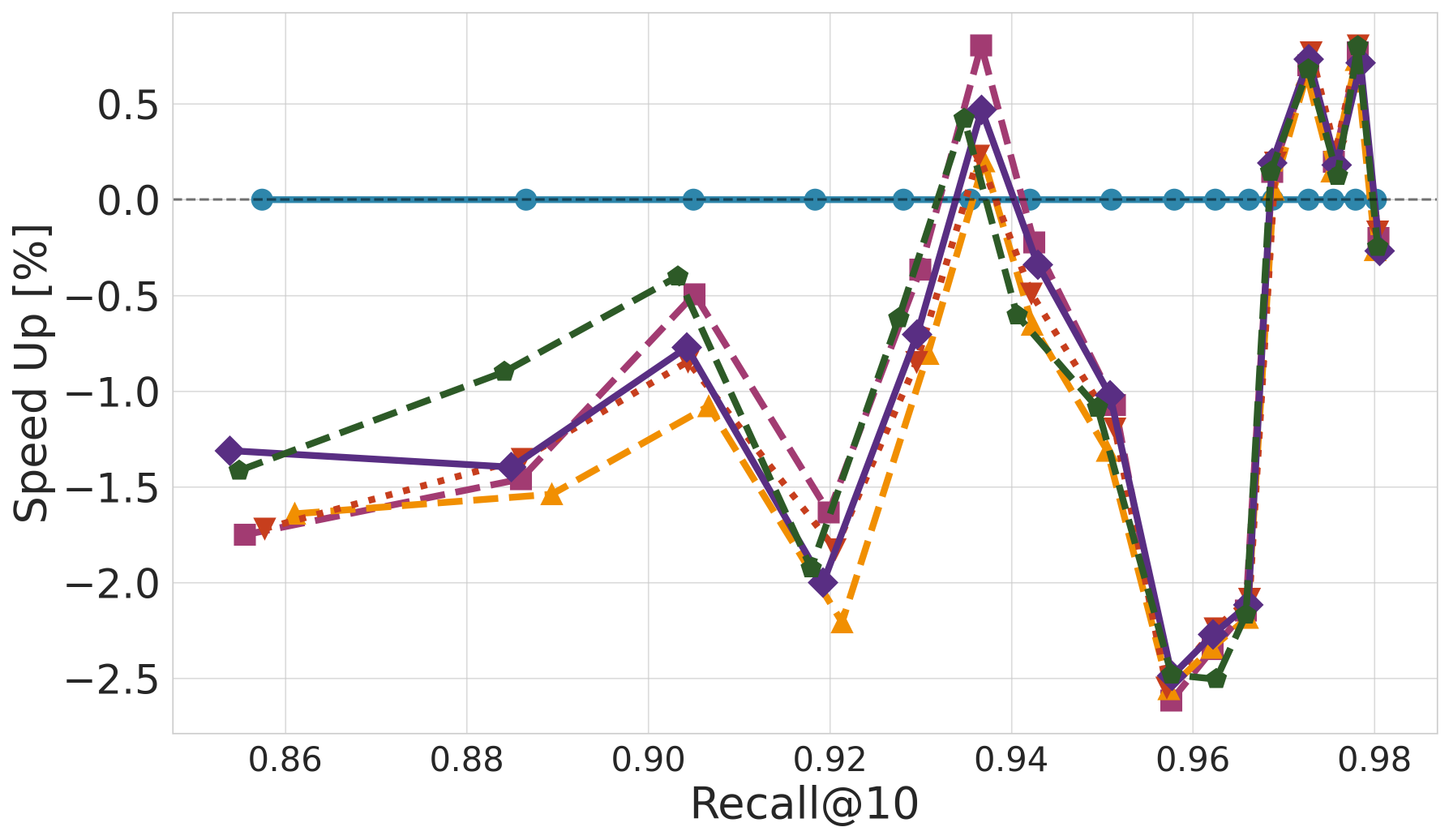}}
    &
    \subcaptionbox{NN-Descent, OpenAI
    1M}[0.24\linewidth]{\includegraphics[width=\linewidth]{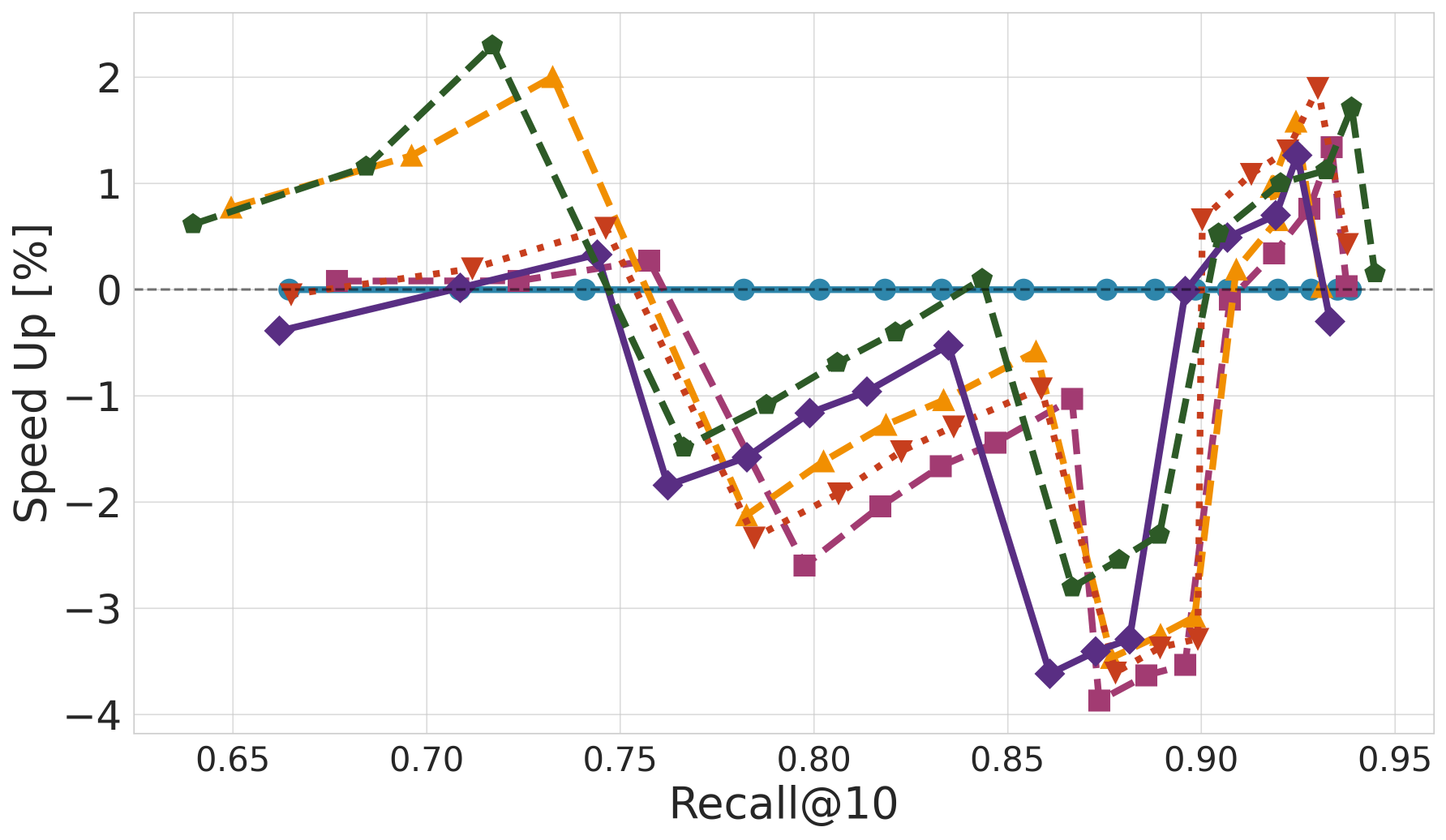}}
    \\
    \subcaptionbox{CAGRA, Yandex T2I
    1M}[0.24\linewidth]{\includegraphics[width=\linewidth]{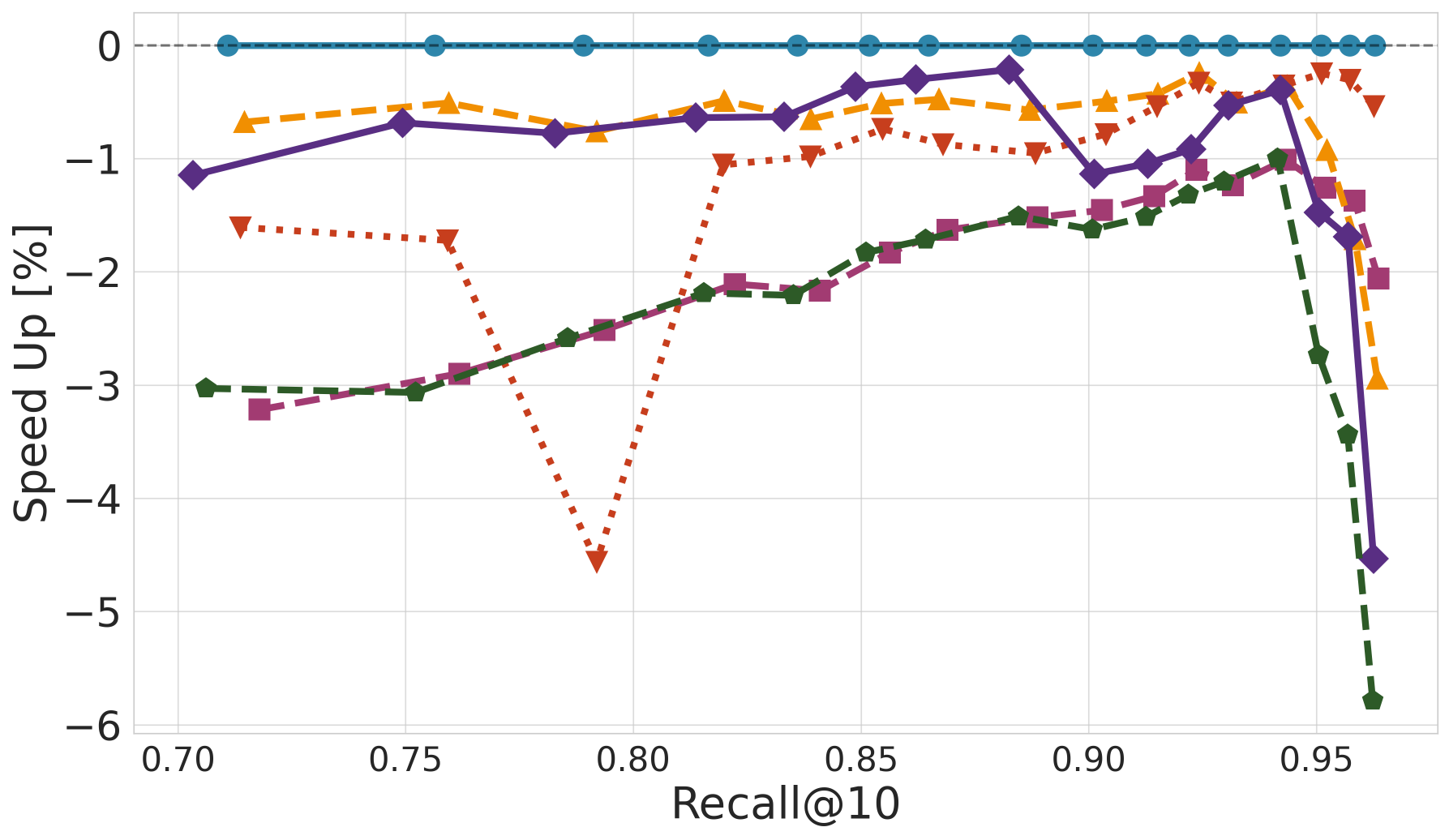}}
    &
    \subcaptionbox{NSG, Yandex T2I
    1M}[0.24\linewidth]{\includegraphics[width=\linewidth]{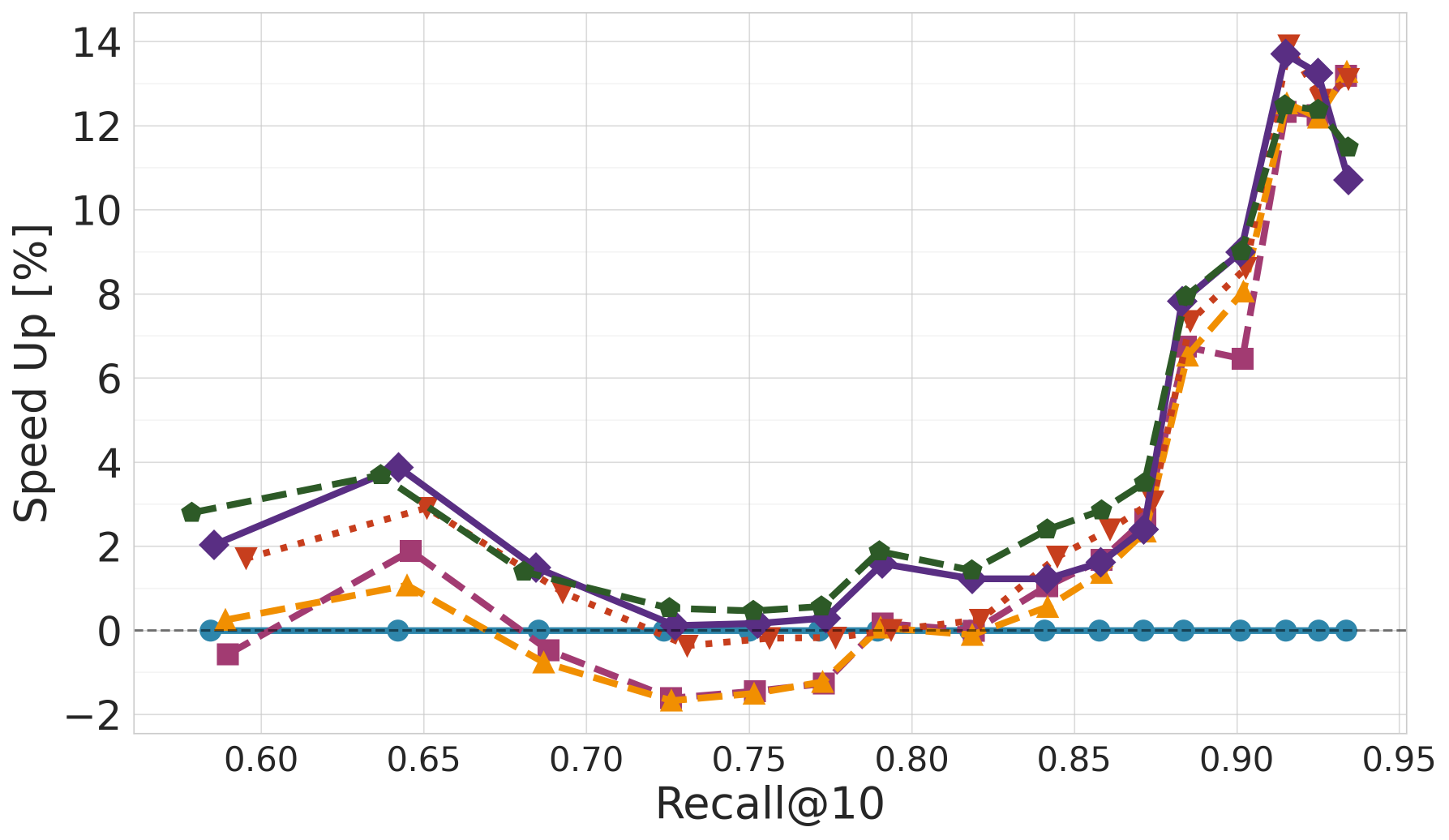}}
    &
    \subcaptionbox{Vamana, Yandex T2I
    1M}[0.24\linewidth]{\includegraphics[width=\linewidth]{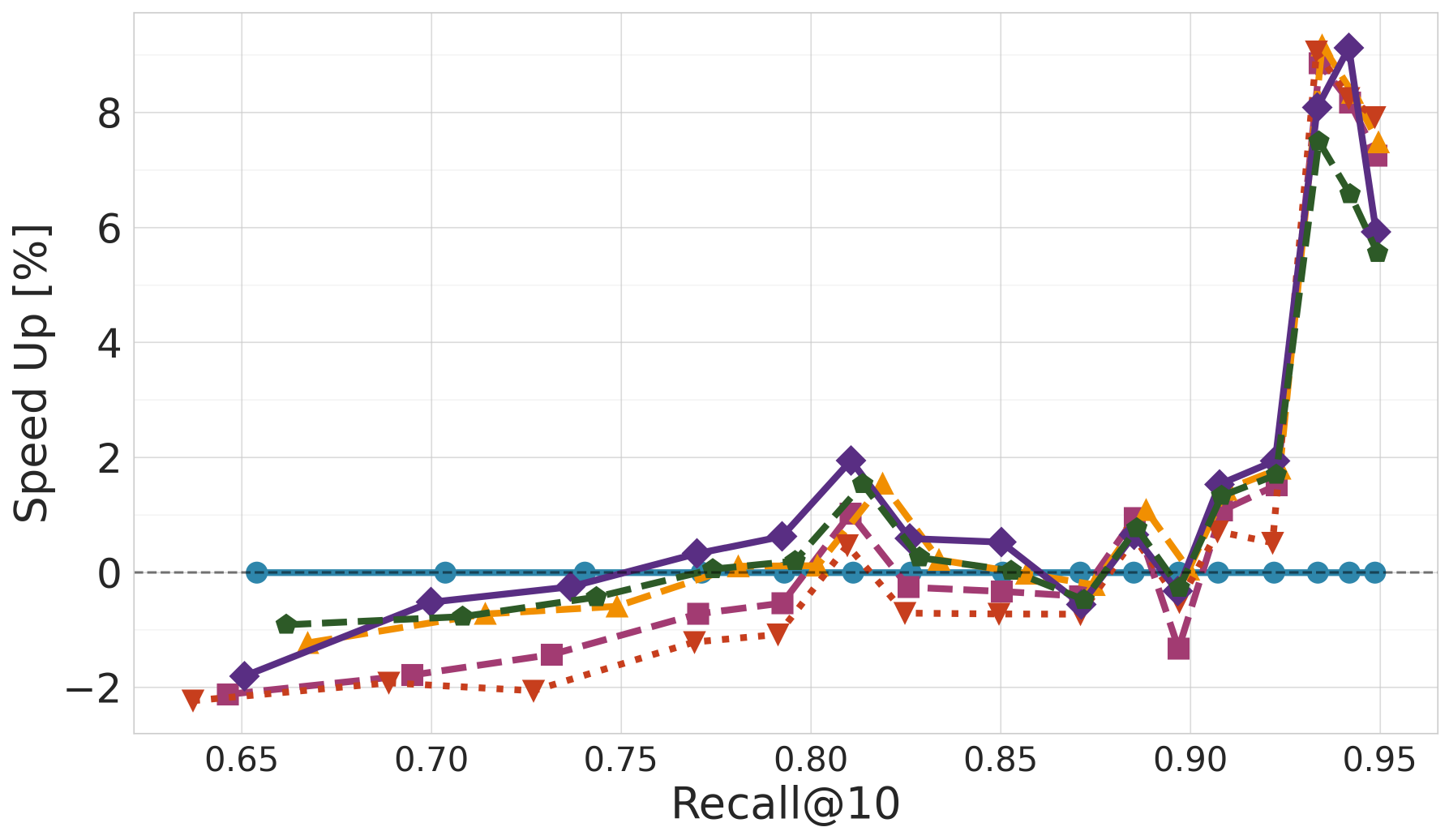}}
    &
    \subcaptionbox{NN-Descent, Yandex T2I
    1M}[0.24\linewidth]{\includegraphics[width=\linewidth]{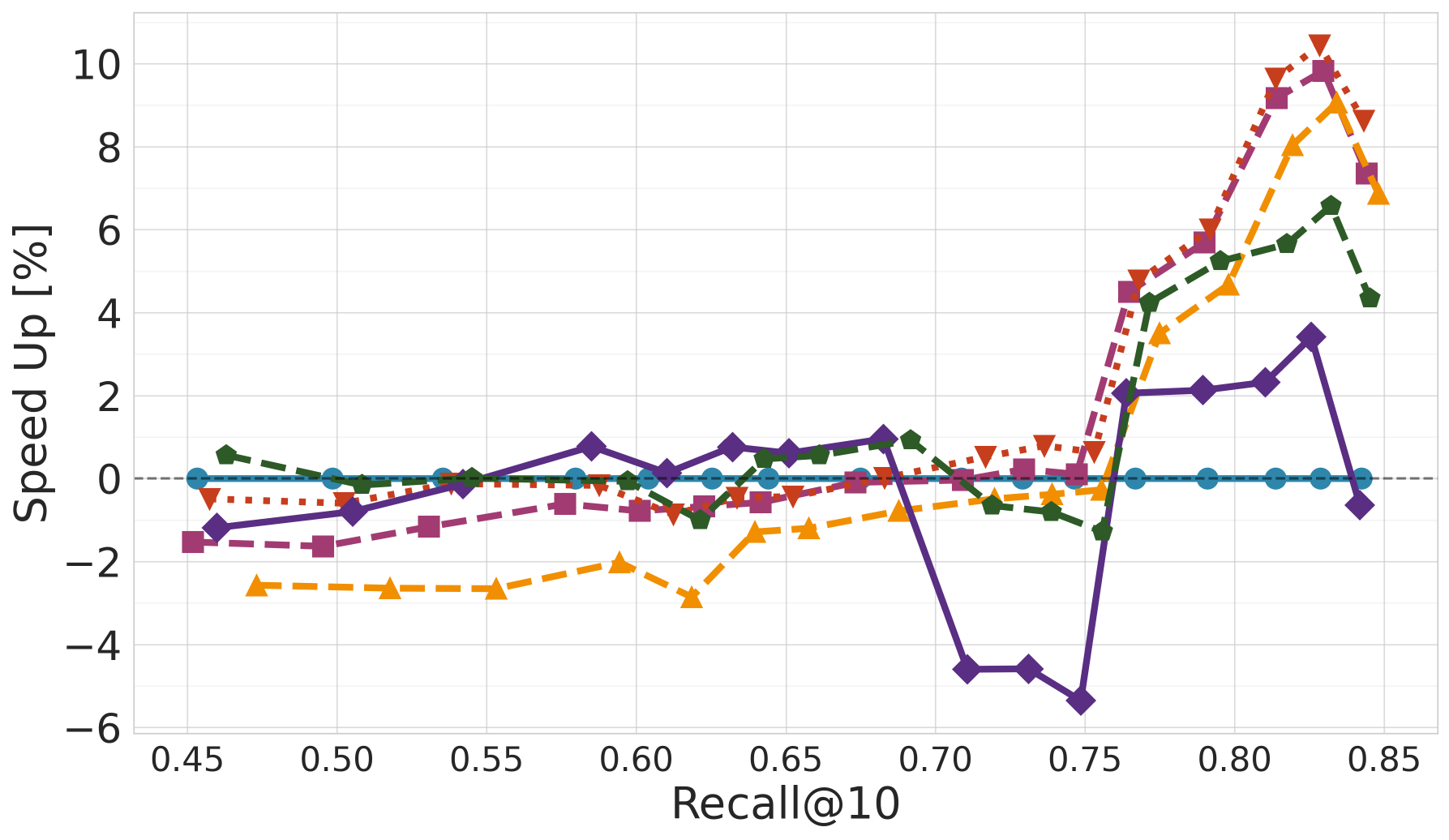}}
    \\
    \subcaptionbox{CAGRA, Deep
    10M}[0.24\linewidth]{\includegraphics[width=\linewidth]{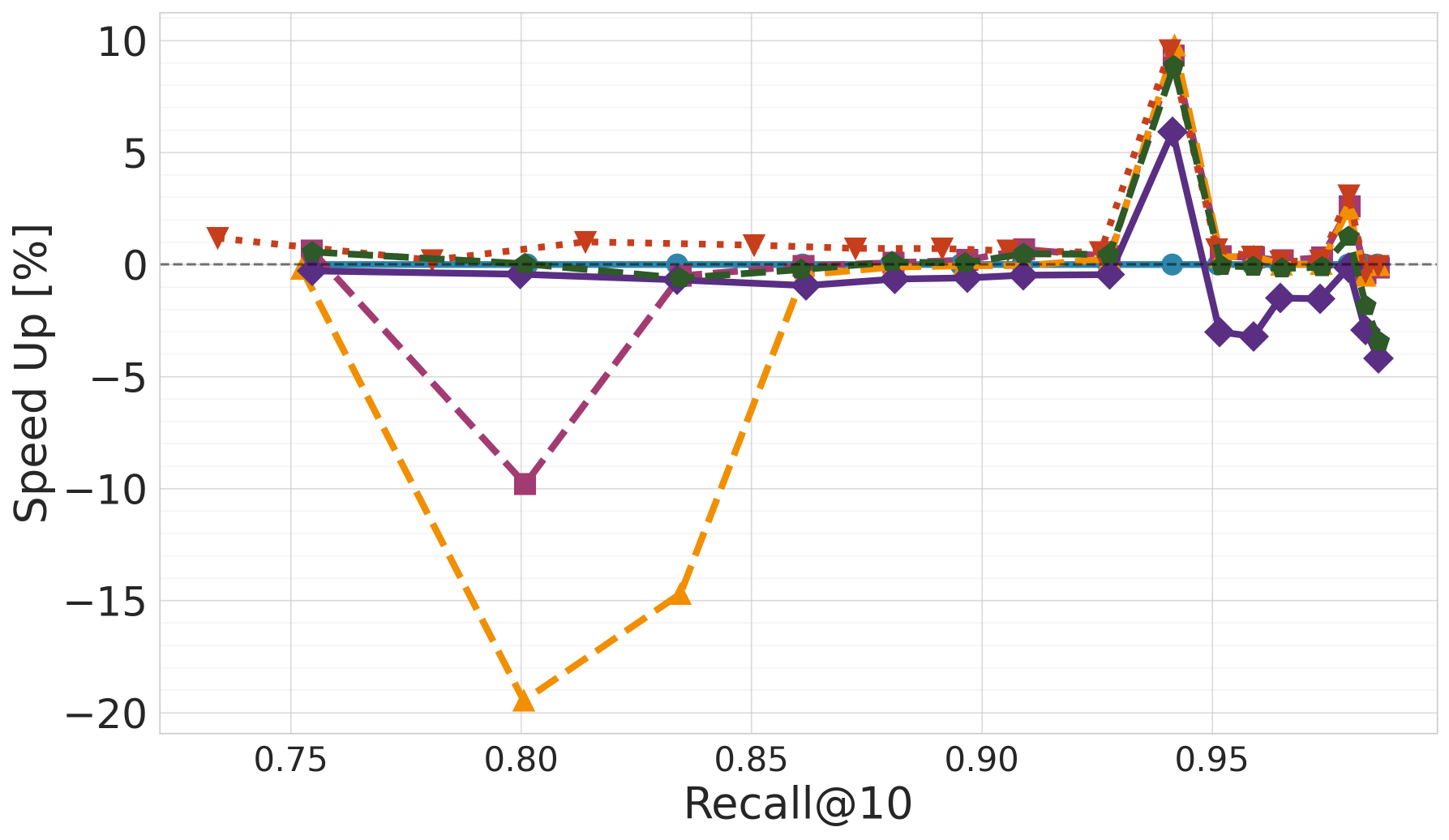}}
    &
    \subcaptionbox{NSG, Deep
    10M}[0.24\linewidth]{\includegraphics[width=\linewidth]{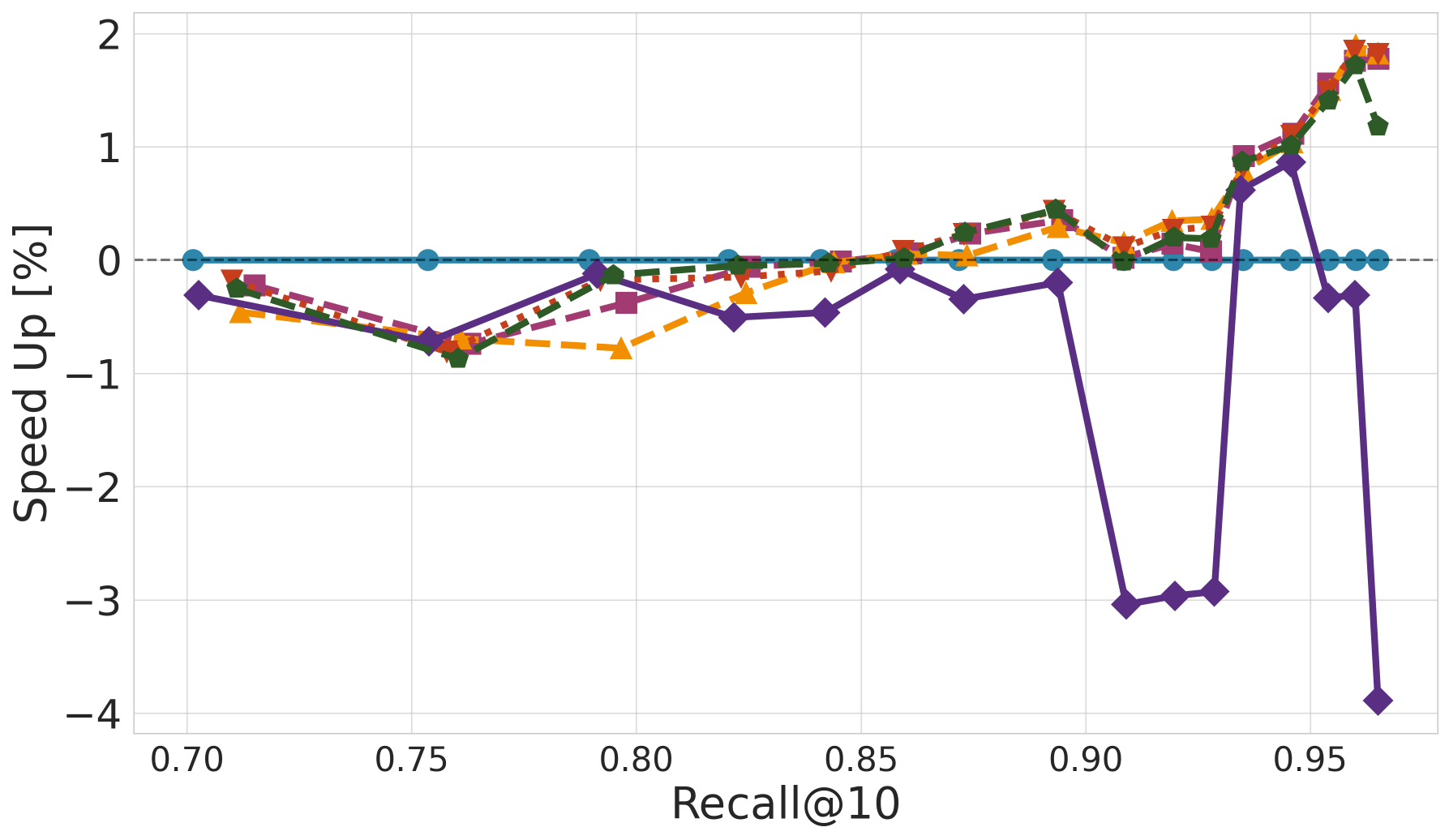}}
    &
    \subcaptionbox{Vamana, Deep
    10M}[0.24\linewidth]{\includegraphics[width=\linewidth]{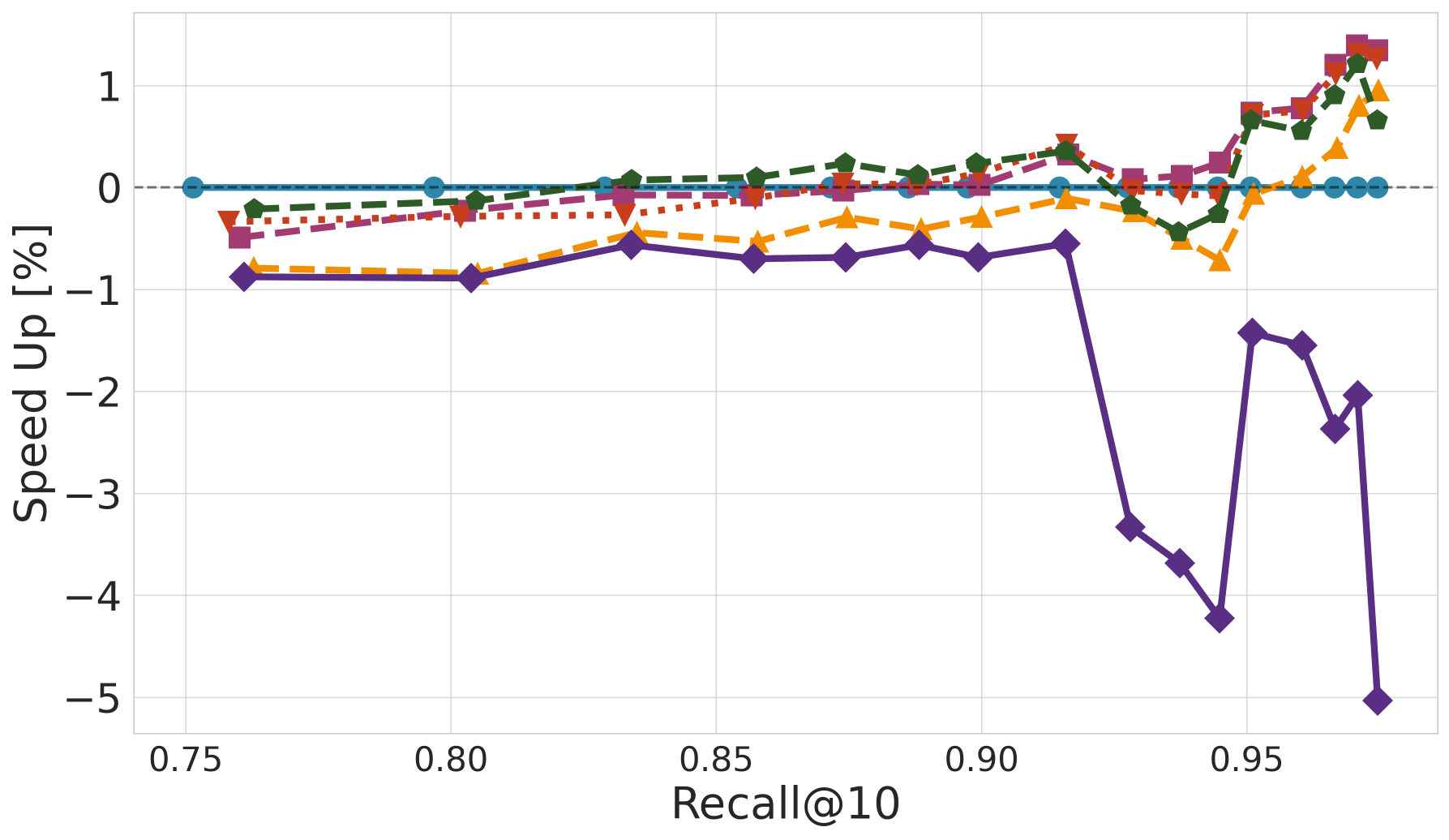}}
    &
    \subcaptionbox{NN-Descent, Deep
    10M}[0.24\linewidth]{\includegraphics[width=\linewidth]{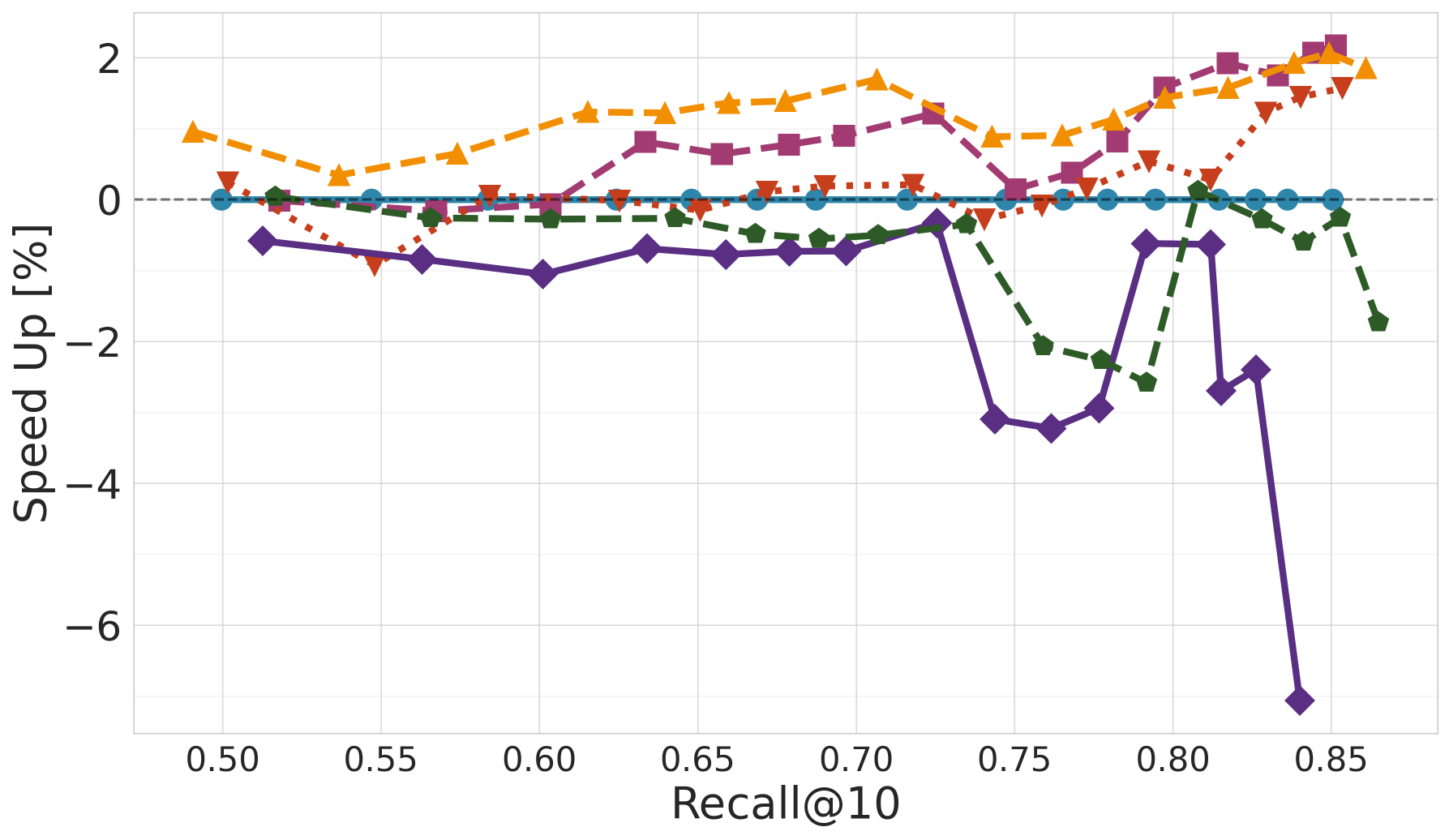}}
  \end{tabular}
  \vspace{0.1in}
  \begin{tabular}{c}
    \includegraphics[width=\textwidth]{fig/reordering_algorithms_legend.pdf}
  \end{tabular}
  \caption{Recall--QPS trade-off with reordering on NVIDIA RTX 6000
  Ada (Part I: traditional and medium-scale datasets).}
  \label{fig:supp-reordering-A6000-part1}
\end{figure*}

\begin{figure*}[h]
  \centering
  \setlength{\tabcolsep}{2pt}
  \renewcommand{\arraystretch}{0}
  \begin{tabular}{cccc}
    \subcaptionbox{CAGRA, Wikipedia
    1M}[0.24\linewidth]{\includegraphics[width=\linewidth]{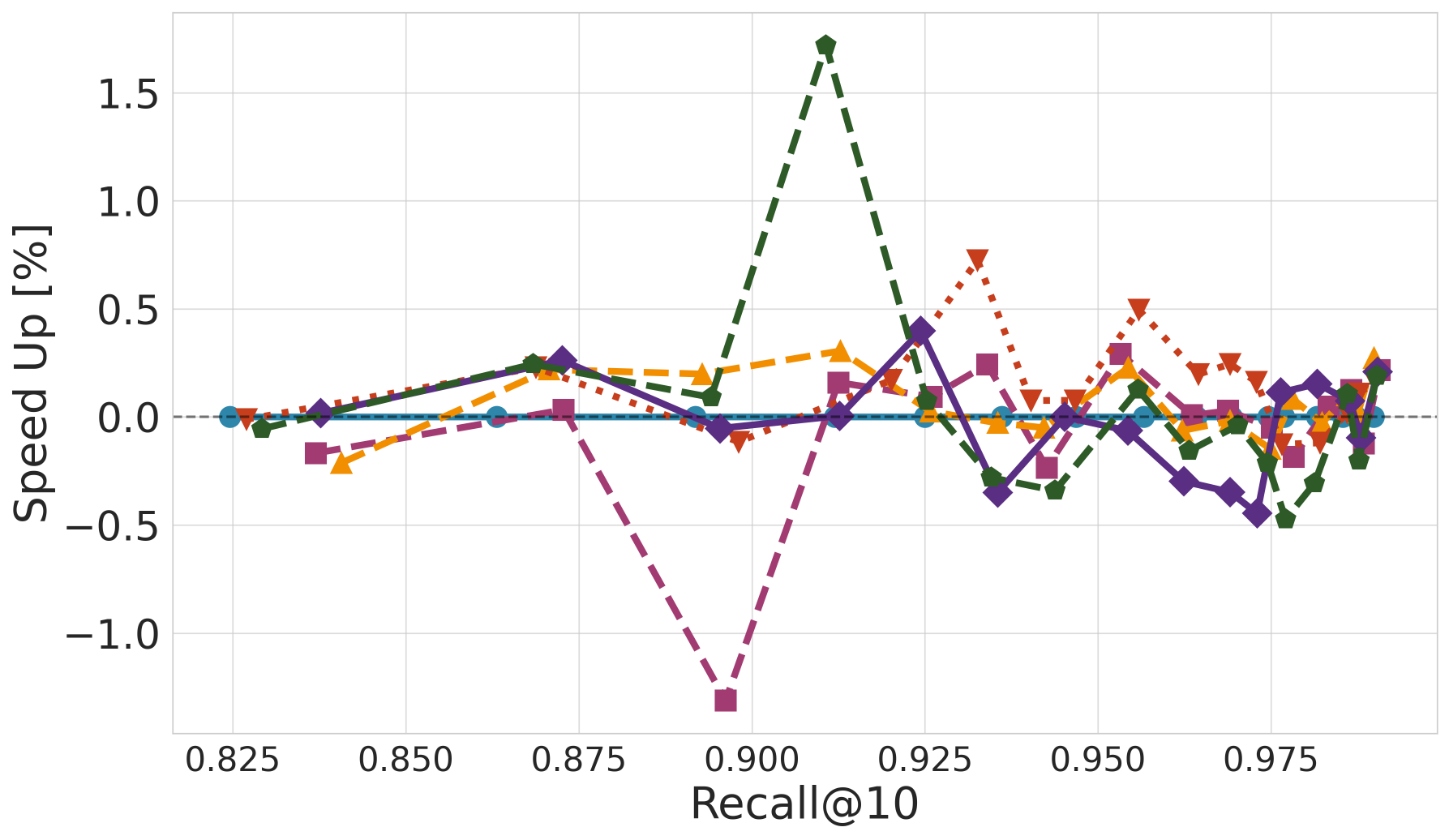}}
    &
    \subcaptionbox{NSG, Wikipedia
    1M}[0.24\linewidth]{\includegraphics[width=\linewidth]{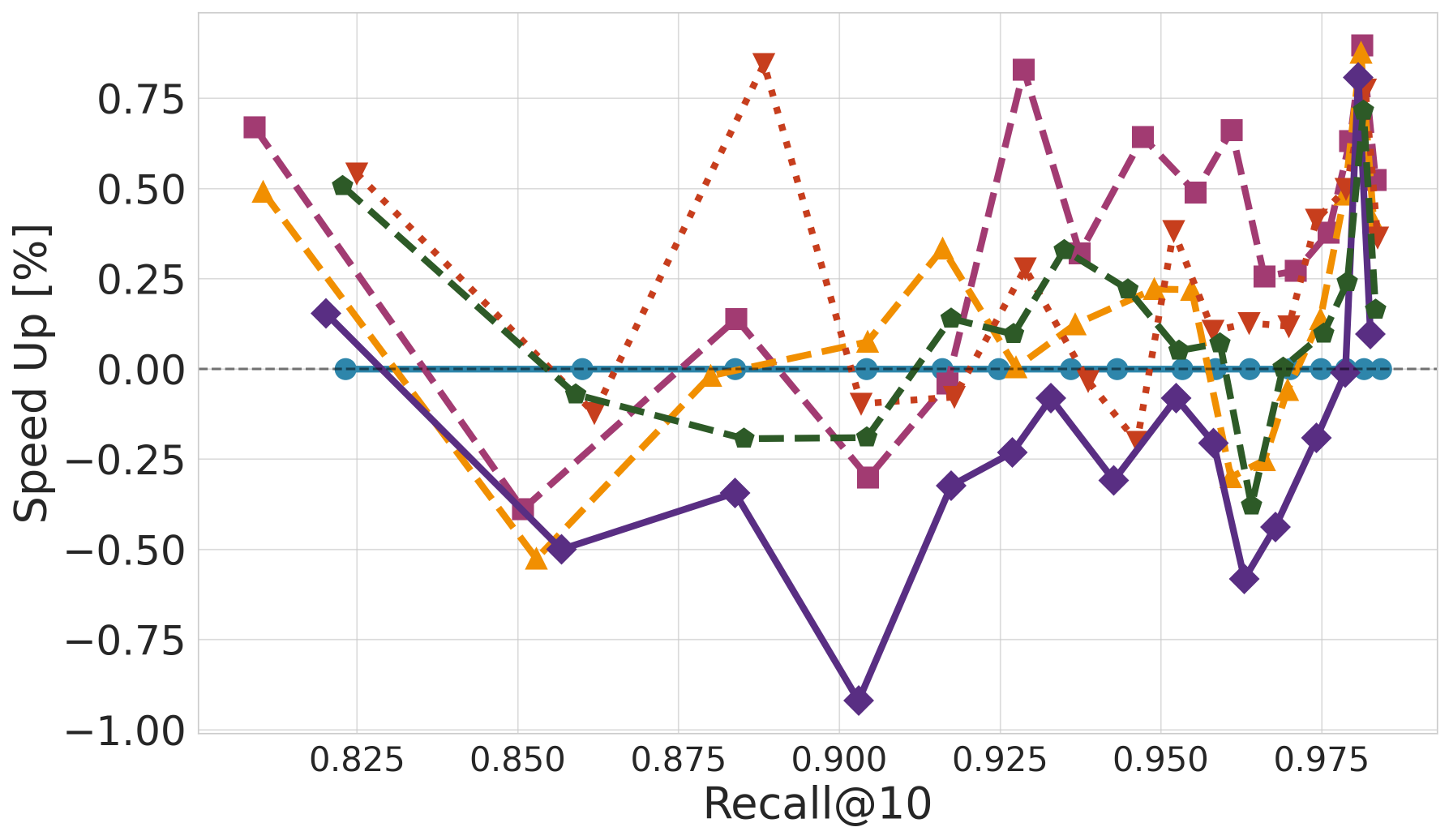}}
    &
    \subcaptionbox{Vamana, Wikipedia
    1M}[0.24\linewidth]{\includegraphics[width=\linewidth]{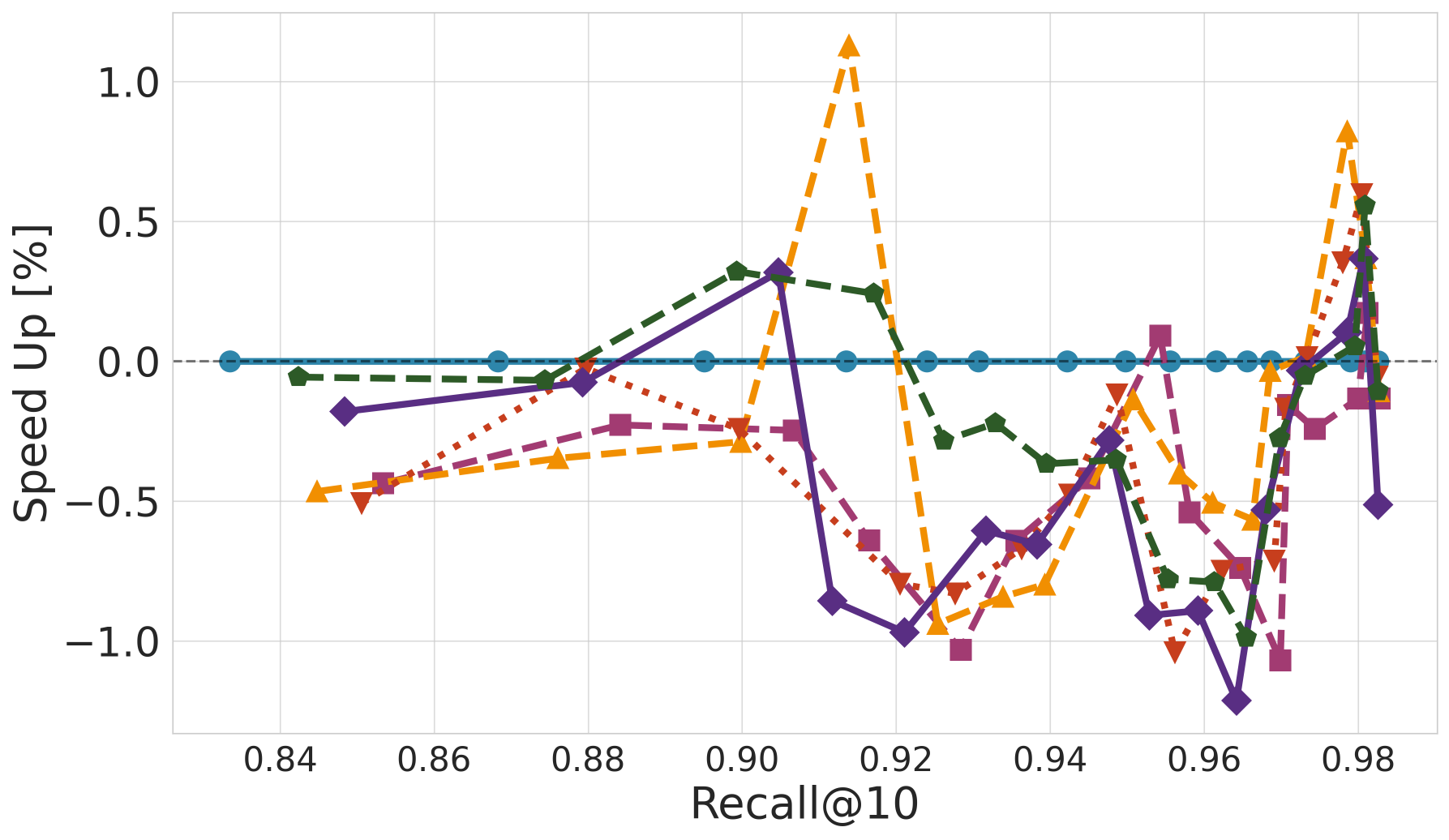}}
    &
    \subcaptionbox{NN-Descent, Wikipedia
    1M}[0.24\linewidth]{\includegraphics[width=\linewidth]{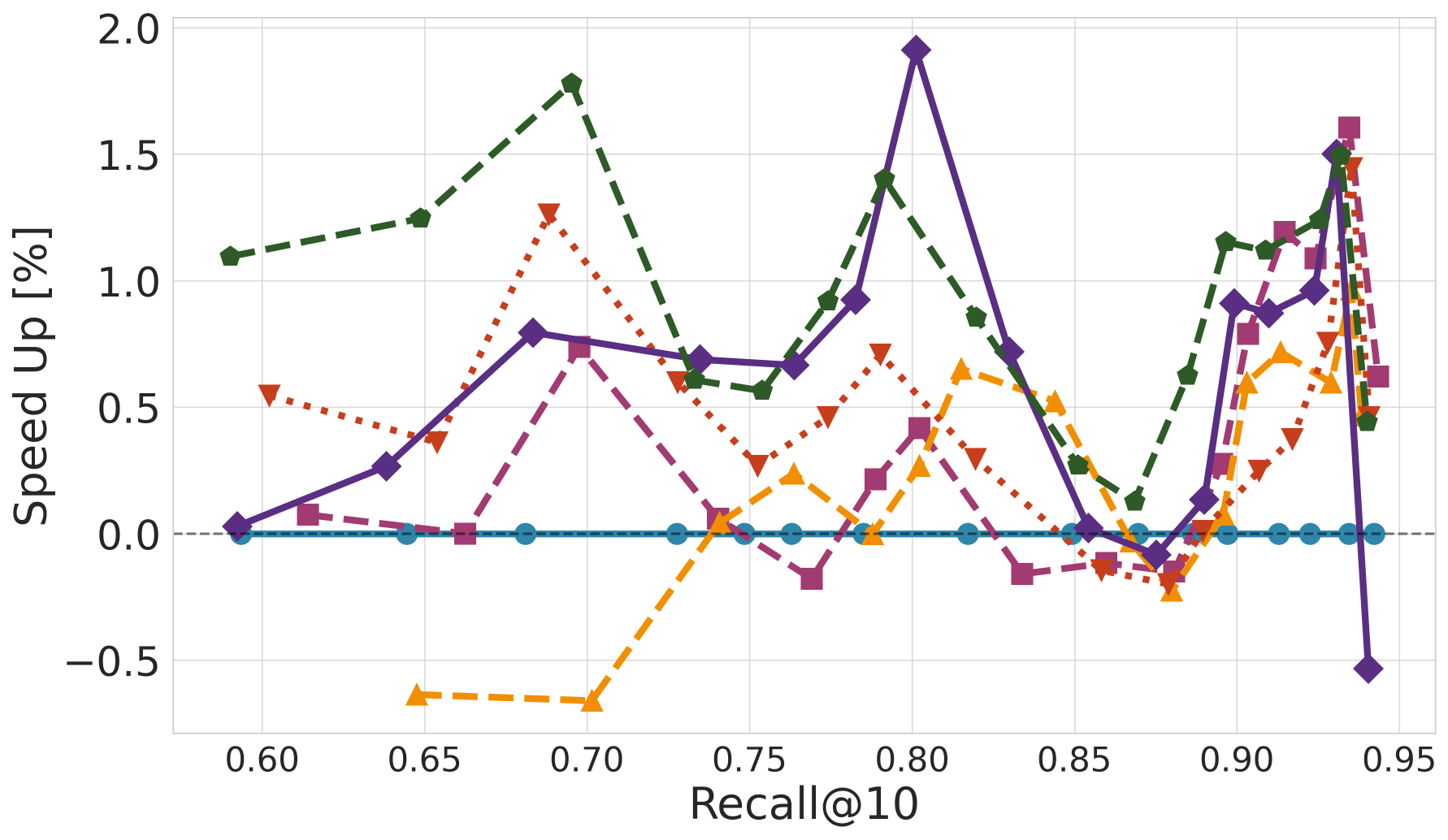}}
    \\
    \subcaptionbox{CAGRA, BioASQ
    1M}[0.24\linewidth]{\includegraphics[width=\linewidth]{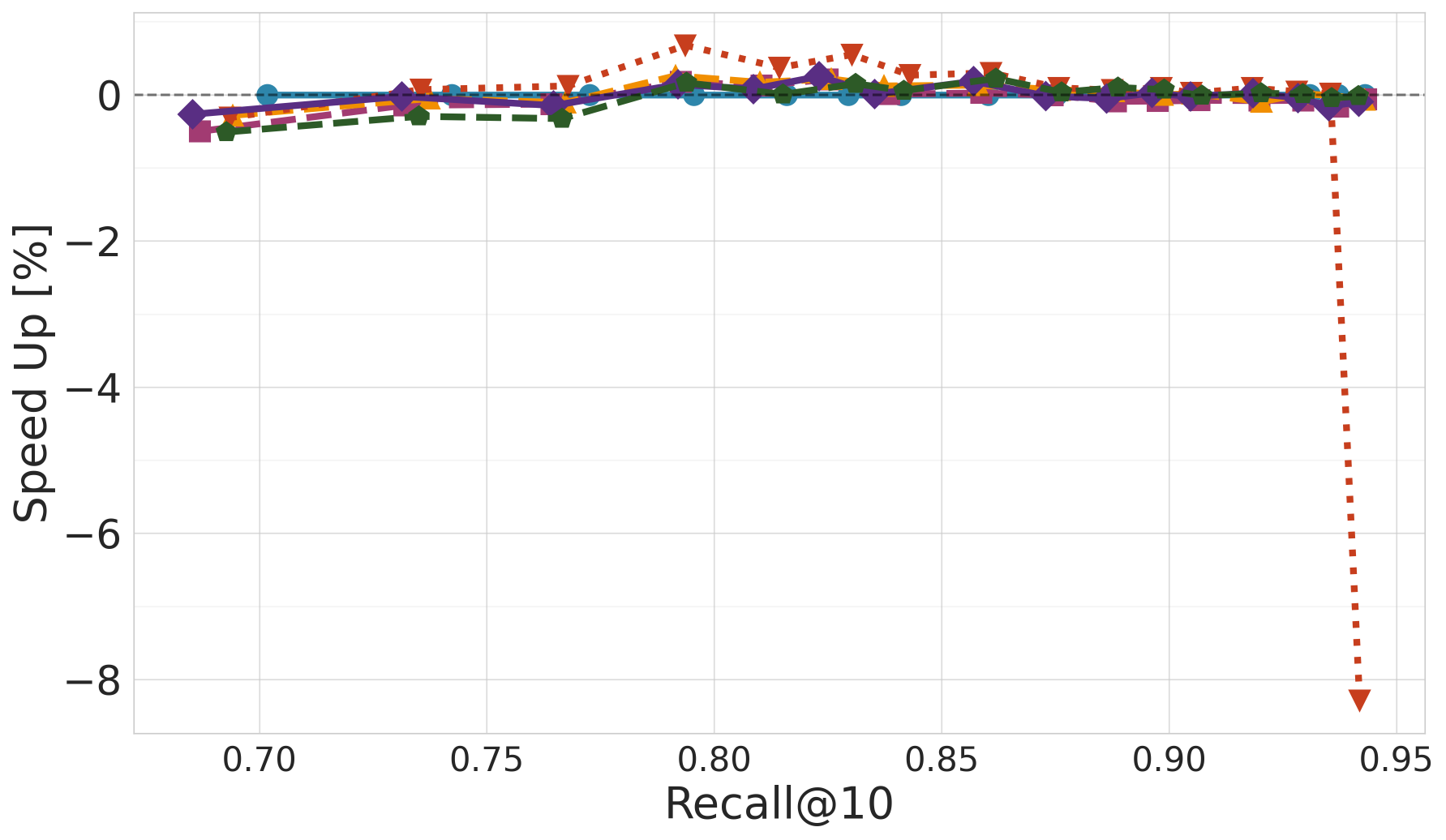}}
    &
    \subcaptionbox{NSG, BioASQ
    1M}[0.24\linewidth]{\includegraphics[width=\linewidth]{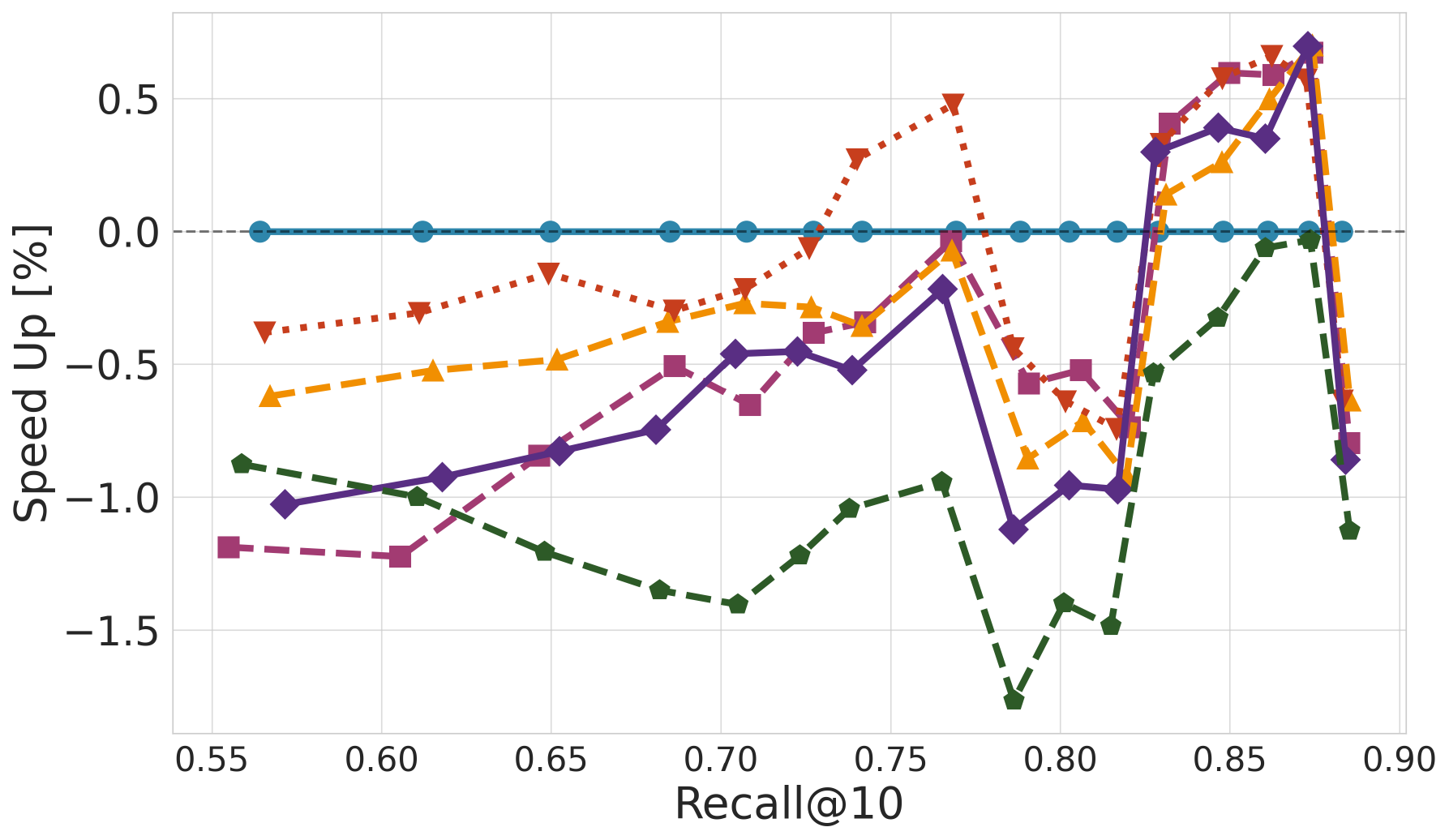}}
    &
    \subcaptionbox{Vamana, BioASQ
    1M}[0.24\linewidth]{\includegraphics[width=\linewidth]{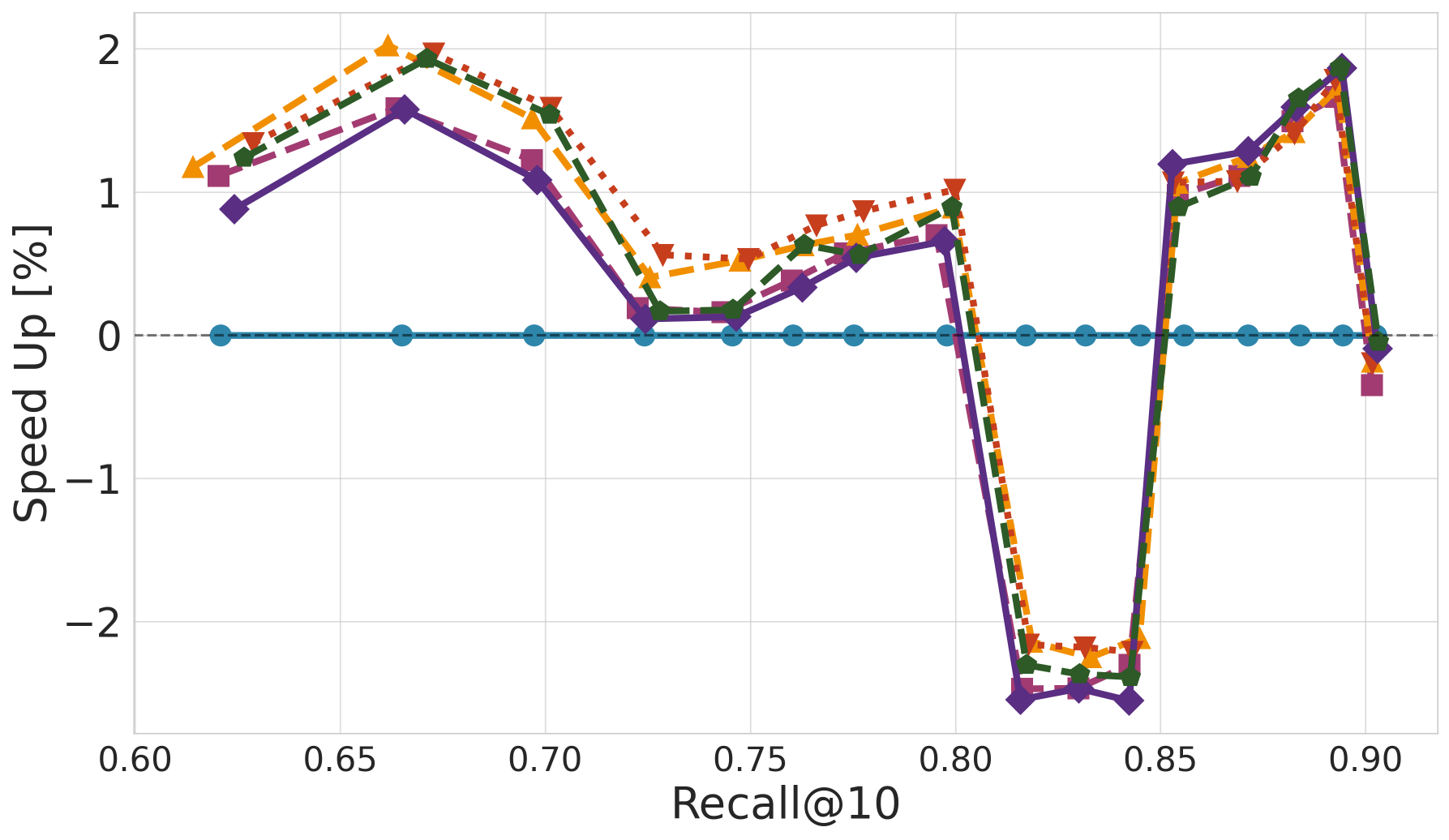}}
    &
    \subcaptionbox{NN-Descent, BioASQ
    1M}[0.24\linewidth]{\includegraphics[width=\linewidth]{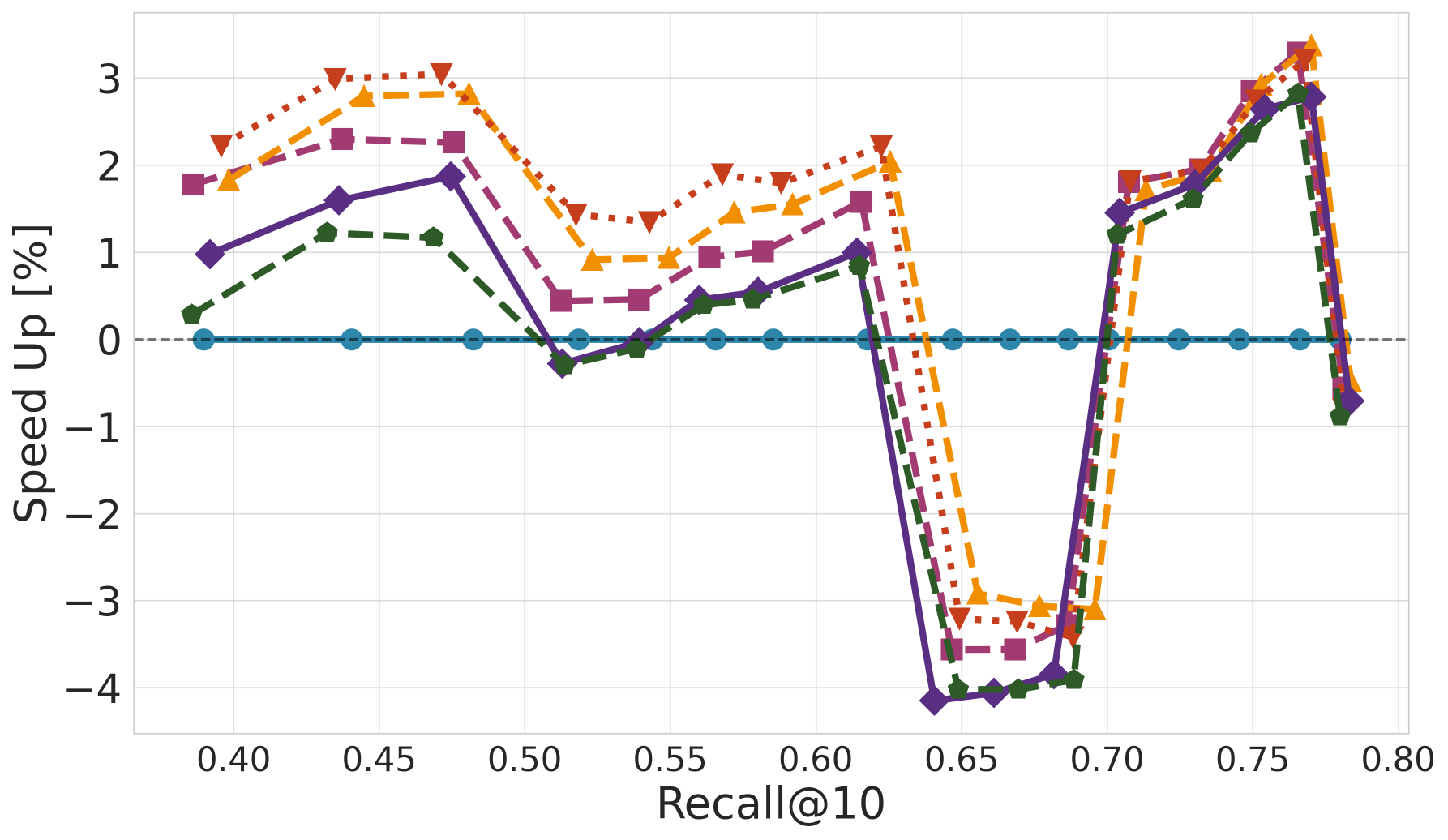}}
    \\
    \subcaptionbox{CAGRA, Wikipedia
    10M}[0.24\linewidth]{\includegraphics[width=\linewidth]{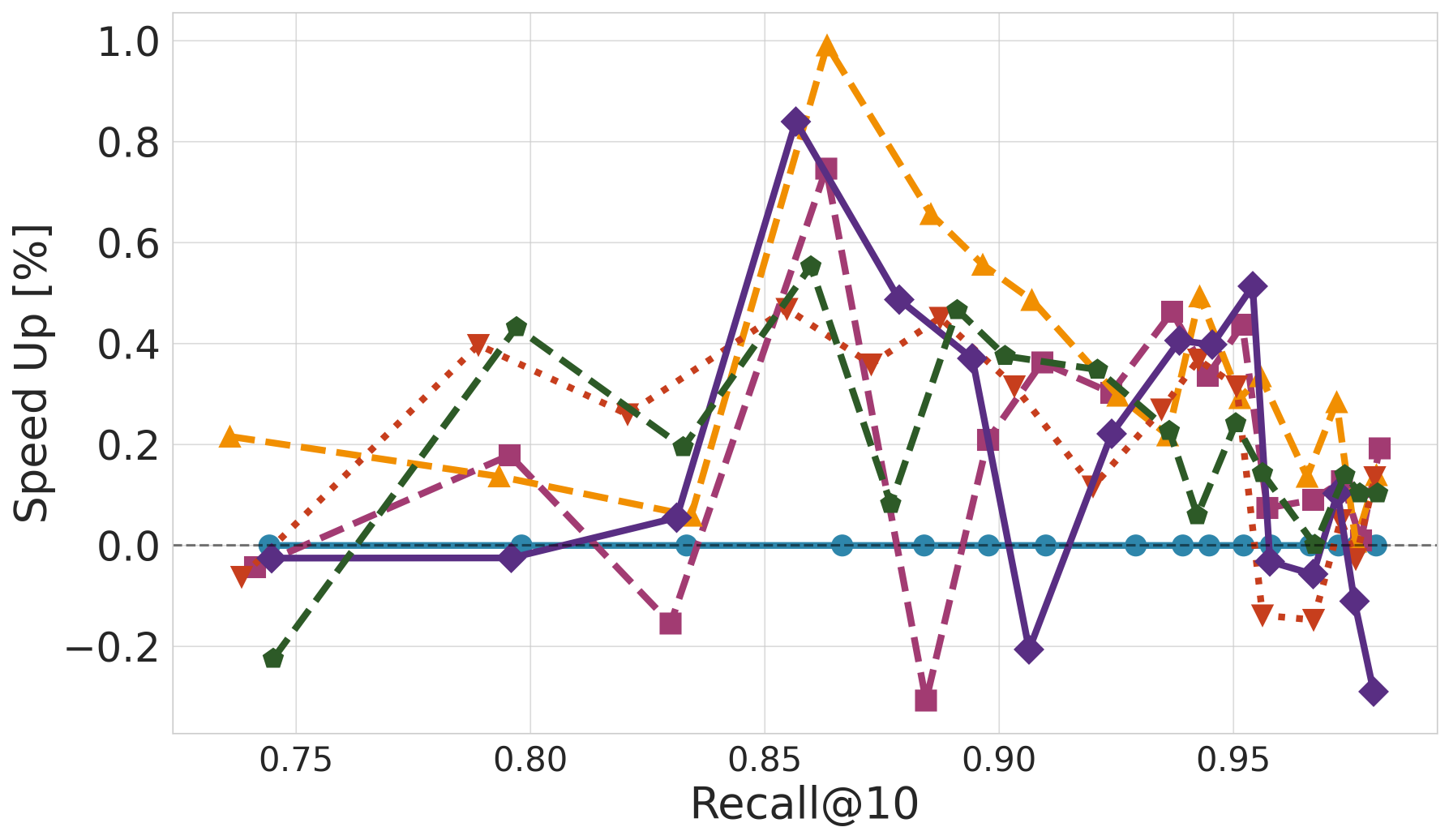}}
    &
    \subcaptionbox{NSG, Wikipedia
    10M}[0.24\linewidth]{\includegraphics[width=\linewidth]{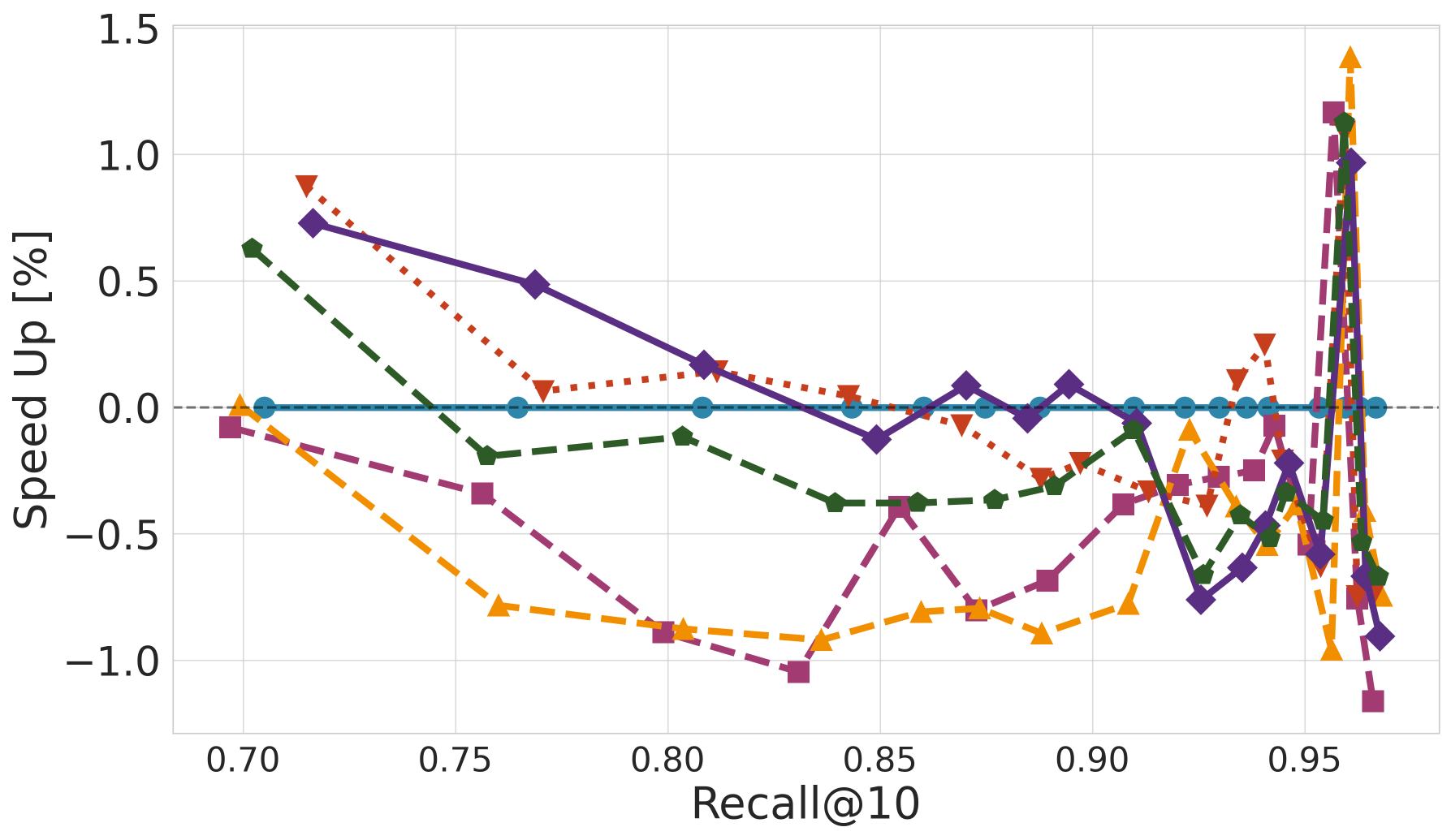}}
    &
    \subcaptionbox{Vamana, Wikipedia
    10M}[0.24\linewidth]{\includegraphics[width=\linewidth]{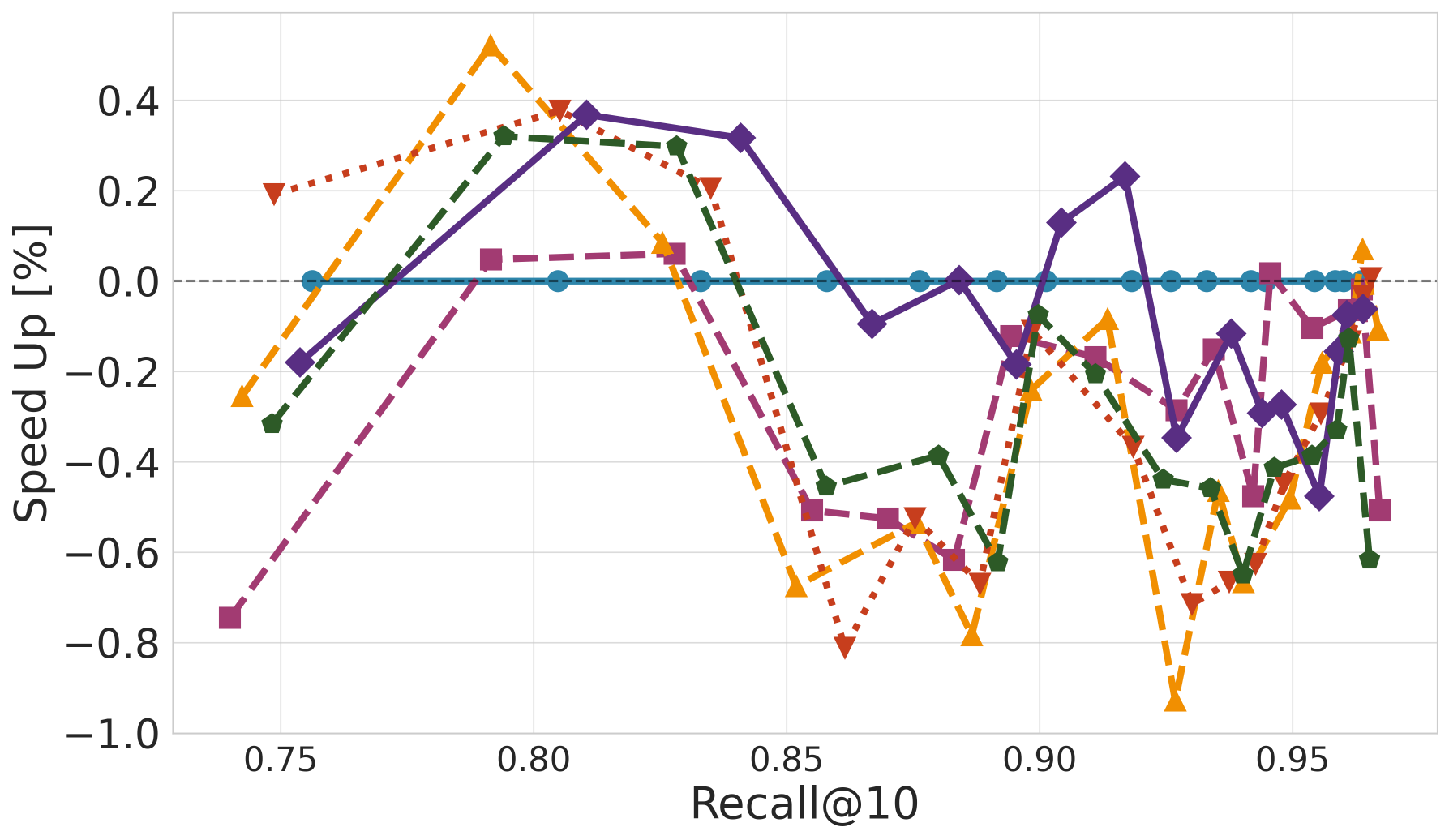}}
    &
    \subcaptionbox{NN-Descent, Wikipedia
    10M}[0.24\linewidth]{\includegraphics[width=\linewidth]{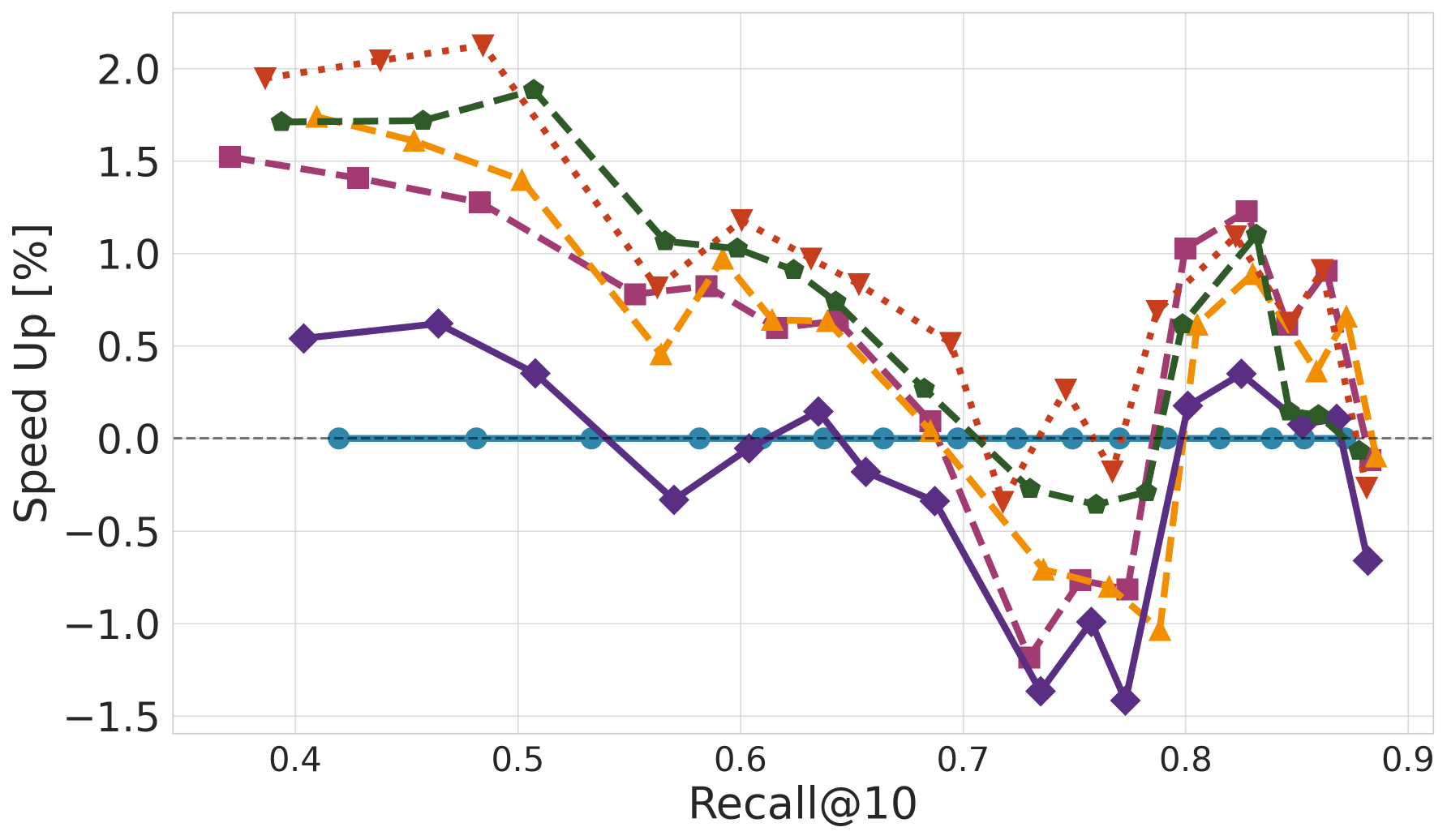}}
    \\
    \subcaptionbox{CAGRA, BioASQ
    10M}[0.24\linewidth]{\includegraphics[width=\linewidth]{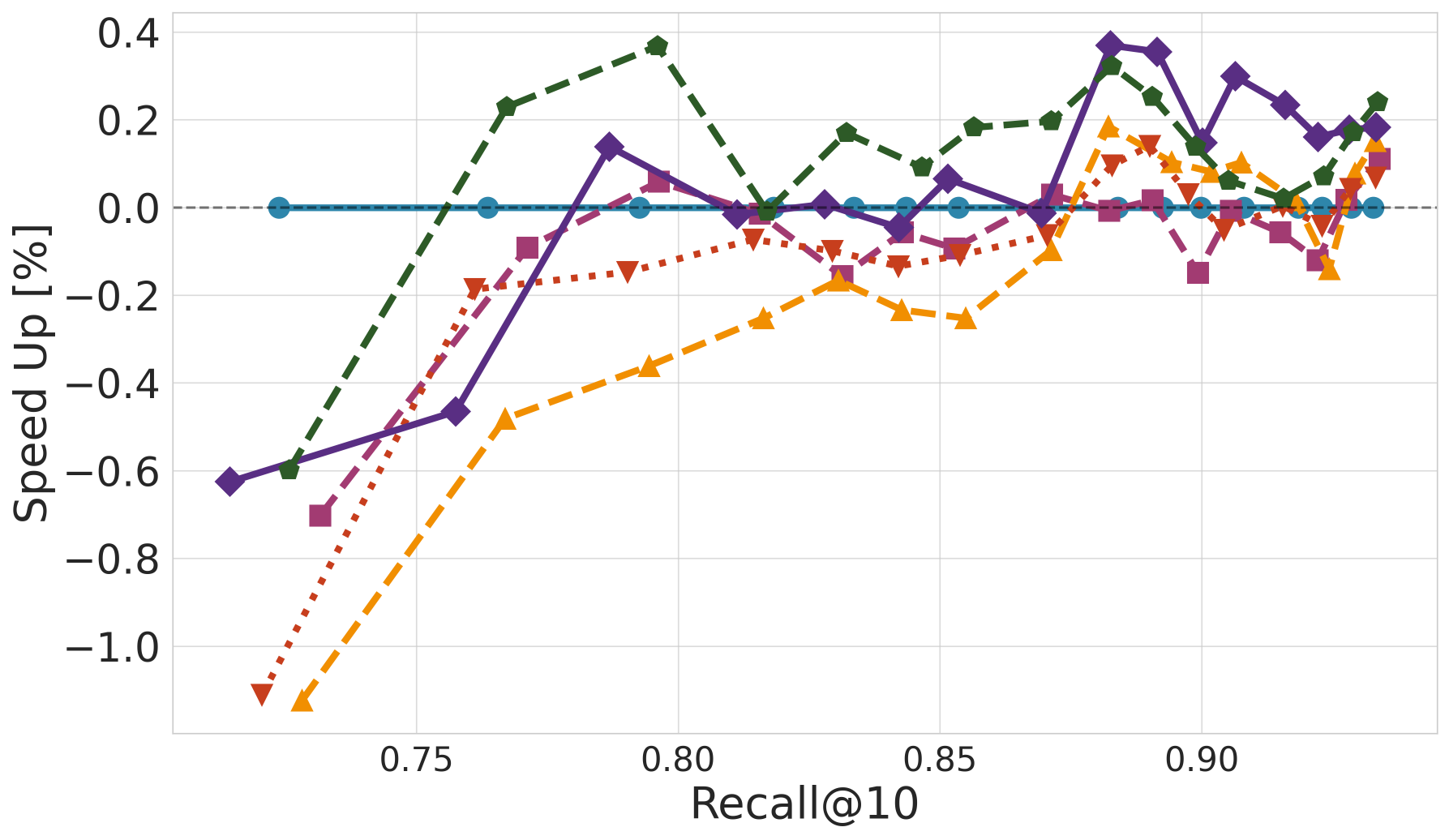}}
    &
    \subcaptionbox{NSG, BioASQ
    10M}[0.24\linewidth]{\includegraphics[width=\linewidth]{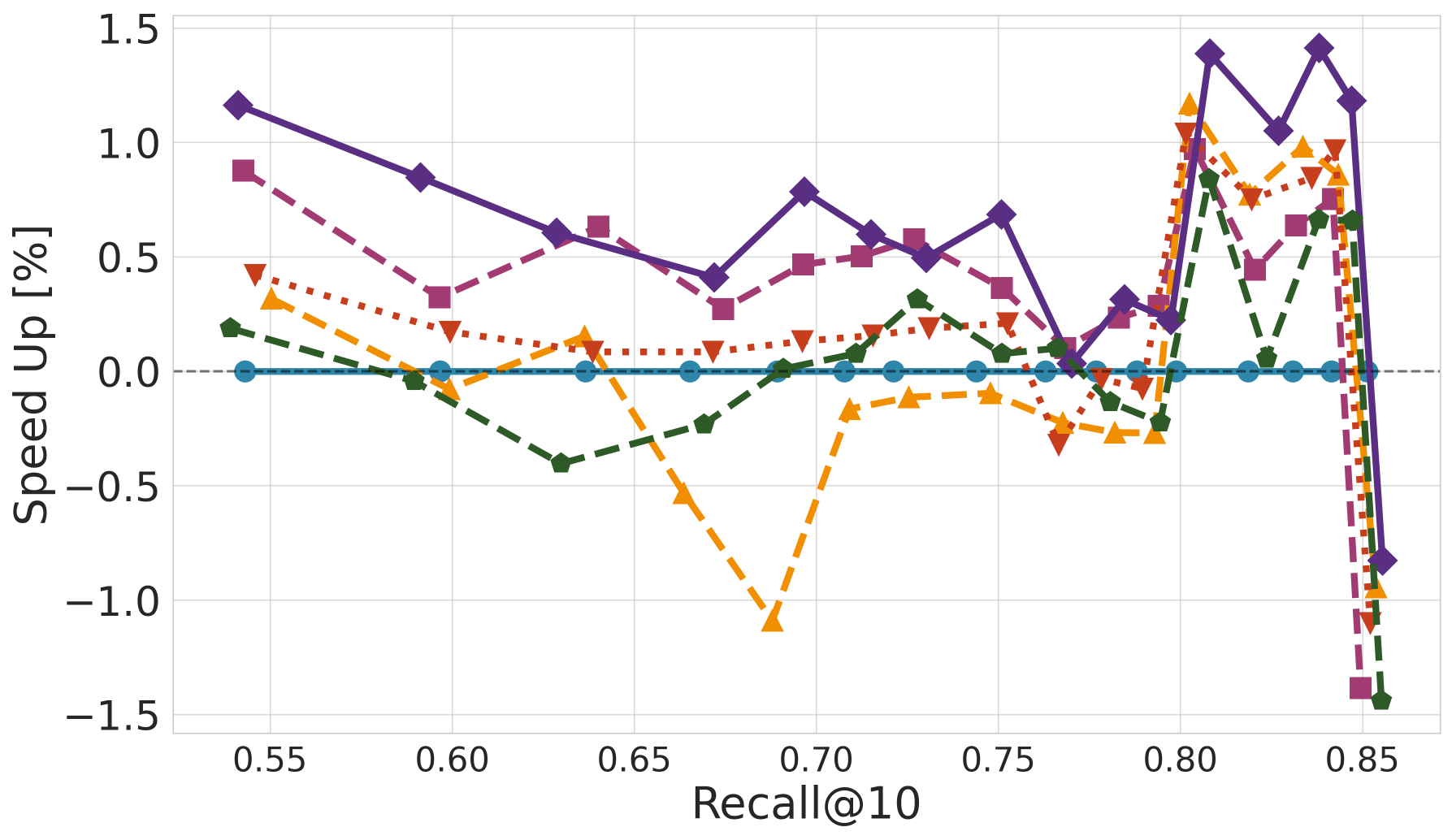}}
    &
    \subcaptionbox{Vamana, BioASQ
    10M}[0.24\linewidth]{\includegraphics[width=\linewidth]{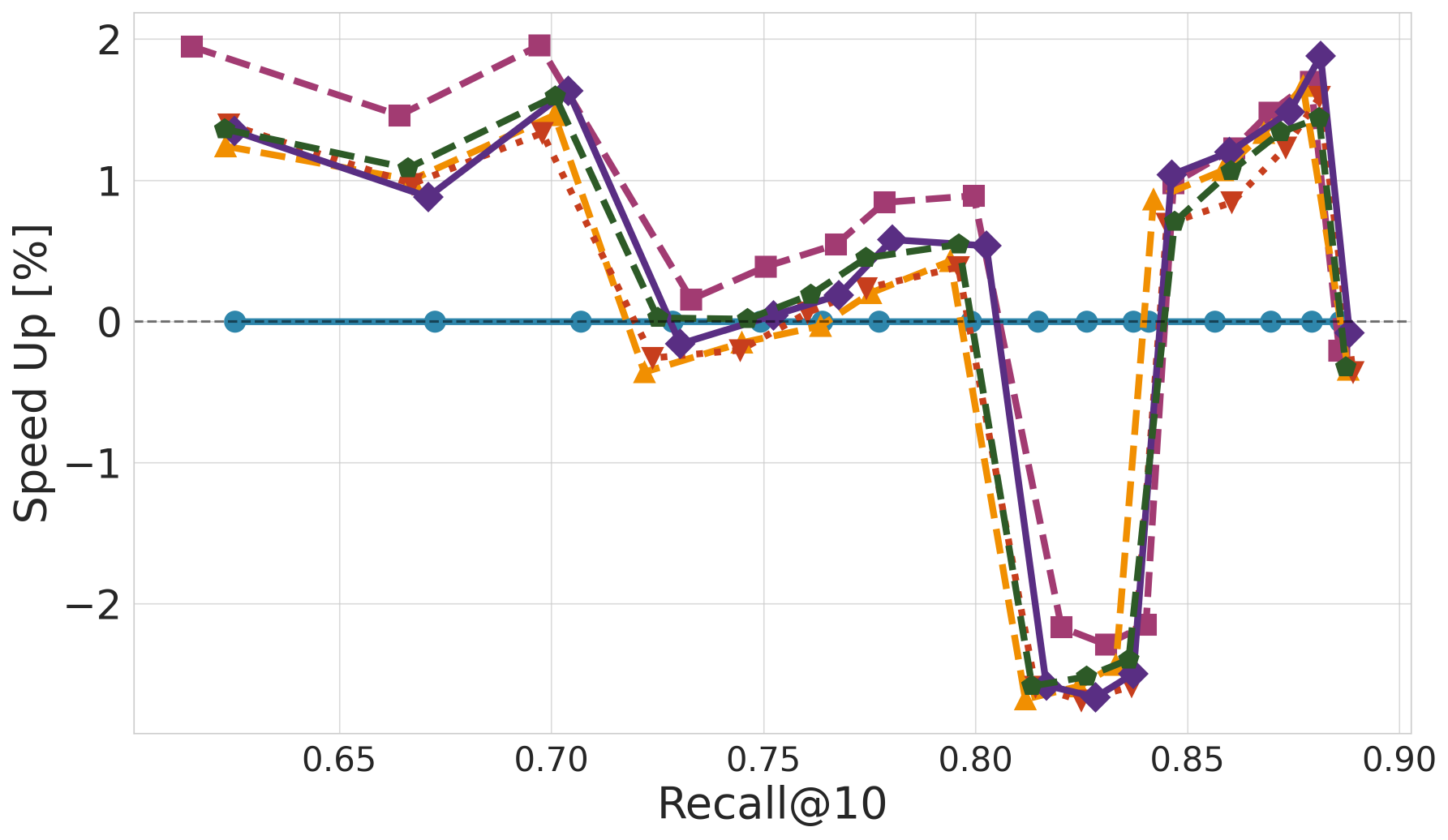}}
    &
    \subcaptionbox{NN-Descent, BioASQ
    10M}[0.24\linewidth]{\includegraphics[width=\linewidth]{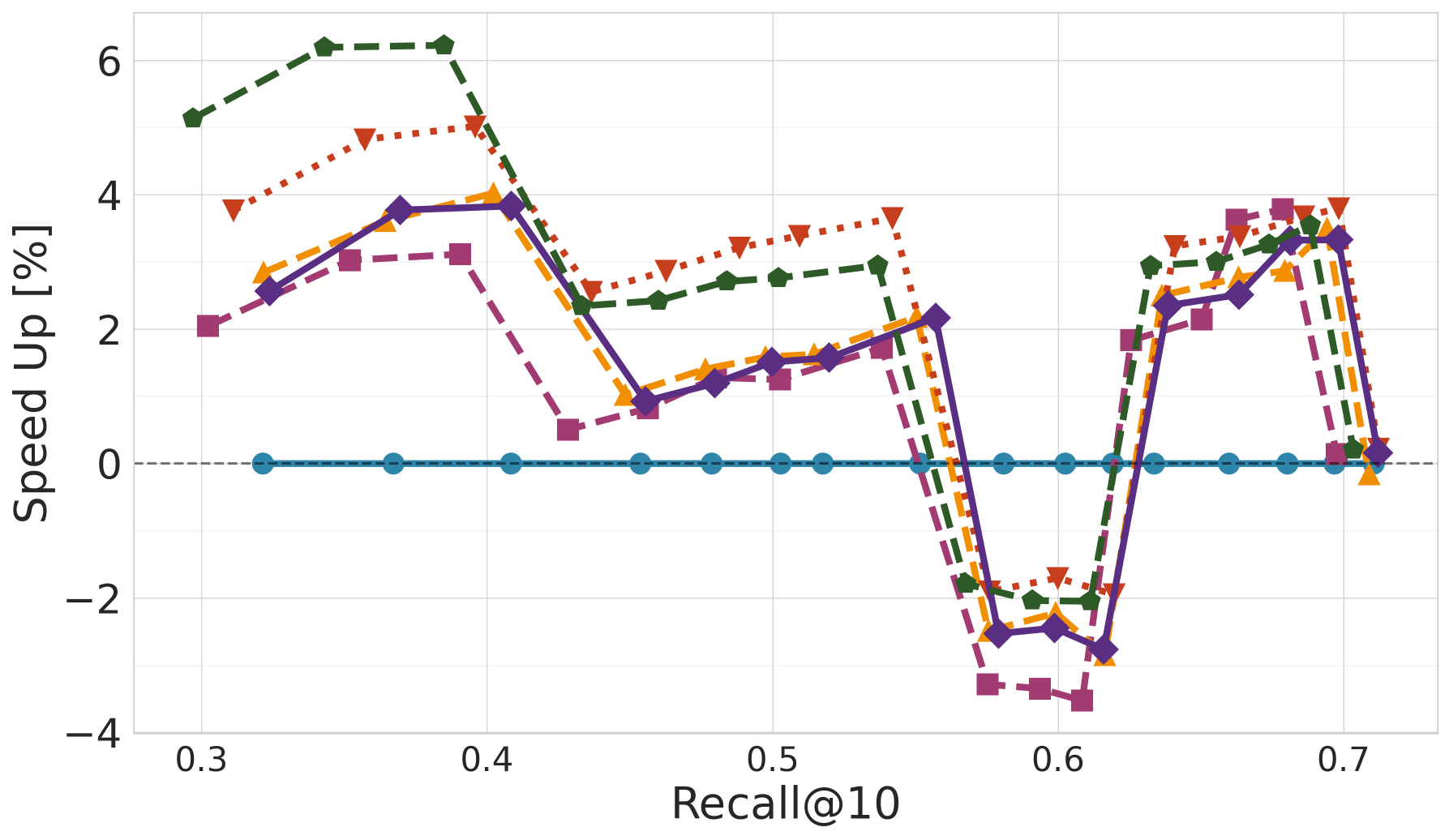}}
    \\
    \subcaptionbox{CAGRA, C4
    5M}[0.24\linewidth]{\includegraphics[width=\linewidth]{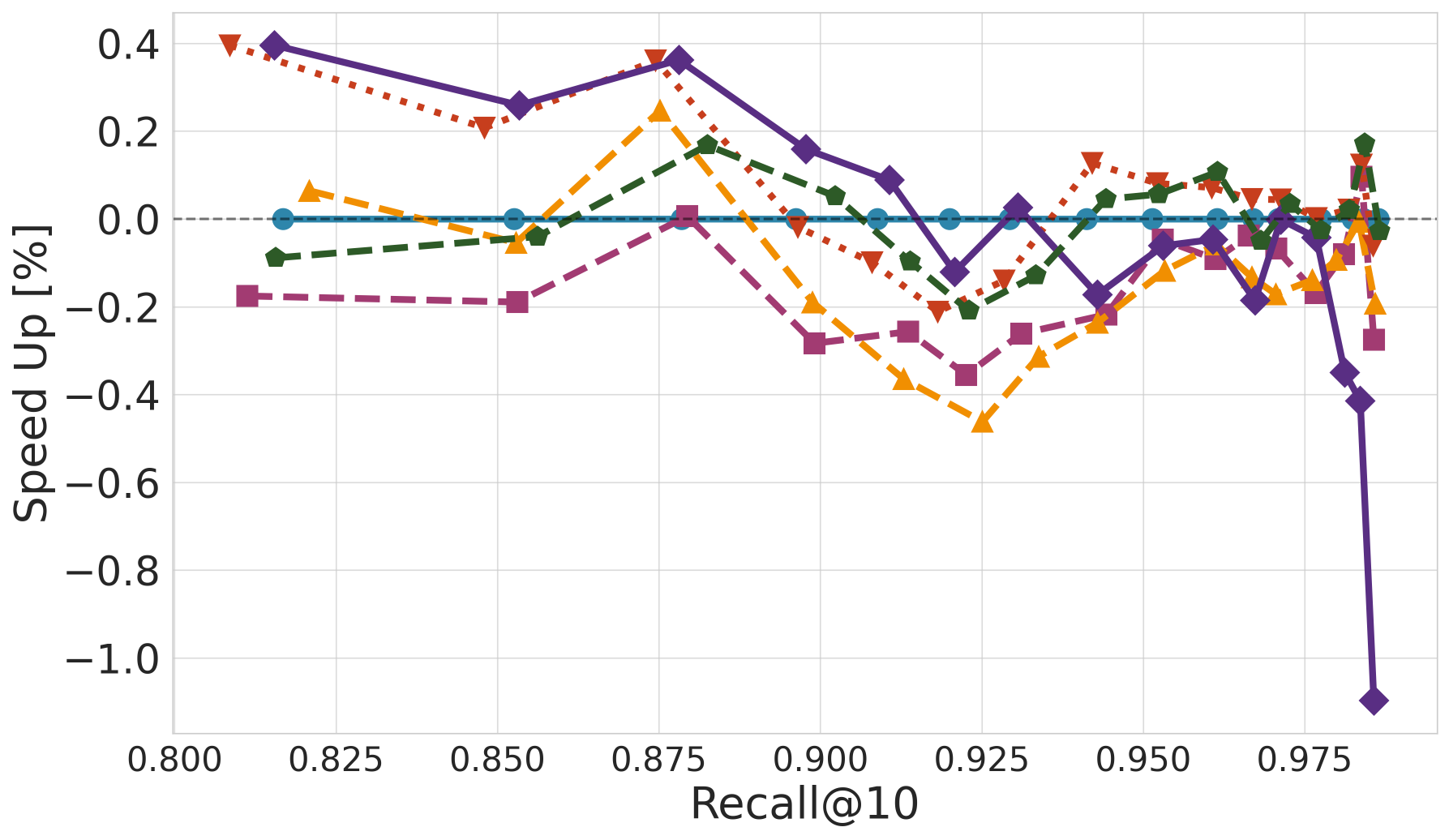}}
    &
    \subcaptionbox{NSG, C4
    5M}[0.24\linewidth]{\includegraphics[width=\linewidth]{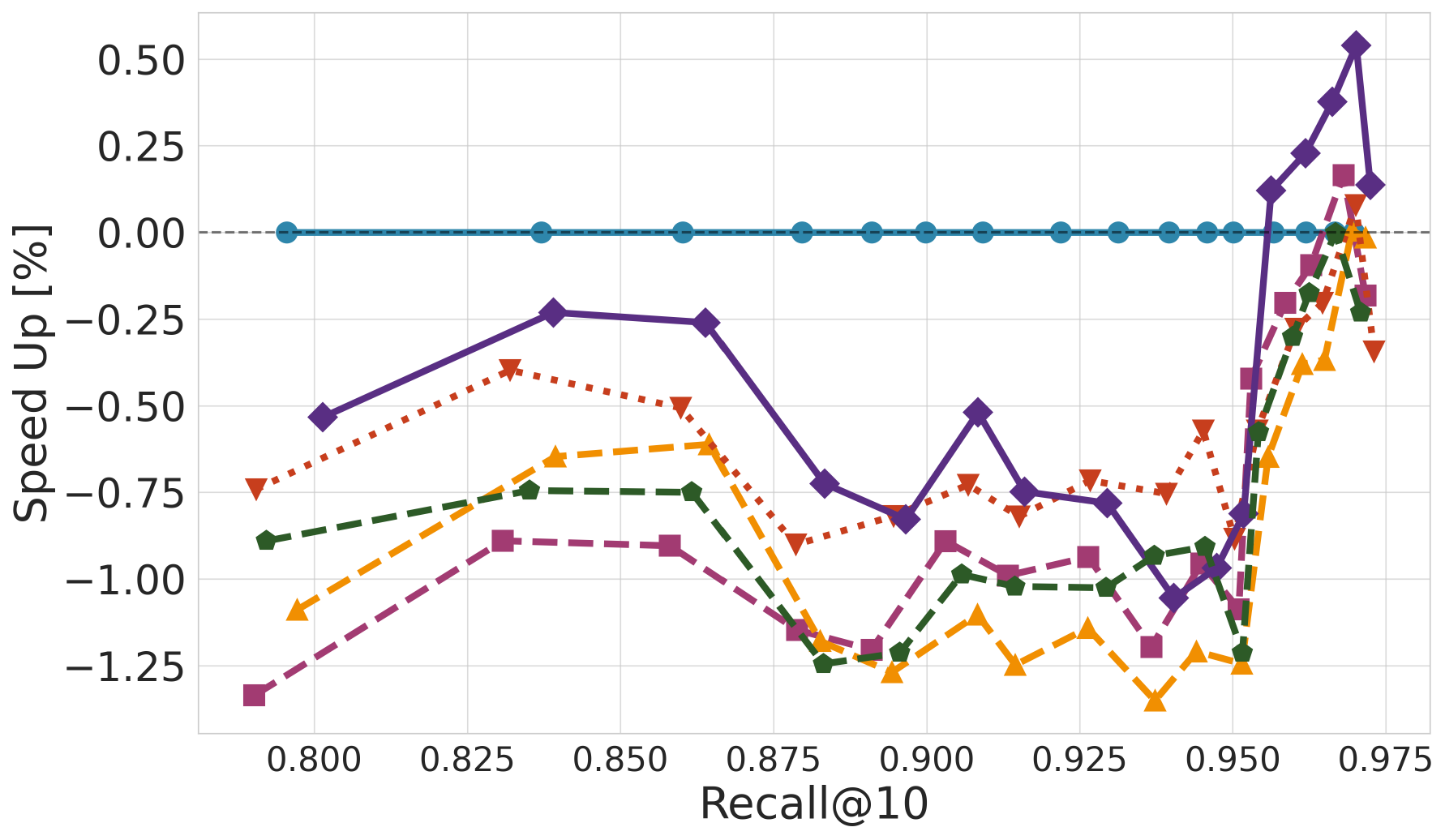}}
    &
    \subcaptionbox{Vamana, C4
    5M}[0.24\linewidth]{\includegraphics[width=\linewidth]{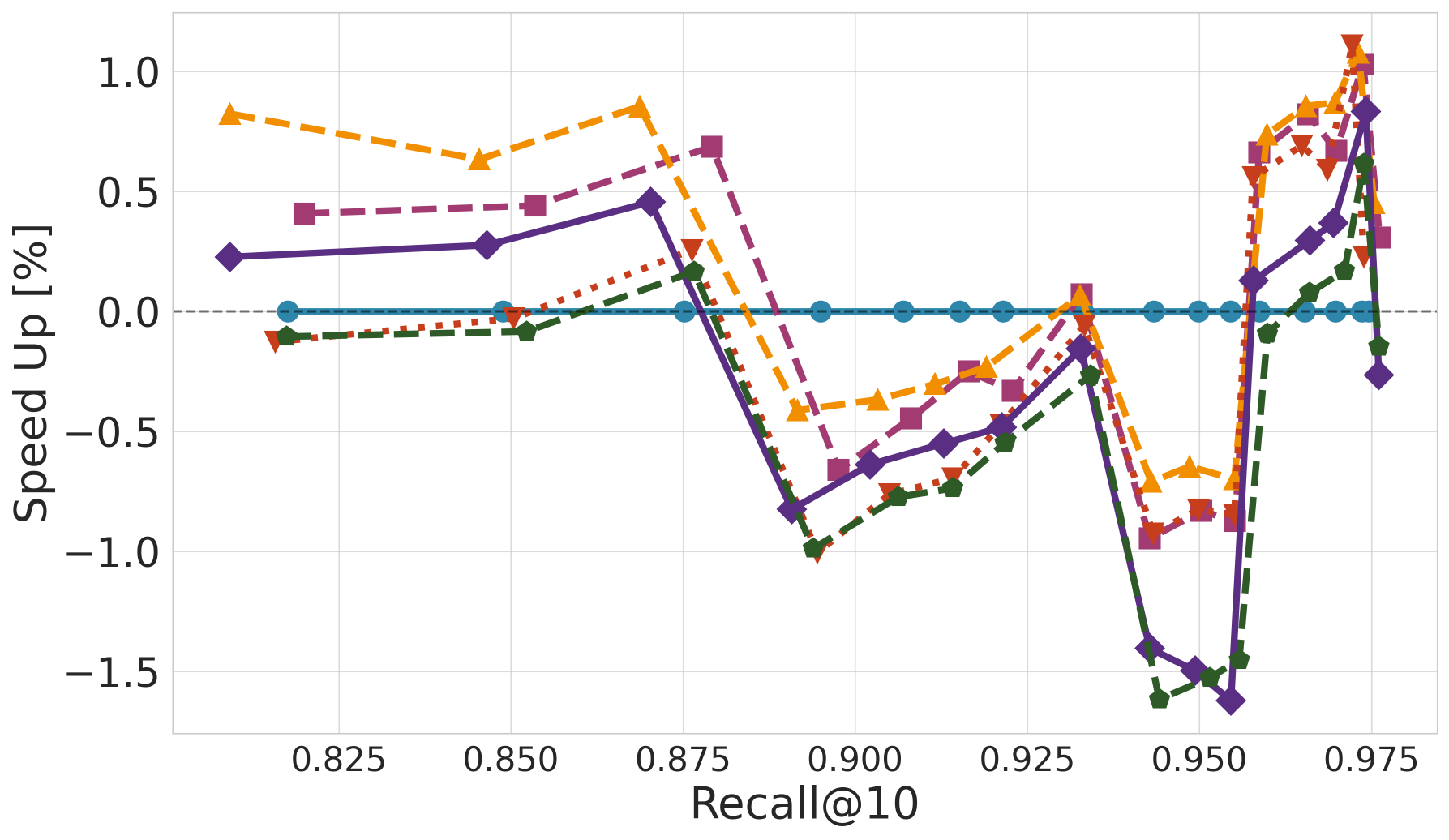}}
    &
    \subcaptionbox{NN-Descent, C4
    5M}[0.24\linewidth]{\includegraphics[width=\linewidth]{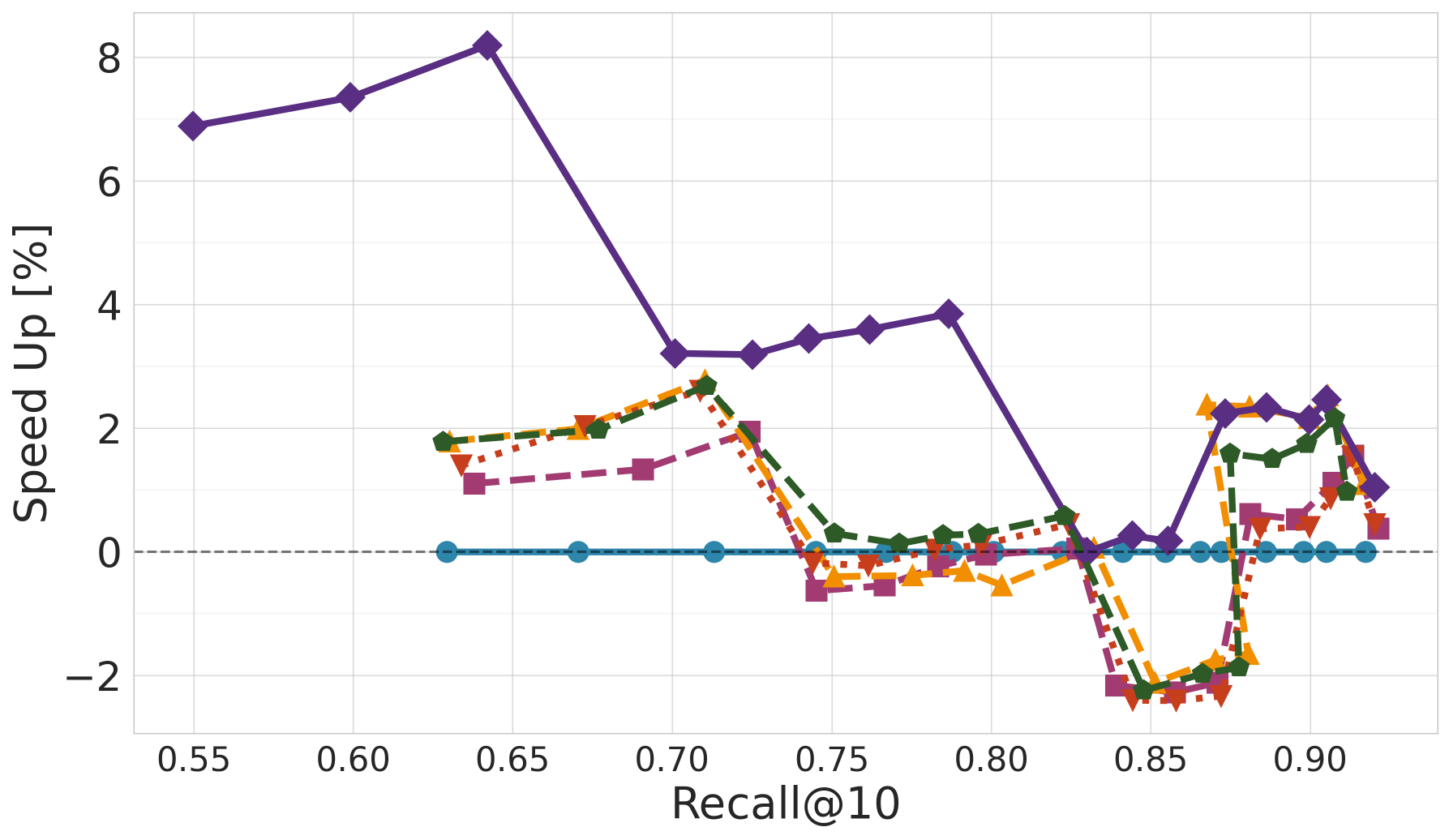}}
    \\
    \subcaptionbox{CAGRA, Deep
    40M}[0.24\linewidth]{\includegraphics[width=\linewidth]{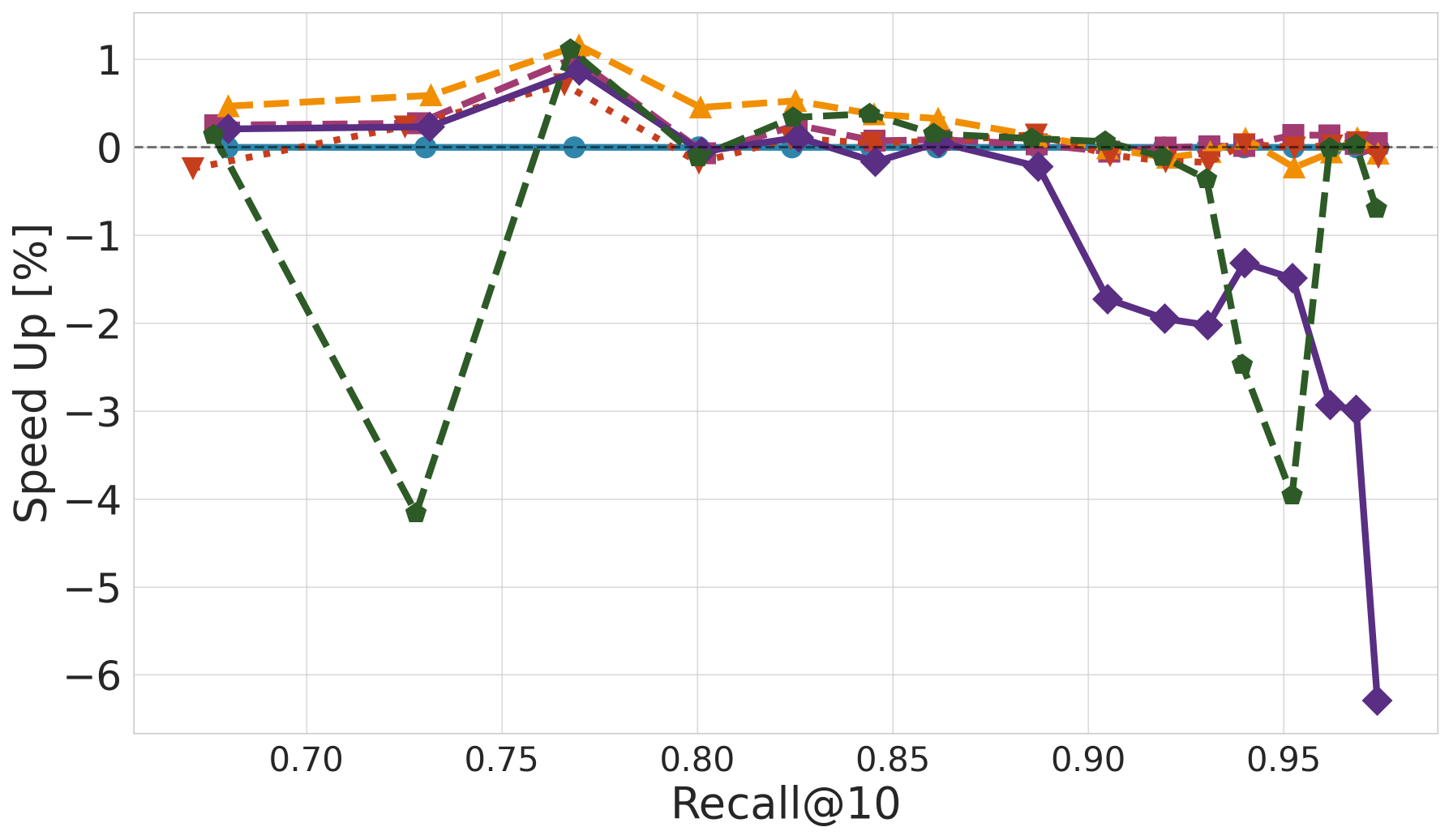}}
    &
    \subcaptionbox{NSG, Deep
    40M}[0.24\linewidth]{\includegraphics[width=\linewidth]{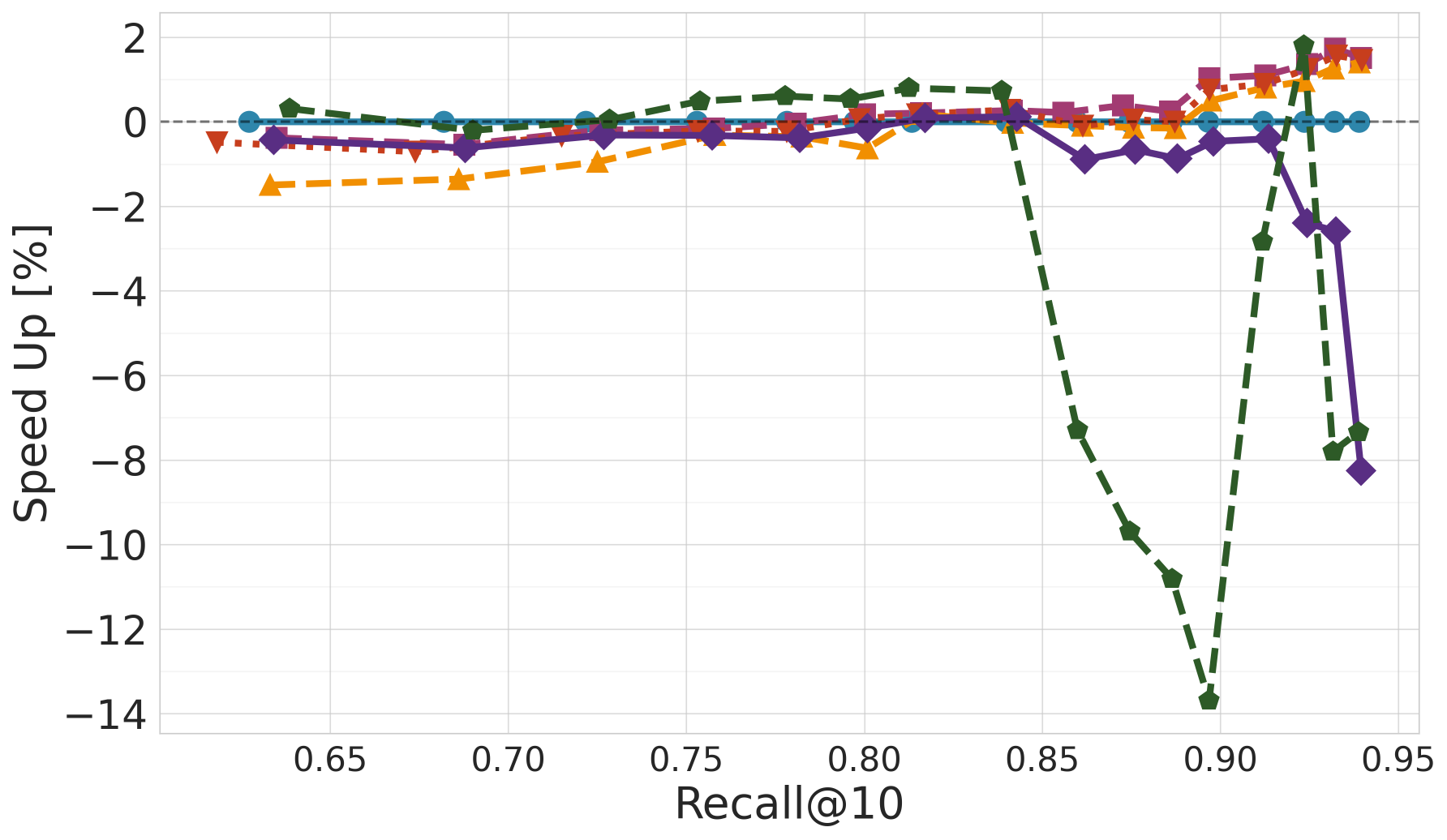}}
    &
    \subcaptionbox{Vamana, Deep
    40M}[0.24\linewidth]{\includegraphics[width=\linewidth]{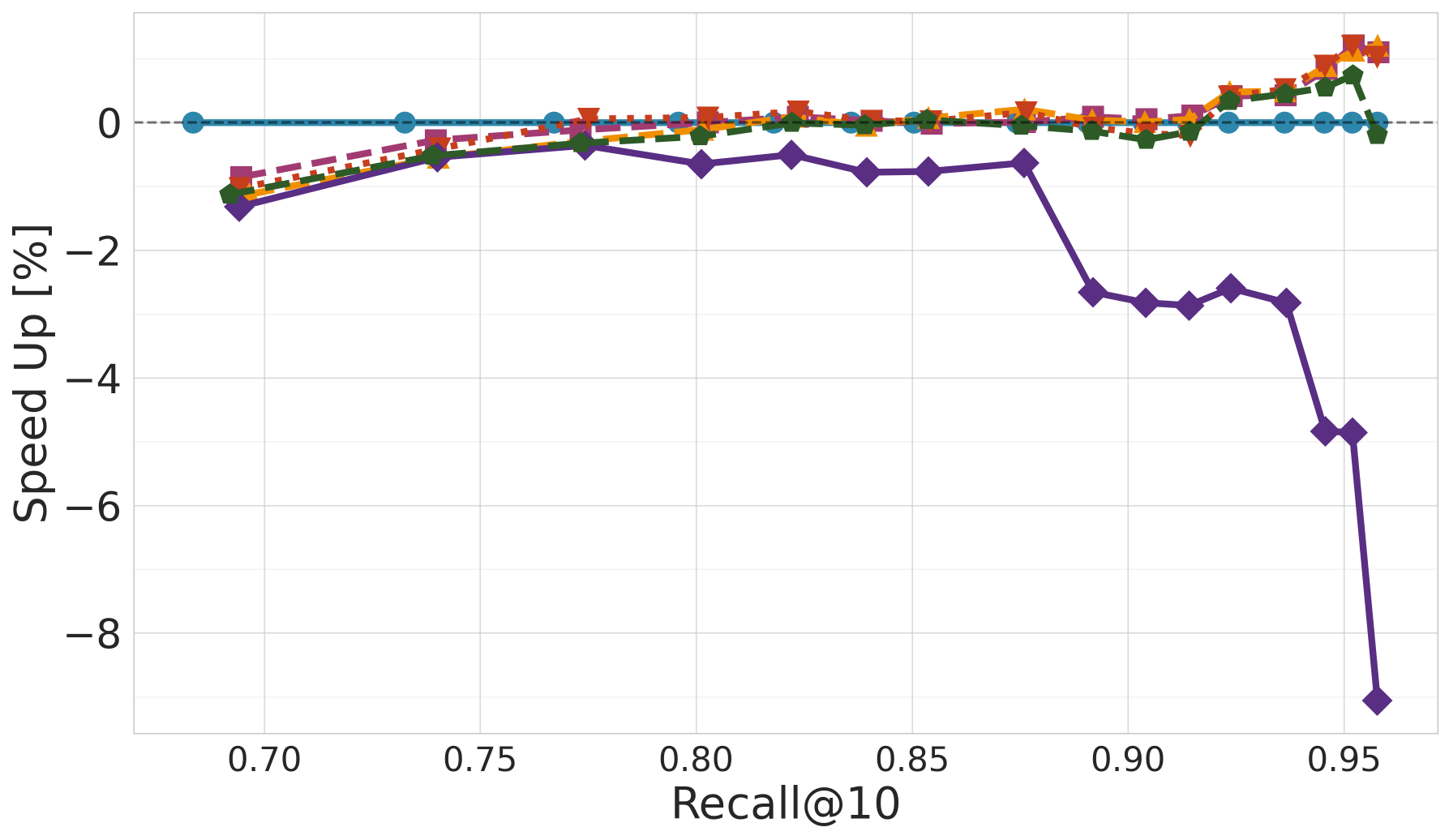}}
    &
    \subcaptionbox{NN-Descent, Deep
    40M}[0.24\linewidth]{\includegraphics[width=\linewidth]{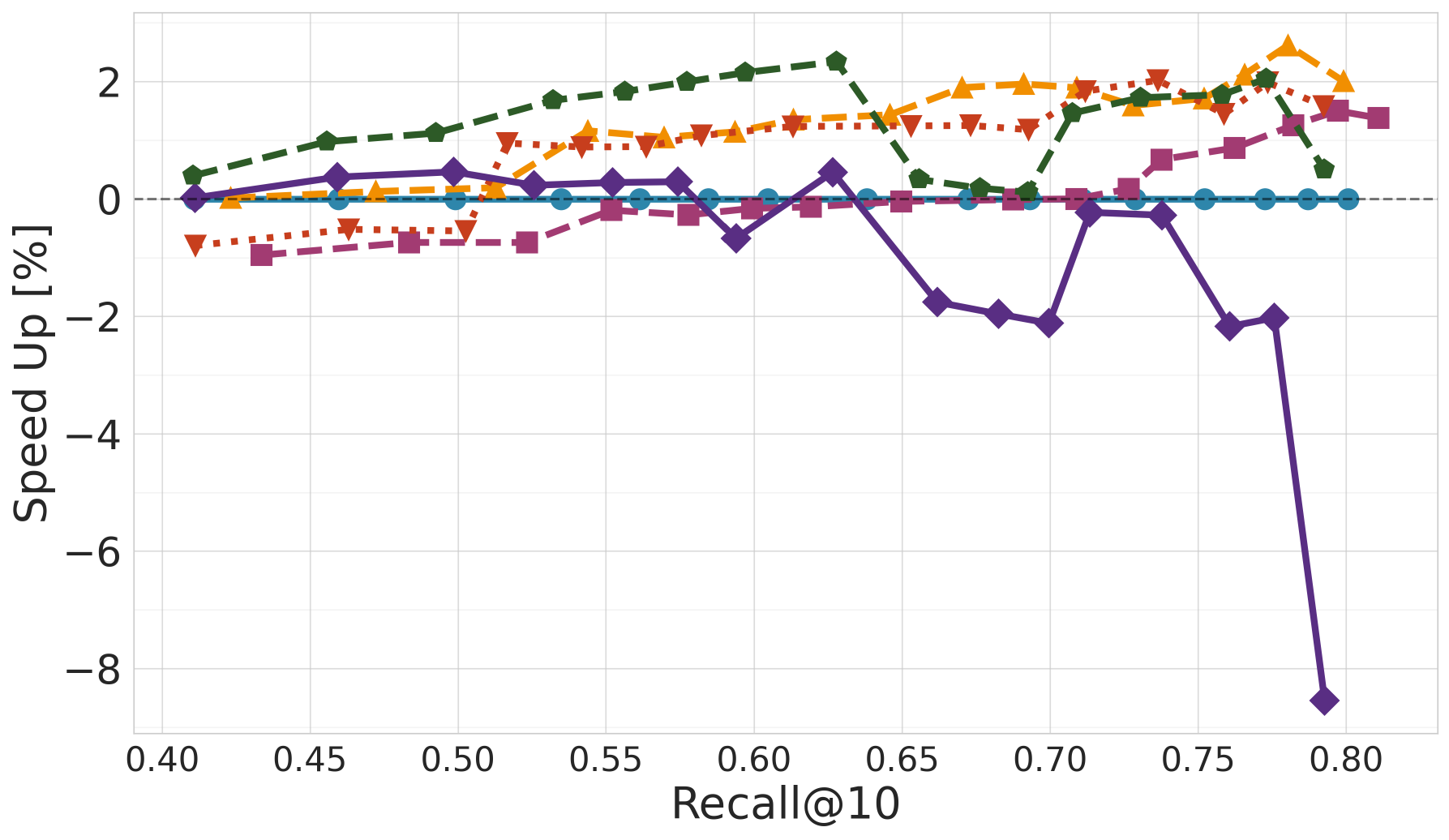}}
  \end{tabular}
  \vspace{0.1in}
  \begin{tabular}{c}
    \includegraphics[width=\textwidth]{fig/reordering_algorithms_legend.pdf}
  \end{tabular}
  \caption{Recall--QPS trade-off with reordering on NVIDIA RTX 6000
  Ada (Part II: large-scale and modern datasets).}
  \label{fig:supp-reordering-A6000-part2}
\end{figure*}

Figures~\ref{fig:supp-reordering-A100-part1},
\ref{fig:supp-reordering-A100-part2},
\ref{fig:supp-reordering-A6000-part1},
and~\ref{fig:supp-reordering-A6000-part2} show the full reordering
evaluation results for all dataset--index combinations on NVIDIA A100
and NVIDIA RTX 6000 Ada. Each row is a dataset and each column is a
graph-based index (CAGRA, NSG, Vamana, NN-Descent).

\section{Reordering Execution Time}
\label{app:reordering-time}
\begin{figure*}[t]
  \centering
    \includegraphics[width=\linewidth]{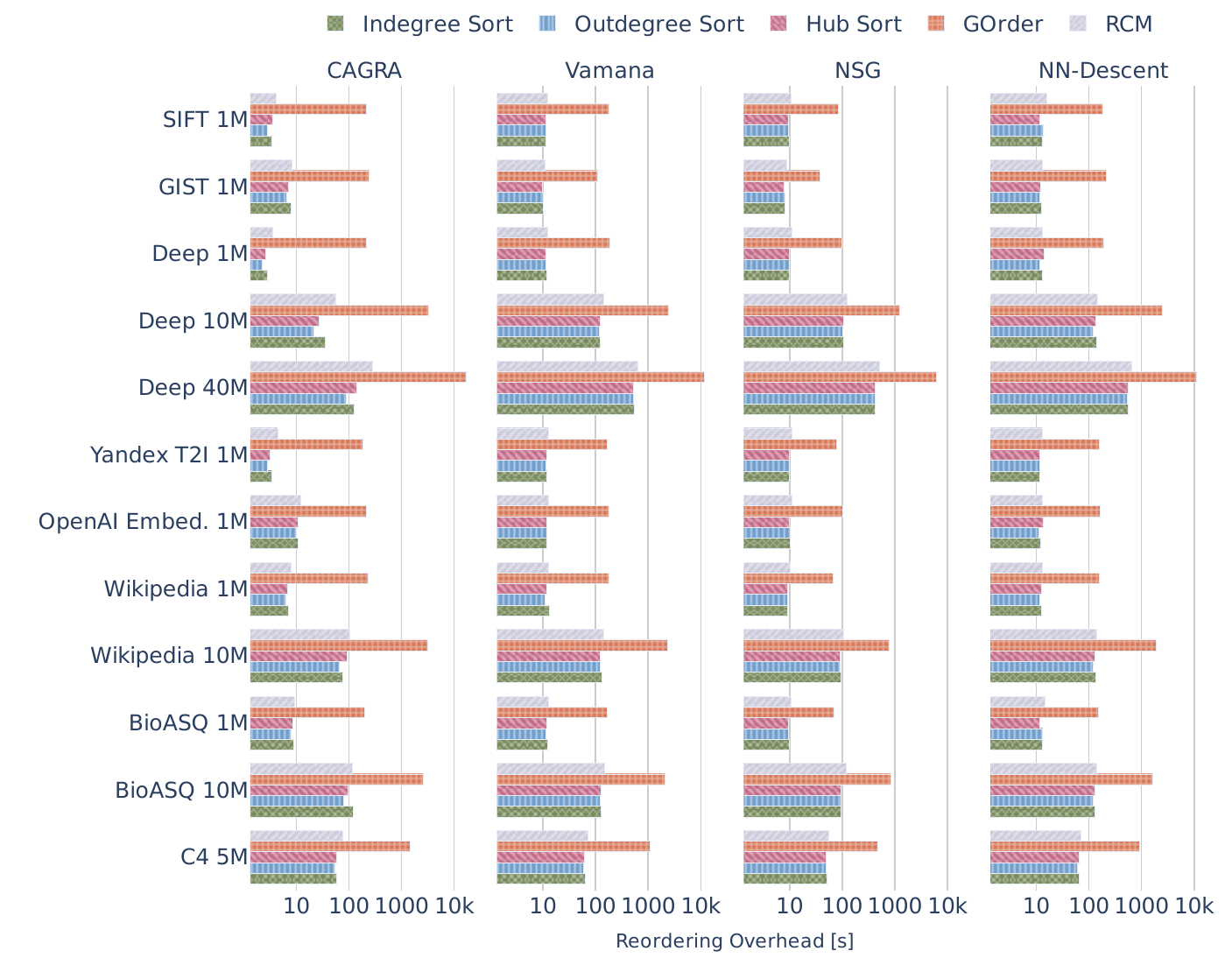}
\caption{Reordering execution time per dataset and graph-based index.
  Each bar shows the wall-clock time (seconds) required to apply a reordering algorithm to the full index.
  GOrder incurs substantially higher overhead than the other methods across all configurations.}
  \label{fig:reorder-time}
\end{figure*}

Our framework assumes \emph{offline} reordering: each reordering algorithm is applied once to the graph index before any queries are issued, and all subsequent queries share the same memory layout.
Under this assumption, the reordering overhead is a one-time preprocessing cost that is amortized across all queries.

\Cref{fig:reorder-time} reports the wall-clock reordering time for each combination of dataset and graph-based index.
The four lightweight methods (Hub Sort, Indegree, Outdegree, and RCM) complete in seconds to at most a few minutes even for the largest datasets, making them practical for most deployment scenarios.
GOrder, by contrast, requires substantially more time: its community-detection-based optimization is computationally expensive, and reordering time grows rapidly with dataset size.

\section{Detailed Hardware Metrics Analysis}
\label{app:hardware-details}

\Cref{tab:hardware-metrics,tab:hardware-metrics-nsg,tab:hardware-metrics-vamana}
present per-configuration measurements of L1/L2 cache hit rates and
DRAM bandwidth utilization at $L=120$ for CAGRA, NSG, and Vamana
respectively, extending the correlation results in Sec. 5.4
of the main paper.
L1 hit rates are near zero ($<0.03$\%) across all configurations and
virtually unaffected by reordering.
L2 hit rates are low ($8$--$20$\%) and change by less than
$1$ percentage point with reordering.
DRAM bandwidth utilization, in contrast, shows gains of several
percentage points that correspond directly to QPS improvements.

\begin{table}[t]
  \centering
  \small
  \setlength{\tabcolsep}{4pt}
  \caption{Hardware metrics (L1/L2 cache hit rate and DRAM bandwidth
    utilization at $L=120$) across datasets and reordering methods:
    \textbf{CAGRA}. NSG and Vamana follow in
    \Cref{tab:hardware-metrics-nsg,tab:hardware-metrics-vamana}.}
  \label{tab:hardware-metrics}
  \begin{tabular}{@{}llccc@{}}
    \toprule
    \multicolumn{5}{c}{\textbf{CAGRA}}\\
    \midrule
    Dataset & Method & L1 (\%) & L2 (\%) & DRAM (\%) \\
    \midrule
    \multirow{4}{*}{SIFT 1M} & Baseline & 0.0152 & 14.9 & 57.6 \\
     & Hub Sort & 0.0153 & 14.9 & 57.5 \\
     & RCM & 0.0157 & 14.9 & 54.0 \\
     & GOrder & 0.0157 & 14.9 & 48.1 \\
    \cmidrule{1-5}
    \multirow{4}{*}{GIST 1M} & Baseline & 0.0073 & 10.1 & 71.5 \\
     & Hub Sort & 0.0073 & 10.1 & 71.7 \\
     & RCM & 0.0073 & 10.1 & 70.9 \\
     & GOrder & 0.0073 & 10.1 & 70.7 \\
    \cmidrule{1-5}
    \multirow{4}{*}{Deep 1M} & Baseline & 0.0178 & 16.0 & 43.7 \\
     & Hub Sort & 0.0177 & 16.1 & 43.6 \\
     & RCM & 0.0183 & 16.0 & 40.7 \\
     & GOrder & 0.0181 & 15.9 & 34.3 \\
    \cmidrule{1-5}
    \multirow{4}{*}{Deep 40M} & Baseline & 0.0156 & 9.1 & 53.4 \\
     & Hub Sort & 0.0158 & 9.0 & 53.3 \\
     & RCM & 0.0158 & 9.0 & 52.0 \\
     & GOrder & 0.0161 & 9.0 & 46.6 \\
    \cmidrule{1-5}
    \multirow{4}{*}{Wikipedia 10M} & Baseline & 0.0078 & 8.8 & 67.1 \\
     & Hub Sort & 0.0079 & 8.9 & 67.3 \\
     & RCM & 0.0078 & 8.9 & 66.9 \\
     & GOrder & 0.0079 & 8.8 & 61.6 \\
    \cmidrule{1-5}
    \multirow{4}{*}{BioASQ 10M} & Baseline & 0.0071 & 9.1 & 77.4 \\
     & Hub Sort & 0.0071 & 9.1 & 77.6 \\
     & RCM & 0.0071 & 9.0 & 77.2 \\
     & GOrder & 0.0072 & 9.1 & 77.5 \\
    \bottomrule
  \end{tabular}
\end{table}

\begin{table}[t]
  \centering
  \small
  \setlength{\tabcolsep}{4pt}
  \caption{Hardware metrics (L1/L2 cache hit rate and DRAM bandwidth
    utilization at $L=120$) across datasets and reordering methods:
    \textbf{NSG}.}
  \label{tab:hardware-metrics-nsg}
  \begin{tabular}{@{}llccc@{}}
    \toprule
    \multicolumn{5}{c}{\textbf{NSG}}\\
    \midrule
    Dataset & Method & L1 (\%) & L2 (\%) & DRAM (\%) \\
    \midrule
    \multirow{4}{*}{SIFT 1M} & Baseline & 0.0214 & 18.0 & 39.4 \\
     & Hub Sort & 0.0221 & 17.1 & 44.2 \\
     & RCM & 0.0223 & 17.1 & 43.1 \\
     & GOrder & 0.0222 & 17.1 & 39.8 \\
    \cmidrule{1-5}
    \multirow{4}{*}{GIST 1M} & Baseline & 0.0109 & 13.4 & 50.2 \\
     & Hub Sort & 0.0109 & 12.7 & 55.3 \\
     & RCM & 0.0109 & 12.8 & 54.8 \\
     & GOrder & 0.0109 & 12.7 & 54.9 \\
    \cmidrule{1-5}
    \multirow{4}{*}{Deep 1M} & Baseline & 0.0240 & 19.5 & 30.2 \\
     & Hub Sort & 0.0246 & 18.7 & 33.5 \\
     & RCM & 0.0249 & 18.7 & 32.2 \\
     & GOrder & 0.0248 & 18.6 & 28.4 \\
    \cmidrule{1-5}
    \multirow{4}{*}{Deep 40M} & Baseline & 0.0198 & 11.0 & 39.7 \\
     & Hub Sort & 0.0206 & 10.0 & 44.0 \\
     & RCM & 0.0207 & 10.0 & 42.9 \\
     & GOrder & 0.0204 & 10.0 & 39.8 \\
    \cmidrule{1-5}
    \multirow{4}{*}{Wikipedia 10M} & Baseline & 0.0111 & 11.0 & 47.6 \\
     & Hub Sort & 0.0112 & 10.3 & 53.6 \\
     & RCM & 0.0111 & 10.3 & 52.2 \\
     & GOrder & 0.0112 & 10.3 & 52.1 \\
    \cmidrule{1-5}
    \multirow{4}{*}{BioASQ 10M} & Baseline & 0.0130 & 12.2 & 66.7 \\
     & Hub Sort & 0.0126 & 11.4 & 70.9 \\
     & RCM & 0.0127 & 11.4 & 70.6 \\
     & GOrder & 0.0128 & 11.3 & 69.5 \\
    \bottomrule
  \end{tabular}
\end{table}

\begin{table}[t]
  \centering
  \small
  \setlength{\tabcolsep}{4pt}
  \caption{Hardware metrics (L1/L2 cache hit rate and DRAM bandwidth
    utilization at $L=120$) across datasets and reordering methods:
    \textbf{Vamana}.}
  \label{tab:hardware-metrics-vamana}
  \begin{tabular}{@{}llccc@{}}
    \toprule
    \multicolumn{5}{c}{\textbf{Vamana}}\\
    \midrule
    Dataset & Method & L1 (\%) & L2 (\%) & DRAM (\%) \\
    \midrule
    \multirow{4}{*}{SIFT 1M} & Baseline & 0.0160 & 17.7 & 47.9 \\
     & Hub Sort & 0.0166 & 16.8 & 52.6 \\
     & RCM & 0.0168 & 16.7 & 50.7 \\
     & GOrder & 0.0169 & 16.7 & 46.2 \\
    \cmidrule{1-5}
    \multirow{4}{*}{GIST 1M} & Baseline & 0.0080 & 12.1 & 62.2 \\
     & Hub Sort & 0.0080 & 11.6 & 66.4 \\
     & RCM & 0.0080 & 11.6 & 66.8 \\
     & GOrder & 0.0080 & 11.6 & 66.3 \\
    \cmidrule{1-5}
    \multirow{4}{*}{Deep 1M} & Baseline & 0.0186 & 18.7 & 36.3 \\
     & Hub Sort & 0.0190 & 17.9 & 40.0 \\
     & RCM & 0.0194 & 17.8 & 37.6 \\
     & GOrder & 0.0194 & 17.8 & 32.8 \\
    \cmidrule{1-5}
    \multirow{4}{*}{Deep 40M} & Baseline & 0.0160 & 10.5 & 46.2 \\
     & Hub Sort & 0.0164 & 9.5 & 50.7 \\
     & RCM & 0.0166 & 9.5 & 49.3 \\
     & GOrder & 0.0168 & 9.5 & 43.4 \\
    \cmidrule{1-5}
    \multirow{4}{*}{Wikipedia 10M} & Baseline & 0.0088 & 10.1 & 56.9 \\
     & Hub Sort & 0.0090 & 9.5 & 61.8 \\
     & RCM & 0.0090 & 9.6 & 61.1 \\
     & GOrder & 0.0090 & 9.5 & 56.2 \\
    \cmidrule{1-5}
    \multirow{4}{*}{BioASQ 10M} & Baseline & 0.0082 & 10.2 & 73.8 \\
     & Hub Sort & 0.0082 & 9.6 & 75.6 \\
     & RCM & 0.0081 & 9.7 & 76.4 \\
     & GOrder & 0.0083 & 9.7 & 75.3 \\
    \bottomrule
  \end{tabular}
\end{table}

\FloatBarrier
\clearpage

\renewcommand{\dbltopfraction}{0.95}
\renewcommand{\dblfloatpagefraction}{0.5}
\renewcommand{\textfraction}{0.05}

\section{Experiments on Randomly Generated Datasets}
\label{app:gaussian-datasets}

\begin{figure*}[b]
  \centering
  \begin{subfigure}[b]{0.48\textwidth}
    \centering
    \includegraphics[width=\textwidth]{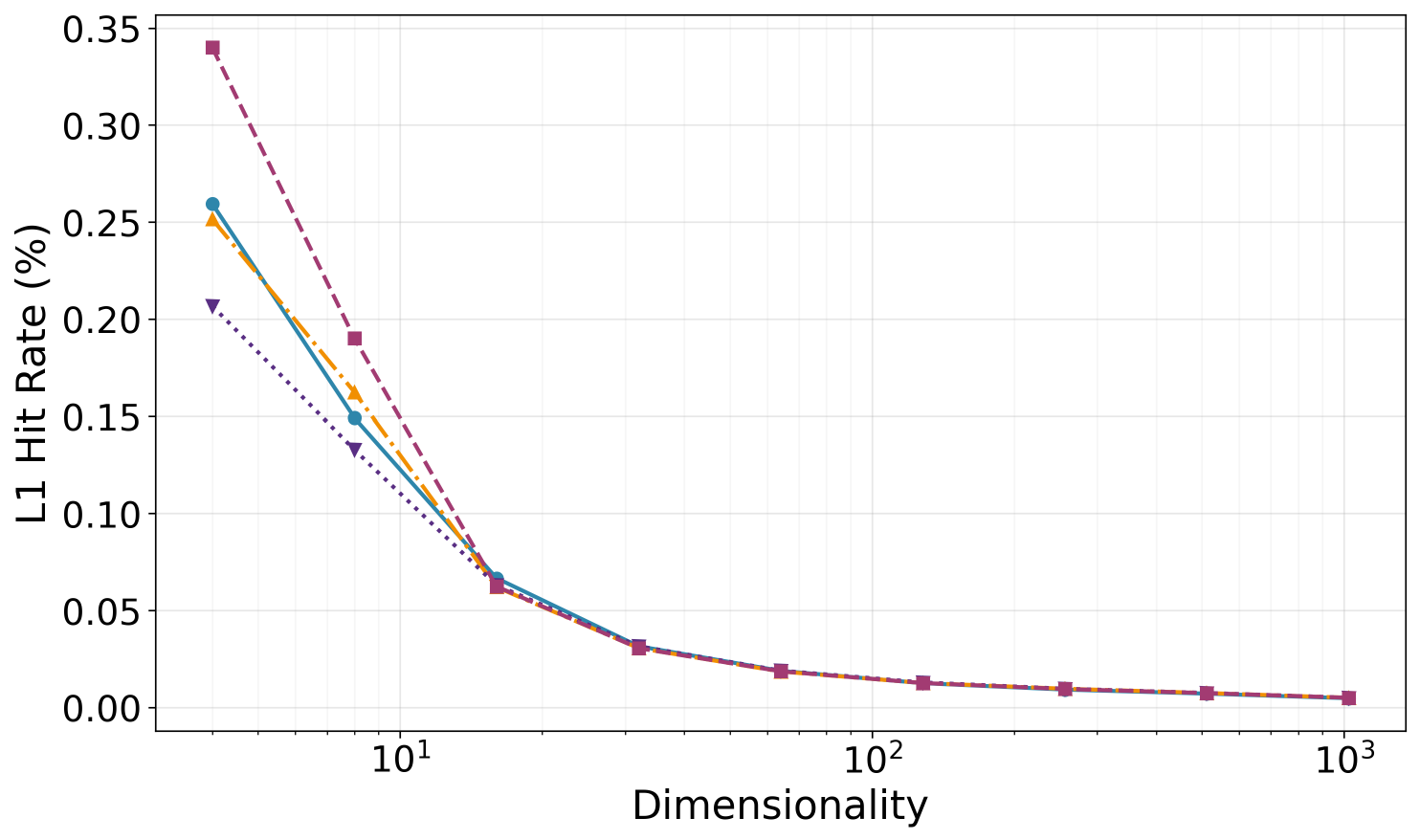}
    \caption{L1 cache hit rate}
    \label{fig:supp-gaussian-l1-hit}
  \end{subfigure}
  \hfill
  \begin{subfigure}[b]{0.48\textwidth}
    \centering
    \includegraphics[width=\textwidth]{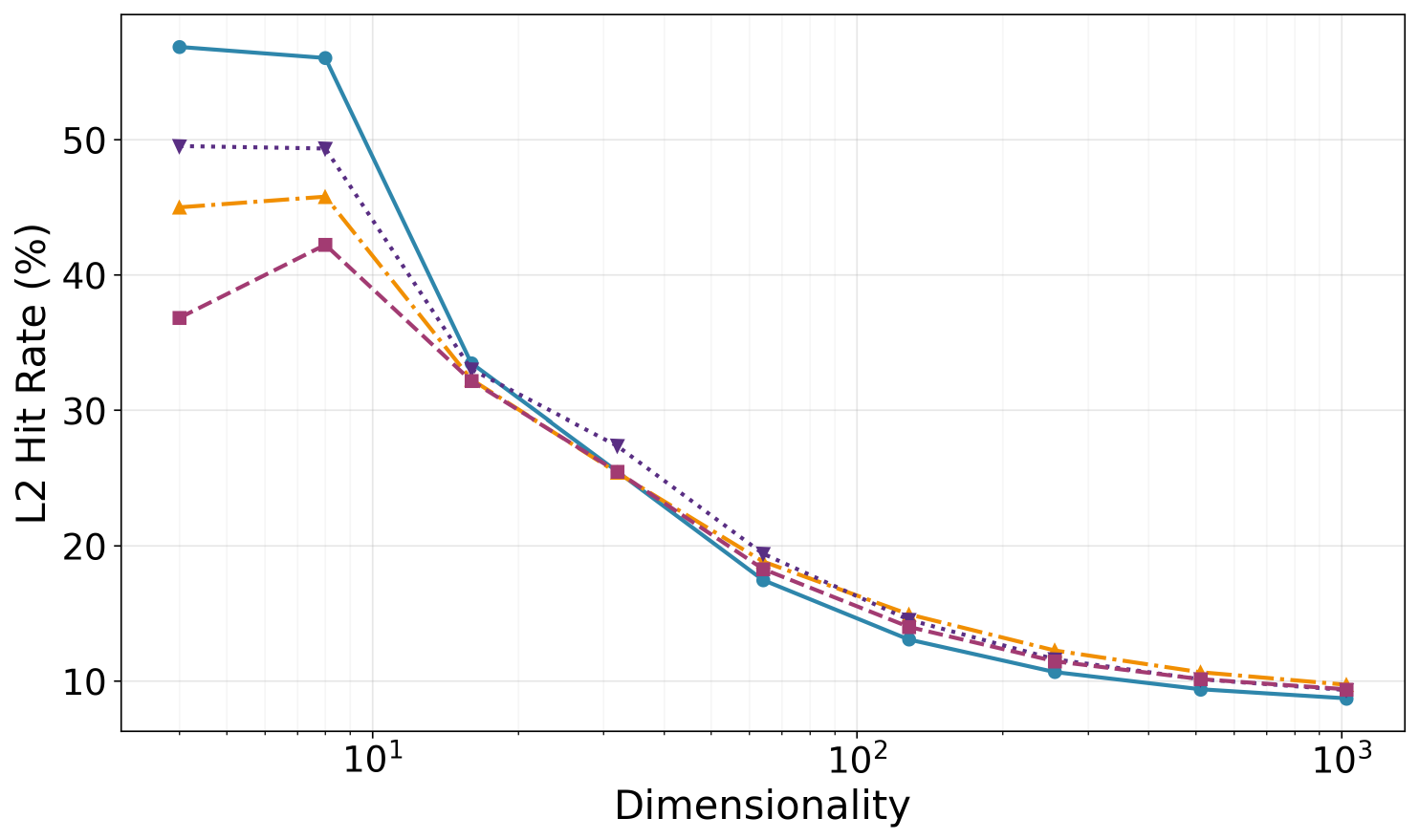}
    \caption{L2 cache hit rate}
    \label{fig:supp-gaussian-l2-hit}
  \end{subfigure}

  \vspace{0.5em}

  \begin{subfigure}[b]{0.48\textwidth}
    \centering
    \includegraphics[width=\textwidth]{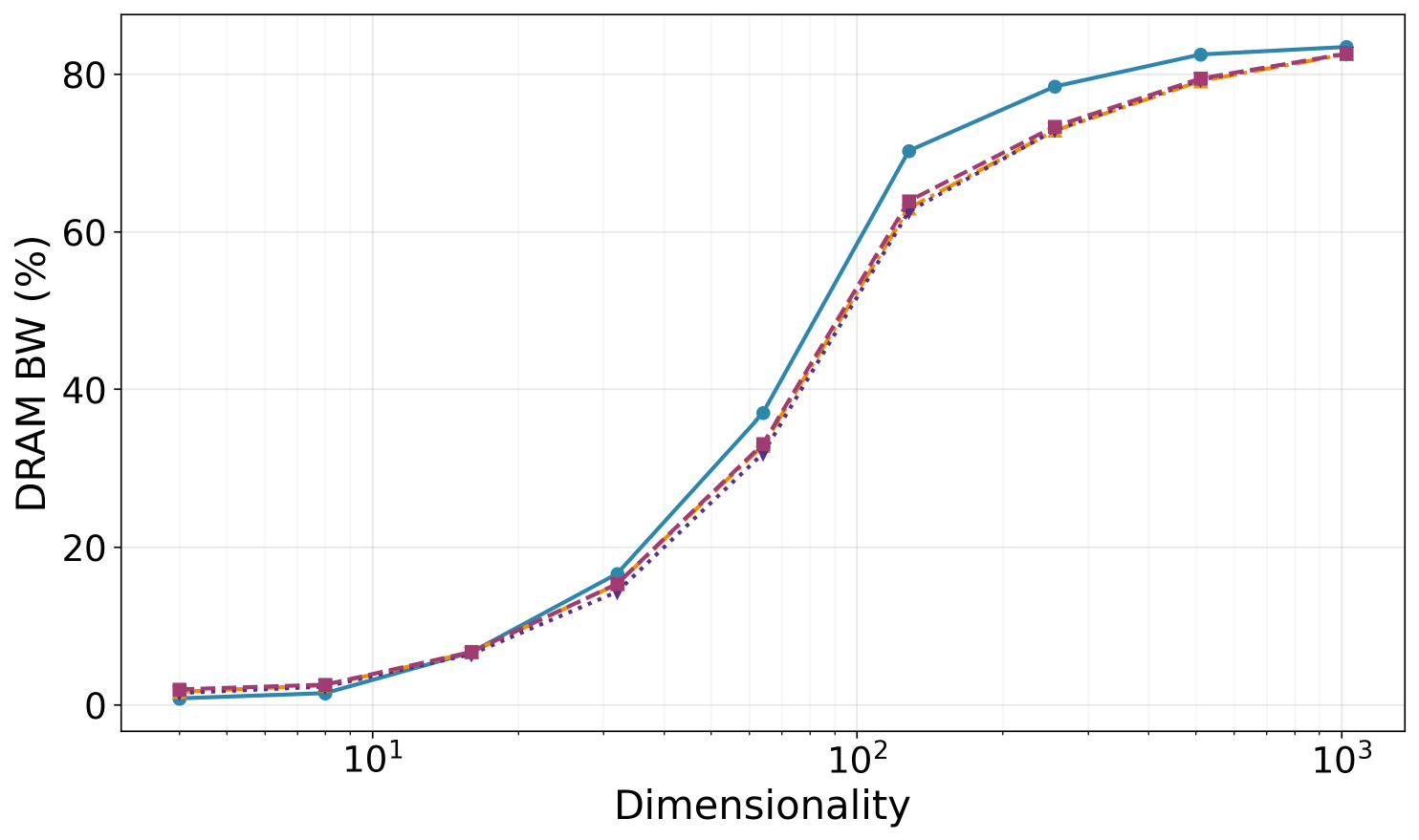}
    \caption{DRAM bandwidth utilization}
    \label{fig:supp-gaussian-dram-bw}
  \end{subfigure}
  \hfill
  \begin{subfigure}[b]{0.48\textwidth}
    \centering
    \vspace{1.0em}
    \includegraphics[width=0.8\textwidth]{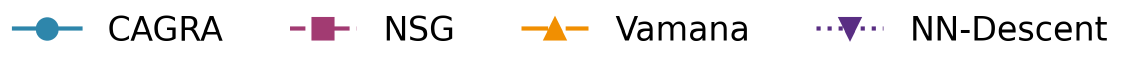}
    \vspace{5.0em}
    \label{fig:supp-gaussian-legend}
  \end{subfigure}
  \caption{Hardware metrics vs.\ vector dimensionality on 1M randomly
    generated vectors ($\mathcal{N}(0,1)$, $d\in\{4,8,16,32,64,128,256,512,1024\}$)
    for four graph-based indices without reordering.}
  \label{fig:supp-gaussian-hardware-metrics}
\end{figure*}

\Cref{fig:supp-gaussian-hardware-metrics} shows how L1/L2 cache hit rates and DRAM bandwidth utilization vary with vector dimensionality.
L1 hit rates remain near zero ($<0.05$\% for $d\geq 64$), and L2 hit rates drop from $>40$\% at $d=4$ to $\approx10$\% at high dimensions.
DRAM bandwidth utilization rises from $<5$\% at $d=4$ to $>80$\% at $d=1024$, confirming that DRAM bandwidth becomes the dominant bottleneck as dimensionality increases.
Our real-world datasets ($d\in [96,1536]$) all fall in this high-dimensional regime.

\end{document}